\documentclass[aps,prd,superscriptaddress,nofootinbib,notitlepage]{revtex4-1}

\usepackage{graphicx}
\usepackage{amsmath}
\usepackage{bm}
\usepackage{slashed}
\usepackage{epsfig}
\usepackage{amsfonts}
\usepackage{epstopdf}
\usepackage{hyperref}
\usepackage[subfigure]{graphfig}
\usepackage{braket}
\usepackage{tikz}
\usepackage{hhline}
\usepackage{soul}
\usepackage{placeins}
\usepackage{caption}

\newcommand{\ben}{\begin{eqnarray}}
\newcommand{\een}{\end{eqnarray}}

\newcommand{\bef}{\begin{figure}[!htp]}
\newcommand{\eef}{\end{figure}}

\newcommand{\msbar}{\overline{\text{MS}}}
\newcommand{\ensem}{a094m358}

\def\be{\begin{equation}}
\def\ee{\end{equation}}
\newcommand{\bea}{\begin{eqnarray}}
\newcommand{\eea}{\end{eqnarray}}

\def\ba{\begin{linenomath*}\begin{equation}}
\def\ea{\end{equation}\end{linenomath*}}

\usepackage{soul}

\hypersetup{colorlinks, linkcolor = [rgb]{0,0.0,1.0}, citecolor = [rgb]{0,0.0,1.0}, urlcolor = [rgb]{0,0.0,1.0}}

\date{\today}

\begin{document}
\title{Towards High-Precision Parton Distributions From Lattice QCD via Distillation}

\newcommand*{\WM}{Department of Physics, William and Mary, Williamsburg, Virginia 23187, USA}\affiliation{\WM}
\newcommand*{\JLAB}{Thomas Jefferson National Accelerator Facility, Newport News, VA 23606, USA}\affiliation{\JLAB}
\newcommand*{\ODU}{Physics Department, Old Dominion University, Norfolk, VA 23529, USA}\affiliation{\ODU}
\newcommand*{\AIX}{Aix Marseille Univ, Université de Toulon, CNRS, CPT, Marseille, France}\affiliation{\AIX}

\author{Colin Egerer}\affiliation{\WM}
\author{Robert G. Edwards}\affiliation{\JLAB}
\author{Christos Kallidonis}\affiliation{\JLAB}
\author{Kostas Orginos}\affiliation{\WM}\affiliation{\JLAB}
\author{Anatoly V. Radyushkin}\affiliation{\ODU}\affiliation{\JLAB}
\author{David G. Richards}\affiliation{\JLAB}
\author{Eloy Romero}\affiliation{\JLAB}
\author{Savvas Zafeiropoulos}\affiliation{\AIX}

\collaboration{On behalf of the \textit{HadStruc Collaboration}}
\begin{abstract}
  We apply the Distillation spatial smearing program to the extraction of the unpolarized isovector valence PDF of the nucleon. The improved volume sampling and control of excited-states afforded by distillation leads to a dramatically improved determination of the requisite Ioffe-time Pseudo-distribution (pITD). The impact of higher-twist effects is subsequently explored by extending the Wilson line length present in our non-local operators to one half the spatial extent of the lattice ensemble considered. The valence PDF is extracted by analyzing both the matched Ioffe-time Distribution (ITD), as well as a direct matching of the pITD to the PDF. Through development of a novel prescription to obtain the PDF from the pITD, we establish a concerning deviation of the pITD from the expected DGLAP evolution of the pseudo-PDF. The presence of DGLAP evolution is observed once more following introduction of a discretization term into the PDF extractions. Observance and correction of this discrepancy further highlights the utility of distillation in such structure studies.
\end{abstract}

\maketitle
\allowdisplaybreaks

\section{Introduction}
Elucidation of the dynamical properties of quarks and gluons and their collective emergent phenomena is the principal charge of contemporary hadronic physics and is a central component of the precision phenomenology program at the Large Hadron Collider (LHC). Crucial for the interpretation and prediction of inclusive and semi-inclusive scattering processes, such as deep inelastic scattering (DIS) and semi-inclusive DIS, are the parton distribution functions (PDFs) which emerge in the QCD factorization~\cite{Collins:1989gx} of such inclusive cross sections. PDFs capture the collinear momentum distribution of partons within a fast moving hadron with $p^\mu=\left(p^+,m^2/2p^+,\mathbf{0}_\perp\right)$, and offer a probabilistic interpretation featured at leading-twist. Despite complementing the growing hadron tomography efforts both theoretically and at upcoming facilities such as the Electron Ion Collider, determination of PDFs remains a high priority as they are often large sources of hadronic error in collider experiments and thereby affect the precision measurements of an array of Standard Model parameters~\cite{Gao:2017yyd}.

The numerical tool of lattice field theory enables the quantitative study of strongly-coupled theories, such as Quantum Chromodynamics (QCD), from first-principles. As PDFs accumulate information on the infrared structure of a hadron, they would seem ideal objects to target in lattice QCD (LQCD) calculations. However, the Euclidean metric of LQCD precludes direct access to the light-like bilinears required to define PDFs and other light-cone distributions. Attempts to circumvent this preclusion date to early efforts to access the hadronic tensor~\cite{Liu:1993cv,Liu:1999ak}, forward Compton amplitude~\cite{Detmold:2005gg}, and light-cone distributions of exclusive processes~\cite{Aglietti:1998ur,Braun:2007wv} from suitably constructed Euclidean correlation functions.

The recent cascade of research into the light-cone structure of hadrons from LQCD followed from X. Ji's proposed connection between matrix elements of \textit{space-like} separated parton bilinears and their light-like counterparts~\cite{Ji:2013dva}. When analyzed with respect to the space-like extent, the resulting distribution of parton longitudinal space-like momenta, or quasi-PDF, can be factorized into the light-cone PDFs $f\left(x,\mu^2\right)$ in the limit of large space-like momenta~\cite{Ji:2014gla}.
Since then, substantial effort has been invested in extracting PDFs from the quasi-distribution formalism~\cite{Alexandrou:2020uyt,Cichy:2019ebf,Izubuchi:2019lyk,Alexandrou:2018eet,Chen:2016utp,Alexandrou:2017dzj,Lin:2017ani,Izubuchi:2019lyk,Fan:2018dxu,Zhang:2018diq,Alexandrou:2017huk,Chen:2017mzz} and the interplay with lattice systematics has been explored~\cite{Alexandrou:2020qtt,Alexandrou:2019lfo}. In the forward limit, quasi-distributions have recently been extended to access distribution amplitudes~\cite{Zhang:2017bzy,Hua:2020gnw,Wang:2019msf} and PDFs of explicit higher-twist~\cite{Bhattacharya:2020cen}. The reader is directed to~\cite{Constantinou:2020pek,Cichy:2018mum} for details on community progress.
An alternative coordinate space interpretation, the pseudo-distribution formalism expounded upon in Sec.~\ref{sec:pseudo}, developed in~\cite{Radyushkin:2017cyf} shares the same matrix elements with quasi-distributions, but its Lorentz-invariant amplitudes factorize in the short-distance space-like regime into PDFs and perturbatively calculable coefficient functions. In this manner, pseudo-distributions are a special case of \textit{Good Lattice Cross Sections}~\cite{Ma:2014jla,Ma:2017pxb}, which has offered complementary information on pion structure~\cite{Sufian:2019bol,Sufian:2020vzb}.

Regardless of the methodology adopted, high-momenta is a requirement in order to access the regime of low momentum fraction.  The remainder of this manuscript is organized as follows. In the interest of self-containment, we begin in Sec.~\ref{sec:pseudo} with a summary of the pseudo-distribution formalism and how its application to certain space-like matrix elements illuminates the forward lightcone structure of hadrons. After introducing the lattice ensemble employed in this work, we proceed in Sec.~\ref{sec:methods} with a recapitulation of the Distillation spatial smearing program, a recent extension of the method to high-momentum observables, and argue why its utilization is essential in such structure calculations. Methods for matrix element and subsequent PDF extraction from our lattice data, especially in light of lattice artifacts, are developed subsequently. Results of these protocols are presented in Sec.~\ref{sec:results}, followed by discussion and concluding remarks in Sec.~\ref{sec:conclusion}.

\section{PDFs and Ioffe-time Pseudo-Distributions\label{sec:pseudo}}
Consider the non-local quark bilinear $\overline{\psi}\left(z\right)\gamma^\alpha\Phi^{(f)}_{\hat{z}}\left(\lbrace z,0\rbrace\right)\psi\left(0\right)$ connected with a straight $z$-separated Wilson line $\Phi^{(f)}_{\hat{z}}\left(\lbrace z,0\rbrace\right)$ in the fundamental representation of SU(3). Lorentz invariance dictates the forward helicity-averaged matrix element of this operator decomposes according to
\begin{align}
    M^\alpha\left(p,z\right)&=\bra{h\left(p\right)}\underbrace{\overline{\psi}\left(z\right)\gamma^\alpha\Phi_{\hat{z}}^{\left(f\right)}\left(\lbrace z,0\rbrace\right)\psi\left(0\right)}_{\mathring{\mathcal{O}}_{\rm WL}^{[\gamma^\mu]}\left(z\right)}\ket{h\left(p\right)}\label{eq:parton-bilinear-had}\\
    &=2p^\alpha\mathcal{M}\left(\nu,z^2\right)+2z^\alpha\mathcal{N}\left(\nu,z^2\right)\label{eq:lorentz-decomp},
\end{align}
with $\nu\equiv p\cdot z$ and $z^2$ the Ioffe-time~\cite{Ioffe:1969kf,Braun:1994jq} and invariant interval, respectively, of the process.
For a fast-moving hadron, the usual unpolarized PDFs are defined with light-cone coordinates where $\alpha=+$, $p^\alpha=\left(p^+,\tfrac{m_h^2}{2p^+},\mathbf{0_\perp}\right)$ and $z^\alpha=\left(0,z^-,\mathbf{0_\perp}\right)$. In this scenario $M^+\left(p,z\right)$ only receives contributions from $\mathcal{M}\left(p^+z^-,0\right)$. Provided the logarithmic singularity that arises for $z^2=0$ is regularized (typically in $\msbar$), $\mathcal{M}\left(p^+z^-,0\right)$ defines the Ioffe-time distribution (ITD)~\cite{Braun:1994jq}:
\be
\mathcal{M}\left(p^+z^-,0\right)_{\mu^2}\equiv\mathcal{Q}\left(\nu,\mu^2\right)=\int_{-1}^1dx\ e^{i\nu x}f_{q/h}\left(x,\mu^2\right),
\label{eq:braun-itd-pdf}
\ee
and obviates the Fourier transform of~\eqref{eq:parton-bilinear-had} to momentum space which defines the conventional PDFs.
Lorentz invariance implies the $\nu$-dependence of $\mathcal{M}\left(p^+z^-,0\right)_{\mu^2}$ can be computed in any frame, and with any choice of $\lbrace z,\alpha\rbrace$ that may be convenient. A particular choice amenable to calculation with lattice QCD is $\alpha=0$, $p^\alpha=(E,\mathbf{0_\perp},p_z)$ and $z^\alpha=(0,\mathbf{0_\perp},z_3)$, which excludes the contamination from the pure higher-twist term $\mathcal{N}\left(\nu,z^2\right)$. The remaining term $\mathcal{M}\left(\nu,z^2\neq0\right)$ is deemed the \textit{Ioffe-time Pseudo-distribution}~\cite{Radyushkin:2017cyf} or pseudo-ITD. In addition to the twist-$2$ contributions, the pseudo-ITD also contains higher-twist contributions $\mathcal{O}\left(z^2\Lambda_{\rm QCD}^2\right)$ that vanish only in the light-cone limit. Furthermore, for all relevant Feynman diagrams~\cite{Radyushkin:2016hsy}, the Fourier transform of the pseudo-ITD with respect to $\nu$ has support only on the canonical parton momentum fraction interval $x\in\left[-1,1\right]$.
The challenge numerically is to extract the leading-twist dependence amongst contributing $\mathcal{O}\left(z^2\right)$ higher-twist terms.

A considerable challenge of using this non-local parton bilinear is the appearance of additional ultraviolet (UV) divergences for space-like separations.
Prior to taking the continuum limit, the UV  divergences must be regularized and removed. 
In perturbation theory, such  divergences appear 
first,  as power-like  $z/a$  ~\cite{Polyakov:1980ca} terms in the gauge-link self-energy 
corrections,  with $a$ being a UV regulator.    
These exponentiate to all orders~\cite{Dotsenko:1979wb,Brandt:1981kf} 
producing  the factor $Z_{\rm link}\left(z_3,a\right)\simeq e^{-A\left|z_3\right|/a}$.
Second, there are logarithmic $\ln (-z^2/a^2)$ UV  corrections
present  both in the link self-energy 
and in the vertex link corrections~\cite{Craigie:1980qs}. 
As these unwanted UV divergences are multiplicative~\cite{Craigie:1980qs,Ishikawa:2017faj,Ji:2017oey,Green:2017xeu}, and independent of the  Ioffe-time,  combined into
an overall  factor $Z_{\rm UV}\left(z_3,a\right)$,
we construct the so-called \textit{reduced} pseudo-ITD~\cite{Radyushkin:2017cyf}
\be
\mathfrak{M}\left(\nu,z^2\right)=\frac{\mathcal{M}\left(\nu,z^2\right)}{\mathcal{M}\left(0,z^2\right)}.
\label{eq:reduced-pITD}
\ee
Since the rest-frame pseudo-ITD $\mathcal{M}\left(0,z^2\right)$ has the same
UV  divergences associated  with the gauge link, by forming the reduced pseudo-ITD we cancel
 the  divergent factors $Z_{\rm link}\left(z_3,a\right)$ and $Z_{\rm UV}\left(z_3,a\right)$,  thereby ensuring a finite continuum limit. 
The choice to construct the RG invariant reduced pseudo-ITD with $\mathcal{M}\left(0,z^2\right)$ is especially \st{curated} straightforward, as $\mathcal{M}\left(0,z^2\right)$ is simply the bare vector charge $Z_V^{-1}$ in the light-cone limit thereby leaving the OPE unaltered. Motivation for the reduced pseudo-ITD also extends to the mitigation of higher-twist $\mathcal{O}\left(z^2\Lambda_{\rm QCD}^2\right)$ effects~\cite{Orginos:2017kos}. An alternative has recently been proposed that makes use of a vacuum matrix element of the space-like parton bilinear~\cite{Braun:2018brg} (see also Ref.~\cite{Li:2020xml}).

Following removal of the UV divergences produced by the space-like Wilson line, the remaining singularities  in~\eqref{eq:reduced-pITD} stem from $\ln (-z^2)$ contributions in QCD. These terms generate the perturbative evolution of the collinear PDFs and complicate the naive $z^2\rightarrow0$ limit. The reduced pseudo-ITD $\mathfrak{M}\left(\nu,z^2\right)$ factorizes in the perturbative small-$z^2$ regime into PDFs with perturbatively calculable hard coefficients. The factorization relationship has been computed to NLO~\cite{Izubuchi:2018srq,Radyushkin:2018cvn,Zhang:2018ggy} and recently NNLO~\cite{Li:2020xml,Chen:2020ody}. The NLO relationship that matches the $\msbar$ ITD $\mathcal{Q}\left(\nu,\mu^2\right)$ to the reduced pseudo-ITD $\mathfrak{M}\left(\nu,z^2\right)$ reads
\begin{equation}
  \mathfrak{M}\left(\nu,z^2\right)=\int_0^1{\rm d}u\ \mathcal{C}\left(u,z^2\mu^2,\alpha_s\left(\mu\right)\right)\mathcal{Q}\left(u\nu,\mu^2\right)
  \quad+\sum_{k=1}^\infty\mathcal{B}_k\left(\nu\right)\left(z^2\right)^k,
  \label{eq:matchingkernel}
\end{equation}
where $\mathcal{Q}\left(\nu,\mu^2\right)$ is the lightcone ITD at a factorization scale $\mu^2$.
The matching kernel
\begin{equation}
  \mathcal{C}\left(u,z^2\mu^2,\alpha_s\left(\mu\right)\right)=\delta\left(1-u\right)
  -\frac{\alpha_s}{2\pi}C_F\left[\ln\left(\frac{e^{2\gamma_E+1}z^2\mu^2}{4}\right)B\left(u\right)+L\left(u\right)\right],
  \label{eq:dglap-match}
\end{equation}
involves a scale-independent kernel $L\left(u\right)=\left[4\frac{\ln\left(1-u\right)}{1-u}-2\left(1-u\right)\right]_+$ that matches the lattice and $\msbar$ regularization schemes, and a scale-dependent kernel that relates the $z^2$ and $\mu^2$ scales through the flavor non-singlet DGLAP evolution kernel $B\left(u\right)=\left[\frac{1+u^2}{1-u}\right]_+$ \mbox{\cite{Dokshitzer:1977sg,Gribov:1972rt,Altarelli:1977zs}.} The factorization is valid in so far as the polynomial corrections $\mathcal{B}_k\left(\nu\right)\left(z^2\right)^k$ can be mitigated. We follow in this work the plus-prescription defined by $\int_0^1{\rm d}u\  G\left(u\right)_+f\left(u x\right)=\int_0^1{\rm d}u\  G\left(u\right)\left[f\left(u x\right)-f\left(x\right)\right]$.

\begin{figure}
  \centering
  \includegraphics[width=0.3\linewidth]{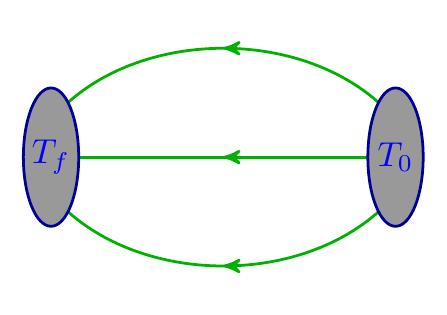}\hspace{1cm}
  \includegraphics[width=0.3 \linewidth]{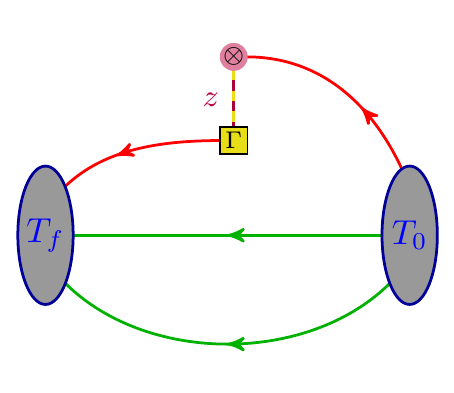}
  \caption{(Left) Two-point and (Right) Wilson-line type three-point correlation functions factorized into distinct elementals (blue), perambulators (green) and generalized perambulators (red).\label{fig:distFact}}
\end{figure}

\section{Numerical Methods\label{sec:methods}}
The pseudo-distribution formalism has been leveraged in several lattice calculations of partonic structure of hadrons, including the valence quark content of the pion~\cite{Joo:2019bzr}, and the unpolarized valence quark~\cite{Karpie:2021pap,Joo:2020spy,Bhat:2020ktg,Joo:2019jct,Orginos:2017kos} and recently gluon~\cite{Fan:2020cpa} contents of the nucleon. Even though each calculation makes use of standard spatial and momentum smearing techniques, considerable statistical fluctuations are met for Ioffe-times in excess of $\nu\gtrsim5$ (and at even smaller values for gluonic matrix elements).

Motivated by the demonstrable success of the union of the distillation paradigm and the momentum smearing idea~\cite{Egerer:2020hnc}, we apply for the first time the distillation spatial smearing program to the extraction of PDFs from lattice QCD. We employ a $349$ configuration isotropic clover ensemble generated by the JLab/W\&M/LANL collaboration~\cite{jlab-wm-lanl}, featuring $2\oplus1$ quark flavors within a $32^3\times64$ lattice volume. The inverse coupling was set to $\beta=6.3$, from which the lattice spacing $a\simeq0.094$ fm was obtained from the Wilson-flow scale $w_0$~\cite{Borsanyi:2012zs}, which yields a pion mass $m_\pi=358$ MeV. The reader is referred to~\cite{Yoon:2016jzj,Yoon:2016dij} for further details of this ensemble, denoted $\ensem$ and summarized in Tab.~\ref{tab:ensem-specs}. A single application of the stout smearing kernel~\cite{Morningstar:2003gk} yields a tadpole-improved tree-level clover coefficient $c_{\rm sw}$ near the value determined non-perturbatively from the Schr{\"o}dinger functional method~\cite{jlab-wm-lanl} a posteriori.
\subsection{Distillation}
The Jacobi smearing kernel $J_{\sigma,n_\sigma}\left(t\right)=\left(1+\sigma\nabla^2\left(t\right)\right)^{n_\sigma}$~\cite{Allton:1993wc} featured in a variety of lattice calculations is but one method with which point-like interpolating fields can be spatially smeared and made more sensitive to confinement scale physics. This low-mode filter can be made explicit by identifying eigenvectors of the three-dimensional gauge-covariant Laplacian
\be
-\nabla^2\left(t\right)\xi^{\left(k\right)}\left(t\right)=\lambda^{\left(k\right)}\left(t\right)\xi^{\left(k\right)}\left(t\right)
\label{eq:dist}
\ee
and ordering solutions according to the eigenvalue magnitude $\lambda^{\left(k\right)}\left(t\right)$. Forming the outer product of equal-time eigenvectors defines the distillation~\cite{Peardon:2009gh} smearing kernel
\be
\Box\left(\vec{x},\vec{y};t\right)_{ab}=\sum_{k=1}^{R_\mathcal{D}}\xi_a^{\left(k\right)}\left(\vec{x},t\right)\xi_b^{\left(k\right)\dagger}\left(\vec{y},t\right),
\label{eq:dist-op}
\ee
where $R_\mathcal{D}$ is the desired distillation space rank and color indices $\lbrace a,b\rbrace$ are made explicit. Two-point correlation functions formed by Wick-contracting quark fields smeared via~\eqref{eq:dist-op} can be recast into a trace over distinct reusable objects constructed within the distillation space, the so-called \textit{elementals} and \textit{perambulators}. The elementals encode the interpolator construction, which in the case of baryons read
\be
\Phi^{\left(i,j,k\right)}_{\alpha\beta\gamma}\left(t\right)=\epsilon^{abc}\left(\mathcal{D}_1\xi^{\left(i\right)}\right)^a\left(\mathcal{D}_2\xi^{\left(j\right)}\right)^b\left(\mathcal{D}_3\xi^{\left(k\right)}\right)^c\left(t\right)S_{\alpha\beta\gamma},
\ee
where $\mathcal{D}_i$ are covariant derivatives, and $S_{\alpha\beta\gamma}$ encode the patterns of subduction of a continuum interpolator across irreducible representations (irreps) of a hypercubic lattice and its associated little groups. Quark propagation between distillation spaces is captured by the perambulators
\be
\tau^{\left(l,k\right)}_{\alpha\beta}\left(T_f,T_0\right)=\xi^{\left(l\right)\dagger}\left(T_f\right)M^{-1}_{\alpha\beta}\left(T_f,T_0\right)\xi^{\left(k\right)}\left(T_0\right),
\ee
with $M$ the Dirac operator. In the case of three-point correlation functions, an additional computational unit appears. The generalized perambulator, or \textit{genprop} for short, carries the same external indices as a standard perambulator, but includes an intermediate operator insertion. In the case of the Wilson-line operator specific to this work, the genprop reads
\begin{equation}
  \Xi_{\alpha\beta}^{\left(l,k\right)}\left(T_f,T_0;\tau,\vec{z}\right)=\sum_{\vec{z}}\xi^{\left(l\right)\dagger}\left(T_f\right)M^{-1}_{\alpha\sigma}\left(T_f,\tau\right)\times 
  \left[\gamma^4\right]_{\sigma\rho}\Phi_{\hat{z}}^{(f)}\left(\lbrace\vec{z},0\rbrace\right)M^{-1}_{\rho\beta}\left(\tau,T_0\right)\xi^{\left(k\right)}\left(T_0\right),
\end{equation}
where the chosen PDF is selected with the Dirac matrix $\Gamma$, and the sum over the Wilson line terminus $\vec{z}$ ensures zero 3-momentum projection. We remark the factorization of correlation functions made manifest by distillation allows for an efficient implementation of the variational method with an extended basis of operators. These factorizations are shown diagrammatically in Fig.~\ref{fig:distFact}.

The viability of the pseudo-distribution formalism hinges on the space-like quark bilinear remaining in a perturbative, or short-distance, regime. To then minimize the impact of polynomial-$z^2$ corrections when mapping the $\nu$-dependence of the reduced pseudo-ITD, the matrix elements~\eqref{eq:parton-bilinear-had} must be isolated in frames of varying external nucleon boosts.
Increasing the 3-momentum of a hadronic interpolator not only leads to exponentially worsening statistical fluctuations, but also reduces the efficacy of spatially-smeared interpolators to overlap onto the desired hadronic states. An improvement program capable of increasing interpolator-state overlaps in boosted frames, now known as \textit{momentum smearing} was first established in~\cite{Bali:2016lva}. Momentum smearing effectively shifts the operator-ground-state overlap peak in momentum space; crucially, this shift also improves the overlap of a boosted interpolator onto neighboring excited-states. In anticipation of an increasingly dense nucleon spectrum in boosted frames, we adopt in this work a modification of the distillation paradigm~\cite{Egerer:2020hnc} which simultaneously leverages the momentum smearing heuristic with improved control over excited-states. In the interest of self-containment, we now summarize this procedure.

The momentum smearing idea is incorporated within distillation by applying spatially-varying phases to a pre-computed eigenvector basis
\be
\tilde{\xi}^{\left(k\right)}\left(\vec{z},t\right)\equiv e^{i\vec{\zeta}\cdot\vec{z}}\xi^{\left(k\right)}\left(\vec{z},t\right),
\ee
where we designate the modified eigenvectors as \textit{phased}. As the eigenvector basis already reflects the periodicity of the spatial lattice, the phase factors introduced in this manner are restricted to allowed lattice momenta. This requirement was not found to be limiting, but rather offered broad improvement of momentum space overlaps for a range of nucleon momenta~\cite{Egerer:2020hnc}. All distillation components must then be reconstructed on each new modified basis. In the interest of compute cycles and storage, we adopt three eigenvector bases: the pre-computed basis~\eqref{eq:dist} for the nucleon at rest and with small $\hat{z}$-momenta ($\left|a_s p_z\right|\leq3\left[2\pi/L_s\right]$), while two additional bases are formed according to
\be
\xi^{\left(k\right)}_\pm\left(\vec{z},t\right)\equiv e^{i\vec{\zeta}_\pm\cdot\vec{z}}\xi^{\left(k\right)}\left(\vec{z},t\right)
\label{eq:phasedEvecs}
\ee
where $\vec{\zeta}_\pm=\pm2\cdot\frac{2\pi}{L}\hat{z}$. These phased bases were found to be sufficient to resolve the ground-state nucleon in $\hat{z}$-boosted frames $\left|a_s p_z\right|>3\left[2\pi/L_s\right]$, respectively~\cite{Egerer:2020hnc}. We employ $R_{\mathcal{D}}=64$ eigenvectors within each basis.

\subsection{Matrix Element Isolation}
\textit{Interpolator Construction.}
The breaking of rotational symmetry by a hypercubic lattice entails baryons appear in lattice calculations according to definite patterns of subduction across the finite number of irreps $\Lambda$ of the octahedral group double-dover $O_h^D$. We elect to use a single spatially local, non-relativistic nucleon interpolating operator constructed according to~\cite{Edwards:2011jj,Dudek:2012ag}. The paradigm established in~\cite{Thomas:2011rh} is adopted herein to ensure our interpolator transforms irreducibly under the appropriate little group irreps. A forthcoming calculation will leverage an extended basis of interpolators to extract these same matrix elements.

\begin{table}[tb]
    \centering
    \begin{tabular}{c|c|c|c|c|c|c}
      \hline\hline
      ID & $a_s$ (fm) & $m_\pi$ (MeV)  & $L_s^3\times N_t$ & $N_{\rm cfg}$ & $N_{\rm srcs}$ & $R_{\mathcal{D}}$ \\
      \hline
      $\ensem$ & $0.094(1)$ & 358(3) & $32^3\times 64$ & 349 & 4 & 64 \\
      \hline\hline
    \end{tabular}
    \caption{Lattice ensemble employed in this work. The number of distinct source positions $N_{\rm srcs}$ per configuration and the distillation space rank $R_{\mathcal{D}}$ are also indicated.\label{tab:ensem-specs}}
\end{table}

The space-like matrix element~\eqref{eq:parton-bilinear-had} is isolated in a flavor isovector combination according to the kinematics highlighted in Sec.~\ref{sec:pseudo}, and requires computation of standard two-point
\begin{equation}
    C_2\left(p_z,T\right)=\langle\mathcal{N}\left(-p_z,T\right)\overline{\mathcal{N}}\left(p_z,0\right)\rangle
    =\sum_n\left|\mathcal{A}_n\right|^2e^{-E_nT}
    \label{eq:2pt}
\end{equation}
and three-point correlation functions featuring the unrenormalized Wilson line operator $\mathring{\mathcal{O}}_{\rm WL}^{[\gamma_4]}\left(z_3,\tau\right)$
\begin{align}
    &C_3\left(p_z,T,\tau;z_3\right)=V_3\ \langle\mathcal{N}\left(-p_z,T\right)\mathring{\mathcal{O}}_{\rm WL}^{[\gamma_4]}\left(z_3,\tau\right)\overline{\mathcal{N}}\left(p_z,0\right)\rangle \nonumber \\ &
    =V_3\sum_{n,n'}\braket{\mathcal{N}|n'}\braket{n|\overline{\mathcal{N}}}\bra{n'}\mathring{\mathcal{O}}_{\rm WL}^{[\gamma_4]}\left(z_3,\tau\right)\ket{n}
     e^{-E_{n'}\left(T-\tau\right)}e^{-E_nT},
    \label{eq:3pt}
\end{align}
where the nucleon interpolating fields $\mathcal{N}$ are smeared with distillation and are separated by a Euclidean time $T$. An explicit momentum projection is performed using the initial points of the Wilson line, thereby leading to an overall spatial volume factor $V_3$ in the forward case. The Wilson line operator is inserted for $0<\tau<T$. These correlators and the factorization manifest through distillation are illustrated in Fig.~\ref{fig:distFact}.

The spectral representations of~\eqref{eq:2pt} and~\eqref{eq:3pt} indicate the desired ground-state matrix element follows from ratios of three-point to two-point correlation functions, which plateau asymptotically for $0\ll\tau\ll T$. The contamination from excited-states is reduced further in this calculation by extracting the matrix elements using the \textit{Summation method}~\cite{Maiani:1987by,Capitani:2012gj}, whereby the time slice $\tau$ on which $\mathring{\mathcal{O}}_{\rm WL}^{[\gamma_4]}\left(z_3\right)$ is introduced is summed over
\be
R\left(p_z,z_3;T\right)=\sum_{\tau=1}^{T-1}\frac{C_3\left(p_z,T,\tau;z_3\right)}{C_2\left(p_z,T\right)}.
\label{eq:sumMat}
\ee
Note any contact terms are explicitly excluded. Excited-states in~\eqref{eq:sumMat} scale as $\exp\left[-\Delta ET\right]$, while in plateau and multi-state fits these effects scale as $\exp\left[-\Delta ET/2\right]$. As the ground-excited state gap $\Delta E$ is generally large at low-momenta, the gains afforded by the summation method over plateau/multi-state fits are modest. However, at high-momenta $\Delta E$ becomes small and the summation method offers considerable suppression of excited-states relative to plateau and multi-state fits. The geometric series resulting from~\eqref{eq:sumMat} depends linearly on the targeted matrix element $M_4\left(p_z,z_3\right)$, for which we implement the fitting functional
\be
R_{\rm fit}\left(p_z,z_3;T\right)=\mathcal{A}+M_4\left(p_z,z_3\right)T+\mathcal{O}\left(e^{-\Delta ET}\right).
\label{eq:L-summ}
\ee
We note in practice, the excited-state term $\mathcal{O}\left(e^{-\Delta ET}\right)$ is found to have no impact on our summation fits and is hence omitted from our results.

\begin{figure}[tb]
    \centering
    \includegraphics[width=.45\linewidth]{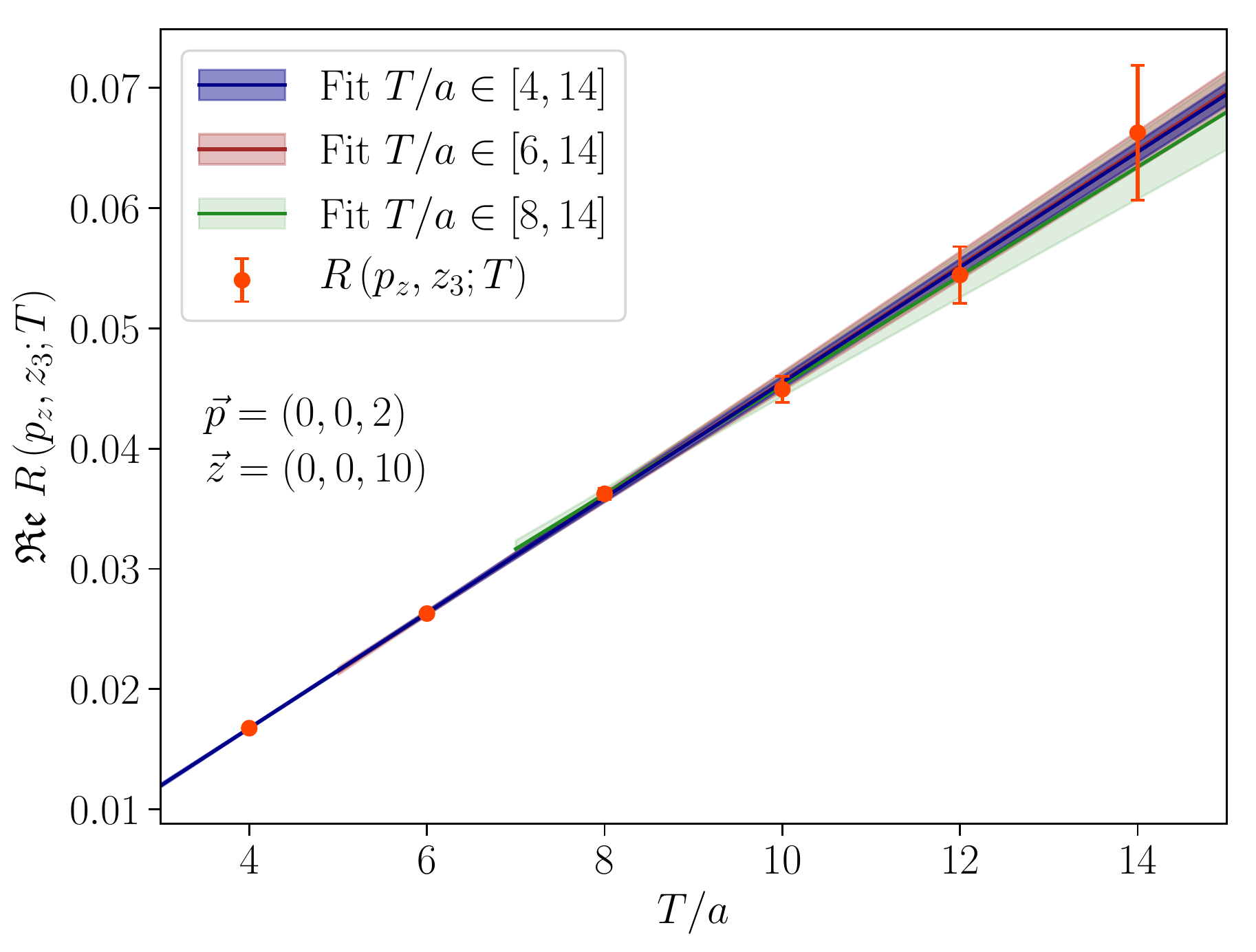}\hfill
    \includegraphics[width=.45\linewidth]{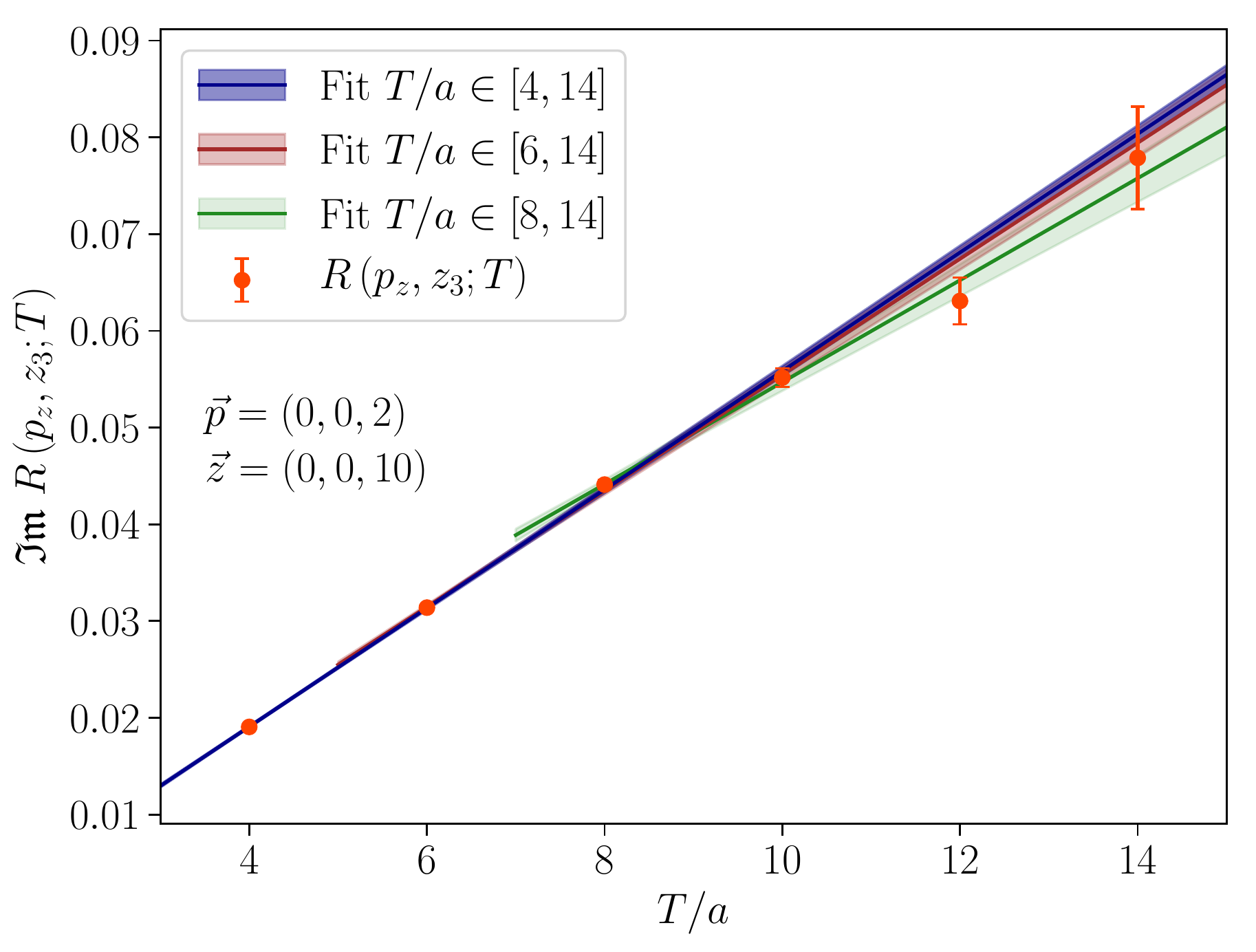}
    \caption{Real (left) and imaginary (right) summation data $R\left(p_3,z_3;T\right)$ for $p_3\simeq0.82\text{ GeV}$ and $z/a=10$, together with the linear fit~\eqref{eq:L-summ} applied for different time series. The slope of each linear fit yields the bare matrix element, which is seen to be consistently determined for varied fitting windows.\label{fig:pz2-z10_SRfits}}
\end{figure}

\begin{figure}[tb]
    \centering
    \includegraphics[width=.45\linewidth]{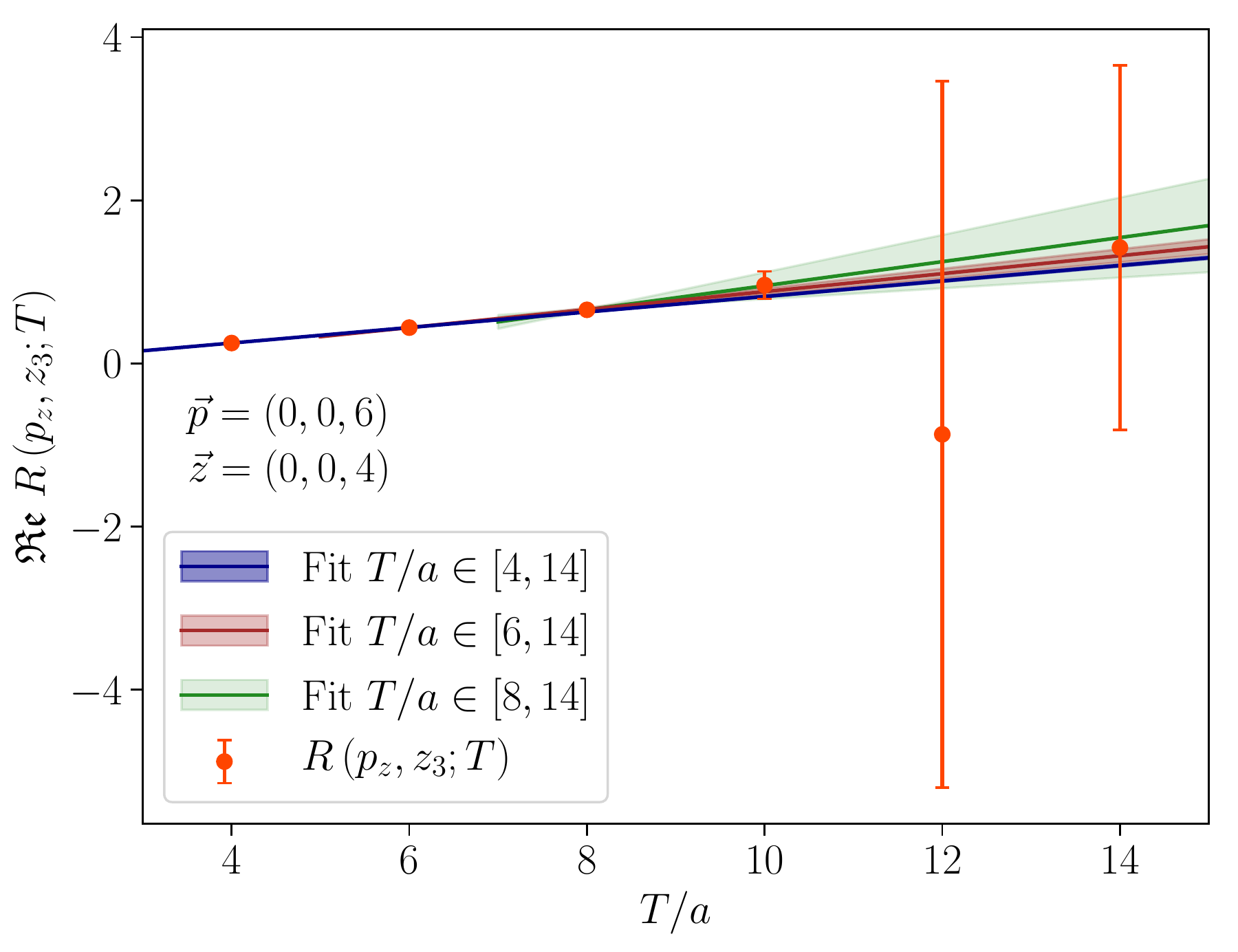}\hfill
    \includegraphics[width=.45\linewidth]{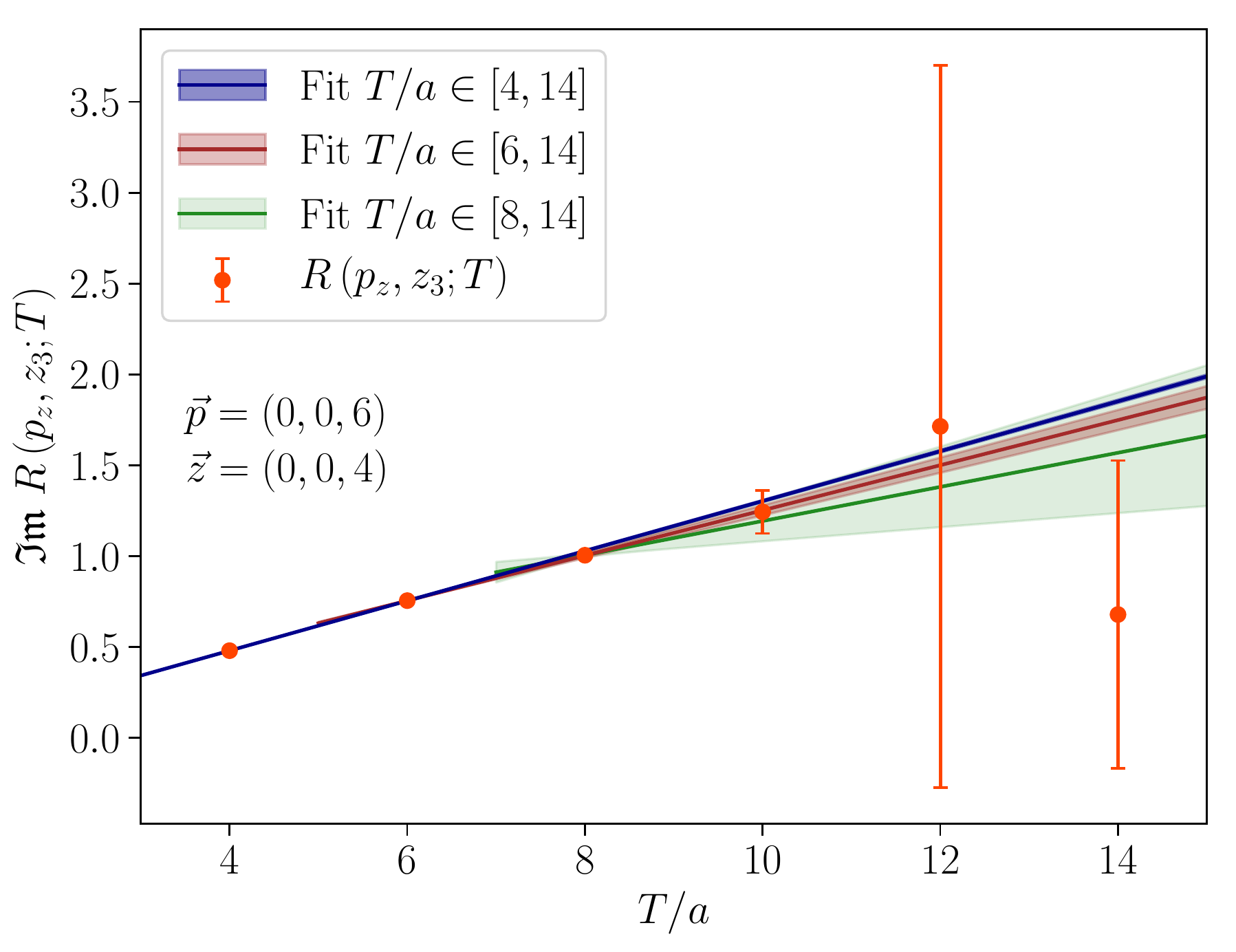}
    \caption{Real (left) and imaginary (right) summation data $R\left(p_3,z_3;T\right)$ for $p_3\simeq2.47\text{ GeV}$ and $z/a=4$, together with the linear fit~\eqref{eq:L-summ} applied for different time series. The slope of each linear fit yields the bare matrix element. Slight tension in the extracted matrix element is observed as the fitting window is altered. Although minor, this stems jointly from the lack of constraint provided by the $T/a=12,14$ data and the greater flexibility afforded to each fit as the minimum $T/a$ is increased.\label{fig:pz6-z4_SRfits}}
\end{figure}

\begin{figure*}[t!]
\subfigure[]{\label{fig:realPITD} \includegraphics[width=0.49\textwidth]{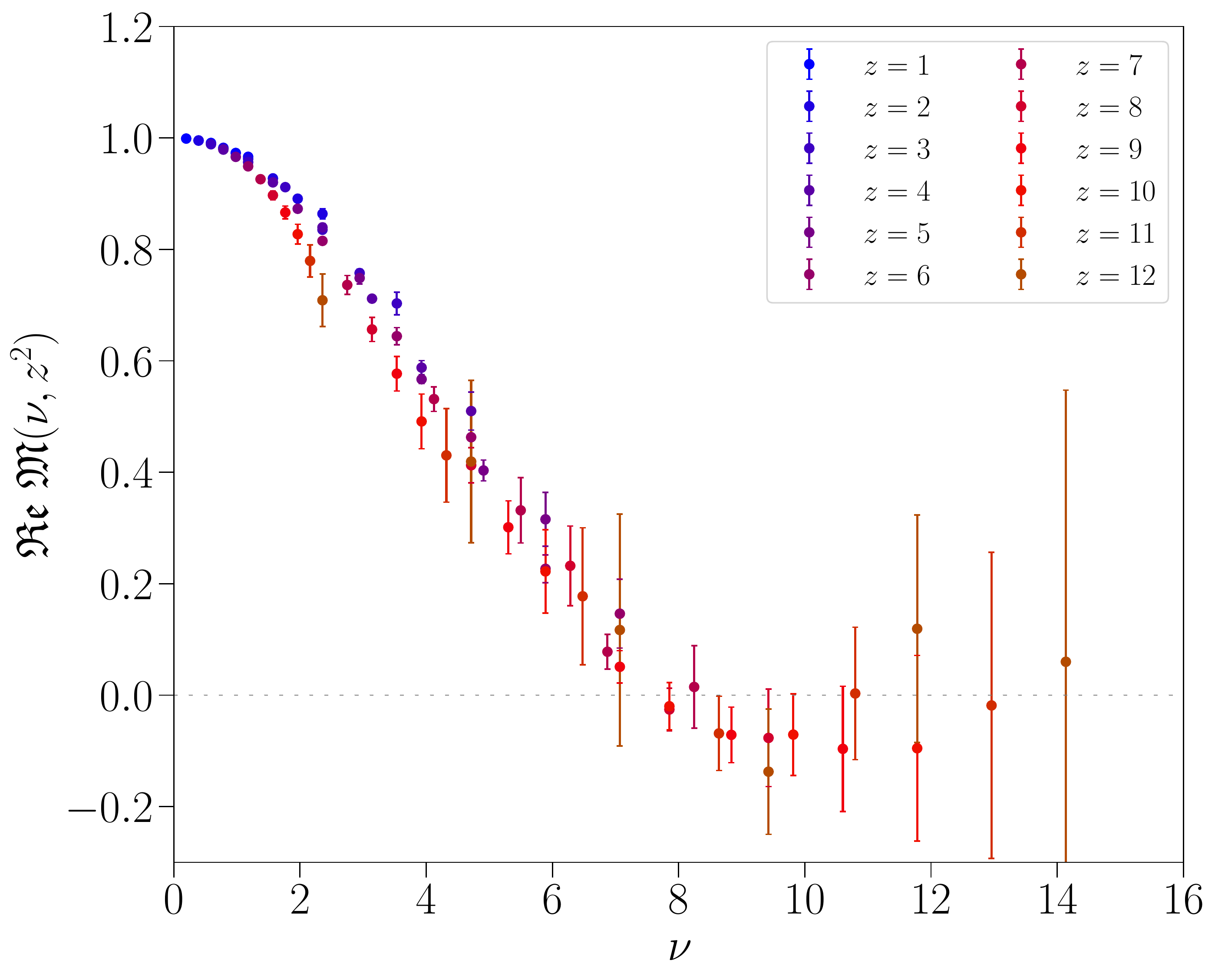}}
\hfill
\subfigure[]{\label{fig:imagPITD} \includegraphics[width=0.49\textwidth]{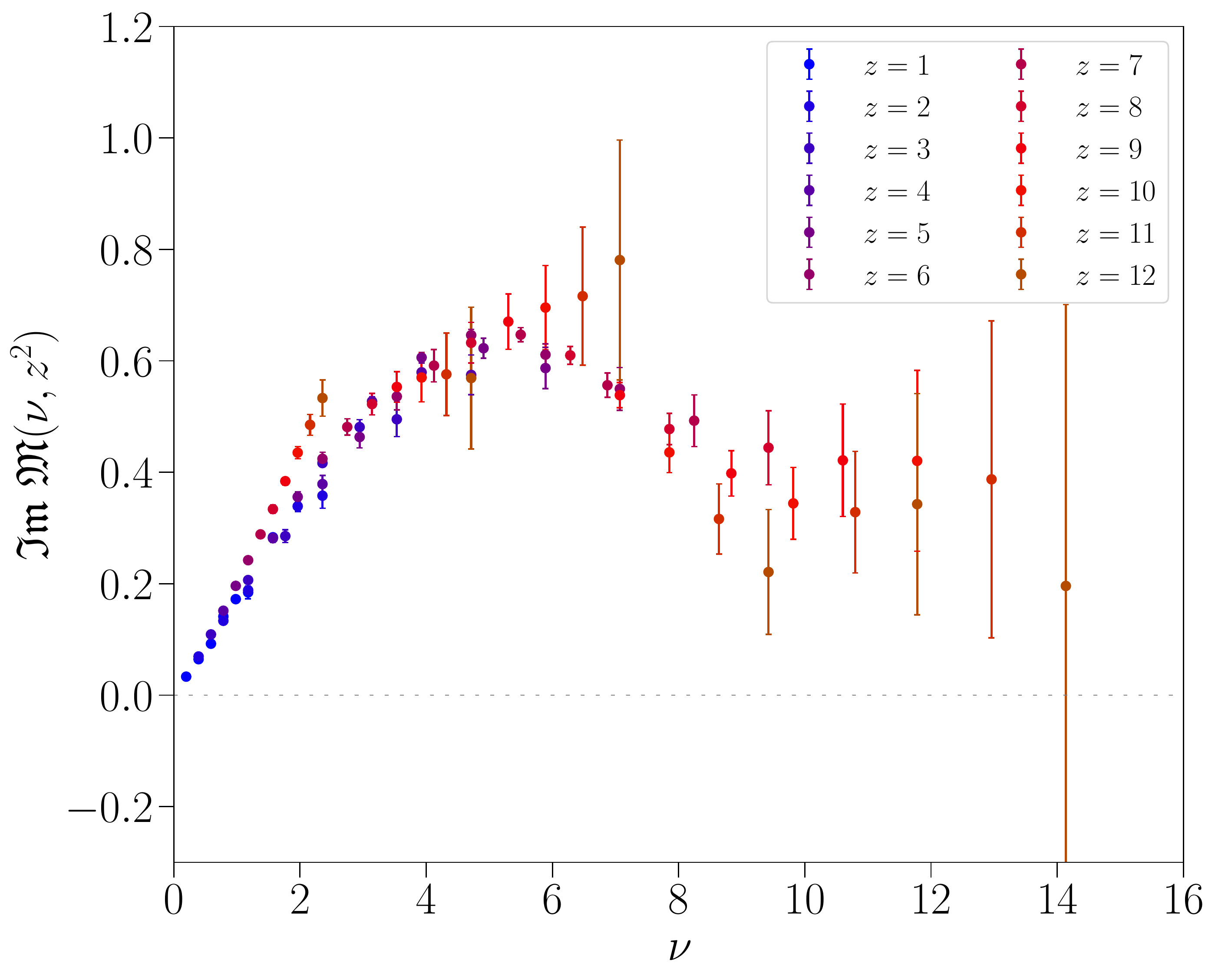}}
\caption{The real (\ref{fig:realPITD}) and the imaginary (\ref{fig:imagPITD}) component of the unpolarized reduced pseudo-ITD on the $\ensem$ ensemble with Wilson line extents $z/a<13$.}
\end{figure*}

The two- and three-point functions are computed on four temporal source origins per configuration with $T/a\in\lbrace4,6,8,10,12,14\rbrace\sim0.38-1.32\text{ fm}$. This number of source-sink separations is chosen to filter out any excited-states that are not captured by the combined effect of distillation and the summation method, and to ensure our linear fits~\eqref{eq:L-summ} do not over fit our data as signal-to-noise problems become unavoidable. We consider nucleon momenta up to $\left|p_z\right|=6\times\left[2\pi/aL\right]\sim2.47\text{ GeV}$ and Wilson line lengths up to $z_3/a=16$, although only $z_3/a\leq12$ will be subsequently used in our analysis.
A representative set of $R\left(p_3,z_3;T\right)$ and applied linear fits are shown in Fig.~\ref{fig:pz2-z10_SRfits} and~\ref{fig:pz6-z4_SRfits}. Repeating the matrix element extraction for all momenta and displacements, the real/imaginary components of the unpolarized reduced pseudo-ITD are given in Fig.~\ref{fig:realPITD} and~\ref{fig:imagPITD}.

\section{Extraction of Unpolarized Nucleon PDFs\label{sec:results}}
\vspace{-2mm}
As PDFs are determined phenomenologically at a factorization scale $\mu^2$ in $\msbar$ to renormalize the associated collinear divergences, the nucleon unpolarized reduced pseudo-ITD shown in Fig~\ref{fig:realPITD} and~\ref{fig:imagPITD} must be matched to a common scale in $\msbar$ prior to any meaningful comparisons. At one-loop and without loss of generality, negating the sign of the $\mathcal{O}\left(\alpha_s\right)$ correction and interchanging the ITD and reduced pseudo-ITD in~\eqref{eq:matchingkernel} one obtains the factorization relationship that matches the reduced pseudo-ITD to the ITD:
\begin{equation}
    \mathcal{Q}\left(\nu,\mu^2\right)=\mathfrak{M}\left(\nu,z^2\right)+\frac{\alpha_sC_F}{2\pi}\int_0^1du\left[\ln\left(\frac{e^{2\gamma_E+1}z^2\mu^2}{4}\right) \times B\left(u\right)+L\left(u\right)\vphantom{\frac{e^a}{1}}\right]\mathfrak{M}\left(u\nu,z^2\right).
    \label{eq:ITD-factor-PITD}
\end{equation}
This relationship describes the evolution of each distinct set of $\mathfrak{M}\left(\nu,z^2\right)$ data at a given $z^2$ to a common scale $\mu^2$ in $\msbar$. Regardless of whether the evolution and matching steps are done separately or in one step, a smooth and continuous description of the reduced pseudo-ITD for each $z^2$ in the interval $\left[0,\nu\right]$ is required. It is common in the literature to find polynomials in Ioffe-time fit to each set of distinct $z^2$ data in order to build $\mathfrak{M}\left(u\nu,z^2\right)$~\cite{Joo:2019jct,Joo:2020spy,Gao:2020ito,Bhat:2020ktg}. Interpolations are also common, and when used have been found to be consistent with polynomial fits~\cite{Joo:2019jct,Bhat:2020ktg}.

A polynomial in $\nu$ is perhaps a dubious choice, as it cannot capture the correct limiting behavior of the ITD at large-$\nu$. To understand this, consider a simple nucleon valence PDF ansatz
\be
f_{q_{\rm v}/N}\left(x\right)=\frac{\Gamma\left(\alpha+\beta+2\right)}{\Gamma\left(\alpha+1\right)\Gamma\left(\beta+1\right)}x^\alpha\left(1-x\right)^\beta.
\label{eq:simple2paramAnsatz}
\ee
The cosine transform of this ansatz with respect to Ioffe-time is given by
\begin{align}
\mathfrak{Re}\ &\mathcal{Q}\left(\nu,\alpha,\beta\right)=\frac{\pi\Gamma\left(2+\alpha+\beta\right)}{2^{1+\alpha+\beta}}
\ _2F_3\left(\frac{1+\alpha}{2},\frac{2+\alpha}{2};\frac{1}{2},\frac{2+\alpha+\beta}{2},\frac{3+\alpha+\beta}{2};-\frac{\nu^2}{4}\right),
\label{eq:itdSimple2ParamAnsatz}
\end{align}
with$\ _2F_3$ a generalized hypergeometric function and $\alpha,\beta>-1$.
In the large Ioffe-time regime~\eqref{eq:itdSimple2ParamAnsatz} behaves as
\be
\mathfrak{Re}\ \mathcal{Q}\left(\nu\right)\simeq\beta\cos\left(\frac{\pi}{2}\alpha\right)\frac{\Gamma\left(\alpha+2\right)}{\nu^{\alpha+2}}-\sin\left(\frac{\pi}{2}\alpha\right)\frac{\Gamma\left(\alpha+1\right)}{\nu^{\alpha+1}}.
\label{eq:itd-large-nu}
\ee
For the real component of the ITD to correspond to a valence PDF with a finite sum rule, the ITD must then vanish for asymptotically large-$\nu$ (i.e. $\alpha>-1$). This suggests the usefulness of a smooth polynomial in $\nu$ extends only so far as interpolating the discrete pseudo-ITD data, and should not be used as a measure of the moments of the pseudo-PDFs given their divergent behavior at large-$\nu$.
This motivates methods to directly extract the PDFs from the reduced pseudo-ITD, thereby obviating the need for a continuous description of $\mathfrak{M}\left(\nu,z^2\right)$ in order to perform the evolution/matching steps. This will be developed in Sections~\ref{ssec:direct} and~\ref{ssec:jacobi}.

\begin{figure}[t]
    \centering
    \includegraphics[width=0.85\linewidth]{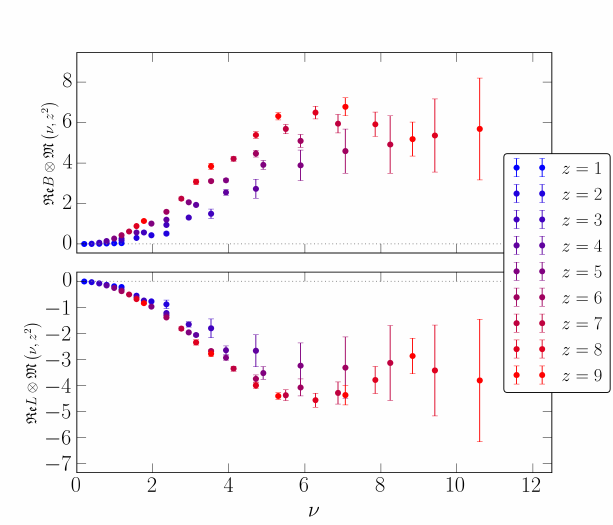}
    \caption{Convolutions needed to evolve (upper) and match (lower) the real component of the reduced pseudo-ITD to a common scale of $2\text{ GeV}$ in $\msbar$. The NLO prefactor $\alpha_sC_F/2\pi$ is included in these data, but omitted from the labels for clarity. The convolutions were performed up to $z/a=16$, but data for $z/a>9$ are generally noisy and not shown.}
    \label{fig:realConvols}
\end{figure}

\begin{figure}[bt]
    \centering
    \includegraphics[width=0.85\linewidth]{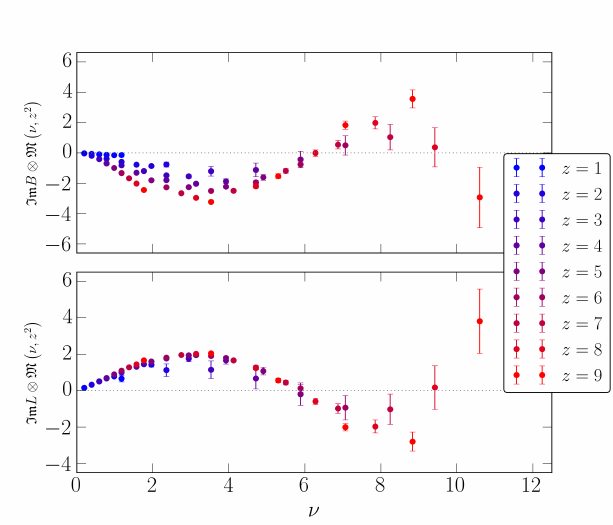}
    \caption{Convolutions needed to (upper) evolve and (lower) match the imaginary component of the reduced pseudo-ITD to a common scale of $2\text{ GeV}$ in $\msbar$. The NLO prefactor $\alpha_sC_F/2\pi$ is included in these data, but omitted from the labels for clarity. The convolutions were performed up to $z/a=16$, but data for $z/a>9$ are generally noisy and not shown.}
    \label{fig:imagConvols}
\end{figure}

To get a handle on the scale dependence of our data and ground the ensuing discussion, we nonetheless start with a provisional sixth order polynomial fit in Ioffe-time to the reduced pseudo-ITD for constant $z^2$:
\be
\mathfrak{M}\left(\nu,z^2\right)=1+\sum_{n=1}^3\left(c_{2n}\,\nu^{2n}\,+\,i\,c_{2n-1}\,\nu^{2n-1}\right).
\label{eq:polyFits}
\ee
The even (odd) powers of the polynomial are applied to jackknife samples of the real (imaginary) component of $\mathfrak{M}\left(\nu,z^2\right)$ given in Fig.~\ref{fig:realPITD} and~\ref{fig:imagPITD}. Higher order polynomials were considered, but were found to be unconstrained by the data. With the polynomial fits in hand, we perform the evolution and scheme conversion convolutions~\eqref{eq:ITD-factor-PITD} in a single step. The matched $\msbar$ scale $\mu=2\text{ GeV}$ was chosen, and the strong coupling $\alpha_s\left(2\text{ GeV}\right)\simeq0.303$ was adopted from LHAPDF6~\cite{Buckley:2014ana}. The scale $\mu=2\text{ GeV}$ corresponds to the reduced pseudo-ITD being evolved to the common scale $z_0^2=4e^{-2\gamma_E-1}\left(2\text{ GeV}\right)^{-2}\simeq0.12\text{ GeV}^{-2}$ or $z_0^{-1}\simeq2.94\text{ GeV}$. On this ensemble $\ensem$, this common scale then equates to $z_0^2/a^2\simeq0.511$. The computed evolution and scheme matching convolutions are depicted in Fig.~\ref{fig:realConvols} and Fig.~\ref{fig:imagConvols} for the real and imaginary components, respectively.
It is curious the evolution and matching convolutions appear to be nearly equal in magnitude but opposite in sign. This feature of the pseudo-distributions has been observed in independent calculations~\cite{Joo:2020spy,Bhat:2020ktg} and hints an NNLO matching relation may not be needed. Nonetheless, a future study will explore to what effect the matching relation~\eqref{eq:matchingkernel}, truncated at NLO in this work, can be improved at NNLO~\cite{Li:2020xml}.

\begin{figure*}[t!]
\subfigure[]{\label{fig:realITD} \includegraphics[width=0.49\textwidth]{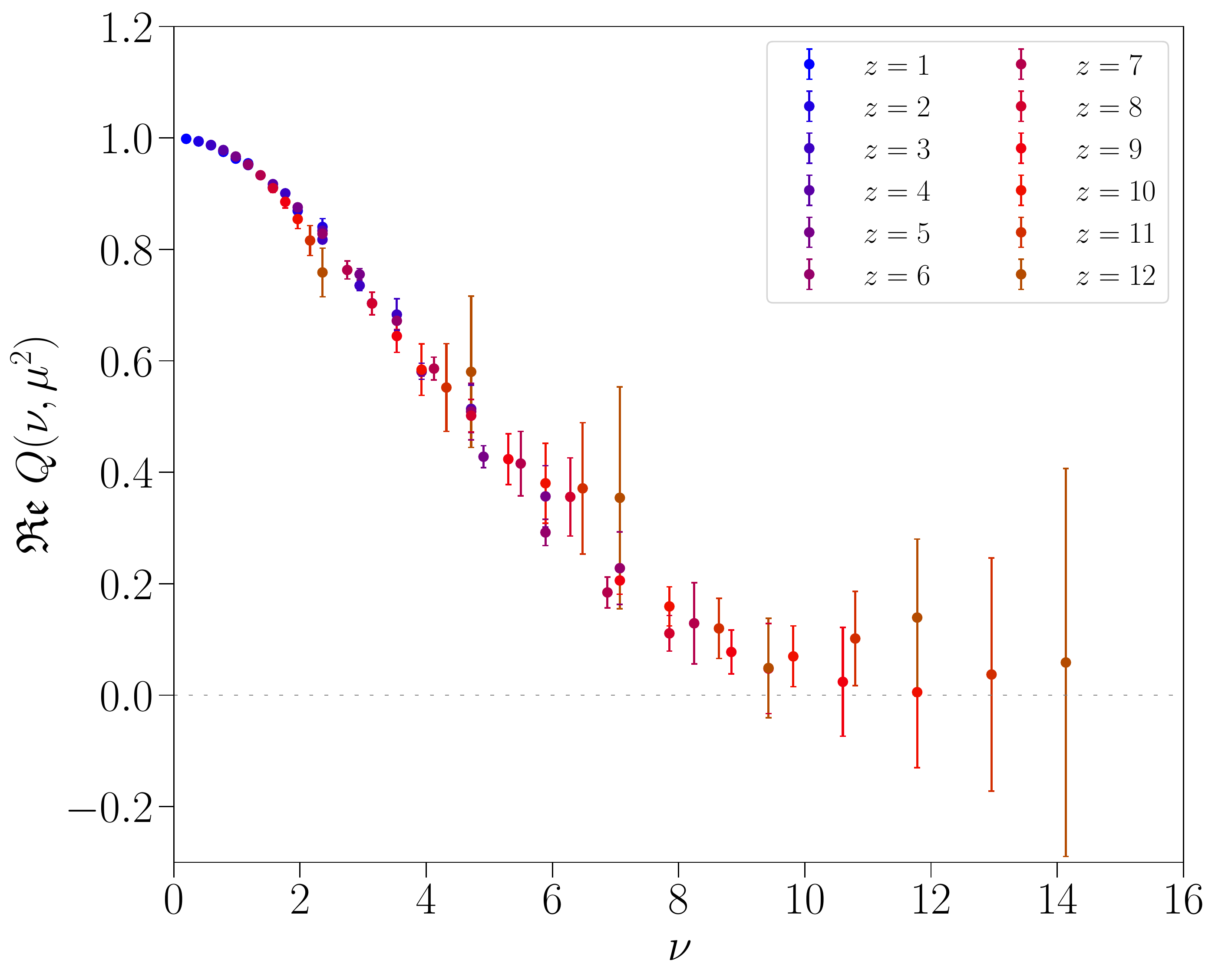}}
\hfill
\subfigure[]{\label{fig:imagITD} \includegraphics[width=0.49\textwidth]{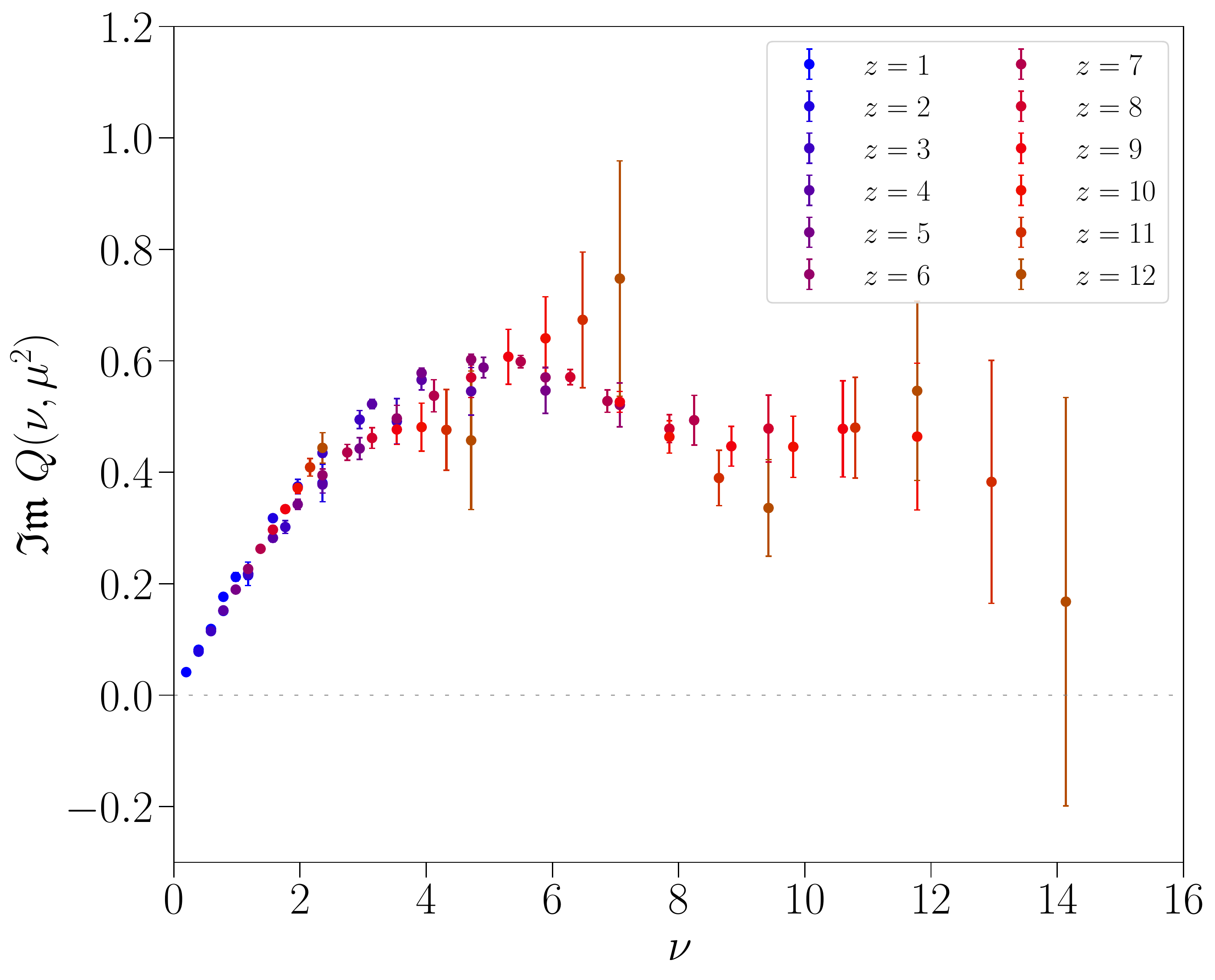}}
\caption{The real (\ref{fig:realITD}) and the imaginary (\ref{fig:imagITD}) component of the unpolarized ITD at a scale of $2\text{ GeV}$ in $\msbar$ obtained from the matching relation~\eqref{eq:ITD-factor-PITD}for the real component and ~\eqref{eq:ITD-factor-PITD} for the imaginary component applied to polynomial fits to the reduced pseudo-ITD data. Data are shown for Wilson line extents $z/a\leq12$; extents $z/a>12$ are considerably uncertain, and thus excluded from our ensuing analysis.}
\end{figure*}

When the scale and scheme conversion are incorporated, we observe in Fig.~\ref{fig:realITD} and Fig.~\ref{fig:imagITD} a dramatic collapse of the reduced pseudo-ITD onto a common curve for $z/a\lesssim10$. The lack of residual $z^2$-dependence is particularly striking in the real component of the ITD, but less so in the imaginary component. This confirms the formation of the reduced ratio~\eqref{eq:reduced-pITD} indeed cancels much of the $z^2$-dependence in the pseudo-ITD, with any remaining at small-$z^2$ ideally described by the coordinate-space DGLAP evolution~\cite{Radyushkin:2017cyf}.

\subsection{An Ill-Posed Inverse}
The Fourier transform relating the $x$-dependence of the PDF to the $\nu$-dependence of the ITD
\be
\mathcal{Q}\left(\nu,\mu^2\right)=\int_{-1}^1dx\ e^{i\nu x}f_{a/h}\left(x,\mu^2\right)
\label{eq:itd-pdf}
\ee
is an ill-posed inverse problem, as the ITD is computed in a discrete and limited range of Ioffe-time. Regularization at this stage is however numerically cheap and more stable relative to a direct matching of the $\lbrace x,\mu^2\rbrace$ dependencies of the PDF to the $\lbrace\nu,z^2\rbrace$ dependencies of the reduced pseudo-ITD.
A direct inversion of~\eqref{eq:itd-pdf} is satisfied by an infinite number of solutions, each of which having little predictive or postdictive credibility. The futility of direct inversions has been demonstrated in a few LCS calculations~\cite{Bhat:2020ktg,Joo:2019jct}, wherein each inversion yielded unstable PDFs with spurious oscillations. The limited range of Ioffe-time accessible to present PDF calculations only compounds the need for refined extraction methods.

This inversion problem is shared with other lattice formalisms that rely on QCD factorization, and indeed the global analysis of inclusive/semi-inclusive processes. Although different in character, this problem pervades the quantitative sciences and even impacts the image reconstruction of black holes~\cite{Akiyama:2019bqs}. Arguably the most serious systematic that must then be confronted in LCS studies is how to reliably extract a targeted distribution, while minimizing numerical artifacts and bias. Numerous sophisticated methods, such as the Backus-Gilbert~\cite{Backus},
maximum entropy~\cite{Asakawa:2000tr},
and Bayesian reconstruction methods have been explored as tools to aid in PDF extractions~\cite{Karpie:2019eiq,Bhat:2020ktg} and other observables more generally~\cite{Liang:2019frk,Hansen:2017mnd}.

A common heuristic to regularize the inverse problem at hand, both in the global fitting of inclusive scattering data~\cite{Accardi:2016qay,Martin:2009iq,Harland-Lang:2014zoa,Hou:2019efy,Ball:2014uwa,Ball:2017nwa} and analogous lattice calculations, is to supply additional information in the form of physically motivated PDF parameterizations. In particular, a parametric form can incorporate the known divergence/convergence of PDFs at small-/large-$x$ and enforce any parton sum rules explicitly. We note parametric frameworks in the literature generally differ in what functional form is used to smoothly connect the two limiting $x$-space regimes.
Inspired by the phenomenological forms of the global fitting community and to establish a benchmark for the alternate extraction methods that follow, we opt to first regularize the inverse problem by parameterizing the valence/plus quark distributions
according to
\begin{align}
    f_{q_{\rm v}/N}\left(x,\mu^2\right)&=f_{q/N}\left(x,\mu^2\right)-f_{\bar{q}/N}\left(x,\mu^2\right)=N_{\rm v}x^\alpha\left(1-x\right)^\beta P\left(x\right) \label{eq:jam-param_valence}\\
    f_{q_+/N}\left(x,\mu^2\right)&=f_{q_{\rm v}/N}\left(x,\mu^2\right)+2f_{\bar{q}/N}\left(x,\mu^2\right)=N_+x^{\alpha_+}\left(1-x\right)^{\beta_+}P\left(x\right),
    \label{eq:jam-param_plus}
\end{align}
where $P\left(x\right)$ is a smooth interpolating polynomial and $N_{\rm v}^{-1}=B\left(\alpha+1,\beta+1\right)+\sum_k\lambda_kB\left(\alpha+1+\frac{k+1}{2},\beta+1\right)$ ensures the valence quark sum rule $\int_0^1dx\ f_{q_{\rm v}/N}\left(x,\mu^2\right)=1$ is satisfied; the normalization of $f_{q_+/N}\left(x,\mu^2\right)$ is not fixed by a sum rule, and is left to float in our fits. Given the limited range of Ioffe-time in our results, we will find the simplest 2-parameter ansatz with $P\left(x\right)=1$ cannot be avoided. Where possible, the bias introduced by this highly-constraining choice will be studied by supplementing $P\left(x\right)$ with additional half-integer powers of $x$: $P\left(x\right)=1+\sum_k\lambda_kx^{\left(k+1\right)/2}$, thereby increasing the flexibility of our parameterizations beyond the nominal PDF behavior $x^\alpha\left(1-x\right)^\beta$.

\begin{figure*}[t!]
\subfigure[]{\label{fig:itd-real-convolC-uncorrelated} \includegraphics[width=0.48\textwidth]{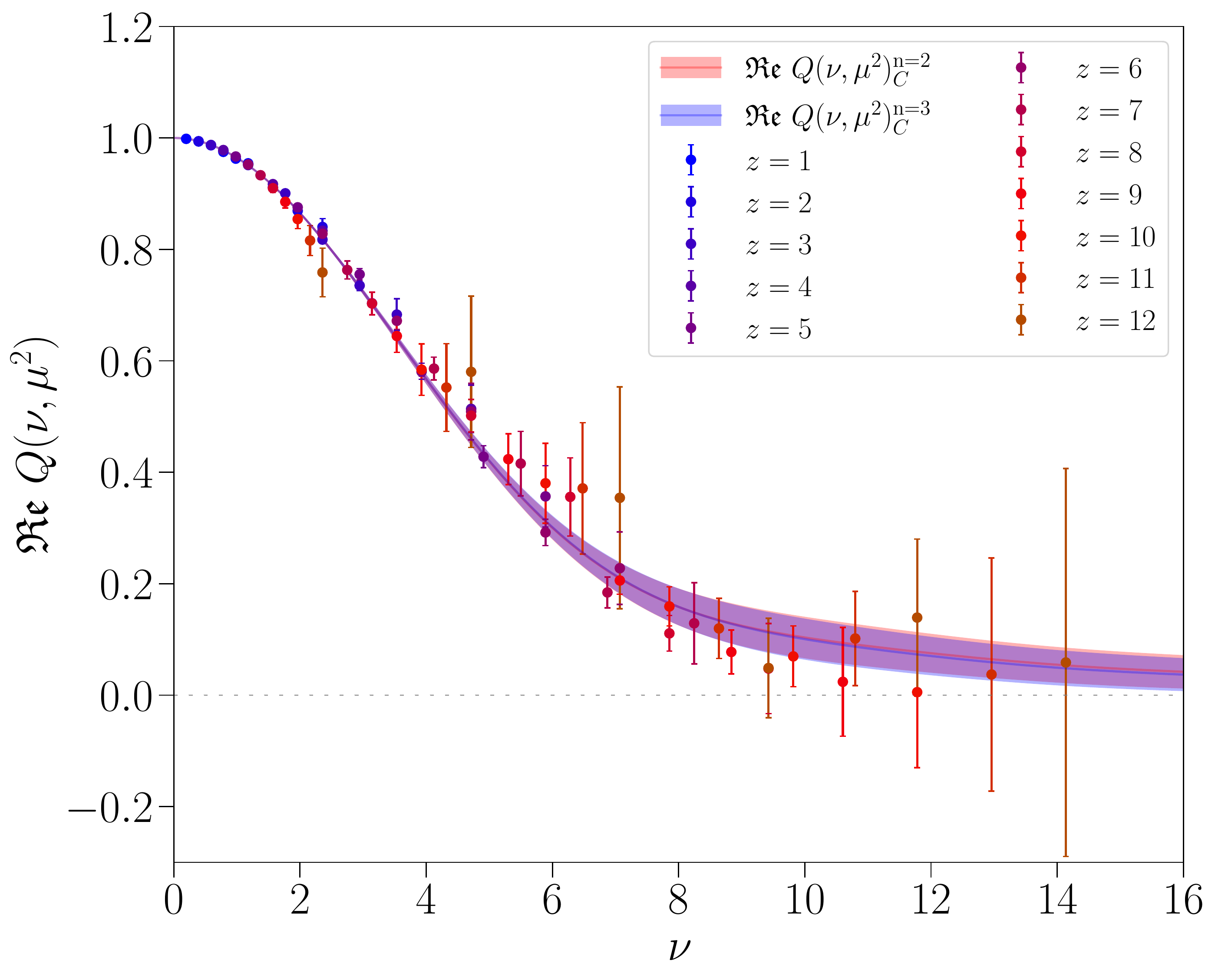}}
\hfill
\subfigure[]{\label{fig:pdf-valence-convolC-uncorrelated} \includegraphics[width=0.459\textwidth]{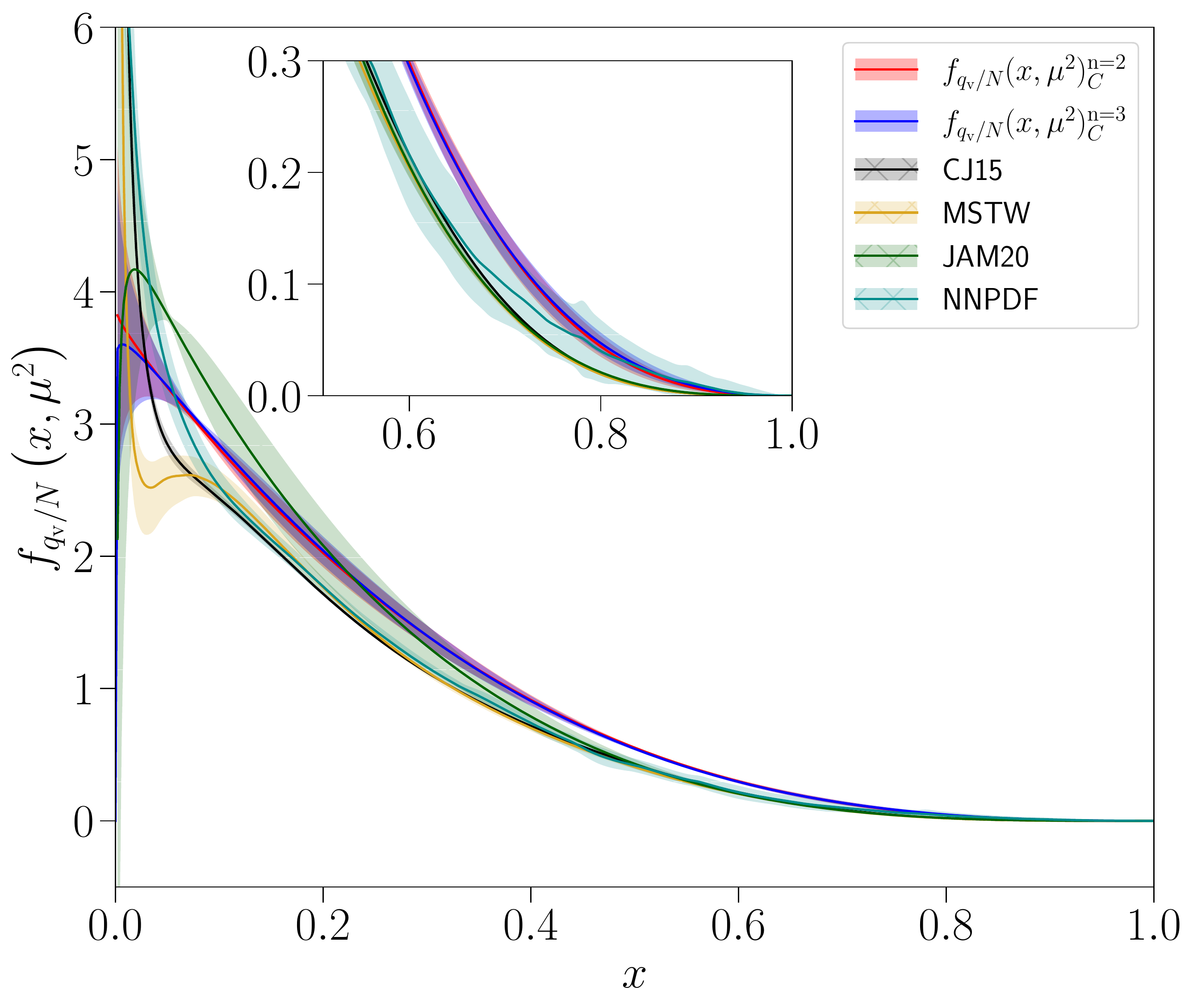}}
\caption{The real component of the matched ITD at $\mu=2\text{ GeV}$ in $\msbar$ (\ref{fig:itd-real-convolC-uncorrelated}) fit by cosine transforms of two- and three-parameter model PDFs~\eqref{eq:cosTransform}. Data has been fit for $z/a\leq12$, and correlations have been neglected.
The resulting PDF parameters and figure of merit are gathered in Tab.~\ref{tab:uncorrelated-fitParams}. The nucleon unpolarized valence quark PDF at $2\text{ GeV}$ in $\msbar$ (\ref{fig:pdf-valence-convolC-uncorrelated}) determined from the uncorrelated cosine transform fits~\eqref{eq:cosTransform} applied to real component of the matched ITD. Comparisons are made with the NLO global analysis of CJ15~\cite{Accardi:2016qay} and JAM20~\cite{Moffat:2021dji}, and the NNLO analyses of MSTW~\cite{Martin:2010db} and NNPDF~\cite{Ball:2017nwa} at the same scale.}
\end{figure*}

\begin{figure*}[t!]
\subfigure[]{\label{fig:itd-imag-convolC-uncorrelated} \includegraphics[width=0.48\textwidth]{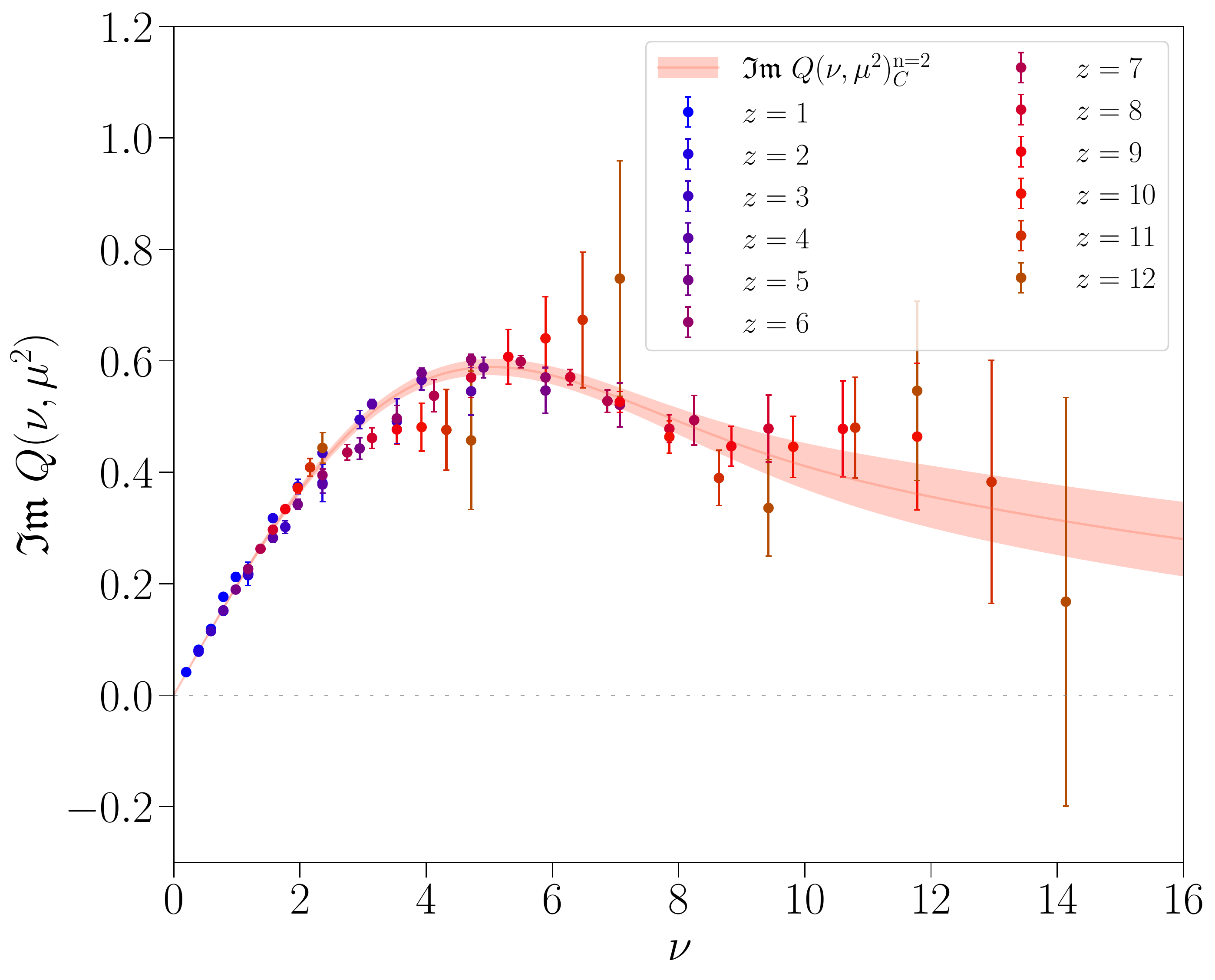}}
\hfill
\subfigure[]{\label{fig:pdf-plus-convolC-uncorrelated} \includegraphics[width=0.459\textwidth]{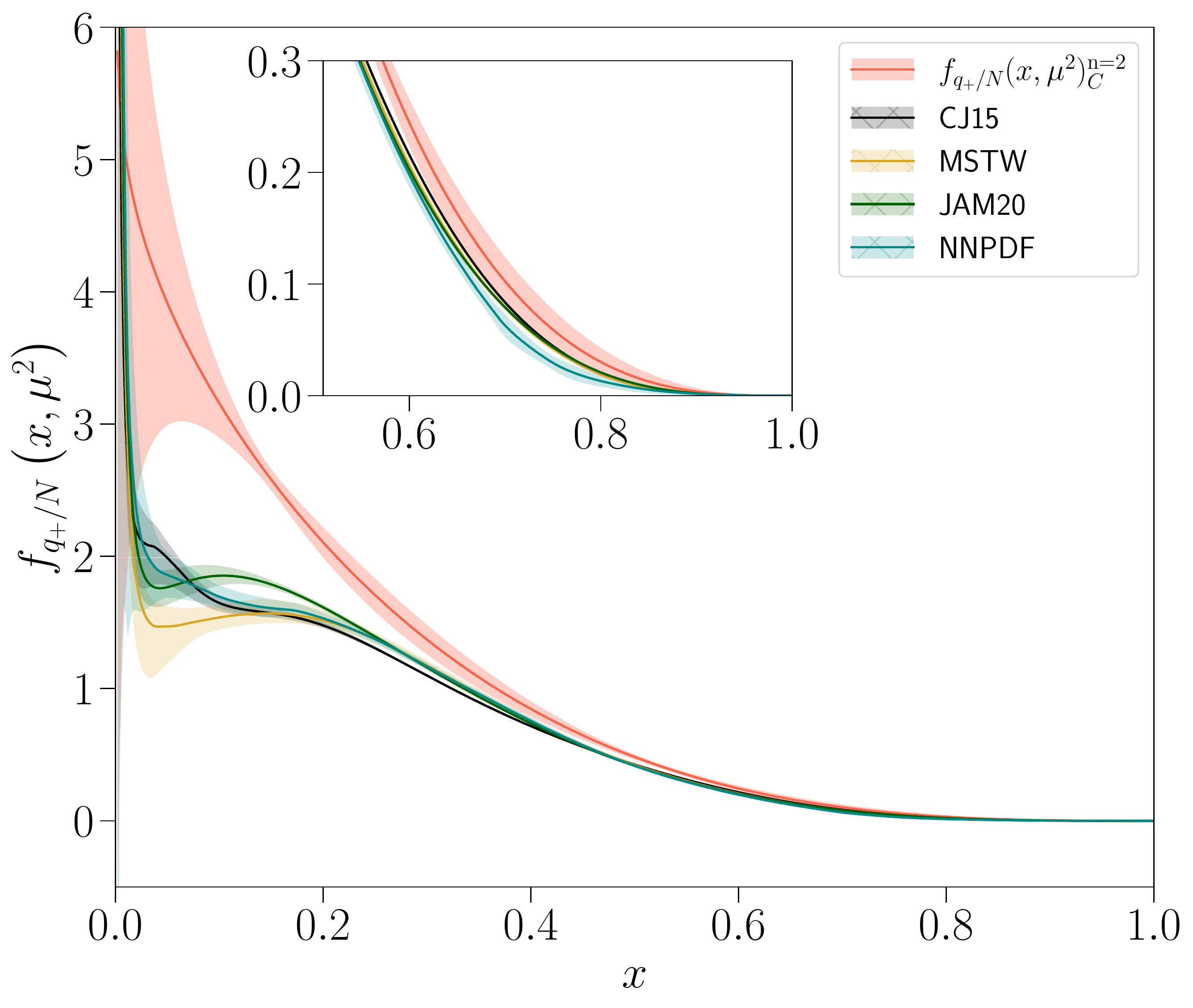}}
\caption{Imaginary component of the matched ITD at $\mu=2\text{ GeV}$ (\ref{fig:itd-imag-convolC-uncorrelated}) in $\msbar$ fit by the sine transform of a two-parameter model PDF~\eqref{eq:sinTransform}. Data has been fit for $z/a\leq12$, and correlations have been neglected. The resulting PDF parameters and figure of merit are gathered in Tab.~\ref{tab:uncorrelated-fitParams}. The nucleon unpolarized plus quark PDF at $2\text{ GeV}$ in $\msbar$ (\ref{fig:pdf-plus-convolC-uncorrelated}) determined from the uncorrelated sine transform fits~\eqref{eq:sinTransform} applied to the imaginary component of the matched ITD. Comparisons are made with the NLO global analysis of CJ15~\cite{Accardi:2016qay} and JAM20~\cite{Moffat:2021dji}, and the NNLO analyses of MSTW~\cite{Martin:2010db} and NNPDF~\cite{Ball:2017nwa} at the same scale. }
\end{figure*}

We start with two- and three-parameter PDF parameterizations, where in the latter we take $P\left(x\right)=1+\delta x$. The cosine/sine transforms of the PDF forms~\eqref{eq:jam-param_valence} and~\eqref{eq:jam-param_plus}
\begin{align}
    &\mathfrak{Re}\ Q\left(\nu,\mu^2\right)=\int_0^1dx\ \cos\left(\nu x\right)f_{q_{\rm v}/N}\left(x,\mu^2\right)\label{eq:cosTransform} \\
    &\mathfrak{Im}\ Q\left(\nu,\mu^2\right)=\int_0^1dx\ \sin\left(\nu x\right)f_{q_+/N}\left(x,\mu^2\right)\label{eq:sinTransform}
\end{align}
are fit to the real/imaginary ITD data using first an uncorrelated least-squares regression
\be
\chi^2=\sum_{\nu_{min}}^{\nu_{max}}\frac{\left[Q\left(\nu,\mu^2\right)-Q_{\rm fit}\left(\nu,\mu^2\right)\right]^2}{\sigma_Q^2},
\label{eq:leastSqrs}
\ee
with $\sigma_{\mathcal{Q}}^2$ the variance of the ITD, and $\lbrace\nu_{\rm min},\nu_{\rm max}\rbrace$ representing potential cuts on the data. These uncorrelated fits include all $z/a\in\lbrace1,\cdots,12\rbrace$ and $ap_z\in\lbrace1,\cdots,6\rbrace\times2\pi/L$. For ease of later reference, this method of extraction is denoted type-{\it C}.
The fits to the real and imaginary components of the ITD are shown in Fig.~\ref{fig:itd-real-convolC-uncorrelated} and Fig.~\ref{fig:itd-imag-convolC-uncorrelated}. The resulting valence and plus quark PDFs are juxtaposed with phenomenological determinations in Fig.~\ref{fig:pdf-valence-convolC-uncorrelated} and Fig.~\ref{fig:pdf-plus-convolC-uncorrelated}. The phenomenological PDFs are three flavor NLO determinations by the CJ~\cite{Accardi:2016qay} and JAM~\cite{Moffat:2021dji} collaborations, and three flavor NNLO determinations of MSTW~\cite{Martin:2010db} and NNPDF~\cite{Ball:2017nwa}.\footnote{LHAPDF set names: CJ15 - {\tt CJ15nlo}, JAM20 - {\tt JAM20-SIDIS\_PDF\_proton\_nlo}, \\ MSTW - {\tt MSTW2008nnlo68cl\_nf4}, NNPDF - {\tt NNPDF31\_nnlo\_pch\_as\_0118\_mc\_164}}

Apart from the $z/a\geq9$ data, such an uncorrelated fit would seem to well describe the $\mathfrak{Re}\ \mathcal{Q}\left(\nu,\mu^2\right)$ data and lead to valence PDFs that feature many structural similarities with phenomenological determinations at the same scale. The statistically consistent figure of merit for the two- and three-parameter fits tabulated in Tab.~\ref{tab:uncorrelated-fitParams}, however indicates the data cannot distinguish between these models. The two-parameter fit to $\mathfrak{Im}\ \mathcal{Q}\left(\nu,\mu^2\right)$ is clearly more heavily constrained by the $z/a\lesssim7$ data, and all but avoids points of the ITD originating from larger separations. Above $x\sim0.4$ the extracted $f_{q_+/N}\left(x,\mu^2\right)_C^{{\rm n}=2}$ result likewise shares structural similarities with the shown phenomenological results. The lack of any large-$\nu$ constraint provided by $\mathfrak{Im}\ \mathcal{Q}\left(\nu,\mu^2\right)$ entails a generally unconstrained fitted PDF in the small-$x$ regime, although this relation is not bijective.

\begin{table*}[]
    \centering
    {
    \begin{tabular}{cccccc}
    \hline
      $N_{\rm param}$ & $N_{{\rm v}/+}$ & $\alpha$ & $\beta$ & $\delta$ & $\chi^2_r$ \\
      \hline
      $2$ & -- & $-0.006(98)$ & $2.754(285)$ & $-$ & $2.183(483)$ \\
      $3$ & -- & $0.019(98)$ & $2.212(291)$ & $-0.737(12)$ & $2.192(490)$\\
      \hline
      $2$ & $3.616(2.260)$ & $-0.077(275)$ & $2.983(606)$ & $-$ & $2.780(806)$ \\
      \hline
    \end{tabular}
    }
    \caption{Unpolarized nucleon valence and plus quark PDF parameters obtained from performing uncorrelated cosine/sine transform fits to the real/imaginary component of the matched ITD at $2\text{ GeV}$ in $\msbar$. Results for the plus quark PDF are only shown for $N_{\rm param}=2$, where the smooth polynomial $P\left(x\right)=1$, as higher numbers of parameters led to uncontrolled fits. The uncorrelated figure of merit is also shown.\label{tab:uncorrelated-fitParams}}
\end{table*}

\subsection{Direct Extraction of PDFs from Reduced Pseudo-ITDs\label{ssec:direct}}
A separate, though in principle equivalent, route to extract PDFs from these data is to directly apply the factorized relationship~\eqref{eq:matchingkernel} having substituted the definition of the ITD~\eqref{eq:itd-pdf}:
\begin{align}
  \mathfrak{M}\left(\nu,z^2\right)=&\int_{-1}^1dx\int_0^1du\ \mathcal{C}\left(u,z^2\mu^2,\alpha_s\left(\mu^2\right)\right)e^{ix\nu}f_{q/N}\left(x,\mu^2\right)
  +\sum_{k=1}^\infty\mathcal{B}_k\left(\nu\right)\left(z^2\right)^k.\label{eq:pITD-PDF-direct-match}
\end{align}
By assuming a PDF parameterization and performing a maximum likelihood regression of the double convolution and  $\mathfrak{M}\left(\nu,z^2\right)$, the introduction of additional systematic errors from the evolution/matching steps and a potentially incorrect functional description of $\mathfrak{M}\left(\nu,z^2\right)$ when interpolating its $\nu$-dependence (e.g. Eq.~\ref{eq:polyFits}) can all be avoided.
The direct matching relationship between the PDFs and the reduced pseudo-ITD is given by
\begin{align}
    \left\lbrace\begin{matrix}\mathfrak{Re}\\ \mathfrak{Im}\end{matrix}\right\rbrace\mathfrak{M}\left(\nu,z^2\right)=\int_0^1dx&\ \left\lbrace\begin{matrix}\mathcal{K}_{\rm v}\left(x\nu,z^2\mu^2\right)f_{q_{\rm v}/N}\left(x,\mu^2\right)\\ \mathcal{K}_+\left(x\nu,z^2\mu^2\right)f_{q_+/N}\left(x,\mu^2\right)\end{matrix}\right\rbrace\ 
    +\sum_{k=1}^\infty\mathcal{B}_k\left(\nu\right)\left(z^2\right)^k,
    \label{eq:pitd-pdf-fit}
\end{align}
where the one-loop kernels that match the $\lbrace x,\mu^2\rbrace$-dependencies of the valence/plus quark PDFs to the reduced pseudo-ITD are given by
\begin{align}
  \mathcal{K}_{\rm v}&\left(x\nu,z^2\mu^2\right)=\cos\left(x\nu\right)-\nonumber 
  \frac{\alpha_sC_F}{2\pi}\left[\ln\left(\frac{e^{2\gamma_E+1}z^2\mu^2}{4}\right)\tilde{B}_{\rm v}\left(x\nu\right)+\tilde{D}_{\rm v}\left(x\nu\right)\right] \\
  \mathcal{K}_+&\left(x\nu,z^2\mu^2\right)=\sin\left(x\nu\right)-
  \frac{\alpha_sC_F}{2\pi}\left[\ln\left(\frac{e^{2\gamma_E+1}z^2\mu^2}{4}\right)\tilde{B}_+\left(x\nu\right)+\tilde{D}_+\left(x\nu\right)\right],
\end{align}
with the Altarelli-Parisi and scheme matching kernels modified to
\begin{align*}
  \tilde{B}_{\rm v}&\left(y\right)=\frac{1-\cos\left(y\right)}{y^2}+\frac{3-4\gamma_E}{2}\cos\left(y\right) 
  +2\sin\left(y\right)\frac{y\text{Si}\left(y\right)-1}{y}+2\cos\left(y\right)\left[\text{Ci}\left(y\right)-\ln\left(y\right)\right] \\
  \tilde{B}_+&\left(y\right)=-\frac{\sin\left(y\right)+y}{y^2}+\frac{3-4\gamma_E}{2}\sin\left(y\right) 
  -2\cos\left(y\right)\frac{y\text{Si}\left(y\right)-1}{y}+2\sin\left(y\right)\left[\text{Ci}\left(y\right)-\ln\left(y\right)\right] \\
  \tilde{D}_{\rm v}&\left(y\right)=-4y\mathfrak{Im}\left[e^{ix}\thinspace_3F_3\left(111;222;-iy\right)
  \right]
  +\left[\cos\left(y\right)\left(1+\frac{2}{y^2}\right)-\frac{2}{y^2}\right] \\
  \tilde{D}_+&\left(y\right)=4y\mathfrak{Re}\left[e^{ix}\thinspace_3F_3\left(111;222;-iy\right)\right] 
  +\left[\sin\left(y\right)\left(1+\frac{2}{y^2}\right)-\frac{2}{y}\right],
\end{align*}
and $\text{Si}\left(y\right)/\text{Ci}\left(y\right)$ are the integral sine/cosine functions and $_3F_3\left(111;222;-iy\right)$ is a generalized Hypergeometric function~\cite{Radyushkin:2019owq,Izubuchi:2018srq}. 
A notable challenge of this direct approach is that accurate computation of the generalized Hypergeometric function requires multi-precision arithmetic, which is computationally inefficient and invariably slows the parameter optimization. 
PDFs extracted by directly performing parametric fits~\eqref{eq:pitd-pdf-fit} to the reduced pseudo-ITD data are denoted type-{\it K}.

\begin{figure*}[t!]
\subfigure[]{\label{fig:uncorr-typeK-valence} \includegraphics[width=0.459\textwidth]{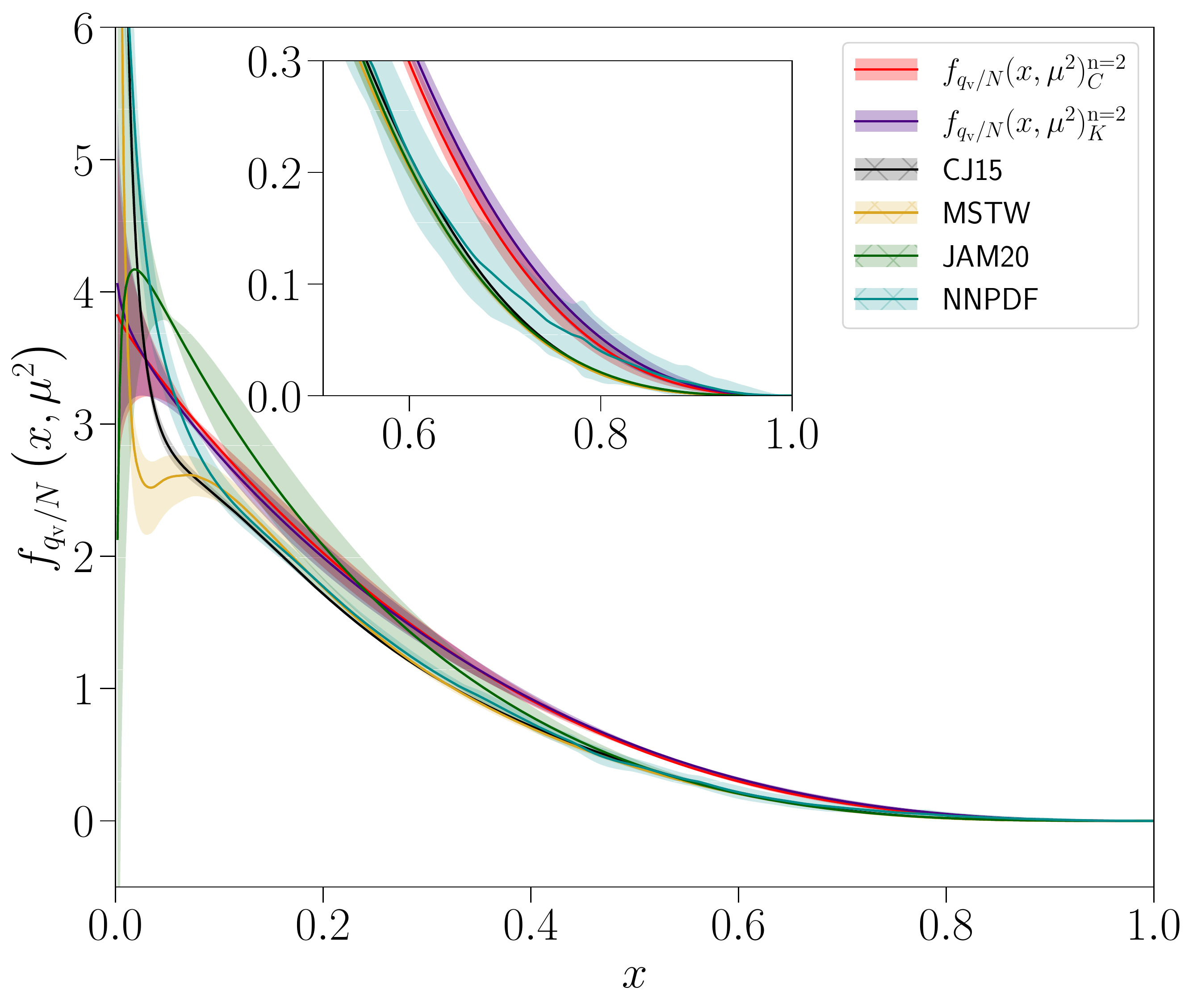}}
\hfill
\subfigure[]{\label{fig:uncorr-typeK-plus} \includegraphics[width=0.459\textwidth]{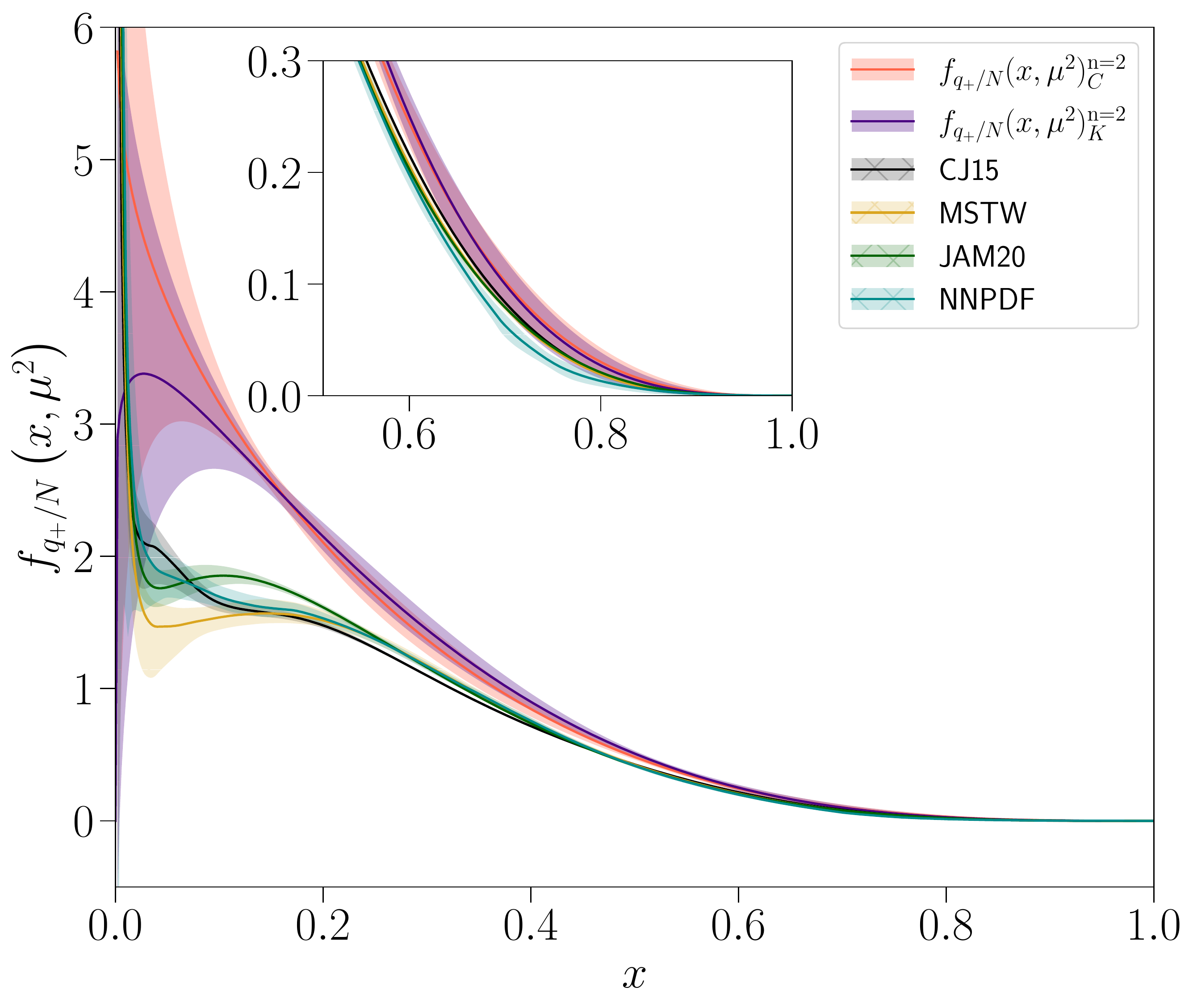}}
\caption{Two-parameter valence (\ref{fig:uncorr-typeK-valence}) and plus (\ref{fig:uncorr-typeK-plus}) quark PDF resulting from type-{\it C} (red) and type-{\it K} (indigo) fits to the unpolarized nucleon ITD and reduced pseudo-ITD, respectively. The direct matching fits are consistent with the cosine/sine transform of the model PDF fit to the ITD.}
\end{figure*}

Simple two-parameter PDFs obtained from uncorrelated type-{\it K} fits are shown in Fig.~\ref{fig:uncorr-typeK-valence} and Fig~\ref{fig:uncorr-typeK-plus}, together with the same phenomenological determinations and the uncorrelated type-{\it C} two-parameter PDF fits.
\begin{table}[t]
  \begin{center}
    \begin{tabular}{ccccc}
      \hline
      $N=2_{{\rm v}/+}$ & $N_{{\rm v}/+}$ & $\alpha$ & $\beta$ & $\chi^2_r$ \\
      \hline
      $2_{\rm v}$ & -- & $-0.030(96)$ & $2.601(277)$ & $7.364(761)$ \\
      $2_+$ & $5.131(3.405)$ & $0.091(299)$ & $3.244(638)$ & $4.536(902)$\\
      \hline
    \end{tabular}
  \end{center}
  \caption{Unpolarized nucleon valence and plus quark PDF parameters obtained from type-{\it K} fits to the real/imaginary component of $\mathfrak{M}\left(\nu,z^2\right)$.\label{tab:uncorrelated-typeK-Params}}
\end{table}
The type-{\it C} and type-{\it K} fits are statistically consistent. This is confirmed by comparing the type-{\it K} fit results in Tab.~\ref{tab:uncorrelated-typeK-Params} to the type-{\it C} results in Tab.~\ref{tab:uncorrelated-fitParams}. However, the central values of the type-{\it K} fits suggest that at small-$x$ the $f_{q_{\rm v}/N}\left(x\right)$ is more divergent and the $f_{q_+/N}\left(x\right)$ is instead convergent for small-$x$ at the scale $\mu=2\text{ GeV}$. The factor of two or three increase in the figure of merit when switching from type-{\it C} to type-{\it K} fits is the first indication of puzzling behavior in $\mathfrak{M}\left(\nu,z^2\right)$. We reiterate the naive two-parameter fits capture the known limiting regimes of the PDFs. The poor figures of merit in the type-{\it K} fits hint that $\mathfrak{M}\left(\nu,z^2\right)$ at this stage apparently does not align well with expectations from the direct matching~\eqref{eq:pitd-pdf-fit}. This is a potentially disastrous conclusion. To gain some insight, we now consider the data correlations.


\begin{figure*}[t!]
\subfigure[]{\label{fig:corr-real-itd} \includegraphics[width=0.462\textwidth]{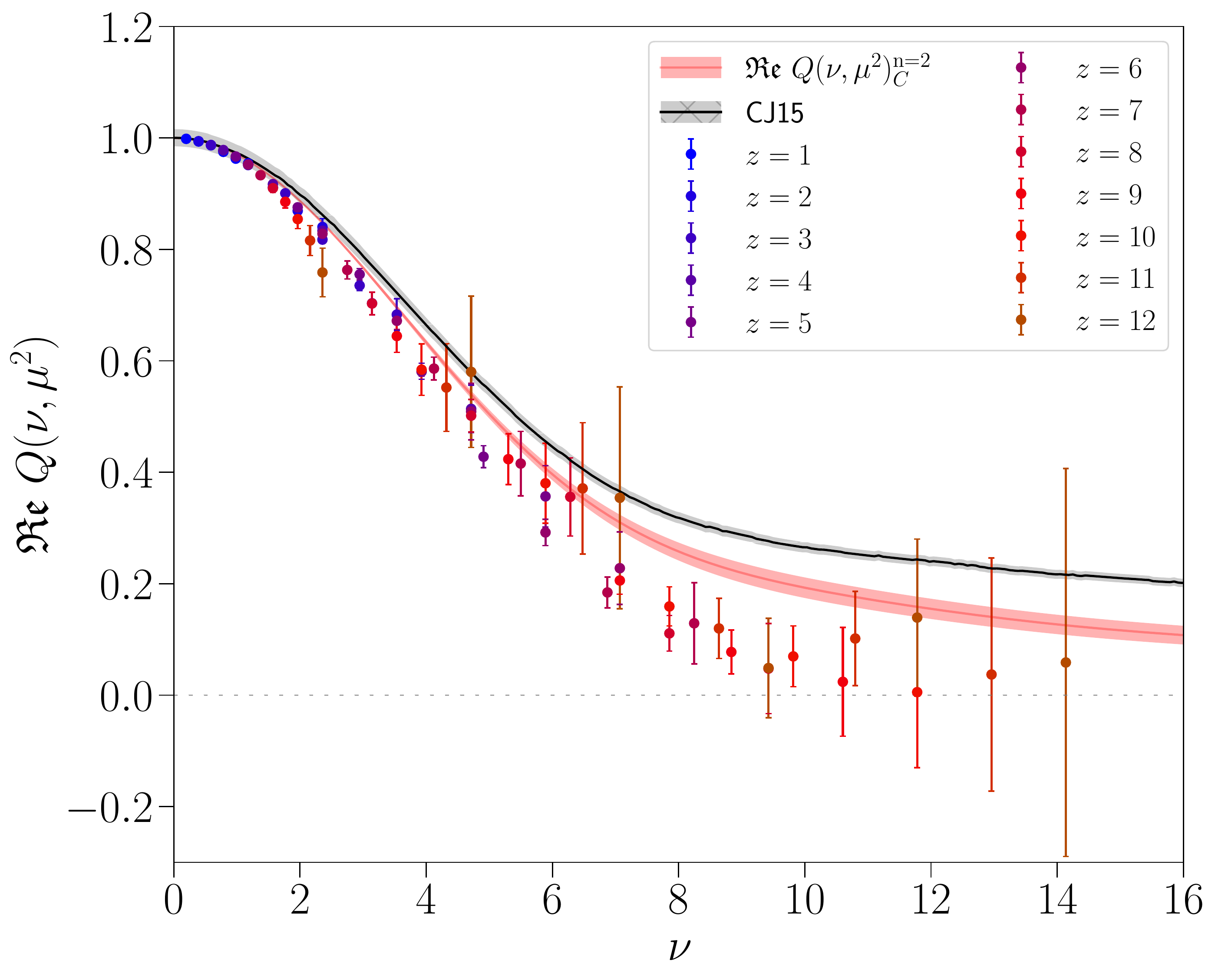}}
\hfill
\subfigure[]{\label{fig:corr-real-itd_zoomed}         \includegraphics[width=0.459\textwidth]{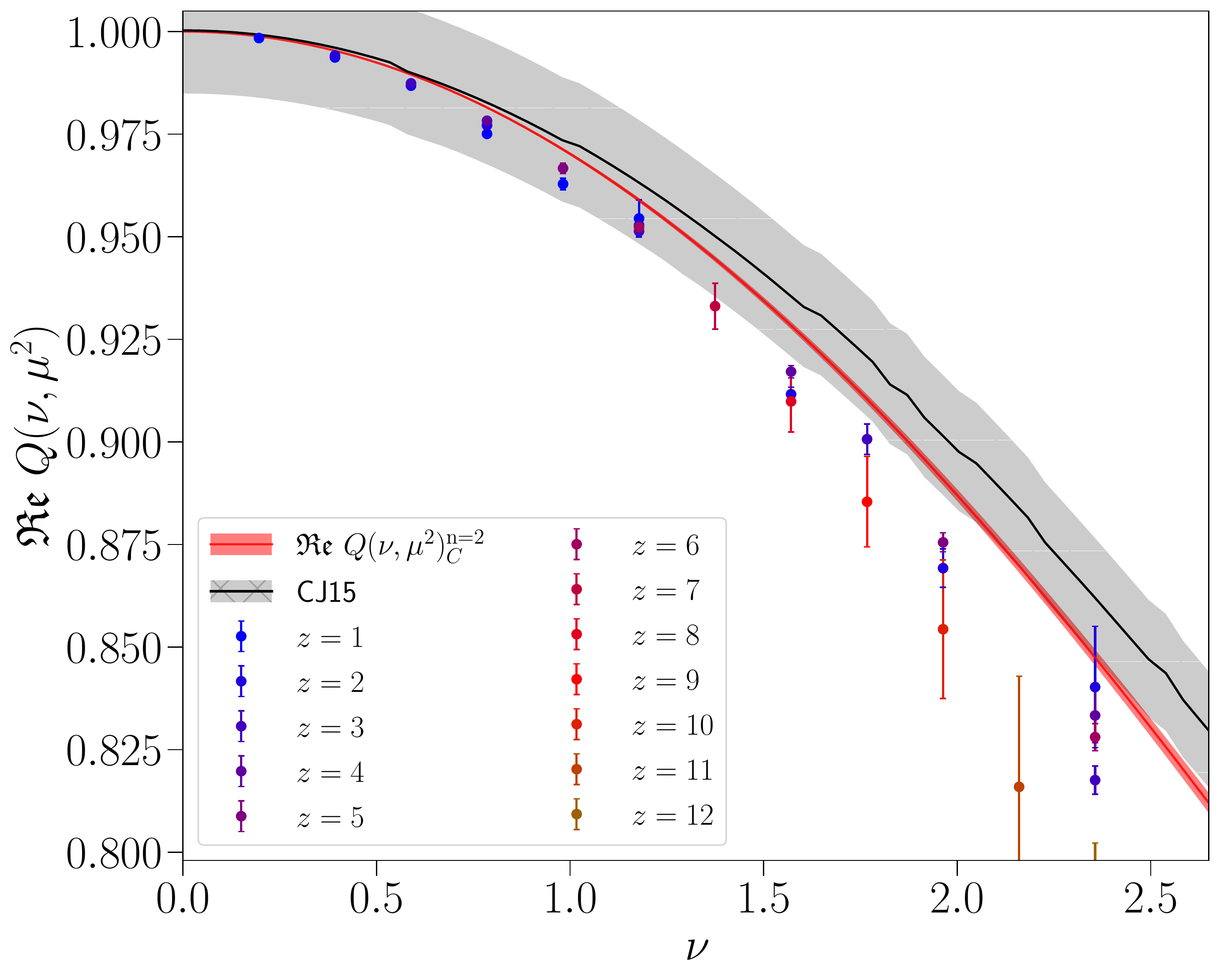}}
\caption{Real component of the matched ITD at $\mu=2\text{ GeV}$ in $\msbar$ (\ref{fig:corr-real-itd}) fit by the cosine transform of a two-parameter model PDF~\eqref{eq:jam-param_valence}. Data have been fit for $z/a\leq12$ and \textit{data correlations have been incorporated}. The fit clearly misses each point of the ITD. The derived CJ15 ITD at the same scale is shown for reference. In (\ref{fig:corr-real-itd_zoomed}) focus is given to the small-$\nu$ region. The correlated two-parameter fit is seen to deviate appreciably from the precise small-$\nu$ data.}
\end{figure*}

\subsubsection{Data Correlation}
The data featured in this work, and indeed any lattice calculation, naturally are correlated.
By beginning this subsection with uncorrelated fits, we highlight that without knowledge or through the simple neglect of data correlations, which appears to be common in the literature, one might incorrectly assume an adequate description of the data has been achieved. These correlations must be taken into account in order to provide a rigorous accounting of mutual fluctuations in the data and thus an agnostic PDF determination.

Simply repeating the two-parameter fit to $\mathfrak{Re}\thinspace\mathcal{Q}\left(\nu,\mu^2\right)$, only this time accounting for the data covariance \textbf{Cov}
\be
\chi_r^2=\sum_{i,j=\nu_{min}}^{\nu_{max}}q_i^\top{\bf Cov}^{-1}_{ij}q_j,
\label{eq:correlatedLeastSqrs}
\ee
with $q_k=\mathcal{Q}\left(\nu,\mu^2\right)_k-\mathcal{Q}_{\rm fit}\left(\nu,\mu^2\right)_k$, we arrive at a much different conclusion shown in Fig.~\ref{fig:corr-real-itd}. The visual discrepancy between the ITD and two-parameter fit is stark, and leads to a correlated figure of merit of $\mathcal{O}\left(40\right)$. Although the fit misses nearly all of the moderate to large-$z$ points, Fig.~\ref{fig:corr-real-itd_zoomed} illustrates the large increase in the figure of merit is primarily due to the slight deviation from the very precise $z/a\lesssim4$ data.

\begin{figure*}[tb!]
     \subfigure[]{\label{fig:real-itd-covariance} \includegraphics[width=0.459\textwidth]{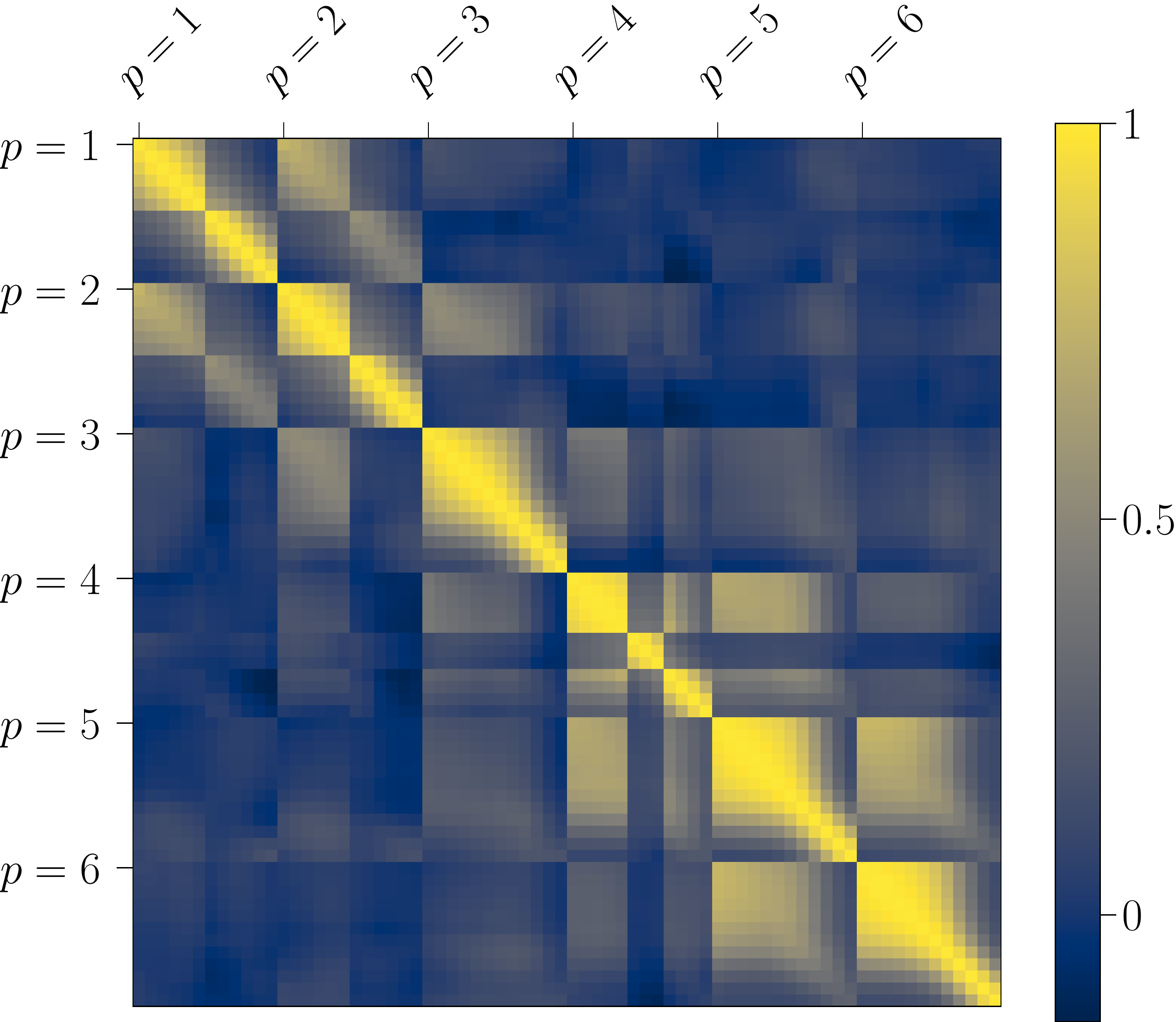}}
     \hfill
     \subfigure[]{\label{fig:imag-itd-covariance} \includegraphics[width=0.459\textwidth]{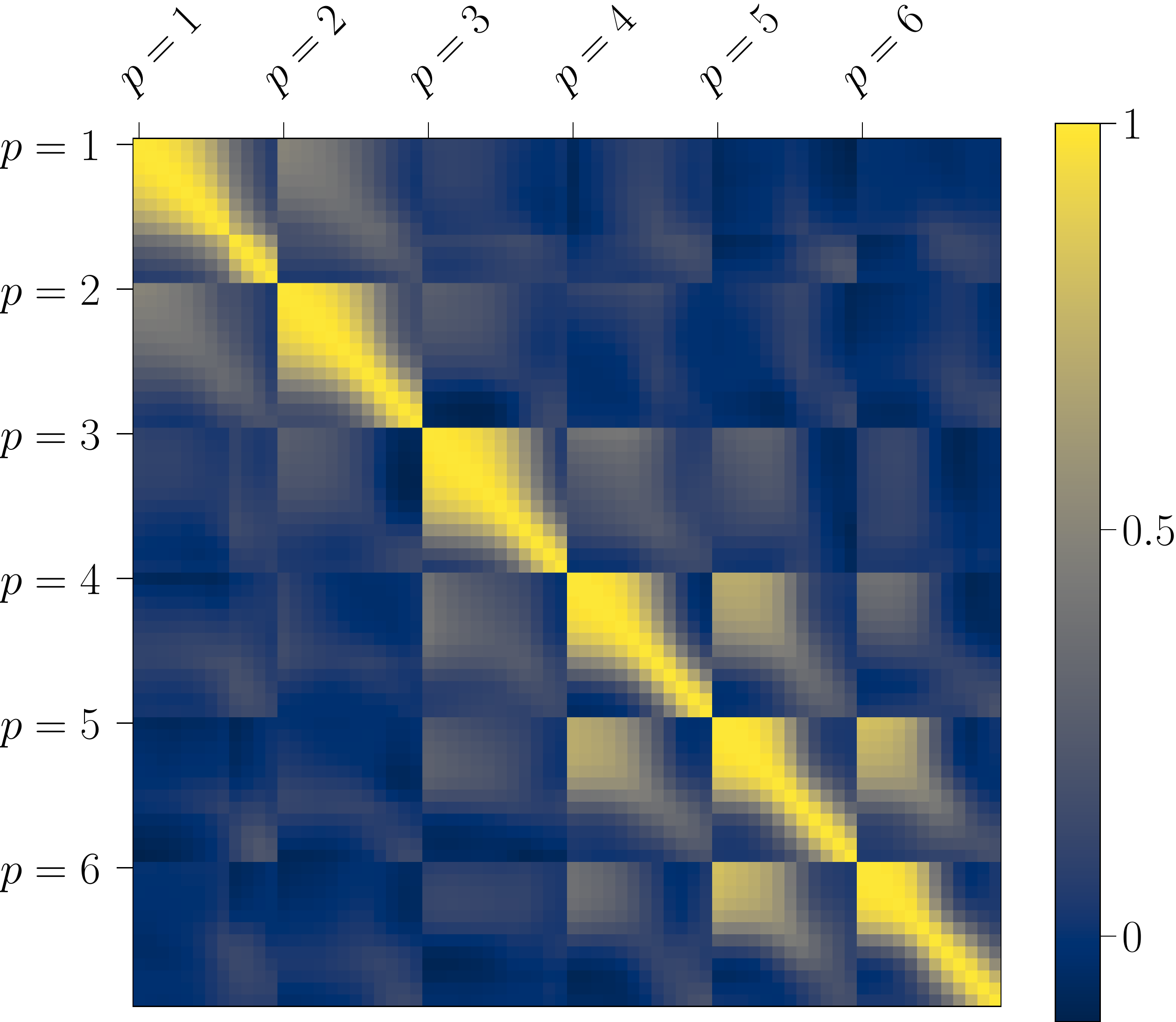}}
     \caption{Data covariance in the real (\ref{fig:real-itd-covariance}) and imaginary (\ref{fig:imag-itd-covariance}) components of the matched ITD at $2\text{ GeV}$, normalized according to ${\rm Cov}_{ij}/\sqrt{{\rm Cov}_{ii}{\rm Cov}_{jj}}$. Within each lattice momentum block, entries are ordered in ascending Wilson line lengths.}
\end{figure*}
Visualizing the data covariance in the real component in Fig.~\ref{fig:real-itd-covariance}, it is clear the low-momentum data $ap_z\leq4\pi/L$ are strongly correlated amongst each other and correlate weakly with the $ap_z\geq8\pi/L$ data; some mild correlation is visible with the $ap_z=6\pi/L$ data with $z/a\leq6$. Within the $ap_z\leq4\pi/L$ channels the strongest correlation can be found in the shortest Wilson line data. These observations provide an explanation for the poor correlated two-parameter fit in Fig.~\ref{fig:corr-real-itd} - the strongest correlation is with the most precise data in our calculation causing any correlated fit to favor the small-$\nu$ data. Indeed strong correlation is also observed amongst the momentum channels $ap_z=\lbrace4,5,6\rbrace\times2\pi/L$, but the signal-to-noise degradation for these high-momentum data minimizes their effect on any fit. It is interesting this delineation corresponds to the transition from an unphased to phased eigenvector basis. The data covariance in the imaginary component, shown in Fig.~\ref{fig:imag-itd-covariance}, shows the strongest correlations within each momentum channel and between adjacent Wilson line lengths (e.g. $z/a=4$ and $z/a=5$). It is then no surprise that a correlated two-parameter PDF parameterization of $\mathfrak{Im}\ \mathcal{Q}\left(\nu,\mu^2\right)$ is also met with a poor figure of merit.
The non-trivial structures of correlation evident in these data are indicative of our simple PDF parameterizations~\eqref{eq:jam-param_valence} and~\eqref{eq:jam-param_plus} being inappropriate for these data.
The above puzzling, and indeed worrisome, conclusions are given a deeper quantitative understanding in the following sections.

\subsection{Classical Orthogonal Polynomials\label{ssec:jacobi}}
The phenomenological parameterizations we have considered thus far are but one way to regulate the ill-posed inverse relation between the ITD/reduced pseudo-ITD and the corresponding PDF.
These parameterizations nevertheless introduce a model dependence into the extracted PDF. Any PDF faithfully reported from a lattice calculation should take into account the space of functions that smoothly connects the $x\rightarrow0$ and $x\rightarrow1$ limits.
As an alternative means to describe the valence/plus quark sectors and minimize model bias, we propose to parameterize the PDFs by a complete basis of classical orthogonal polynomials~\cite{Karpie:2021pap}. The leveraging of orthogonal polynomials to obtain an unknown distribution is not unique to this work. The approach we adopt parallels efforts to extract PDFs from phenomenological fits of inclusive processes~\cite{Harland-Lang:2014zoa,Dulat:2015mca}, as well as distribution amplitudes~\cite{Bali:2018spj,Bali:2019dqc,Segovia:2013eca} and inelastic scattering cross sections~\cite{Fukaya:2020wpp} from matrix elements calculated in lattice QCD.

Consider the Jacobi (hypergeometric) polynomials
\begin{align}
    P_n^{\left(\alpha,\beta\right)}&\left(z\right)=\frac{\Gamma\left(\alpha+n+1\right)}{n!\Gamma\left(\alpha+\beta+n+1\right)}\times
    \sum_{j=0}^n{n\choose j}\frac{\Gamma\left(\alpha+\beta+n+j+1\right)}{\Gamma\left(\alpha+j+1\right)}\left(\frac{z-1}{2}\right)^j,
    \label{eq:jacobi-orig}
\end{align}
which for $\alpha,\beta>-1$ form a basis of orthogonal polynomials on the interval $\left[-1,1\right]$ with respect to the metric $\left(1-z\right)^\alpha\left(1+z\right)^\beta$. Under the mapping $z\mapsto1-2x$ the shifted Jacobi polynomials
\be
\Omega_n^{\left(\alpha,\beta\right)}\left(x\right)=\sum_{j=0}^n\omega_{n,j}^{\left(\alpha,\beta\right)}x^j
\label{eq:jacobi-poly}
\ee
form a complete basis of orthogonal polynomials on the interval $\left[0,1\right]$ with respect to the metric $x^\alpha\left(1-x\right)^\beta$. In Eq.~\ref{eq:jacobi-poly} we have defined
\begin{align}
    \omega_{n,j}^{\left(\alpha,\beta\right)}&=\frac{\Gamma\left(\alpha+n+1\right)}{n!\Gamma\left(\alpha+\beta+n+1\right)} 
   {n\choose j}\frac{\left(-1\right)^j\Gamma\left(\alpha+\beta+n+j+1\right)}{\Gamma\left(\alpha+j+1\right)}.
\label{eq:omega-nj}
\end{align}
As the set of polynomials $\lbrace\Omega_n^{\left(\alpha,\beta\right)}\rbrace$ span $x\in\left[0,1\right]$, a PDF can be expressed generically as
\be
f_{a/h}\left(x\right)=x^\alpha\left(1-x\right)^\beta\sum_{n=0}^\infty C_{a,n}^{\left(\alpha,\beta\right)}\Omega_n^{\left(\alpha,\beta\right)}\left(x\right),
\label{eq:func_jacobiParam}
\ee
with expansion coefficients $C_{a,n}^{\left(\alpha,\beta\right)}$. The parameters $\lbrace\alpha,\beta\rbrace$ lose their familiar characterization of the $x\rightarrow0/x\rightarrow1$ PDF behaviors in place of delineating between different choices of bases. The expansion in Jacobi polynomials in Eq.~\ref{eq:func_jacobiParam} is thus entirely generic and model-independent. However, the series of Jacobi polynomials must in practice be truncated at some finite order, $n$. The bias then introduced may be studied by fixing the order of truncation and determining the optimum $\lbrace\alpha,\beta\rbrace$, or tuning $\lbrace\alpha,\beta\rbrace$ to capture generic properties of a PDF and subsequently optimize the order of truncation - we will adopt the former.

Our strategy to parameterize the reduced pseudo-ITD using a set of $\Omega_n^{\left(\alpha,\beta\right)}\left(x\right)$ will be met by similar numerical difficulties as the type-{\tt K} fits discussed above. The numerical effort is lessened by considering a Taylor series expansion in $\nu$ for fixed separations $z^2$ of the direct matching kernels $\mathcal{K}_{{\rm v}/+}\left(x\nu,z^2\mu^2\right)$ and Eq.~\ref{eq:func_jacobiParam}. The contribution of an $n^{\text{th}}$-order Jacobi polynomial $\Omega_n^{\left(\alpha,\beta\right)}\left(x\right)$ to $\mathfrak{Re}\ \mathfrak{M}\left(\nu,z^2\right)$ and $\mathfrak{Im}\ \mathfrak{M}\left(\nu,z^2\right)$ is given by
\begin{align*}
    &\sigma_n^{\left(\alpha,\beta\right)}\left(\nu,z^2\mu^2\right)=\mathfrak{Re}\int_0^1dx\ \mathcal{K}_{\rm v}\left(x\nu,z^2\mu^2\right) 
   x^\alpha\left(1-x\right)^\beta\Omega_n^{\left(\alpha,\beta\right)}\left(x\right) \\ 
     &\eta_n^{\left(\alpha,\beta\right)}\left(\nu,z^2\mu^2\right)=\mathfrak{Im}\int_0^1dx\ \mathcal{K}_+\left(x\nu,z^2\mu^2\right)
   x^\alpha\left(1-x\right)^\beta\Omega_n^{\left(\alpha,\beta\right)}\left(x\right).
\end{align*}
Expanding the direct matching kernels $\mathcal{K}_{{\rm v}/+}\left(x\nu,z^2\mu^2\right)$ in even/odd powers of $\nu$ one finds
\begin{align}
    \sigma_n^{\left(\alpha,\beta\right)}\left(\nu,z^2\mu^2\right)&=\sum_{j=0}^n\sum_{k=0}^\infty\frac{\left(-1\right)^k}{\left(2k\right)!}c_{2k}\left(z^2\mu^2\right)
    \omega_{n,j}^{\left(\alpha,\beta\right)}B\left(\alpha+2k+j+1,\beta+1\right)\nu^{2k} \label{eq:sigmaN} \\
  \eta_n^{\left(\alpha,\beta\right)}\left(\nu,z^2\mu^2\right)&=\sum_{j=0}^n\sum_{k=0}^\infty\frac{\left(-1\right)^k}{\left(2k+1\right)!}c_{2k+1}\left(z^2\mu^2\right)
  \omega_{n,j}^{\left(\alpha,\beta\right)}B\left(\alpha+2k+j+2,\beta+1\right)\nu^{2k+1}\label{eq:etaN},
\end{align}
where
\be
c_n\left(z^2\mu^2\right)=1-\frac{\alpha_sC_F}{2\pi}\left[\gamma_n\ln\left(\frac{e^{2\gamma_E+1}}{4}z^2\mu^2\right)+d_n\right]
\ee
and the constants $\gamma_n$ and $d_n$ are the leading order moments of the Altarelli-Parisi and scheme matching kernels derived in~\cite{Karpie:2018zaz}, respectively. The sum over $k$ is to be performed to assure convergence for a given value of $\nu$ - we have identified $k_{max}=75$ as providing more than adequate numerical precision, in reasonable computation time.
With the above definitions, the leading-twist valence and plus quark PDFs describe the reduced pseudo-ITD components according to
\begin{align}
    &\mathfrak{Re}\ \mathfrak{M}^{lt}\left(\nu,z^2\right)=\sum_{n=0}^\infty\sigma_n^{\left(\alpha,\beta\right)}\left(\nu,z^2\mu^2\right)C^{lt\left(\alpha,\beta\right)}_{{\rm v},n}\label{eq:lt-sigma} \\
    &\mathfrak{Im}\ \mathfrak{M}^{lt}\left(\nu,z^2\right)=\sum_{n=0}^\infty\eta_n^{\left(\alpha,\beta\right)}\left(\nu,z^2\mu^2\right)C_{+,n}^{lt\left(\alpha,\beta\right)}\label{eq:lt-eta},
\end{align}
where the $C_{{\rm v}/+,n}^{lt\ \left(\alpha,\beta\right)}$ are the Jacobi polynomial expansion coefficients.
\begin{figure*}[t!]
\subfigure[]{\label{fig:sigma0n} \includegraphics[width=0.459\textwidth]{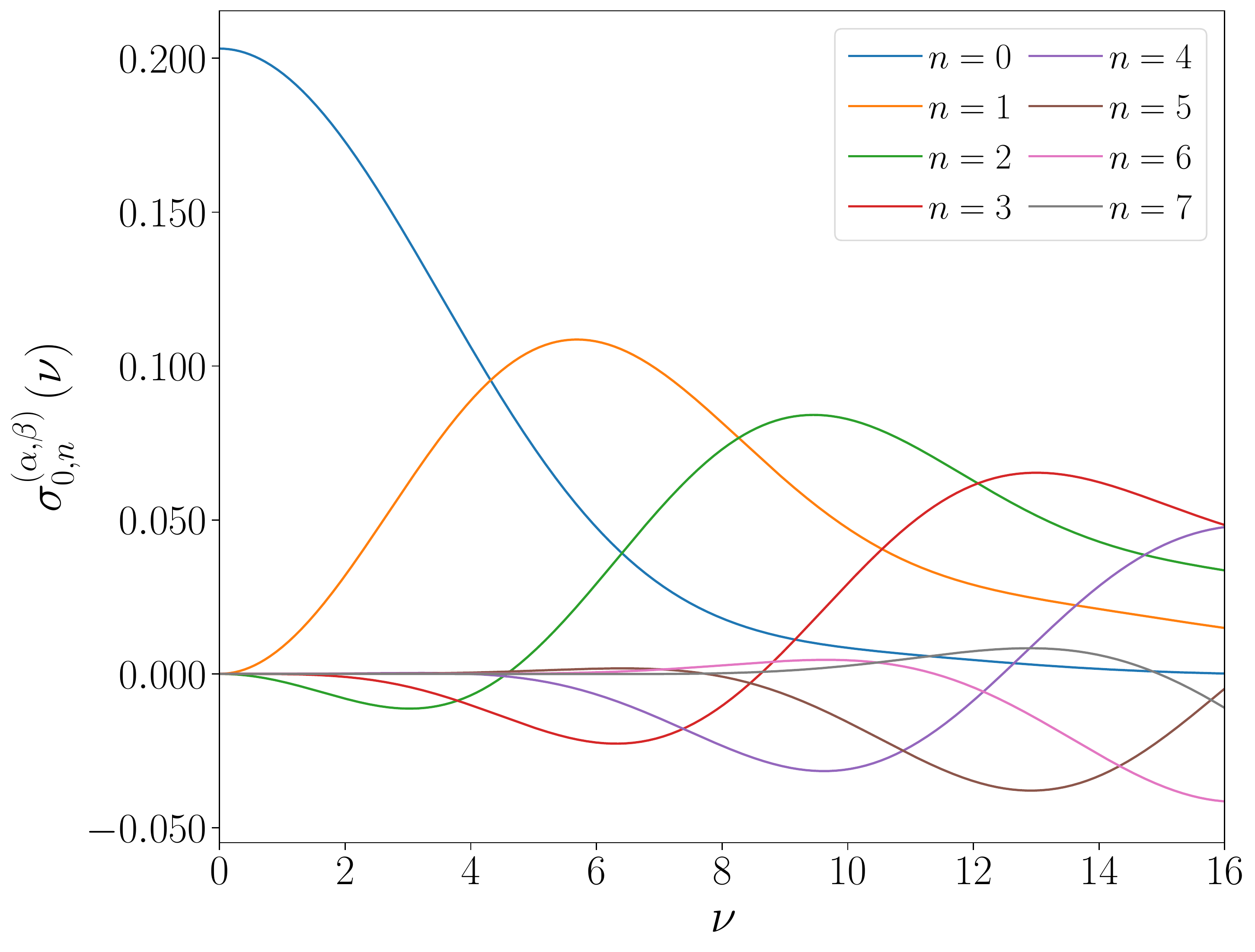}}
\hfill
\subfigure[]{\label{fig:eta0n} \includegraphics[width=0.459\textwidth]{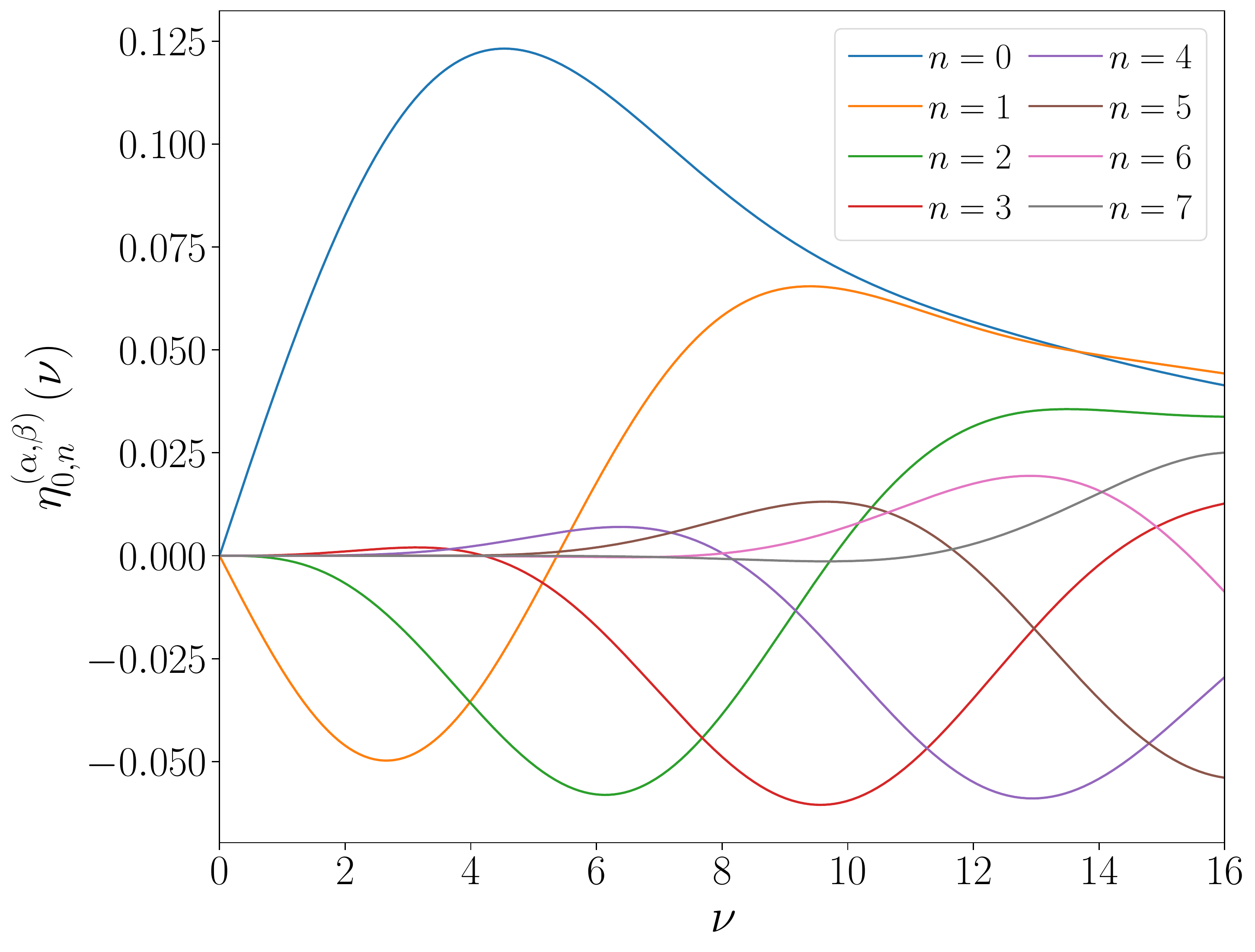}}
\caption{The eight lowest-order $\sigma_{0,n}^{\left(\alpha,\beta\right)}$ (\ref{fig:sigma0n}) and $\eta_{0,n}^{\left(\alpha,\beta\right)}$ (\ref{fig:eta0n}) polynomials for an arbitrarily chosen basis $\alpha=0.125$ and $\beta=2.85$. Each polynomial features an extremum in a range of Ioffe-time accessible in our lattice calculation, and asymptotically approaches zero for $\nu\rightarrow\infty$.}
\end{figure*}

The reduced pseudo-ITD is subject to discretization errors that vanish in the continuum limit, and higher-twist effects that survive the continuum limit. A reliable determination of the leading-twist PDF in the continuum then depends on parameterization and removal of these effects. As the Fourier transform in $\nu$ of the reduced pseudo-ITD only has support on the momentum fraction interval $x\in\left[-1,1\right]$~\cite{Radyushkin:2016hsy}, any contaminating effects must also have support only in this interval and can be parameterized by the same basis of Jacobi polynomials. Any corrections by construction must be functions of $\nu^2$ in the real component, and $\nu$ in the imaginary component. Since the space-like matrix element~\eqref{eq:parton-bilinear-had} is on-shell, at a single lattice spacing we may account for contaminating discretization and higher-twist effects of $\mathcal{O}\left(a/z\right)$ and $\mathcal{O}\left(z^2\Lambda_{\rm QCD}^2\right)^n$. The latter are the expected polynomial corrections to the reduced pseudo-ITD  factorization~\eqref{eq:matchingkernel}, while the discretization correction should scale based on parity. Namely, the real (imaginary) component of $\mathfrak{M}\left(\nu,z^2\right)$ is even (odd) in $z$, so any discretization effect should behave in this manner as well. We will find this to be especially subtle in our data, motivating the present designation of $\mathcal{O}\left(a/z\right)$.

The contaminating $x$-space distributions are of the same form in~\eqref{eq:func_jacobiParam} with distinct expansion coefficients. The coefficients of the corrections are denoted $C^{\thinspace{\rm corr}\thinspace\left(\alpha,\beta\right)}_{\tau,n}$ with $\tau=\lbrace{\rm v},+\rbrace$ indicating whether the effect arises in the valence/plus quark PDFs. The choice of basis $\lbrace\alpha,\beta\rbrace$ in~\eqref{eq:lt-sigma} and~\eqref{eq:lt-eta} may equally be utilized to quantify these distributions.
Supposing, for simplicity, these effects enter at tree-level, their contributions to the reduced pseudo-ITD signals with $\sigma_{0,n}^{\left(\alpha,\beta\right)}\equiv\sigma_n^{\left(\alpha,\beta\right)}\left(\nu,z^2\mu^2\right)\mid_{\alpha_s=0}$ and $\eta_{0,n}^{\left(\alpha,\beta\right)}\equiv\eta_n^{\left(\alpha,\beta\right)}\left(\nu,z^2\mu^2\right)\mid_{\alpha_s=0}$ are then
\begin{align}
    &\mathfrak{Re}\ \mathfrak{M}^{\rm corr}\left(\nu,z^2\right)=\kappa_{\rm corr}\sum_{n=1}^\infty\sigma_{0,n}^{\left(\alpha,\beta\right)}C_{{\rm v},n}^{corr\ \left(\alpha,\beta\right)}\label{eq:reJacobiCorr} \\
    &\mathfrak{Im}\ \mathfrak{M}^{\rm corr}\left(\nu,z^2\right)=\kappa_{\rm corr}\sum_{n=0}^\infty\eta_{0,n}^{\left(\alpha,\beta\right)}C_{+,n}^{corr\ \left(\alpha,\beta\right)},\label{eq:imJacobiCorr}
\end{align}
where $\kappa_{\rm corr}$ is a dimensionless parameter constructed from the dimensionful parameters of the calculation which describes the scaling of each correction (e.g. $\kappa_{\rm corr}=a/z$).

Visualizing the Taylor-expanded matching kernels~\eqref{eq:sigmaN} and~\eqref{eq:etaN} at tree-level across a range of Ioffe-times in Fig.~\ref{fig:sigma0n} and Fig.~\ref{fig:eta0n}, it is seen the polynomials $\sigma_{0,n}^{\left(\alpha,\beta\right)},\eta_{0,n}^{\left(\alpha,\beta\right)}$ reach an extremum in Ioffe-time commensurate with the polynomial order and asymptotically approach zero. This conveniently reflects the correct large-$\nu$ behavior of the ITD in the same limit (see Eq.~\ref{eq:itd-large-nu}).
Since $\mathfrak{M}\left(0,z^2\right)=1$ by construction, all corrections must vanish at zero Ioffe-time. Of the Jacobi polynomial expanded corrections, only $\sigma_{0,0}^{\left(\alpha,\beta\right)}\left(0\right)\neq0$ (blue curve of Fig.~\ref{fig:sigma0n}). The corrections to the real component of $\mathfrak{M}\left(\nu,z^2\right)$~\eqref{eq:reJacobiCorr} must then be restricted to order $n\geq1$.

\begin{widetext}
The complete functional forms we apply to the real/imaginary components of the reduced pseudo-ITD are:
\begin{align}
  &\mathfrak{Re}\thinspace\mathfrak{M}_{\text{fit}}\left(\nu,z^2\right)=\sum_{n=0}^\infty\sigma_n^{\left(\alpha,\beta\right)}\left(\nu,z^2\mu^2\right)C_{{\rm v},n}^{lt\thickspace\left(\alpha,\beta\right)}+\left(\frac{a}{z}\right)\sum_{n=1}^\infty\sigma_{0,n}^{\left(\alpha,\beta\right)}\left(\nu\right)C_{{\rm v},n}^{az\thinspace\left(\alpha,\beta\right)}+z^2\Lambda_{\text{QCD}}^2\sum_{n=1}^\infty\sigma_{0,n}^{\left(\alpha,\beta\right)}\left(\nu\right)C_{{\rm v},n}^{t4\thinspace\left(\alpha,\beta\right)}\label{eq:bigReJacobiFit} \\
  &\qquad\qquad\qquad\qquad\qquad\qquad\qquad\qquad\qquad\qquad\qquad\qquad+z^4\Lambda_{\text{QCD}}^4\sum_{n=1}^\infty\sigma_{0,n}^{\left(\alpha,\beta\right)}\left(\nu\right)C_{{\rm v},n}^{t6\thinspace\left(\alpha,\beta\right)} \nonumber \end{align}
  
  \begin{align}
  &\mathfrak{Im}\thinspace\mathfrak{M}_{\text{fit}}\left(\nu,z^2\right)=\sum_{n=0}^\infty\eta_n^{\left(\alpha,\beta\right)}\left(\nu,z^2\mu^2\right)C_{+,n}^{lt\thickspace\left(\alpha,\beta\right)}+\left(\frac{a}{z}\right)\sum_{n=0}^\infty\eta_{0,n}^{\left(\alpha,\beta\right)}\left(\nu\right)C_{+,n}^{az\thinspace\left(\alpha,\beta\right)}+z^2\Lambda_{\text{QCD}}^2\sum_{n=0}^\infty\eta_{0,n}^{\left(\alpha,\beta\right)}\left(\nu\right)C_{+,n}^{t4\thinspace\left(\alpha,\beta\right)}\label{eq:bigImJacobiFit} \\
  &\qquad\qquad\qquad\qquad\qquad\qquad\qquad\qquad\qquad\qquad\qquad\qquad+z^4\Lambda_{\text{QCD}}^4\sum_{n=0}^\infty\eta_{0,n}^{\left(\alpha,\beta\right)}\left(\nu\right)C_{+,n}^{t6\thinspace\left(\alpha,\beta\right)} \nonumber.
\end{align}
\end{widetext}
The leading-twist ($C^{lt}_{\tau,n}$) and discretization ($C^{az}_{\tau,n}$) corrections are accompanied by twist-4 ($C^{t4}_{\tau,n}$) and twist-6 ($C^{t6}_{\tau,n}$) corrections. The twist-6 corrections will almost certainly not be constraining, as their effect will be large beyond the range of Ioffe-time for which we have statistically clean data ($\nu\sim10$); they are included nonetheless for exploration purposes. We note higher-twist corrections in a lattice calculation must arise in even powers of $z$, as odd powers are not hypercubic invariants.

In the fits we perform according to~\eqref{eq:bigReJacobiFit} and~\eqref{eq:bigImJacobiFit}, we elect to fix the order of truncation for the leading-twist and each type of correction, and numerically search for the optimal $\lbrace\alpha,\beta,C^{\rm corr}_{\tau,n}\rbrace$. Treating each fitted parameter as non-linear in a maximum likelihood fit leads to wildly unstable results. The way forward is to recognize $\alpha,\beta$ are the only fitted parameters that are truly non-linear; the correction coefficients $C^{\rm corr}_{\tau,n}$ are all linear. A maximum likelihood fit of the posterior distribution of the linear terms is then Gaussian and cheap to obtain.

\subsubsection{PDF Results with Jacobi Polynomials}
The reduced pseudo-ITD fits of Eq.~\ref{eq:bigReJacobiFit} and Eq.~\ref{eq:bigImJacobiFit} are implemented with the help of the Variable Projection (VarPro) algorithm~\cite{Golub} for separable non-linear optimization problems. This reduces the dimension of the non-linear optimization from $d=N_{\rm corr}+2$ to $d=2$, where $N_{\rm corr}$ are the number of linear correction coefficients. In our case, minimization is performed in the $d=2$ Jacobi polynomial basis $\lbrace\alpha,\beta\rbrace$, and any correction terms $C^{corr\left(\alpha,\beta\right)}_{\tau,n}$ are solved for exactly in terms of the non-linear basis functions $\lbrace\sigma_n^{\left(\alpha,\beta\right)},\sigma_{0,n}^{\left(\alpha,\beta\right)},\eta_n^{\left(\alpha,\beta\right)},\eta_{0,n}^{\left(\alpha,\beta\right)}\rbrace$. We note without VarPro, optimizations with $N_{\rm corr}\geq4$ are numerically unstable, regardless of the type of correction included in the model.

Care needs to be taken as correction terms are included in Eqns.~\ref{eq:bigReJacobiFit} and~\ref{eq:bigImJacobiFit}, as physical insight can quickly be replaced with over fitting. The first sensible restriction to impose is for all $x$-space corrections $\mathcal{O}\left(a/z\right)$, $\mathcal{O}\left(z^2\Lambda_{\rm QCD}^2\right)$ and $\mathcal{O}\left(z^4\Lambda_{\rm QCD}^4\right)$ to be sub-leading relative to the leading-twist PDF. It would be alarming to obtain, say, a twist-$4$ contribution that is larger than the leading-twist PDF, given that the short-distance factorization of the pseudo-distributions implies leading-twist dominance. Such disastrous scenarios are avoided with several Bayesian constraints of a Gaussian form. So as to allow the reduced pseudo-ITD to dictate the best fit results, all Bayesian priors are fixed to zero. The hierarchy we desire is realized with the following prior widths:
\begin{itemize}
\item Leading-twist:\\
$\delta C^{lt\left(\alpha,\beta\right)}_{\tau,0}=1.1$,
$\delta C^{lt\left(\alpha,\beta\right)}_{\tau,1}=0.75$,
$\delta C^{lt\left(\alpha,\beta\right)}_{\tau,2}=0.5$,\\
$\delta C^{lt\left(\alpha,\beta\right)}_{\tau,3}=0.25$,
$\delta C^{lt\left(\alpha,\beta\right)}_{\tau,4}=0.125$,
$\delta C^{lt\left(\alpha,\beta\right)}_{\tau,5}=0.1$,\\
$\delta C^{lt\left(\alpha,\beta\right)}_{\tau,6}=0.05$,
$\delta C^{lt\left(\alpha,\beta\right)}_{\tau,7}=0.025$
\item Corrections:\\
$\delta C^{corr\left(\alpha,\beta\right)}_{\tau,n\in\mathbb{Z}_3}=0.25$, $\delta C^{corr\left(\alpha,\beta\right)}_{\tau,n=3,4,5}=0.125$,
$\delta C^{corr\left(\alpha,\beta\right)}_{\tau,n=6,7}=0.1$.
\end{itemize}
The validity of the entire Jacobi polynomial parameterization is guaranteed using shifted log-normally distributed priors to ensure $\alpha,\beta>-1$. In practice, the log-normal prior on beta is shifted to $\beta=0$ to secure $\beta>0$ and hence convergent PDFs at $x=1$.

 \begin{figure*}[b]
        \centering
        \subfigure[]{\includegraphics[width=0.49\textwidth]{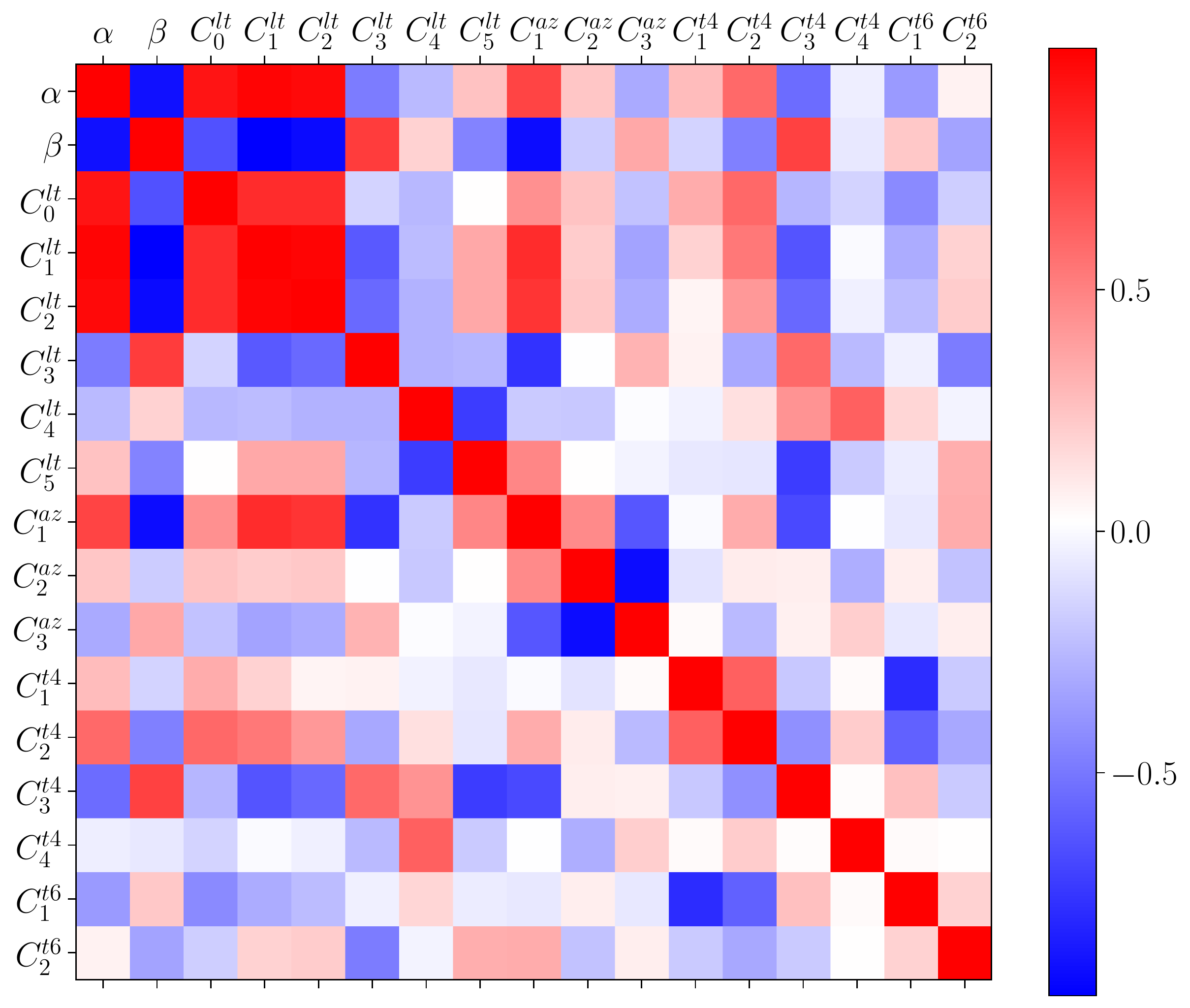}\label{fig:bigJacobiReal}}
        \subfigure[]{\includegraphics[width=0.49\textwidth]{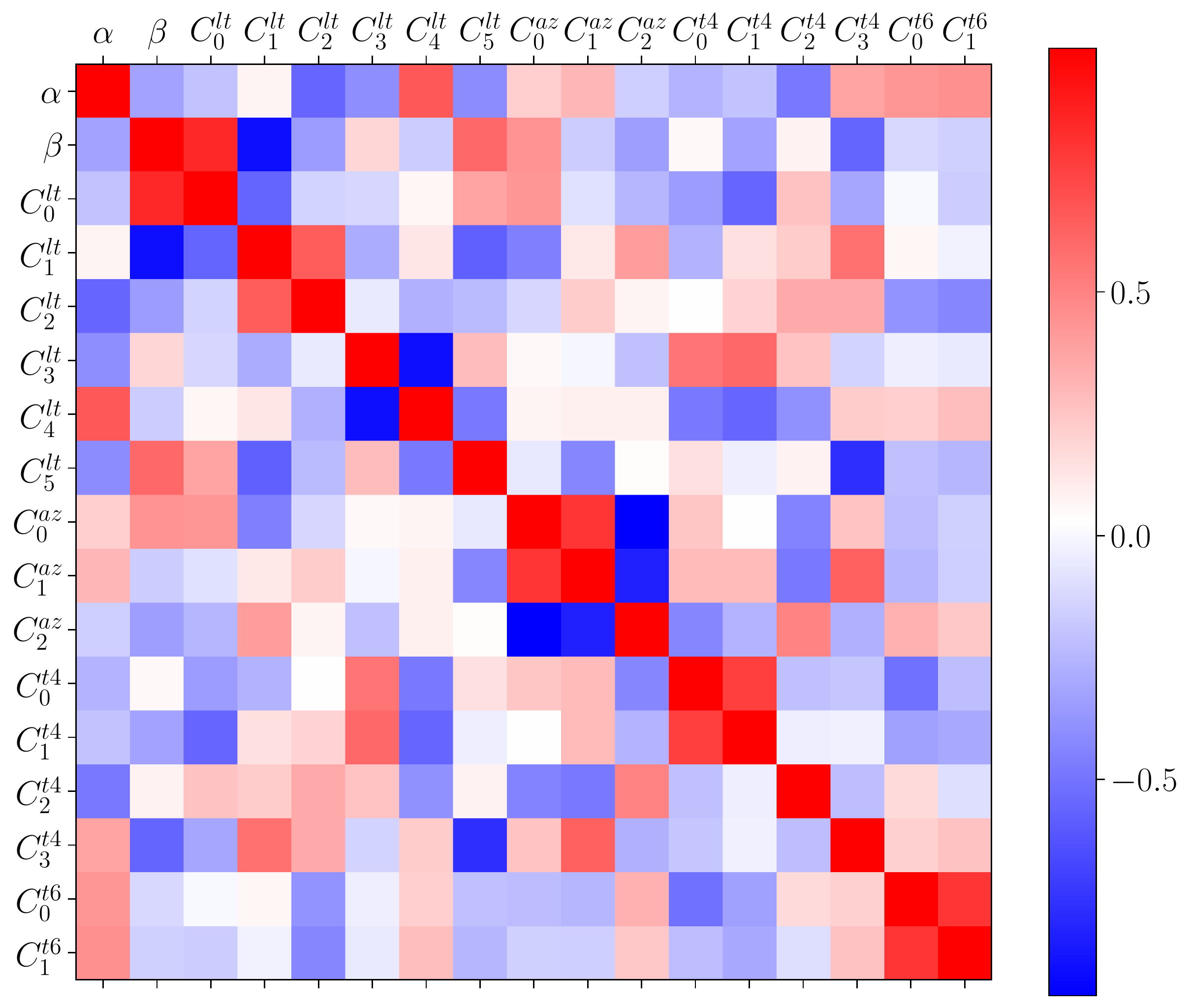}\label{fig:bigJacobiImag}}
        \caption{Parameter covariances in Jacobi polynomial fits with $\left[n_{lt},n_{az},n_{t4},n_{t6}\right]=\left[6342\right]$ to the real (a) and imaginary (b) components of the unpolarized reduced pseudo-ITD for $z/a\leq12$. Entries are normalized according to ${\rm Cov}_{ij}/\sqrt{{\rm Cov}_{ii}{\rm Cov}_{jj}}$.\label{fig:bigJacobiFits}}
    \end{figure*}

As leading-twist and correction terms are added, the question becomes at which order each series of Jacobi polynomials should be truncated. We address this by scanning over all possible combinations of truncation orders for $n_{lt}\in\lbrace3,4,5,6\rbrace$, $n_{az}\in\mathbb{Z}_4$, $n_{t4}\in\mathbb{Z}_5$, $n_{t6}\in\mathbb{Z}_3$, where $n_*$ are the orders of truncation in the fits~\eqref{eq:bigReJacobiFit} and~\eqref{eq:bigImJacobiFit}. Figure~\ref{fig:bigJacobiFits} illustrates the covariances of $\alpha,\beta$ and each linear correction term $C^{corr\left(\alpha,\beta\right)}_{\tau,n}$ in fits to the real~\eqref{eq:bigReJacobiFit} and imaginary~\eqref{eq:bigImJacobiFit} reduced pseudo-ITD components for Wilson line lengths $z/a\leq12$.
The covariance of each pair of fitted parameters is estimated via jackknife resampling
\be
    {\rm Cov}_{ij}\simeq\frac{N-1}{N}\sum_{n=1}^N\left(\mathfrak{f}_{n,i}-\bar{\mathfrak{f}}_i\right)\left(\mathfrak{f}_{n,j}-\bar{\mathfrak{f}}_j\right),
\ee
where fit parameters associated with jackknife sample $n$ are denoted by $\mathfrak{f}_{n,k}$, with jackknife average $\bar{\mathfrak{f}}_k$.
Without observing the quality of agreement between each fit and the reduced pseudo-ITD, it is clear several parameters correlate weakly or not at all with other parameters in the fit. This implies these weakly correlated parameters are not well-constrained by the data, and their removal will not affect the information content of the fit. For instance, the real component fit parameter covariances, shown in Fig.~\ref{fig:bigJacobiReal}, suggest the leading-twist expansion coefficients $C^{lt}_{n_{lt}}$ are constrained by the data for $n_{lt}\leq3$, while $C^{lt}_4,C^{lt}_5$ weakly correlate with the remaining parameters. The discretization, twist-$4$ and twist-$6$ corrections exhibit mild correlation for $C^{az}_1,C^{t4}_2,C^{t4}_3,C^{t6}_1$, with the remaining correction parameters largely unconstrained. In the imaginary component, the fit parameter covariances shown in Fig.~\ref{fig:bigJacobiImag} suggest a more nuanced pattern of correlation. Several leading-twist Jacobi polynomials appear to be well-constrained by the data, while the relative correlation between the $C^{az}$ and $C^{t4},C^{t6}$ parameters is increased relative to the corresponding entries in Fig.~\ref{fig:bigJacobiReal}.
\begin{table*}[b]
    \begin{center}
      \begin{tabular}{c|cccc|cc}
        $\lbrace n_{lt},n_{az},n_{t4},n_{t6}\rbrace_{{\rm v}/+}$ & $\lbrace4,1,3,2\rbrace_{\rm v}$ & $\lbrace4,0,3,2\rbrace_{\rm v}$ & $\lbrace3,3,1,0\rbrace_+$ & $\lbrace3,0,1,0\rbrace_+$ & $\lbrace6,3,4,2\rbrace_{\rm v}$ & $\lbrace6,3,4,2\rbrace_+$\\
        \hline
        $\alpha$ & $-0.209(147)$ & $-0.376(37)$ & $-0.328(20)$ & $-0.331(31)$ & $-0.264(117)$ & $-0.326(20)$ \\
        $\beta$ & $1.330(415)$ & $2.032(496)$ & $2.361(243)$ & $3.227(297)$ & $1.438(404)$ & $2.051(260)$ \\
        \hline
        $C^{lt}_{\tau,0}$ & $1.606(257)$ & $1.340(165)$ & $2.041(108)$ & $1.156(83)$ & $1.489(213)$ & $1.954(107)$ \\
        $C^{lt}_{\tau,1}$ & $0.427(752)$ & $0.335(261)$ & $0.123(248)$ & $0.161(243)$ & $0.174(620)$ & $0.404(213)$ \\
        $C^{lt}_{\tau,2}$ & $-0.880(409)$ & $-0.125(100)$ & $-0.464(121)$ & $0.700(98)$ & $-1.002(301)$ & $-0.241(118)$ \\
        $C^{lt}_{\tau,3}$ & $-0.675(122)$ & $-0.651(140)$ & $-$ & $-$ & $-0.568(118)$ & $-0.018(79)$ \\
        $C^{lt}_{\tau,4}$ & $-$ & $-$ & $-$ & $-$ & $0.089(28)$ & $0.020(27)$ \\
        $C^{lt}_{\tau,5}$ & $-$ & $-$ & $-$ & $-$ & $0.020(12)$ & $-0.023(10)$ \\
        \hline
        $C^{az}_{\tau,0}$ & $-$ & $-$ & $-0.001(43)$ & $-$ & $-$ & $0.054(35)$ \\
        $C^{az}_{\tau,1}$ & $-0.279(48)$ & $-$ & $-0.338(39)$ & $-$ & $-0.226(53)$ & $-0.219(46)$ \\
        $C^{az}_{\tau,2}$ & $-$ & $-$ & $0.434(74)$ & $-$ & $0.209(67)$ & $0.283(67)$ \\
        $C^{az}_{\tau,3}$ & $-$ & $-$ & $-$ & $-$ & $-0.164(48)$ & $-$ \\
        \hline
        $C^{t4}_{\tau,0}$ & $-$ & $-$ & $0.170(28)$ & $0.391(46)$ & $-$ & $0.185(47)$ \\
        $C^{t4}_{\tau,1}$ & $0.052(53)$ & $-0.090(52)$ & $-$ & $-$ & $0.060(50)$ & $0.032(68)$ \\
        $C^{t4}_{\tau,2}$ & $-0.371(106)$ & $-0.112(77)$ & $-$ & $-$ & $-0.341(93)$ & $-0.200(79)$ \\
        $C^{t4}_{\tau,3}$ & $-0.407(122)$ & $0.274(99)$ & $-$ & $-$ & $-0.397(131)$ & $0.076(29)$ \\
        $C^{t4}_{\tau,4}$ & $-$ & $-$ & $-$ & $-$ & $0.088(30)$ & $-$ \\
        \hline
        $C^{t6}_{\tau,0}$ & $-$ & $-$ & $-$ & $-$ & $-$ & $-0.067(34)$ \\
        $C^{t6}_{\tau,1}$ & $-0.045(37)$ & $0.011(39)$ & $-$ & $-$ & $-0.045(36)$ & $-0.079(53)$ \\
        $C^{t6}_{\tau,2}$ & $0.228(52)$ & $0.397(84)$ & $-$ & $-$ & $0.227(53)$ & $-$ \\
        \hline
        $\chi^2_r$ & $2.620(345)$ & $45.68(1.72)$ & $2.845(387)$ & $123.16(2.73)$ & $2.809(374)$ & $3.110(431)$ \\
        \hline
      \end{tabular}
  \end{center}
  \caption{Various Jacobi polynomial fits to the real and imaginary components of the unpolarized reduced pseudo-ITD for $z/a\leq12$. Each column represents distinct orders of truncation in the Jacobi polynomial expansions to the leading-twist, discretization, twist-$4$ and twist-$6$ corrections. The real and imaginary component fits were found to have the highest likelihoods of describing the data with truncation orders $\lbrace4,1,3,2\rbrace_{\rm v}$ and $\lbrace3,3,1,0\rbrace_+$, respectively. The dramatic effect even a single discretization term has on each fit is shown in the columns $\lbrace4,0,3,2\rbrace_{\rm v}$ and $\lbrace3,0,1,0\rbrace_+$.\label{tab:jacobiFitResults}}
\end{table*}

This exercise demonstrates an important point. Although the VarPro implementation of the Jacobi polynomial fits allows for arbitrarily many leading-twist and correction coefficients, the reduced pseudo-ITD data simply do not contain enough information to constrain so many parameters.
One should then expect the likelihood function is maximized for the real component of the reduced pseudo-ITD with truncation orders $n_{lt}\sim3$ and $n_{az}\sim1-2$, and $n_{lt}\sim3$ and $n_{az}\sim3$ for the imaginary component.

By scanning over the order of truncation for the leading-twist and correction terms parameterized by Jacobi polynomials,
we find the likelihood of the functional~\eqref{eq:bigReJacobiFit} to describe $\mathfrak{Re}\ \mathfrak{M}\left(\nu,z^2\right)$ with $z/a\leq12$ to be maximized for $\lbrace n_{lt},n_{az},n_{t4},n_{t6}\rbrace_{\rm v}=\lbrace4,1,3,2\rbrace_{\rm v}$. Likewise, the likelihood of the functional~\eqref{eq:bigImJacobiFit} to describe $\mathfrak{Im}\ \mathfrak{M}\left(\nu,z^2\right)$ with $z/a\leq12$ is maximized for $\lbrace n_{lt},n_{az},n_{t4},n_{t6}\rbrace_+=\lbrace3,3,1,0\rbrace_+$. The fit results for each and their respective figures-of-merit are given in Tab.~\ref{tab:jacobiFitResults}.

\begin{figure*}[t]
    \centering
    \includegraphics[width=\linewidth]{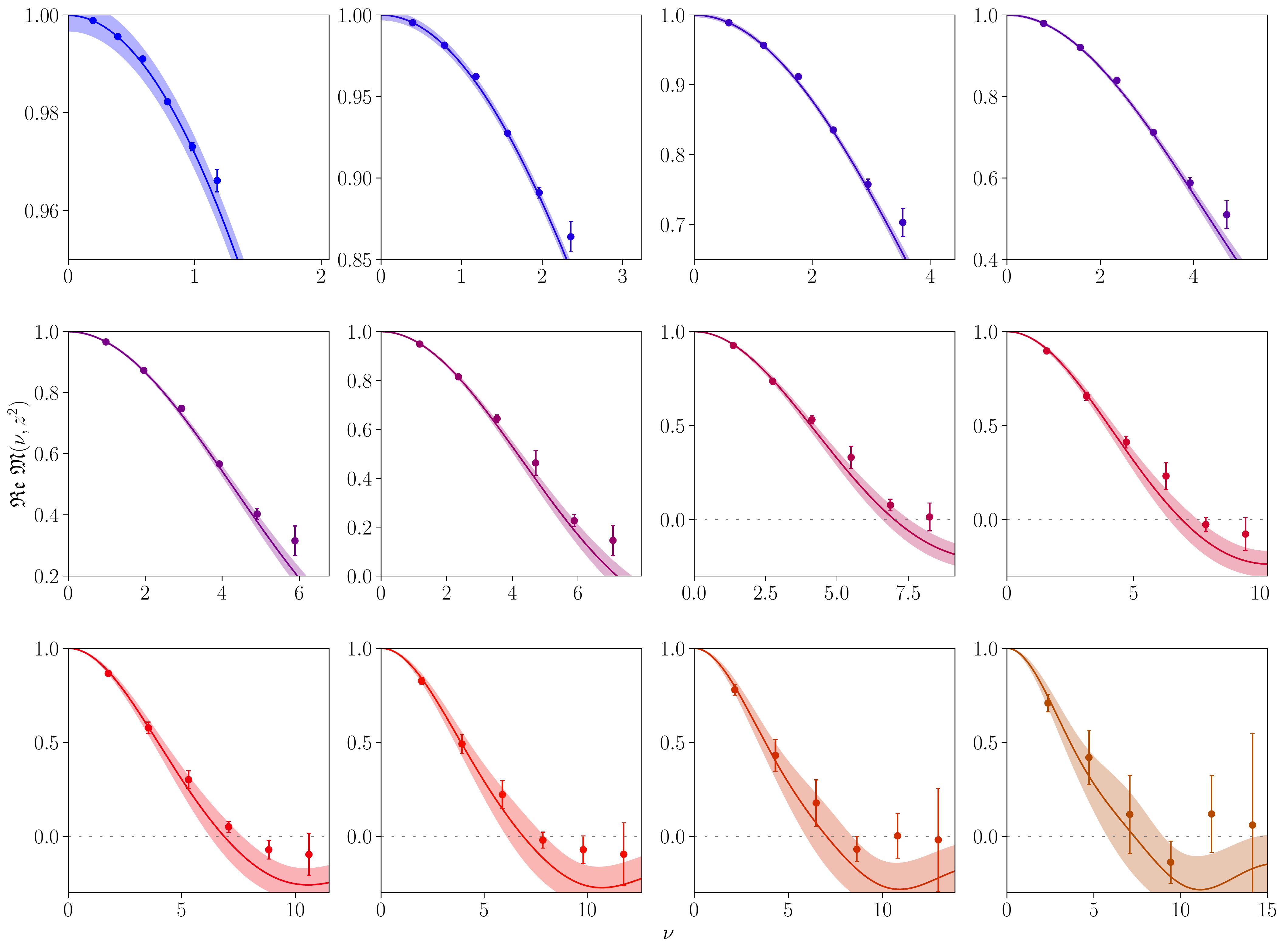}
    \caption{Fit to the real component of the unpolarized reduced pseudo-ITD where the leading-twist, discretization, twist-$4$, and twist-$6$ corrections have been expanded in Jacobi polynomials up to order $\lbrace n_{lt},n_{az},n_{t4},n_{t6}\rbrace_{\rm v}=\lbrace4,1,3,2\rbrace_{\rm v}$. Starting from the upper left panel and traversing horizontally, the leading-twist plus corrections are shown for each $z/a\leq12$.\label{fig:bestValencePITD}}
\end{figure*}

\begin{figure*}[t]
    \centering
    \includegraphics[width=\linewidth]{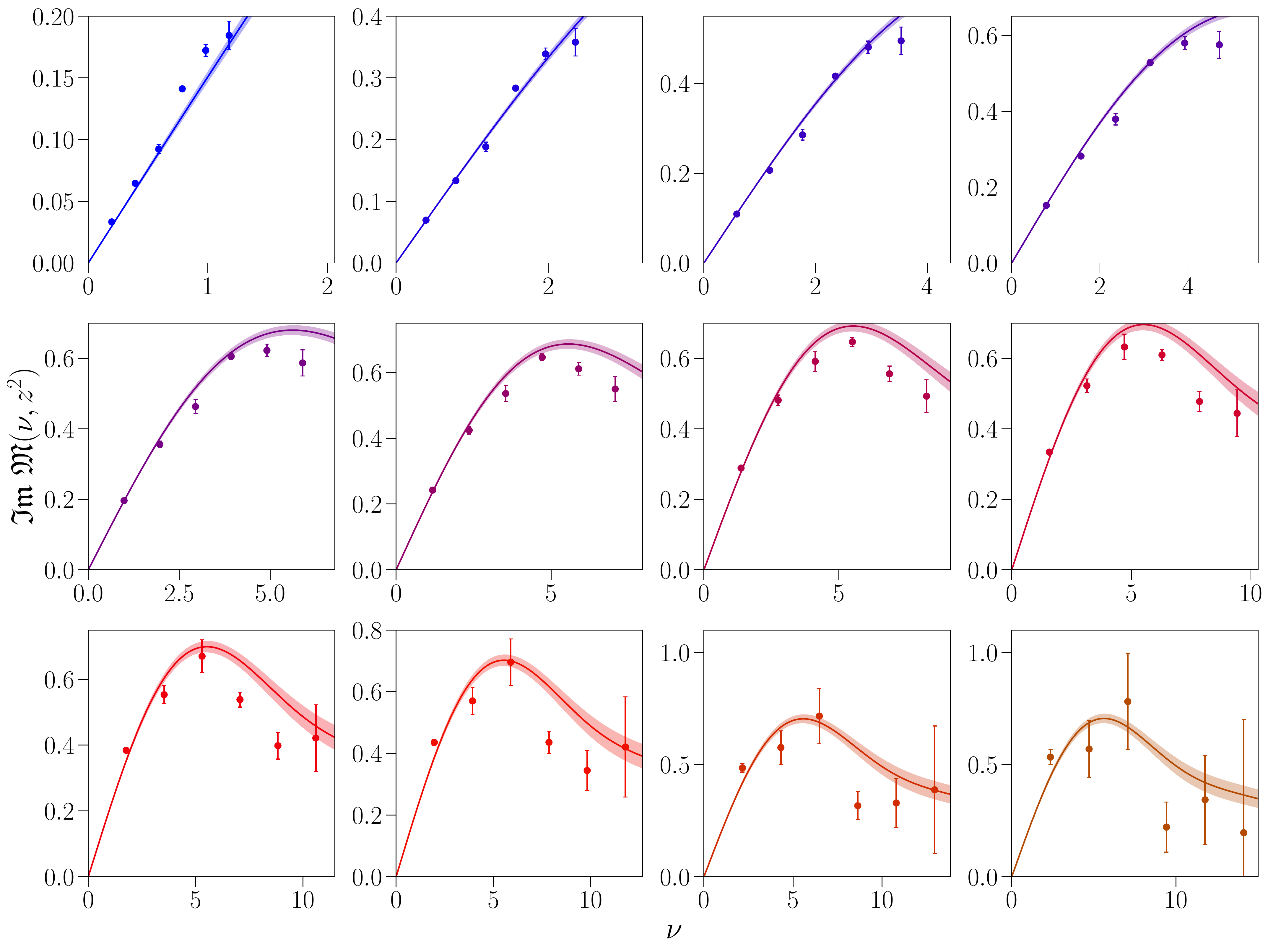}
    \caption{Fit to the imaginary component of the unpolarized reduced pseudo-ITD where the leading-twist, discretization, twist-$4$, and twist-$6$ corrections have been expanded in Jacobi polynomials up to order $\lbrace n_{lt},n_{az},n_{t4},n_{t6}\rbrace_+=\lbrace3,3,1,0\rbrace_+$. Starting from the upper left panel and traversing horizontally, the leading-twist plus corrections are shown for each $z/a\leq12$.\label{fig:bestPlusPITD}}
\end{figure*}

The Jacobi polynomial fits to order $\lbrace n_{lt},n_{az},n_{t4},n_{t6}\rbrace_{\rm v}=\lbrace4,1,3,2\rbrace_{\rm v}$ and
$\lbrace n_{lt},n_{az},n_{t4},n_{t6}\rbrace_+=\lbrace3,3,1,0\rbrace_+$ applied to the real and imaginary components of $\mathfrak{M}\left(\nu,z^2\right)$ are presented in Fig.~\ref{fig:bestValencePITD} and Fig.~\ref{fig:bestPlusPITD}, respectively.

\begin{figure*}[t!]
\subfigure[]{\label{fig:bestValenceHeatMap} \includegraphics[width=0.459\textwidth]{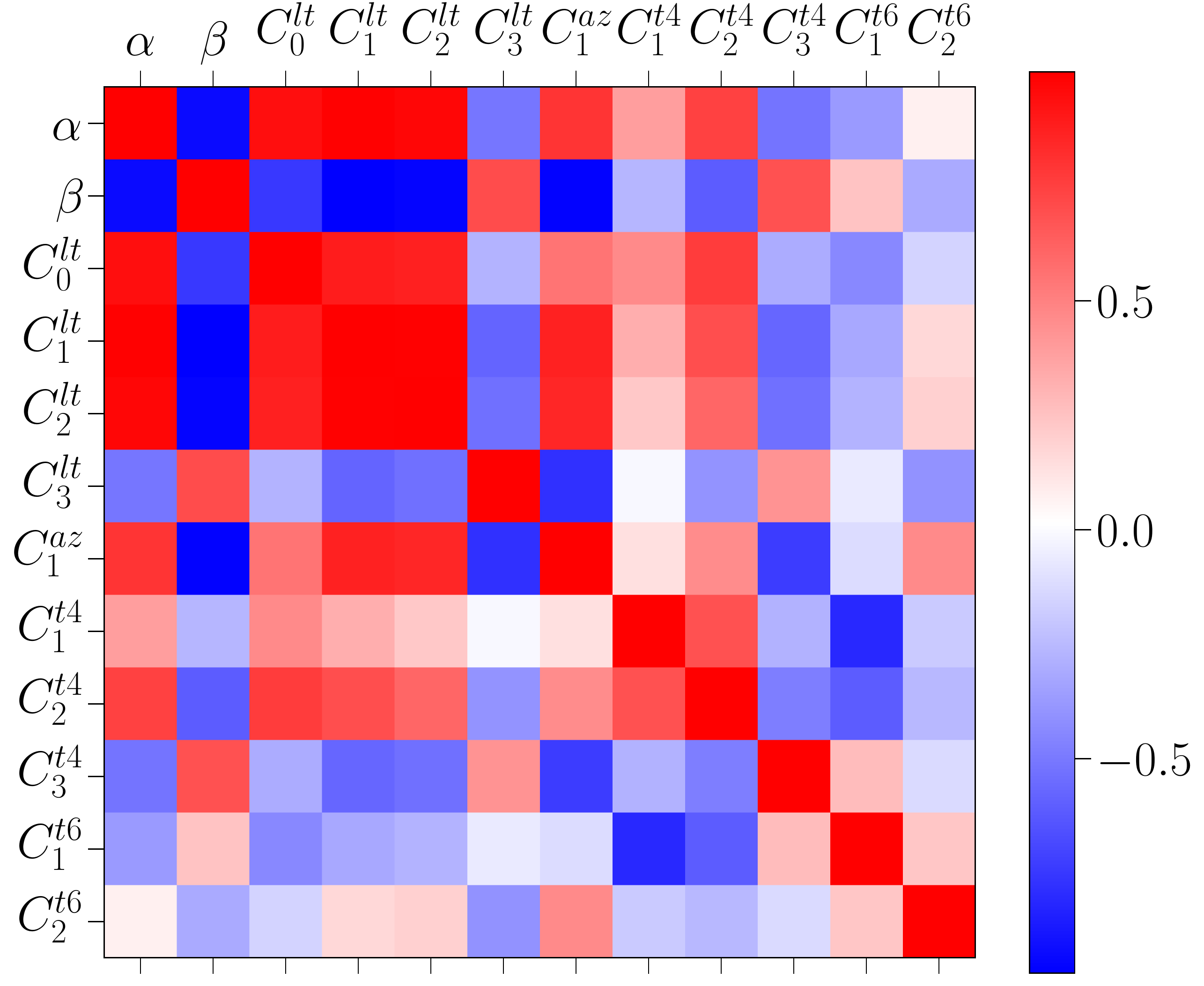}}
\hfill
\subfigure[]{\label{fig:bestPlusHeatMap} \includegraphics[width=0.459\textwidth]{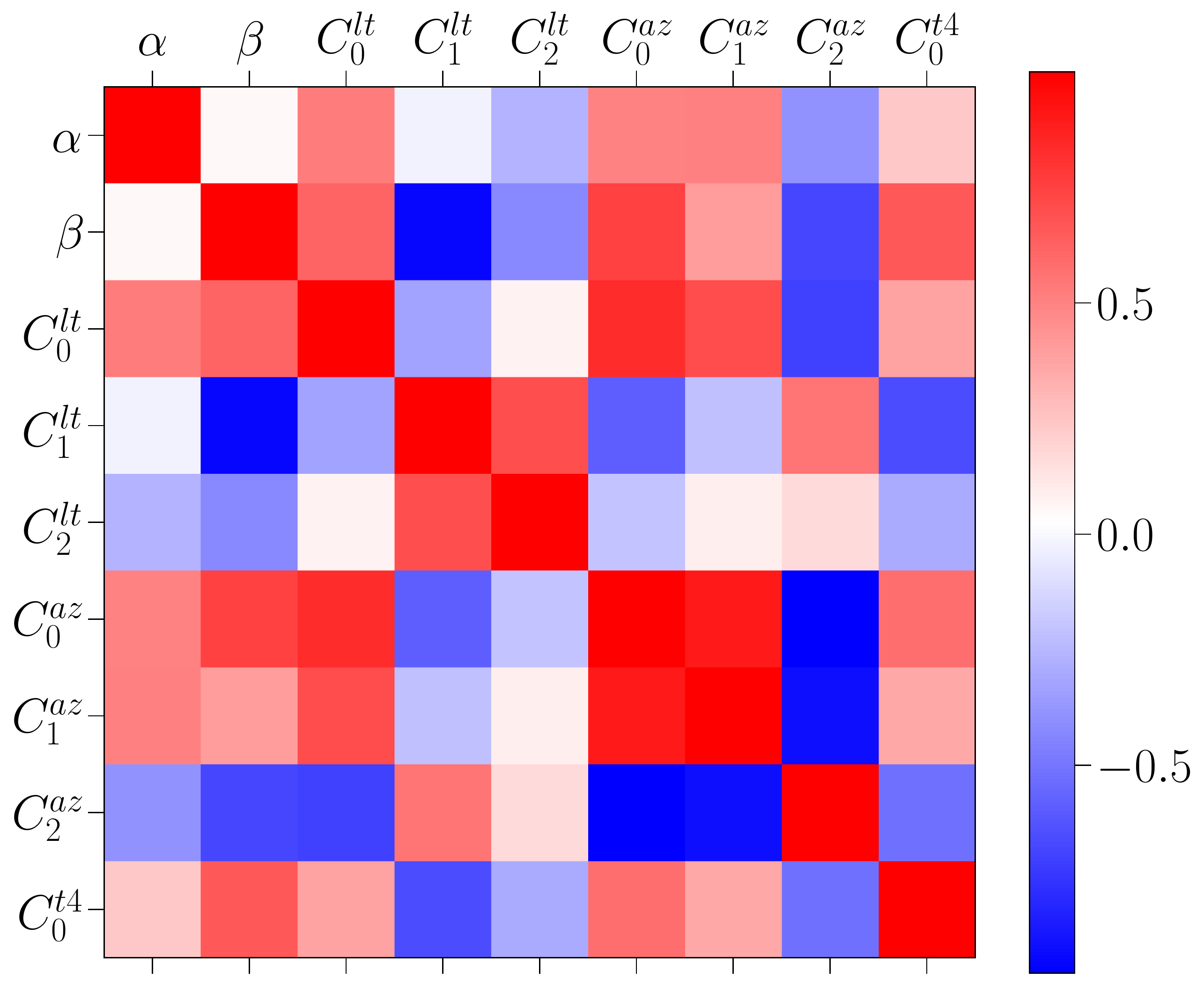}}
\caption{Parameter covariances of the optimal Jacobi polynomial fit to the real (\ref{fig:bestValenceHeatMap}) and the imaginary (\ref{fig:bestPlusHeatMap}) component  of the unpolarized reduced pseudo-ITD for $z/a\leq12$ with truncation orders $\lbrace n_{lt},n_{az},n_{t4},n_{t6}\rbrace_{\rm v}=\lbrace4,1,3,2\rbrace_{\rm v}$ and $\lbrace n_{lt},n_{az},n_{t4},n_{t6}\rbrace_+=\lbrace3,3,1,0\rbrace_+$. Entries are normalized according to ${\rm Cov}_{ij}/\sqrt{{\rm Cov}_{ii}{\rm Cov}_{jj}}$.}
\end{figure*}

Considering first the real component fit, each set of $\mathfrak{Re}\ \mathfrak{M}\left(\nu,z^2\right)$ for $z/a\leq8$ are well represented by the expansion in Jacobi polynomials. The main exception is the highest momentum point $ap_z=6\times\left(2\pi/L\right)\sim2.47\text{ GeV}$. The $\mathfrak{Re}\ \mathfrak{M}\left(\nu,z^2\right)$ data for $z/a>8$ are also reasonably well described, however the highest two momenta are seen to deviate. This behavior is not surprising despite the twist-$4$ and twist-$6$ corrections, which capture large-$z^2$ deviations, as the highest momentum data are subject to loss of signal in both the two- and three-point functions.
The associated fit parameter covariances shown in Fig.~\ref{fig:bestValenceHeatMap} demonstrate the leading-twist, discretization and twist-4 corrections are well constrained by the $\mathfrak{Re}\ \mathfrak{M}\left(\nu,z^2\right)$ data; as expected, the twist-6 corrections are only weakly constrained.


\begin{figure*}[t!]
\subfigure[]{\label{fig:jacobiBestRealITD} \includegraphics[width=0.48\textwidth]{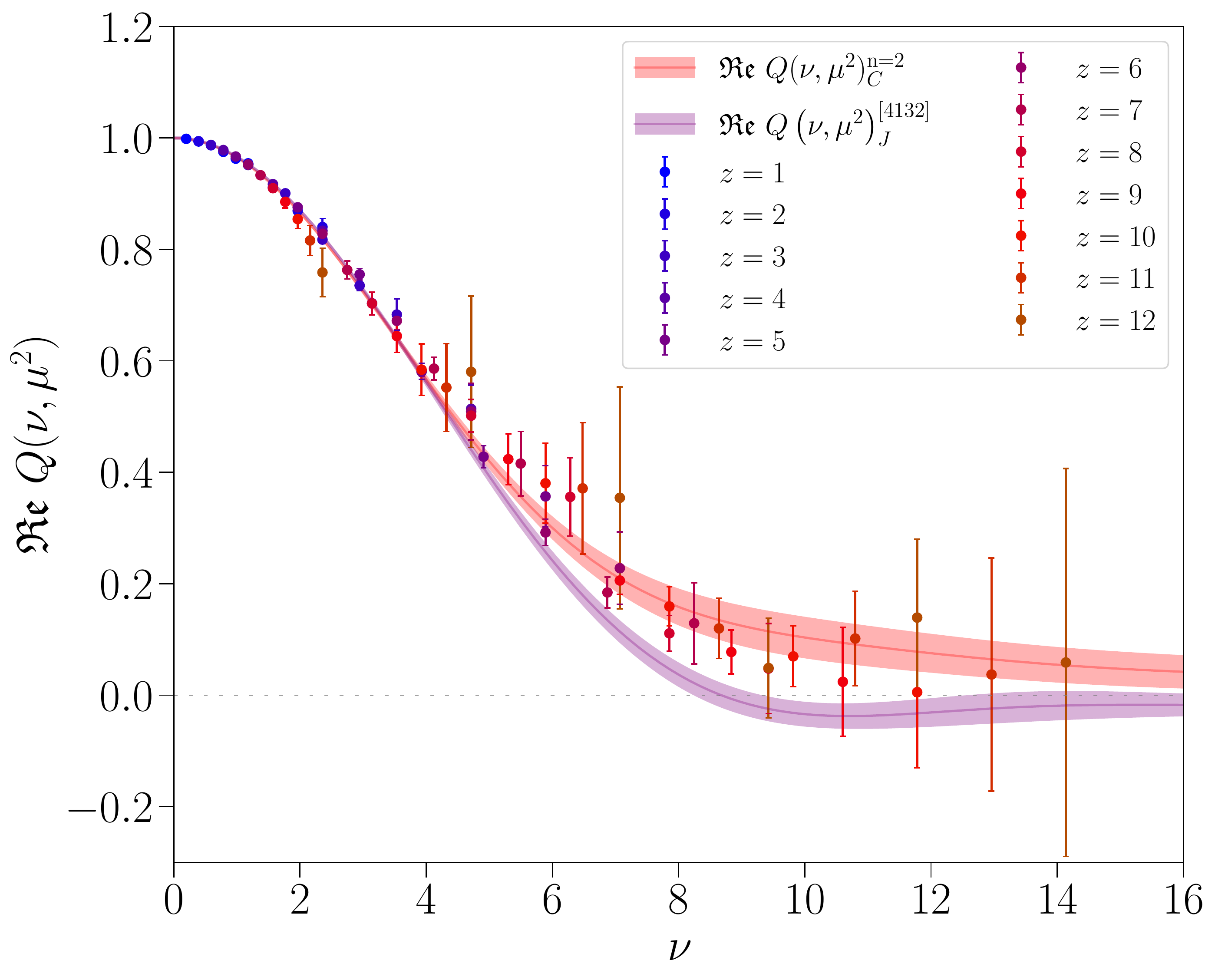}}
\hfill
\subfigure[]{\label{fig:jacobiBestValencePDF} \includegraphics[width=0.459\textwidth]{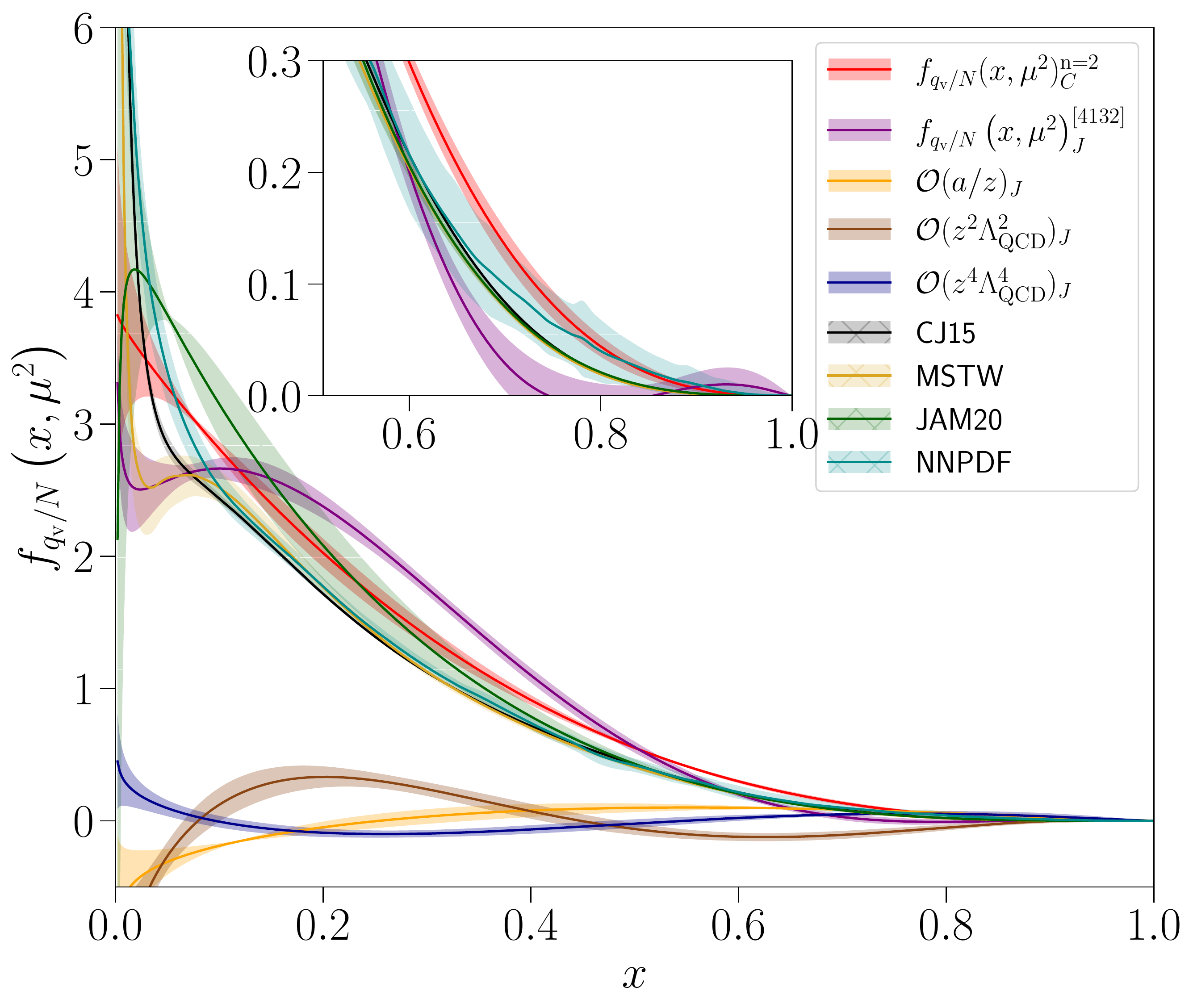}}
\caption{The leading-twist real ITD (purple) (\ref{fig:jacobiBestRealITD}) at $2\text{ GeV}$ derived from the Jacobi polynomial expansion of the reduced pseudo-ITD for $z/a\leq12$ with $\lbrace n_{lt},n_{az},n_{t4},n_{t6}\rbrace_{\rm v}=\lbrace4,1,3,2\rbrace_{\rm v}$. The result is compared with the uncorrelated $2$-parameter phenomenological form of Eq.~\ref{eq:jam-param_valence} shown in red. The valence quark leading-twist PDF (purple) (\ref{fig:jacobiBestValencePDF})  obtained from the $\lbrace n_{lt},n_{az},n_{t4},n_{t6}\rbrace_{\rm v}=\lbrace4,1,3,2\rbrace_{\rm v}$ Jacobi polynomial expansion of the reduced pseudo-ITD. The $a/z$ (orange), twist-$4$ (brown), and twist-$6$ (navy) $x$-space distributions are also shown and seen to be sub-leading. The distributions are compared with the uncorrelated $2$-parameter phenomenological fit of Eq.~\ref{eq:jam-param_valence} (red), as well as the NLO global analyses CJ15~\cite{Accardi:2016qay} and JAM20~\cite{Moffat:2021dji}, and the NNLO analyses of MSTW~\cite{Martin:2010db} and NNPDF~\cite{Ball:2017nwa} at the same scale.}
\end{figure*}

The resultant leading-twist $f_{q_{\rm v}/N}$ PDF and $x$-space distributions corresponding to the $\lbrace n_{lt},n_{az},n_{t4},n_{t6}\rbrace_{\rm v}=\lbrace4,1,3,2\rbrace_{\rm v}$ Jacobi polynomial fit are gathered in Fig.~\ref{fig:jacobiBestValencePDF}. As expected, the corrections in $x$-space are sub-leading to the leading-twist PDF. The Jacobi-parameterized leading-twist PDF, however, features many structural differences with the included phenomenological PDFs and the uncorrelated two-parameter PDF fit. Most evident is the softer approach to $x=1$.
Due to the valence quark sum rule, this enhances the low- to moderate-$x$ region and leads to further tension with the phenomenological results. By evaluating the cosine transform of the pure leading-twist component, we see in Fig.~\ref{fig:jacobiBestRealITD} that the $z/a\gtrsim7$ ITD data deviate successively further from the derived leading-twist ITD shown in purple. Whereas the uncorrelated two-parameter ITD fit shown in red attempts to capture all the $z/a\leq12$ data and indeed the unwanted impact from higher-twist effects, the Jacobi polynomial parameterization has effectively isolated and removed these polynomial-$z^2$ effects, leaving the pure leading-twist contribution.



The quality of the Jacobi polynomial fit to the imaginary component of the reduced pseudo-ITD shown in Fig.~\ref{fig:bestPlusPITD} is more puzzling. The $z/a\leq4$ appear reasonably well represented by the expansion in Jacobi polynomials, but by $z/a=5$ it is evident the data for a given $z^2$ segregate into two distinct groups - one for lattice momenta $p_{\rm latt}\in\lbrace1,2,3\rbrace$ and another for $p_{\rm latt}\in\lbrace4,5,6\rbrace$. This distinction coincides with the switch from an unphased eigenvector basis to the phased bases $\vec{\zeta}_\pm$ defined in Eq.~\ref{eq:phasedEvecs}.

\begin{figure*}[t!]
\subfigure[]{\label{fig:jacobiBestImagITD} \includegraphics[width=0.48\textwidth]{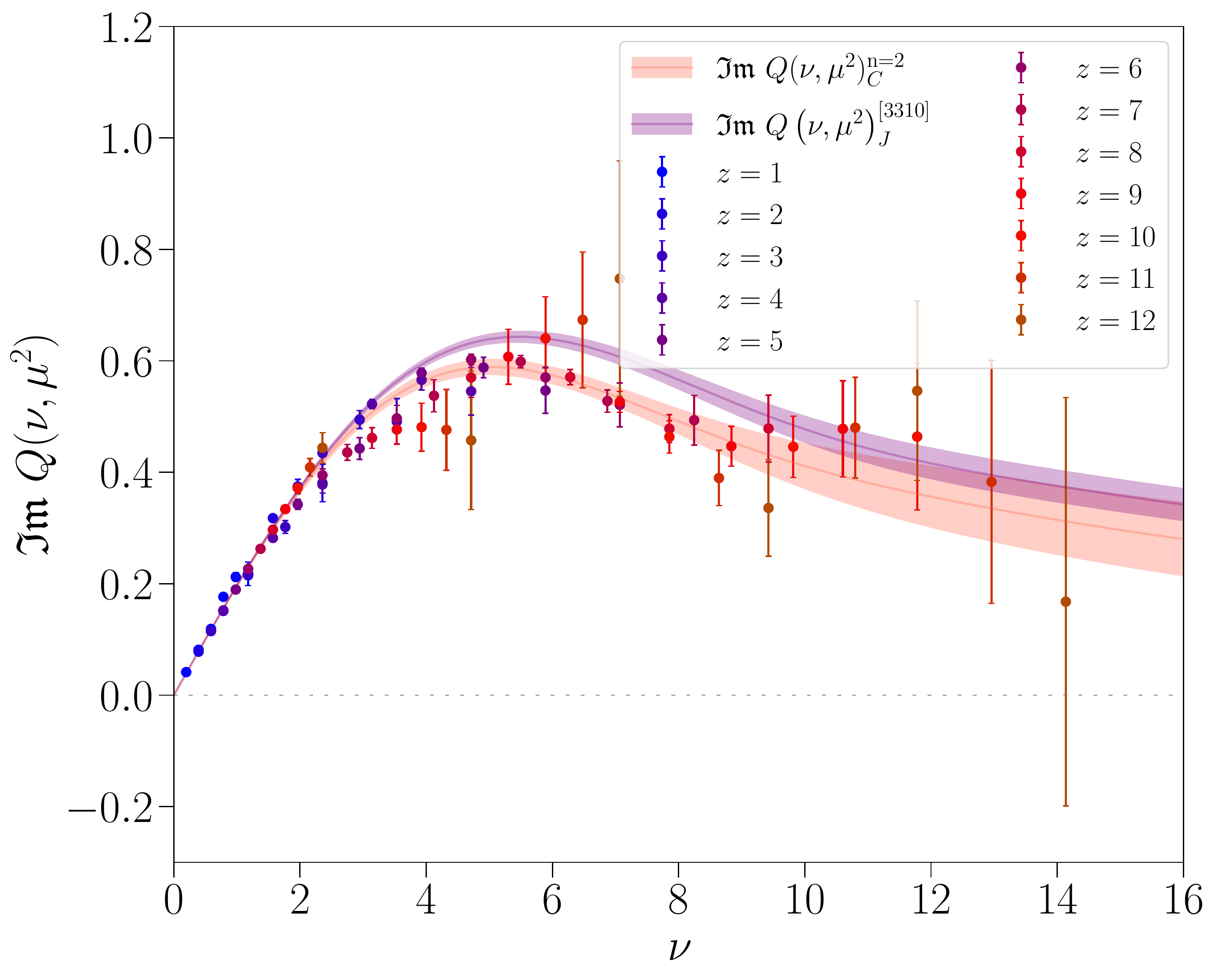}}
\hfill
\subfigure[]{\label{fig:jacobiBestPlusPDF} \includegraphics[width=0.459\textwidth]{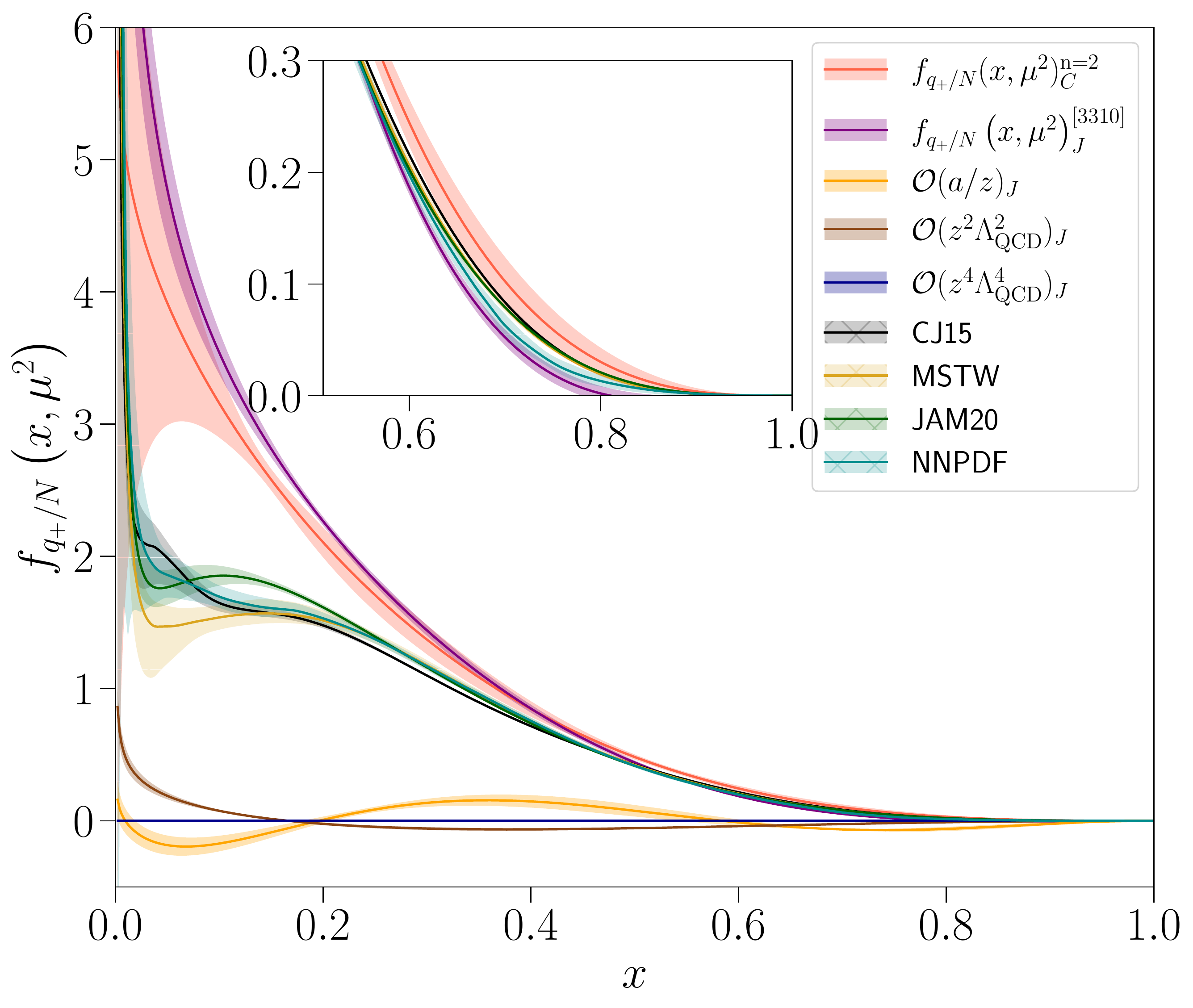}}
\caption{The plus quark leading-twist PDF (purple) in (\ref{fig:jacobiBestPlusPDF}) obtained from the $\lbrace n_{lt},n_{az},n_{t4},n_{t6}\rbrace_+=\lbrace3,3,1,0\rbrace_+$ Jacobi polynomial expansion of the reduced pseudo-ITD. The $a/z$ (orange) and twist-$4$ (brown) $x$-space distributions are also shown and seen to be sub-leading. The distributions are compared with the uncorrelated $2$-parameter phenomenological fit of Eq.~\ref{eq:jam-param_plus} (red), as well as the NLO global analyses CJ15~\cite{Accardi:2016qay} and JAM20~\cite{Moffat:2021dji}, and the NNLO analyses of MSTW~\cite{Martin:2010db} and
NNPDF~\cite{Ball:2017nwa} at the same scale. The imaginary component of the leading-twist ITD (purple) at $2\text{ GeV}$ (\ref{fig:jacobiBestImagITD}) derived from the Jacobi polynomial expansion of the reduced pseudo-ITD for $z/a\leq12$ with $\lbrace n_{lt},n_{az},n_{t4},n_{t6}\rbrace_+=\lbrace3,3,1,0\rbrace_+$. The result is compared with the uncorrelated $2$-parameter phenomenological form of Eq.~\ref{eq:jam-param_plus} shown in red.}
\end{figure*}

The fit parameter covariances shown in Fig.~\ref{fig:bestPlusHeatMap} demonstrate a milder constraint of the first and second order leading-twist Jacobi polynomials compared to the best fit of the real component. The discretization and twist-4 corrections are also seen to be well constrained by the data. The resultant leading-twist $f_{q_+/N}$ PDF and $x$-space distributions corresponding to the $\lbrace n_{lt},n_{az},n_{t4},n_{t6}\rbrace_+=\lbrace3,3,1,0\rbrace_+$ Jacobi polynomial fit are illustrated in Fig.~\ref{fig:jacobiBestPlusPDF}. As in the real component fit, the corrections are sub-leading to the leading-twist PDF which in this case are in agreement with the NNPDF result~\cite{Ball:2017nwa} for $x\geq0.5$. At small values of $x$, the leading-twist PDF parameterized by Jacobi polynomials is generally consistent with the two-parameter uncorrelated PDF fit. The sine transform of the pure leading-twist component is shown in Fig.~\ref{fig:jacobiBestImagITD} together with the $\mathfrak{Im}\ \mathcal{Q}\left(\nu,\mu^2\right)$ data at $2\text{ GeV}$. Unlike the real component of the derived leading-twist ITD in Fig.~\ref{fig:jacobiBestRealITD}, the derived imaginary component of the leading-twist ITD does not agree with the $\mathfrak{Im}\ \mathcal{Q}\left(\nu,\mu^2\right)$ data for any of the $ap_z\gtrsim4\pi/L$ data with $z/a\gtrsim7$. As the imaginary component of $\mathfrak{M}\left(\nu,z^2\right)$ is optimally fit with three discretization corrections and only one higher-twist term, the $\lbrace n_{lt},n_{az},n_{t4},n_{t6}\rbrace_+=\lbrace3,3,1,0\rbrace_+$ fit would suggest the imaginary component of the ITD is susceptible to less higher-twist effects in exchange for greater discretization effects. This is a tenuous conclusion, however, in light of the segregation of the $\mathfrak{Im}\ \mathfrak{M}\left(\nu,z^2\right)$ data into two distinct clusters, a low- and high-momentum set, for large Wilson line lengths. A future study exploring the side effects of phased distillation is warranted.

By far the biggest indicator of a reasonable description of the reduced pseudo-ITD data via Jacobi polynomials is a discretization term. Repeating the above Jacobi polynomial fits but leaving out any discretization corrections, namely $\lbrace n_{lt},n_{az},n_{t4},n_{t6}\rbrace_{\rm v}=\lbrace4,0,3,2\rbrace_{\rm v}$ and $\lbrace n_{lt},n_{az},n_{t4},n_{t6}\rbrace_+=\lbrace3,0,1,0\rbrace_+$, the correlated figures of merit once more inflate to unacceptable values (see Tab.~\ref{tab:jacobiFitResults}). This same conclusion is reached when cuts on momentum and Wilson line lengths are made. Since the discretization term we have included is of $\mathcal{O}\left(a/z\right)$, its effect is most pronounced at short distances. This is precisely the regime wherein the short distance factorization~\eqref{eq:ITD-factor-PITD}, or equivalently~\eqref{eq:pITD-PDF-direct-match}, is applicable. This motivates a more detailed look at the short-distance behavior of the computed reduced pseudo-ITD.

\section{On the Numerical Consistency with DGLAP}
The one-loop matching relationship between the ITD and the reduced pseudo-ITD~\eqref{eq:ITD-factor-PITD} implies that {$\mathcal{Q}\left(\nu,\mu^2\right)=\mathfrak{M}\left(\nu,z^2\right)$} at tree-level. The scatter that exists for a given $z^2$ should ideally be compensated at $\mathcal{O}\left(\alpha_s\right)$ by the \mbox{$\ln z^2$-dependence}  produced by the DGLAP evolution, up to large-$z^2$ higher-twist corrections.
In this section we study  the \mbox{$z^2$-dependence} of $\mathfrak{M}\left(\nu,z^2\right)$ more closely, and
investigate whether the observed dependence is numerically consistent with DGLAP, thus yielding a truly $z^2$-independent ITD.

We begin by focusing on the real component of the reduced pseudo-ITD. The dependence of $\mathfrak{Re}\ \mathfrak{M}\left(\nu,z^2\right)$ on the invariant space-like interval $z^2$ can be most easily visualized by parameterizing the valence pseudo-PDF $\mathcal{P}_{\rm v}\left(x,z^2;\alpha,\beta\right)$ by a simple two-parameter phenomenological form
\be
\mathcal{P}_{\rm v}\left(x,z^2;\alpha,\beta\right)=\frac{\Gamma\left(2+\alpha+\beta\right)}{\Gamma\left(1+\alpha\right)\Gamma\left(1+\beta\right)}x^\alpha\left(1-x\right)^\beta,
\label{eq:model-pPDF}
\ee
and fitting its cosine-transform to $\mathfrak{Re}\ \mathfrak{M}\left(\nu,z^2\right)$ separately for each $z^2$. In order to more readily expose the role of the Altarelli-Parisi kernel, we impose the added restriction $\beta=3$. This choice not only captures the naive $x\rightarrow1$ behavior of the nucleon's valence quark PDF~\cite{Brodsky:1973kr}, but also forces $\alpha$ to reflect any $z$-dependence in the reduced distribution; further, this value of $\beta$ is in statistical agreement with those obtained from the uncorrelated ITD fits (see Tab.~\ref{tab:uncorrelated-fitParams}).

\begin{figure*}[t]
  \centering
  \includegraphics[width=\linewidth]{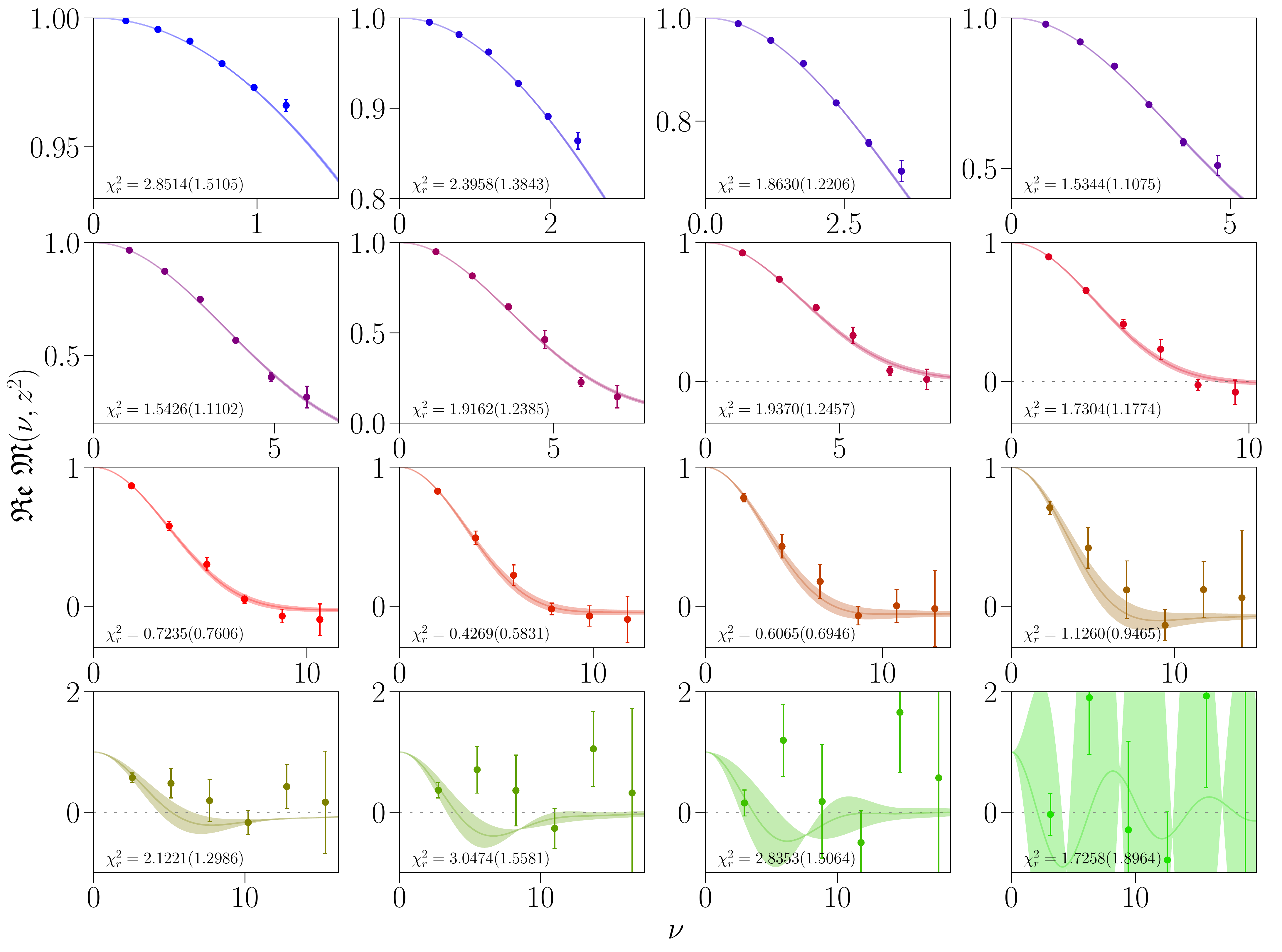}
  \caption{Cosine transform of the model pseudo-PDF in Eq.~\ref{eq:model-pPDF} fit separately to $\mathfrak{Re}\ \mathfrak{M}\left(\nu,z^2\right)$ for distinct $z^2$; data correlations have been included in each fit. Starting from the upper left panel and traversing horizontally, $z/a$ increases from unity. The correlated figure of merit for each separate fit is also indicated.\label{fig:real-pITD_pseudoPDF-fits}}
\end{figure*}

\begin{figure}[tb]
    \includegraphics[width=0.45\linewidth]{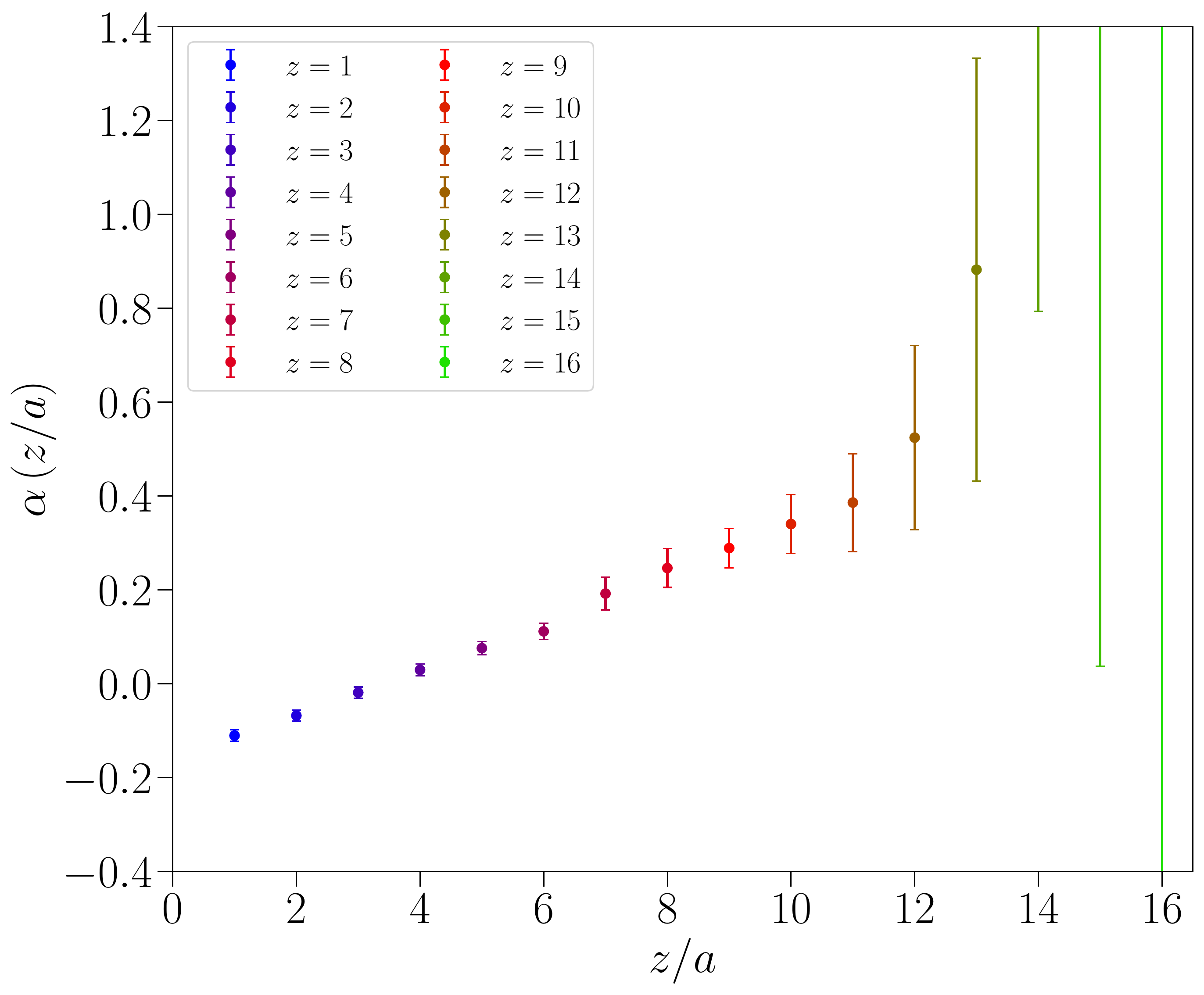}
  \caption{The fitted value of $\alpha$ as a function of $z/a$ resulting from the cosine-transform
  of the model pseudo-PDF in Eq.~\ref{eq:model-pPDF} fit to $\mathfrak{Re}\ \mathfrak{M}\left(\nu,z^2\right)$. The decrease of $\alpha$ with $z/a$ is in agreement with expectations from the Altarelli-Parisi evolution of the pseudo-PDF. This dependence is however clearly linear.
  \label{fig:pITD_pPDF-fits_alpha-vs-z}}
\end{figure}

Figure~\ref{fig:real-pITD_pseudoPDF-fits} illustrates the cosine-transform of the model valence pseudo-PDF~\eqref{eq:model-pPDF} fit separately to each $z^2$ of the real component of the reduced pseudo-ITD.
The cosine-transforms of $\mathcal{P}_{\rm v}\left(x,z^2;\alpha,3\right)$ are seen to describe $\mathfrak{Re}\ \mathfrak{M}\left(\nu,z^2\right)$ quite well for $z/a\lesssim10$, with the greatest tension seen for the highest momentum point for each separation. The fits for $z/a\geq13$ are also shown for completeness, but are clearly noise dominated. Also noteworthy, the highest figures-of-merit are observed for the smallest separations, with a somewhat monotonic reduction until $z/a\simeq11$.
The dependence of the fitted values of $\alpha$ on the separation $z/a$ is visualized for $\mathfrak{Re}\ \mathfrak{M}\left(\nu,z^2\right)$ in Fig.~\ref{fig:pITD_pPDF-fits_alpha-vs-z}. 
As a function of $z/a$, $\alpha$ decreases with the Wilson line length, matching expectations from the Altarelli-Parisi evolution of the pseudo-PDF. However, it is clear $\mathfrak{Re}\  \mathfrak{M}\left(\nu,z^2\right)$ depends linearly on $z/a$ for $z/a\lesssim12$, most notably for small-$z$.

This manifest lack of $\ln z^2$ behavior of $\mathfrak{Re}\ \mathfrak{M}\left(\nu,z^2\right)$ at short distances immediately suggests tension with the presumed DGLAP evolution of the pseudo-PDF. To determine if this $z^2$-dependence in $\mathfrak{Re}\ \mathfrak{M}\left(\nu,z^2\right)$ is nevertheless numerically consistent with DGLAP, the one-loop matching relationship between the reduced pseudo-ITD and ITD is applied. In the ideal scenario where the $z^2$-dependence of $\mathfrak{Re}\ \mathfrak{M}\left(\nu,z^2\right)$ is exactly described by DGLAP, the matched ITD will be independent of the interval $z^2$ up to polynomial corrections for large-$z^2$. Rather than perform the matching step to a common scale in $\msbar$ using a smooth polynomial in Ioffe-time (e.g. Eq.~\ref{eq:polyFits}) as was done in Sec.~\ref{sec:results}, we leverage the cosine-transform of the model pseudo-PDF~\eqref{eq:model-pPDF} as the smooth and continuous description of the reduced pseudo-ITD data. That is, we perform the matching of $\mathfrak{Re}\ \mathfrak{M}\left(\nu,z^2\right)$ to a common scale in $\msbar$ according to
\be
\mathfrak{Re}\ \mathcal{Q}\left(\nu,\mu^2\right)=\mathfrak{Re}\ \mathfrak{M}\left(\nu,z^2\right)+\frac{\alpha_sC_F}{2\pi}\int_0^1du\ \mathfrak{P}\left(u\nu,z^2;\alpha,\beta=3\right)
\left[\ln\left(\frac{e^{2\gamma_E+1}}{4}z^2\mu^2\right)B\left(u\right)+L\left(u\right)\right]
\label{eq:pitd-itd_match_w_pPDF-fits},
\ee
where $\mathfrak{P}\left(u\nu,z^2;\alpha,\beta=3\right)$ is the cosine-transform of the model pseudo-PDF $\mathcal{P}_{\rm v}\left(x,z^2;\alpha,\beta=3\right)$ expressed in a closed form 
by a generalized hypergeometric function
\be
\mathfrak{P}\left(\nu,z^2;\alpha,\beta=3\right)=\thinspace_2F_3\left(\frac{1+\alpha}{2},\frac{2+\alpha}{2};\frac{1}{2},\frac{5+\alpha}{2},\frac{6+\alpha}{2};-\frac{\nu^2}{4}\right).
\label{eq:pseudoPDF-2F3}
\ee
For an explicit, albeit crude, conversion to $\msbar$, we set $\alpha=0.2$ in Eq.~\ref{eq:pseudoPDF-2F3}.

\begin{figure*}[tb]
    \centering
    \includegraphics[width=.925\linewidth]{{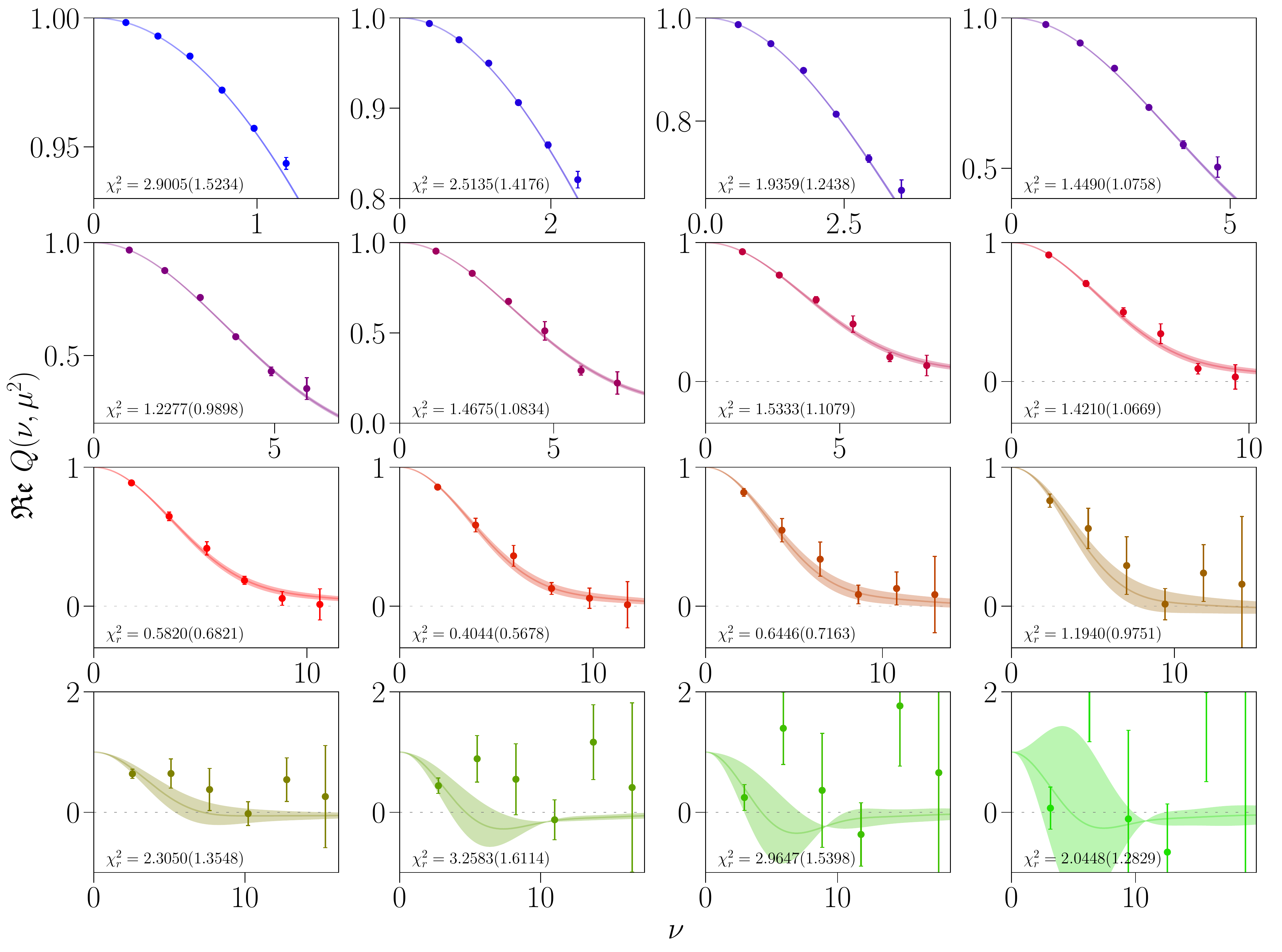}}
    \caption{Cosine transform of the two-parameter model PDF, with the same functional form as~\eqref{eq:model-pPDF}, fit separately to each $z^2$ of the matched ITD. The ITD was obtained using~\eqref{eq:pseudoPDF-2F3} for the evolution/matching step. Data correlations have been included in each fit. Starting from the upper left panel and traversing horizontally, $z/a$ increases from unity. The correlated figure of merit for each separate fit is also indicated.\label{fig:real-ITD_PDF-fits_dglap_check}}
\end{figure*}

\begin{figure}[bt]
    \centering
    \includegraphics[width=.45\linewidth]{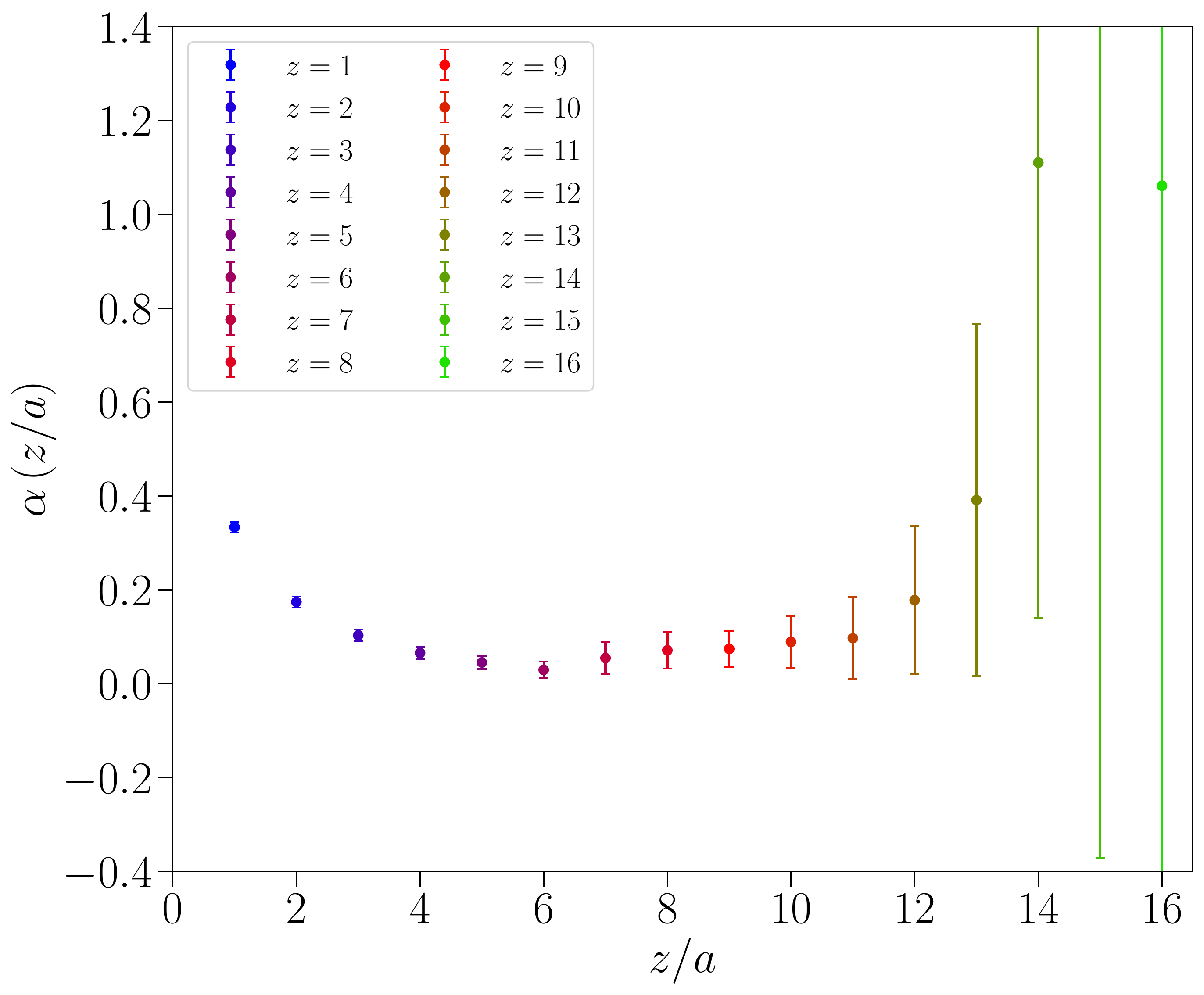}
    \caption{The fitted values of $\alpha$ from the cosine-transform of the two-parameter PDF functional form~\eqref{eq:model-pPDF} fit to each $z^2$ of the matched ITD. The latter was obtained using~\eqref{eq:pseudoPDF-2F3} for the evolution/matching step. The values of $\alpha$ are statistically constant for $4\lesssim z/a\lesssim11$, with sharp deviations for small-$z/a$.\label{fig:real-ITD_PDF-fits_alpha-vs-z_dglap_check}}
\end{figure}

Our strategy to expose any $z$-dependence in the ITD $\mathcal{Q}\left(\nu,\mu^2\right)$ remains identical to the reduced distribution above. The resultant matched ITD at $2\text{ GeV}$ in $\msbar$ is once more fit using the two-parameter form in Eq.~\ref{eq:model-pPDF} independently for each $z^2$ and with $\beta=3$. The parameterized distribution in this case is of course no longer the valence pseudo-PDF, but rather the valence PDF itself. As illustrated in Fig.~\ref{fig:real-ITD_PDF-fits_dglap_check}, each $z^2$ of the matched ITD is well described by the simple two-parameter form. The poorest figures-of-merit are again observed for the smallest ($z/a\lesssim3$) and largest ($z/a\gtrsim13$) separations.
The dependence of the fitted values of $\alpha$ on the separation $z/a$ for $\mathfrak{Re}\ \mathcal{Q}\left(\nu,\mu^2\right)$ is illustrated in Fig.~\ref{fig:real-ITD_PDF-fits_alpha-vs-z_dglap_check}. For $4\lesssim z/a\lesssim11$ the fitted value of $\alpha$ is observed to be independent of $z/a$ and hence numerically consistent with DGLAP in said interval. Remarkably, however, the values of $\alpha$ for the shortest separations, namely $z/a\lesssim4$, deviate increasingly from this constancy as $z/a\rightarrow1$. A subsequent analysis of the imaginary component of both the reduced pseudo-ITD and matched ITD arrived at a similar conclusion, but has been omitted for brevity.


\subsection{Jacobi Polynomial Corrections - Discretization Effects}
The findings above rigorously demonstrate the reduced pseudo-ITD is numerically inconsistent with DGLAP in the small-$z$ regime. Whether matching the reduced pseudo-ITD to the light-cone ITD or directly to the light-cone PDF, the presence of the Altarelli-Parisi evolution kernel should in principle capture and remove the $\ln z^2$ scatter that theoretically exists in $\mathfrak{M}\left(\nu,z^2\right)$ for small-$z$. As $\mathfrak{M}\left(\nu,z^2\right)$ was found to depend only linearly on the separation $z$ (Fig.~\ref{fig:pITD_pPDF-fits_alpha-vs-z}), the Altarelli-Parisi kernel effectively introduces a $\ln z^2$-dependence into the small-$z$ ITD and thus explains the dependence of $\alpha$ on $z/a$ in Fig.~\ref{fig:real-ITD_PDF-fits_alpha-vs-z_dglap_check}. Despite this concerning conclusion, a broad subset of $\mathfrak{M}\left(\nu,z^2\right)$ remains consistent with DGLAP: The statistically constant value of $\alpha\left(z/a\right)$ observed in the ITD fits in the interval $4\lesssim z/a\lesssim11$ (e.g. Fig.~\ref{fig:real-ITD_PDF-fits_alpha-vs-z_dglap_check}) validates the nice collapse of the $\mathfrak{M}\left(\nu,z^2\right)$ data onto a common curve (Fig.~\ref{fig:realITD}) when matched to a common scale in $\msbar$.

\begin{figure}[bt]
    \centering
    \includegraphics[width=0.45\linewidth]{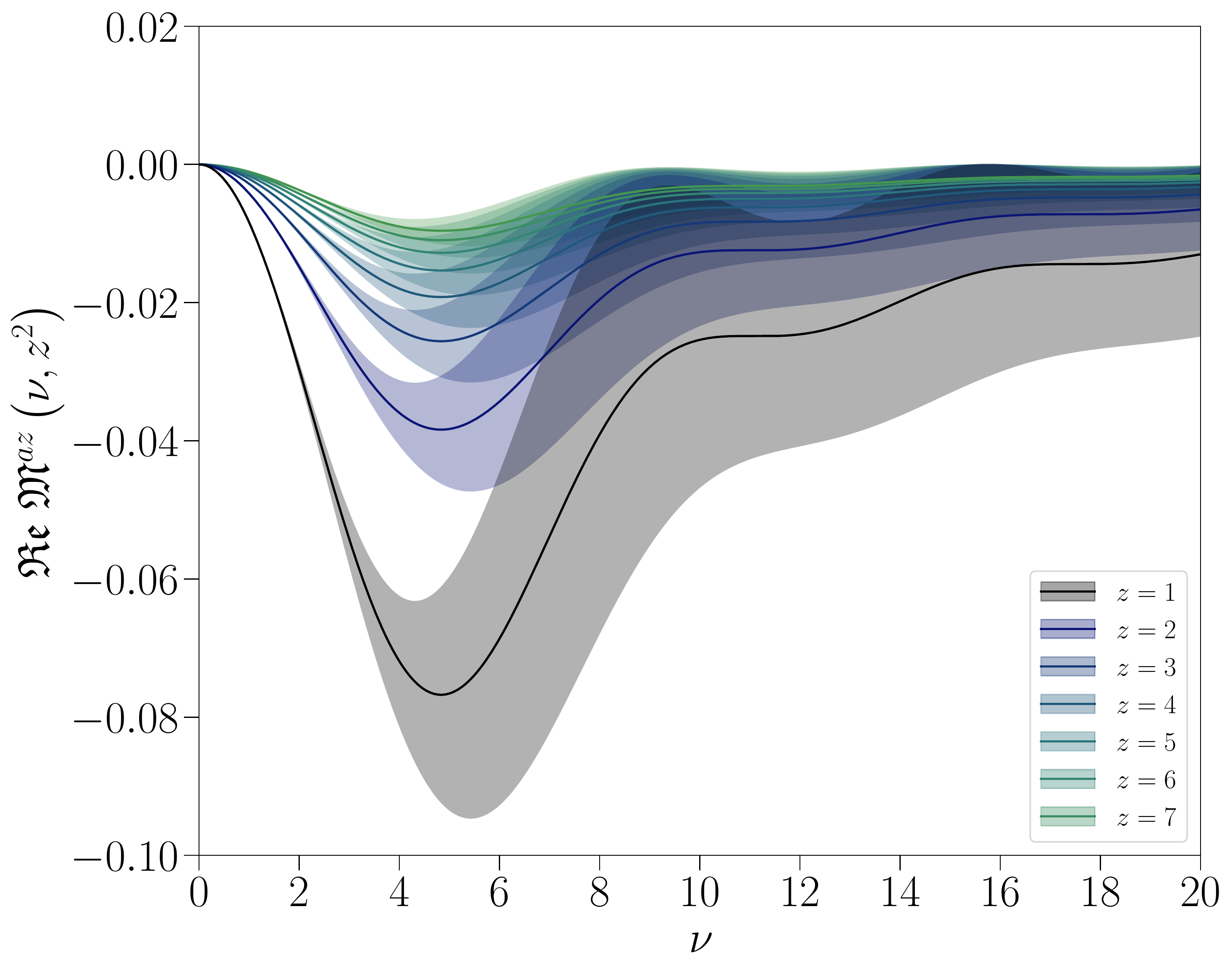}
    \includegraphics[width=0.45\linewidth]{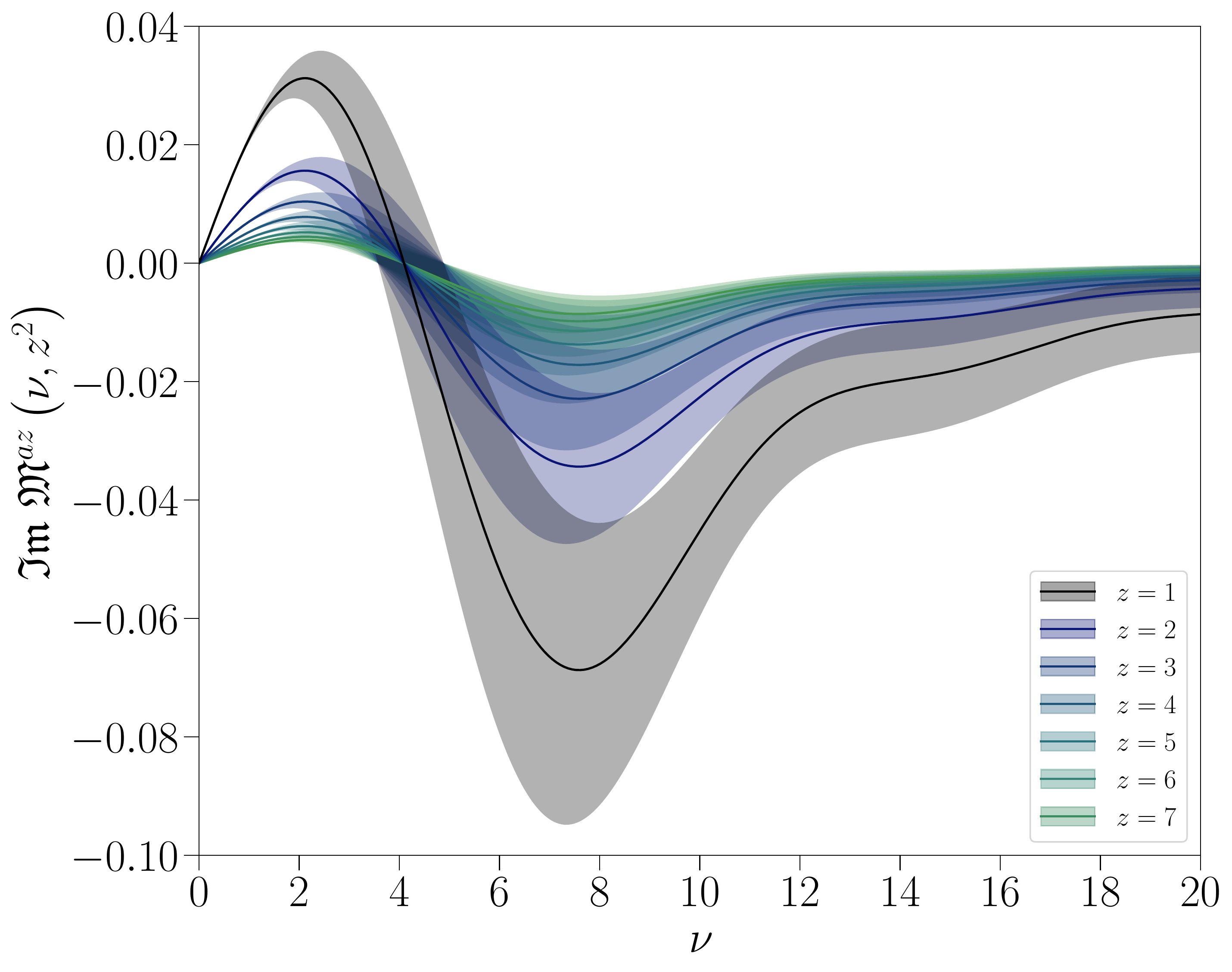}
    \caption{Visualization of the discretization effects determined by the optimal Jacobi polynomial fits $\lbrace n_{lt},n_{az},n_{t4},n_{t6}\rbrace_{\rm v}=\lbrace4,1,3,2\rbrace_{\rm v}$ (left) and $\lbrace n_{lt},n_{az},n_{t4},n_{t6}\rbrace_+=\lbrace3,3,1,0\rbrace_+$ (right) for $z/a\leq7$.}
    \label{fig:jac-corrections}
\end{figure}

To gain further insight into the regions wherein DGLAP is not respected, we return to the optimal Jacobi polynomial fits $\lbrace n_{lt},n_{az},n_{t4},n_{t6}\rbrace_{\rm v}=\lbrace4,1,3,2\rbrace_{\rm v}$ and $\lbrace n_{lt},n_{az},n_{t4},n_{t6}\rbrace_+=\lbrace3,3,1,0\rbrace_+$. The reader is reminded Sec.~\ref{sec:results} concluded with the realization that a suitable description of $\mathfrak{M}\left(\nu,z^2\right)$ was only possible with the nominal inclusion of an $\mathcal{O}\left(a/z\right)$ correction in Eq.~\ref{eq:bigReJacobiFit} and Eq.~\ref{eq:bigImJacobiFit}. The discretization effect parameterized by each of these fits are given by
\be
\mathfrak{M}^{az}\left(\nu,z^2\right)=\frac{a}{z}\times
\begin{cases}
  C_{{\rm v},1}^{az\ \left(\alpha,\beta\right)}\sigma_{0,1}^{\left(\alpha,\beta\right)}\left(\nu\right)&\quad\text{for }\lbrace4,1,3,2\rbrace_{\rm v} \\
  \sum_{n=0}^2C_{+,n}^{az\ \left(\alpha,\beta\right)}\eta_{0,n}^{\left(\alpha,\beta\right)}\left(\nu\right)&\quad\text{for }\lbrace3,3,1,0\rbrace_+
\end{cases}
\label{eq:discretization-pieces},
\ee
and visualized in Fig.~\ref{fig:jac-corrections}. The discretization effect $\mathfrak{Re}\ \mathfrak{M}^{az}\left(\nu,z^2\right)$ is seen to be strictly negative in the interval of Ioffe-time in which $\mathfrak{M}\left(\nu,z^2\right)$ has been computed. By comparison, the discretization effect $\mathfrak{Im}\ \mathfrak{M}^{az}\left(\nu,z^2\right)$ involves three Jacobi polynomials and suggests the $\mathfrak{Im}\ \mathfrak{M}\left(\nu,z^2\right)$ data are subject to a discretization effect that is opposite in sign at small and large values of Ioffe-time.

\begin{figure}[bt]
  \centering
  \includegraphics[width=0.55\linewidth]{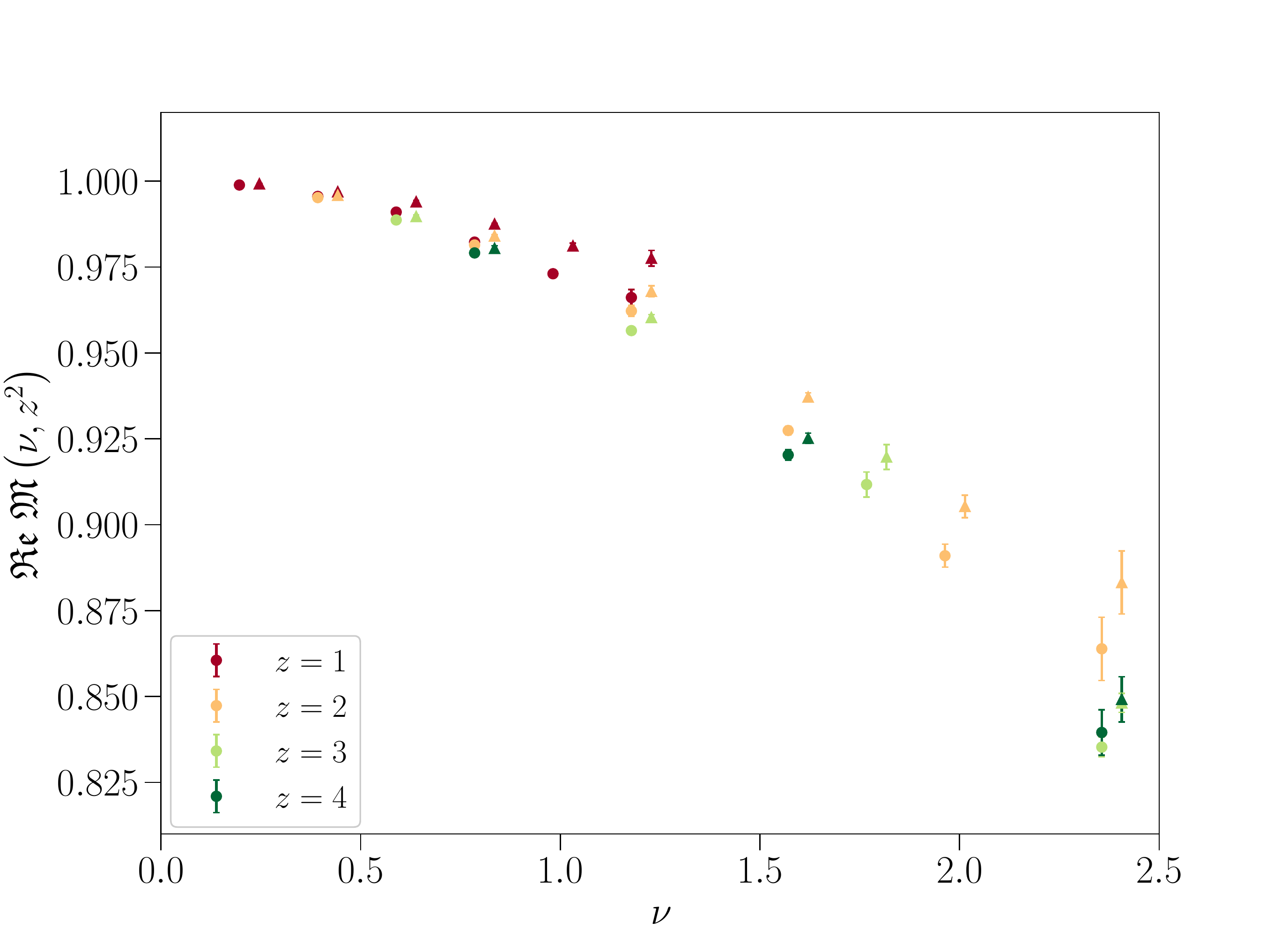}
  \caption{Juxtaposition of the raw $\mathfrak{Re}\ \mathfrak{M}\left(\nu,z^2\right)$ and discretization corrected $\mathfrak{Re}\ \mathfrak{M}\thinspace'\!\left(\nu,z^2\right)$ distributions, represented by circles and wedges respectively. The $\mathfrak{Re}\ \mathfrak{M}\thinspace'\!\left(\nu,z^2\right)$ data is shifted horizontally for legibility.\label{fig:raw_disc-correct_compare}}
\end{figure}

We now justify the necessity of the $\mathcal{O}\left(a/z\right)$ discretization correction by considering the removal of the $\mathfrak{Re}\ \mathfrak{M}^{az}\left(\nu,z^2\right)$ effect from the computed $\mathfrak{Re}\ \mathfrak{M}\left(\nu,z^2\right)$ data, which we denote $\mathfrak{Re}\ \mathfrak{M}\thinspace'\!\left(\nu,z^2\right)\equiv\mathfrak{Re}\ \mathfrak{M}\left(\nu,z^2\right)-\mathfrak{Re}\ \mathfrak{M}^{az}\left(\nu,z^2\right)$. Based on the left panel of Fig.~\ref{fig:jac-corrections}, the removal should shift the small-$z$ points of $\mathfrak{Re}\ \mathfrak{M}\left(\nu,z^2\right)$ to larger values, with the largest impact for $\nu\sim4.5$. Figure~\ref{fig:raw_disc-correct_compare} juxtaposes the original $\mathfrak{Re}\ \mathfrak{M}\left(\nu,z^2\right)$ and discretization corrected $\mathfrak{Re}\ \mathfrak{M}\thinspace'\!\left(\nu,z^2\right)$ in the interval $\nu\in\left[0,2.5\right]$. Although the differences are numerically small, at small Ioffe-times $\mathfrak{Re}\ \mathfrak{M}\thinspace'\!\left(\nu,z^2\right)$ is noticeably larger than the uncorrected reduced pseudo-ITD. The importance of removing this discretization effect is quantitatively discerned by repeating the DGLAP investigation for $\mathfrak{Re}\ \mathfrak{M}\thinspace'\!\left(\nu,z^2\right)$.

\begin{figure}[tb]
    \centering
    \includegraphics[width=0.45\linewidth]{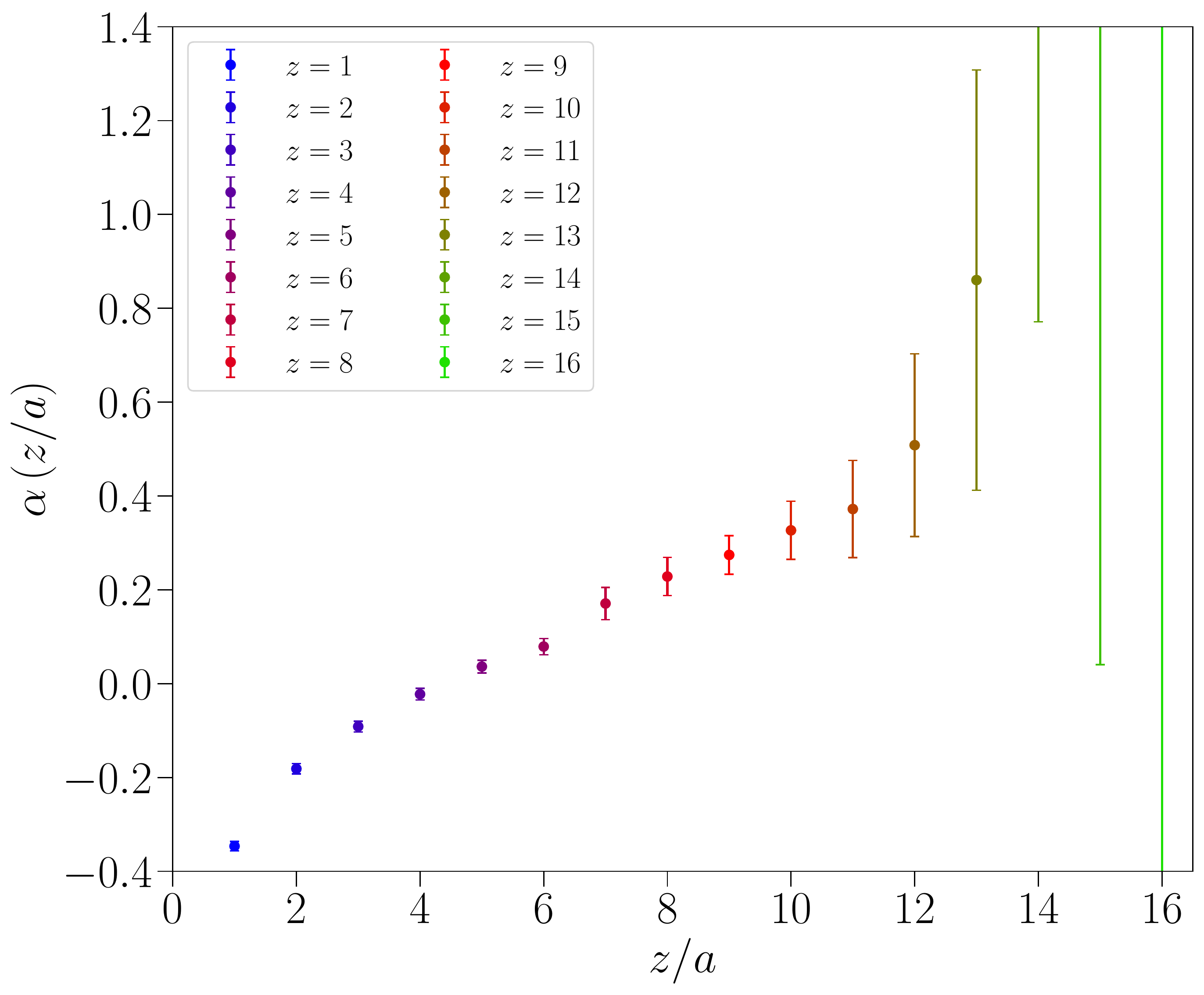}
    \includegraphics[width=0.45\linewidth]{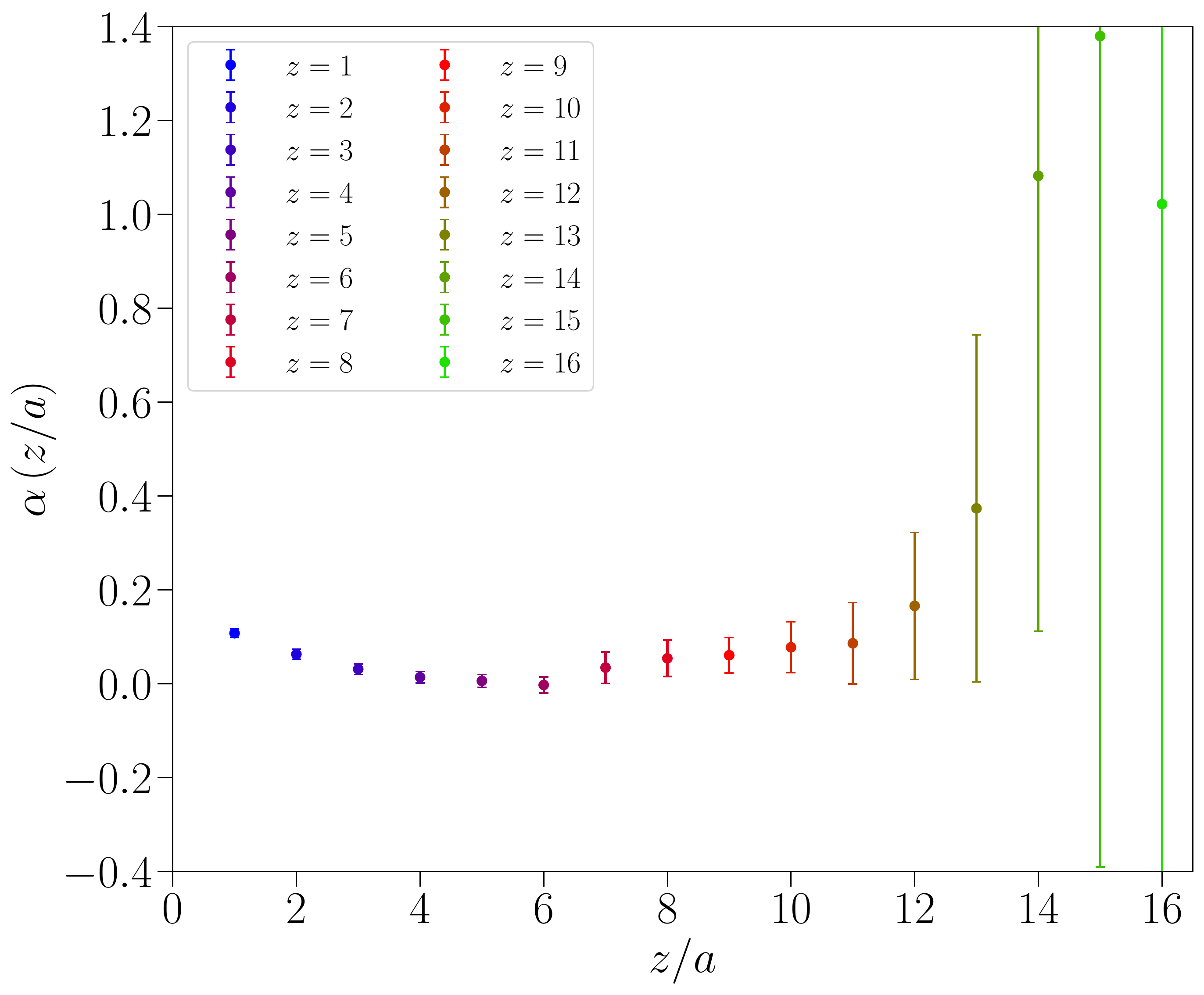}
    \caption{(Left) The variation of $\alpha$ with $z/a$ resulting from the cosine-transform of the model pseudo-PDF in Eq.~\ref{eq:model-pPDF} fit to $\mathfrak{Re}\ \mathfrak{M}\thinspace'\!\left(\nu,z^2\right)$ for each $z^2$. The discretization effect captured by the optimal Jacobi polynomial expansion $\lbrace n_{lt},n_{az},n_{t4},n_{t6}\rbrace_{\rm v}=\lbrace4,1,3,2\rbrace_{\rm v}$ is subtracted from $\mathfrak{Re}\ \mathfrak{M}\left(\nu,z^2\right)$ prior to performing each fit. (Right) The variation of $\alpha$ with $z/a$ resulting from the cosine-transform of the two-parameter PDF form in Eq.~\ref{eq:model-pPDF} fit to the discretization corrected and matched ITD $\mathfrak{Re}\ \mathcal{Q}'\left(\nu,z^2\right)$ for each $z^2$. The discretization corrected ITD is considerably more independent of the interval $z^2$.\label{fig:disc-corrected-pITD_pPDF-ITD_PDF-fits_alpha-vs-z}}
\end{figure}

Parameterizing the discretization corrected valence pseudo-PDF $\mathcal{P}'_{\rm v}\left(x,z^2\right)$ with the two-parameter form in Eq.~\ref{eq:model-pPDF} and fitting its cosine-transform to $\mathfrak{Re}\ \mathfrak{M}\thinspace'\!\left(\nu,z^2\right)$ with $\beta=3$, the $z$-dependence of $\mathfrak{Re}\ \mathfrak{M}\thinspace'\!\left(\nu,z^2\right)$ is once more reflected in the variation of $\alpha$ with $z/a$. As illustrated in the left panel of Fig.~\ref{fig:disc-corrected-pITD_pPDF-ITD_PDF-fits_alpha-vs-z}, $\alpha$ now varies non-linearly with $z/a$ for $z/a\lesssim4$ and linearly for $4\lesssim z/a\lesssim11$. Whether this markedly distinct $z$-dependence (c.f. Fig.~\ref{fig:pITD_pPDF-fits_alpha-vs-z}) is numerically consistent with DGLAP is once again checked by performing the matching to a common scale in $\msbar$ using Eq.~\ref{eq:pseudoPDF-2F3}, and repeating the two-parameter fits to the discretization corrected ITD $\mathfrak{Re}\ \mathcal{Q}'\left(\nu,\mu^2\right)$ for each $z^2$ and with $\beta=3$. The resulting fitted values of $\alpha$ are presented in the right panel of Fig.~\ref{fig:disc-corrected-pITD_pPDF-ITD_PDF-fits_alpha-vs-z}. Relative to the $z$-dependence of the uncorrected ITD shown in Fig.~\ref{fig:real-ITD_PDF-fits_alpha-vs-z_dglap_check}, the variation of $\alpha$ with $z/a$ is considerably more constant for $z/a\lesssim11$. In other words, the ITD is seen to fall into better agreement with DGLAP in the short-distance regime following removal of the $\mathcal{O}\left(a/z\right)$ effect.
That the optimal Jacobi polynomial fits $\lbrace n_{lt},n_{az},n_{t4},n_{t6}\rbrace_{\rm v}=\lbrace4,1,3,2\rbrace_{\rm v}$ and $\lbrace n_{lt},n_{az},n_{t4},n_{t6}\rbrace_+=\lbrace3,3,1,0\rbrace_+$ provide the best description of $\mathfrak{M}\left(\nu,z^2\right)$ can now be quantitatively explained by the compensating effect the $\mathcal{O}\left(a/z\right)$ term provides. The poor quality of the correlated phenomenological fits to the matched ITD, as well as the correlated Jacobi polynomial fits to $\mathfrak{M}\left(\nu,z^2\right)$ without any corrections, are a direct result of attempting to fit a singular function in $z$ to data that do not exhibit singular behavior. By excluding $z/a\lesssim4$ and $z/a\gtrsim11$, the short-distance tension and any large-$z$ polynomial effects can be removed yielding reduced pseudo-ITD or matched ITD data that are well in line with theoretical expectations. Such cuts are common in the literature, however their nominal effect is to neglect deviating behavior.

Although the DGLAP investigation has been shown for the real component of the reduced pseudo-ITD, the considerable reduction in the correlated figure-of-merit when discretization effects are included in fits to $\mathfrak{Im}\ \mathfrak{M}\left(\nu,z^2\right)$ ($\lbrace n_{lt},n_{az},n_{t4},n_{t6}\rbrace_+=\lbrace3,3,1,0\rbrace_+$ versus $\lbrace n_{lt},n_{az},n_{t4},n_{t6}\rbrace_+=\lbrace3,0,1,0\rbrace_+$ in Tab.~\ref{tab:jacobiFitResults}) indicates the imaginary component of the raw reduced pseudo-ITD likewise deviates from expectations of DGLAP at short-distances. The central question left for future research is the origin of this discretization effect.


\section{Conclusions\label{sec:conclusion}}
In this work we presented in detail the first lattice calculation of the unpolarized nucleon PDF in the distillation framework, showing its efficacy and emphasizing in detail the advantages of this smearing technique. We have observed that distillation can yield a pseudo-distribution of superior quality, addressing one of the principal components limiting the competitiveness of contemporary lattice QCD calculations of PDFs with respect to precision phenomenological extractions.
For the unpolarized PDF, the impact of this lattice calculation on global fits was found to be rather marginal. The quality of data, however, hints that a future calculation of a distribution less constrained by experiment may benefit from the use of distillation. By validating the distillation method in the unpolarized nucleon PDF, our study opens new avenues of synergy between lattice and phenomenology in the spirit of ~\cite{DelDebbio:2020rgv,Bringewatt:2020ixn,Cichy:2019ebf}. We performed a careful study of the correct extraction of the matrix elements that yield the pseudo-ITD and we analyzed in detail its real and imaginary components. This was followed by a discussion of correlations of lattice data and how the commonly used uncorrelated fits can yield erroneous results and can severely flaw the extraction of the PDFs. We regulate the ill-posed inverse problem that relates the ITD to the PDF with the use of Jacobi polynomials, and stress the necessity of removing discretization errors as well as higher twist effects in order to ensure a trustworthy extraction of the PDF. The quality of the results herein bolsters confidence that distillation is the natural path to be pursued in these types of hadron structure calculations. We soon plan to extend these analyses to include the extrapolation to the physical pion mass and the continuum limit.

\FloatBarrier

\subsection{Acknowledgements}
We would like to thank all the members of the Hadstruc collaboration for fruitful and stimulating exchanges. 
This work is supported by Jefferson
Science Associates, LLC under U.S. DOE Contract \#DE-AC05-06OR23177.
KO was supported in part by U.S.  DOE grant \mbox{
  \#DE-FG02-04ER41302} and in part by the Center for Nuclear Femtography grants \#C2-2020-FEMT-006, \#C2019-FEMT-002-05.
    AR was supported in part by U.S. DOE Grant
\mbox{\#DE-FG02-97ER41028. }
CE was supported in
part by the U.S. Department of Energy under Contract
No. DEFG02-04ER41302, a Department of Energy Office
of Science Graduate Student Research fellowship, through
the U.S. Department of Energy, Office of Science, Office of
Workforce Development for Teachers and Scientists, Office
of Science Graduate Student Research (SCGSR) program,
and a Jefferson Science Associates graduate fellowship.
The SCGSR program is administered by the Oak Ridge
Institute for Science and Education (ORISE) for the DOE.
ORISE is managed by ORAU under Contract No. DE-
SC0014664.
We would also like to thank the Texas Advanced Computing Center (TACC) at the University of Texas at Austin for providing HPC resources
on Frontera~\cite{frontera} that have contributed to the results in this paper. 
 We acknowledge the facilities of the USQCD Collaboration used for this research in part, which are funded by the Office of Science of the U.S. Department of Energy. This work was performed in part using computing
facilities at the College of William and Mary which were provided by
contributions from the National Science Foundation (MRI grant
PHY-1626177), and the Commonwealth of Virginia Equipment Trust Fund.
The authors acknowledge William \& Mary Research Computing for providing computational resources and/or technical support that have contributed to the results reported within this paper. This work used the Extreme Science and Engineering Discovery Environment (XSEDE), which is supported by National Science Foundation grant number ACI-1548562~\cite{xsede}.
 In addition, this work used resources at
NERSC, a DOE Office of Science User Facility supported by the Office
of Science of the U.S. Department of Energy under Contract
\#DE-AC02-05CH11231, as well as resources of the Oak Ridge Leadership Computing Facility at the Oak Ridge National Laboratory, which is supported by the Office of Science of the U.S. Department of Energy under Contract No. \mbox{\#DE-AC05-00OR22725}. In addition, this work was made possible using results obtained  at NERSC, a DOE Office of Science User Facility supported by the Office of Science of the U.S. Department of Energy under Contract \mbox{\#DE-AC02-05CH11231}, as well as resources of the Oak Ridge Leadership Computing Facility (ALCC and INCITE) at the Oak Ridge National Laboratory, which is supported by the Office of Science of the U.S. Department of Energy under Contract No. \mbox{\#DE-AC05-00OR22725}.  The software libraries used on these machines were Chroma~\cite{Edwards:2004sx}, QUDA ~\cite{Clark:2009wm,Babich:2010mu}, QDP-JIT~\cite{Winter:2014dka} and QPhiX~\cite{Joo:2013lwm,optimising} developed with  
 support from the U.S. Department of Energy, Office of Science, Office of Advanced Scientific Computing Research and Office of Nuclear Physics, Scientific Discovery through Advanced Computing (SciDAC) program, and of the U.S. Department of Energy Exascale Computing Project.

We acknowledge PRACE (Partnership for Advanced Computing in Europe) for awarding us access to the high performance computing system Marconi100 at CINECA (Consorzio Interuniversitario per il Calcolo Automatico dell’Italia Nord-orientale) under the grant Pra21-5389.
Results were obtained also by using Piz Daint at Centro Svizzero di Calcolo Scientifico (CSCS), via the project with id s994. We thank the staff of CSCS for access to the computational resources and for their constant support. This work also benefited from access to the Jean Zay supercomputer at the Institute for Development and Resources in Intensive Scientific Computing (IDRIS) in Orsay, France under project A0080511504.


\FloatBarrier
\bibliographystyle{/home/colin/texmf/tex/latex/commonstuff/bibtex/bst/revtex/apsrev4-1}
\bibliography{srcs}

\begin{thebibliography}{108}%
\makeatletter
\providecommand \@ifxundefined [1]{%
 \@ifx{#1\undefined}
}%
\providecommand \@ifnum [1]{%
 \ifnum #1\expandafter \@firstoftwo
 \else \expandafter \@secondoftwo
 \fi
}%
\providecommand \@ifx [1]{%
 \ifx #1\expandafter \@firstoftwo
 \else \expandafter \@secondoftwo
 \fi
}%
\providecommand \natexlab [1]{#1}%
\providecommand \enquote  [1]{``#1''}%
\providecommand \bibnamefont  [1]{#1}%
\providecommand \bibfnamefont [1]{#1}%
\providecommand \citenamefont [1]{#1}%
\providecommand \href@noop [0]{\@secondoftwo}%
\providecommand \href [0]{\begingroup \@sanitize@url \@href}%
\providecommand \@href[1]{\@@startlink{#1}\@@href}%
\providecommand \@@href[1]{\endgroup#1\@@endlink}%
\providecommand \@sanitize@url [0]{\catcode `\\12\catcode `\$12\catcode
  `\&12\catcode `\#12\catcode `\^12\catcode `\_12\catcode `\%12\relax}%
\providecommand \@@startlink[1]{}%
\providecommand \@@endlink[0]{}%
\providecommand \url  [0]{\begingroup\@sanitize@url \@url }%
\providecommand \@url [1]{\endgroup\@href {#1}{\urlprefix }}%
\providecommand \urlprefix  [0]{URL }%
\providecommand \Eprint [0]{\href }%
\providecommand \doibase [0]{http://dx.doi.org/}%
\providecommand \selectlanguage [0]{\@gobble}%
\providecommand \bibinfo  [0]{\@secondoftwo}%
\providecommand \bibfield  [0]{\@secondoftwo}%
\providecommand \translation [1]{[#1]}%
\providecommand \BibitemOpen [0]{}%
\providecommand \bibitemStop [0]{}%
\providecommand \bibitemNoStop [0]{.\EOS\space}%
\providecommand \EOS [0]{\spacefactor3000\relax}%
\providecommand \BibitemShut  [1]{\csname bibitem#1\endcsname}%
\let\auto@bib@innerbib\@empty
\bibitem [{\citenamefont {Collins}\ \emph {et~al.}(1989)\citenamefont
  {Collins}, \citenamefont {Soper},\ and\ \citenamefont
  {Sterman}}]{Collins:1989gx}%
  \BibitemOpen
  \bibfield  {author} {\bibinfo {author} {\bibfnamefont {J.~C.}\ \bibnamefont
  {Collins}}, \bibinfo {author} {\bibfnamefont {D.~E.}\ \bibnamefont {Soper}},
  \ and\ \bibinfo {author} {\bibfnamefont {G.~F.}\ \bibnamefont {Sterman}},\
  }\bibfield  {title} {\enquote {\bibinfo {title} {{Factorization of Hard
  Processes in QCD}},}\ }\href {\doibase 10.1142/9789814503266_0001} {\bibfield
   {journal} {\bibinfo  {journal} {Adv. Ser. Direct. High Energy Phys.}\
  }\textbf {\bibinfo {volume} {5}},\ \bibinfo {pages} {1} (\bibinfo {year}
  {1989})},\ \Eprint {http://arxiv.org/abs/hep-ph/0409313}
  {arXiv:hep-ph/0409313} \BibitemShut {NoStop}%
\bibitem [{\citenamefont {Gao}\ \emph {et~al.}(2018)\citenamefont {Gao},
  \citenamefont {Harland-Lang},\ and\ \citenamefont {Rojo}}]{Gao:2017yyd}%
  \BibitemOpen
  \bibfield  {author} {\bibinfo {author} {\bibfnamefont {J.}~\bibnamefont
  {Gao}}, \bibinfo {author} {\bibfnamefont {L.}~\bibnamefont {Harland-Lang}}, \
  and\ \bibinfo {author} {\bibfnamefont {J.}~\bibnamefont {Rojo}},\ }\bibfield
  {title} {\enquote {\bibinfo {title} {{The Structure of the Proton in the LHC
  Precision Era}},}\ }\href {\doibase 10.1016/j.physrep.2018.03.002} {\bibfield
   {journal} {\bibinfo  {journal} {Phys. Rept.}\ }\textbf {\bibinfo {volume}
  {742}},\ \bibinfo {pages} {1} (\bibinfo {year} {2018})},\ \Eprint
  {http://arxiv.org/abs/1709.04922} {arXiv:1709.04922 [hep-ph]} \BibitemShut
  {NoStop}%
\bibitem [{\citenamefont {Liu}\ and\ \citenamefont {Dong}(1994)}]{Liu:1993cv}%
  \BibitemOpen
  \bibfield  {author} {\bibinfo {author} {\bibfnamefont {K.-F.}\ \bibnamefont
  {Liu}}\ and\ \bibinfo {author} {\bibfnamefont {S.-J.}\ \bibnamefont {Dong}},\
  }\bibfield  {title} {\enquote {\bibinfo {title} {{Origin of difference
  between anti-d and anti-u partons in the nucleon}},}\ }\href {\doibase
  10.1103/PhysRevLett.72.1790} {\bibfield  {journal} {\bibinfo  {journal}
  {Phys. Rev. Lett.}\ }\textbf {\bibinfo {volume} {72}},\ \bibinfo {pages}
  {1790} (\bibinfo {year} {1994})},\ \Eprint
  {http://arxiv.org/abs/hep-ph/9306299} {arXiv:hep-ph/9306299} \BibitemShut
  {NoStop}%
\bibitem [{\citenamefont {Liu}(2000)}]{Liu:1999ak}%
  \BibitemOpen
  \bibfield  {author} {\bibinfo {author} {\bibfnamefont {K.-F.}\ \bibnamefont
  {Liu}},\ }\bibfield  {title} {\enquote {\bibinfo {title} {{Parton degrees of
  freedom from the path integral formalism}},}\ }\href {\doibase
  10.1103/PhysRevD.62.074501} {\bibfield  {journal} {\bibinfo  {journal} {Phys.
  Rev. D}\ }\textbf {\bibinfo {volume} {62}},\ \bibinfo {pages} {074501}
  (\bibinfo {year} {2000})},\ \Eprint {http://arxiv.org/abs/hep-ph/9910306}
  {arXiv:hep-ph/9910306} \BibitemShut {NoStop}%
\bibitem [{\citenamefont {Detmold}\ and\ \citenamefont
  {Lin}(2006)}]{Detmold:2005gg}%
  \BibitemOpen
  \bibfield  {author} {\bibinfo {author} {\bibfnamefont {W.}~\bibnamefont
  {Detmold}}\ and\ \bibinfo {author} {\bibfnamefont {C.~J.~D.}\ \bibnamefont
  {Lin}},\ }\bibfield  {title} {\enquote {\bibinfo {title} {{Deep-inelastic
  scattering and the operator product expansion in lattice QCD}},}\ }\href
  {\doibase 10.1103/PhysRevD.73.014501} {\bibfield  {journal} {\bibinfo
  {journal} {Phys. Rev. D}\ }\textbf {\bibinfo {volume} {73}},\ \bibinfo
  {pages} {014501} (\bibinfo {year} {2006})},\ \Eprint
  {http://arxiv.org/abs/hep-lat/0507007} {arXiv:hep-lat/0507007} \BibitemShut
  {NoStop}%
\bibitem [{\citenamefont {Aglietti}\ \emph {et~al.}(1998)\citenamefont
  {Aglietti}, \citenamefont {Ciuchini}, \citenamefont {Corbo}, \citenamefont
  {Franco}, \citenamefont {Martinelli},\ and\ \citenamefont
  {Silvestrini}}]{Aglietti:1998ur}%
  \BibitemOpen
  \bibfield  {author} {\bibinfo {author} {\bibfnamefont {U.}~\bibnamefont
  {Aglietti}}, \bibinfo {author} {\bibfnamefont {M.}~\bibnamefont {Ciuchini}},
  \bibinfo {author} {\bibfnamefont {G.}~\bibnamefont {Corbo}}, \bibinfo
  {author} {\bibfnamefont {E.}~\bibnamefont {Franco}}, \bibinfo {author}
  {\bibfnamefont {G.}~\bibnamefont {Martinelli}}, \ and\ \bibinfo {author}
  {\bibfnamefont {L.}~\bibnamefont {Silvestrini}},\ }\bibfield  {title}
  {\enquote {\bibinfo {title} {{Model independent determination of the light
  cone wave functions for exclusive processes}},}\ }\href {\doibase
  10.1016/S0370-2693(98)01138-1} {\bibfield  {journal} {\bibinfo  {journal}
  {Phys. Lett. B}\ }\textbf {\bibinfo {volume} {441}},\ \bibinfo {pages} {371}
  (\bibinfo {year} {1998})},\ \Eprint {http://arxiv.org/abs/hep-ph/9806277}
  {arXiv:hep-ph/9806277} \BibitemShut {NoStop}%
\bibitem [{\citenamefont {Braun}\ and\ \citenamefont
  {M\"uller}(2008)}]{Braun:2007wv}%
  \BibitemOpen
  \bibfield  {author} {\bibinfo {author} {\bibfnamefont {V.}~\bibnamefont
  {Braun}}\ and\ \bibinfo {author} {\bibfnamefont {D.}~\bibnamefont
  {M\"uller}},\ }\bibfield  {title} {\enquote {\bibinfo {title} {{Exclusive
  processes in position space and the pion distribution amplitude}},}\ }\href
  {\doibase 10.1140/epjc/s10052-008-0608-4} {\bibfield  {journal} {\bibinfo
  {journal} {Eur. Phys. J. C}\ }\textbf {\bibinfo {volume} {55}},\ \bibinfo
  {pages} {349} (\bibinfo {year} {2008})},\ \Eprint
  {http://arxiv.org/abs/0709.1348} {arXiv:0709.1348 [hep-ph]} \BibitemShut
  {NoStop}%
\bibitem [{\citenamefont {Ji}(2013)}]{Ji:2013dva}%
  \BibitemOpen
  \bibfield  {author} {\bibinfo {author} {\bibfnamefont {X.}~\bibnamefont
  {Ji}},\ }\bibfield  {title} {\enquote {\bibinfo {title} {{Parton Physics on a
  Euclidean Lattice}},}\ }\href {\doibase 10.1103/PhysRevLett.110.262002}
  {\bibfield  {journal} {\bibinfo  {journal} {Phys. Rev. Lett.}\ }\textbf
  {\bibinfo {volume} {110}},\ \bibinfo {pages} {262002} (\bibinfo {year}
  {2013})},\ \Eprint {http://arxiv.org/abs/1305.1539} {arXiv:1305.1539
  [hep-ph]} \BibitemShut {NoStop}%
\bibitem [{\citenamefont {Ji}(2014)}]{Ji:2014gla}%
  \BibitemOpen
  \bibfield  {author} {\bibinfo {author} {\bibfnamefont {X.}~\bibnamefont
  {Ji}},\ }\bibfield  {title} {\enquote {\bibinfo {title} {{Parton Physics from
  Large-Momentum Effective Field Theory}},}\ }\href {\doibase
  10.1007/s11433-014-5492-3} {\bibfield  {journal} {\bibinfo  {journal} {Sci.
  China Phys. Mech. Astron.}\ }\textbf {\bibinfo {volume} {57}},\ \bibinfo
  {pages} {1407} (\bibinfo {year} {2014})},\ \Eprint
  {http://arxiv.org/abs/1404.6680} {arXiv:1404.6680 [hep-ph]} \BibitemShut
  {NoStop}%
\bibitem [{\citenamefont {Alexandrou}\ \emph
  {et~al.}(2021{\natexlab{a}})\citenamefont {Alexandrou}, \citenamefont
  {Constantinou}, \citenamefont {Hadjiyiannakou}, \citenamefont {Jansen},\ and\
  \citenamefont {Manigrasso}}]{Alexandrou:2020uyt}%
  \BibitemOpen
  \bibfield  {author} {\bibinfo {author} {\bibfnamefont {C.}~\bibnamefont
  {Alexandrou}}, \bibinfo {author} {\bibfnamefont {M.}~\bibnamefont
  {Constantinou}}, \bibinfo {author} {\bibfnamefont {K.}~\bibnamefont
  {Hadjiyiannakou}}, \bibinfo {author} {\bibfnamefont {K.}~\bibnamefont
  {Jansen}}, \ and\ \bibinfo {author} {\bibfnamefont {F.}~\bibnamefont
  {Manigrasso}},\ }\bibfield  {title} {\enquote {\bibinfo {title} {{Flavor
  decomposition for the proton helicity parton distribution functions}},}\
  }\href {\doibase 10.1103/PhysRevLett.126.102003} {\bibfield  {journal}
  {\bibinfo  {journal} {Phys. Rev. Lett.}\ }\textbf {\bibinfo {volume} {126}},\
  \bibinfo {pages} {102003} (\bibinfo {year} {2021}{\natexlab{a}})},\ \Eprint
  {http://arxiv.org/abs/2009.13061} {arXiv:2009.13061 [hep-lat]} \BibitemShut
  {NoStop}%
\bibitem [{\citenamefont {Cichy}\ \emph {et~al.}(2019)\citenamefont {Cichy},
  \citenamefont {Del~Debbio},\ and\ \citenamefont {Giani}}]{Cichy:2019ebf}%
  \BibitemOpen
  \bibfield  {author} {\bibinfo {author} {\bibfnamefont {K.}~\bibnamefont
  {Cichy}}, \bibinfo {author} {\bibfnamefont {L.}~\bibnamefont {Del~Debbio}}, \
  and\ \bibinfo {author} {\bibfnamefont {T.}~\bibnamefont {Giani}},\ }\bibfield
   {title} {\enquote {\bibinfo {title} {{Parton distributions from lattice
  data: the nonsinglet case}},}\ }\href {\doibase 10.1007/JHEP10(2019)137}
  {\bibfield  {journal} {\bibinfo  {journal} {JHEP}\ }\textbf {\bibinfo
  {volume} {10}},\ \bibinfo {pages} {137} (\bibinfo {year} {2019})},\ \Eprint
  {http://arxiv.org/abs/1907.06037} {arXiv:1907.06037 [hep-ph]} \BibitemShut
  {NoStop}%
\bibitem [{\citenamefont {Izubuchi}\ \emph {et~al.}(2019)\citenamefont
  {Izubuchi}, \citenamefont {Jin}, \citenamefont {Kallidonis}, \citenamefont
  {Karthik}, \citenamefont {Mukherjee}, \citenamefont {Petreczky},
  \citenamefont {Shugert},\ and\ \citenamefont {Syritsyn}}]{Izubuchi:2019lyk}%
  \BibitemOpen
  \bibfield  {author} {\bibinfo {author} {\bibfnamefont {T.}~\bibnamefont
  {Izubuchi}}, \bibinfo {author} {\bibfnamefont {L.}~\bibnamefont {Jin}},
  \bibinfo {author} {\bibfnamefont {C.}~\bibnamefont {Kallidonis}}, \bibinfo
  {author} {\bibfnamefont {N.}~\bibnamefont {Karthik}}, \bibinfo {author}
  {\bibfnamefont {S.}~\bibnamefont {Mukherjee}}, \bibinfo {author}
  {\bibfnamefont {P.}~\bibnamefont {Petreczky}}, \bibinfo {author}
  {\bibfnamefont {C.}~\bibnamefont {Shugert}}, \ and\ \bibinfo {author}
  {\bibfnamefont {S.}~\bibnamefont {Syritsyn}},\ }\bibfield  {title} {\enquote
  {\bibinfo {title} {{Valence parton distribution function of pion from fine
  lattice}},}\ }\href {\doibase 10.1103/PhysRevD.100.034516} {\bibfield
  {journal} {\bibinfo  {journal} {Phys. Rev. D}\ }\textbf {\bibinfo {volume}
  {100}},\ \bibinfo {pages} {034516} (\bibinfo {year} {2019})},\ \Eprint
  {http://arxiv.org/abs/1905.06349} {arXiv:1905.06349 [hep-lat]} \BibitemShut
  {NoStop}%
\bibitem [{\citenamefont {Alexandrou}\ \emph
  {et~al.}(2018{\natexlab{a}})\citenamefont {Alexandrou}, \citenamefont
  {Cichy}, \citenamefont {Constantinou}, \citenamefont {Jansen}, \citenamefont
  {Scapellato},\ and\ \citenamefont {Steffens}}]{Alexandrou:2018eet}%
  \BibitemOpen
  \bibfield  {author} {\bibinfo {author} {\bibfnamefont {C.}~\bibnamefont
  {Alexandrou}}, \bibinfo {author} {\bibfnamefont {K.}~\bibnamefont {Cichy}},
  \bibinfo {author} {\bibfnamefont {M.}~\bibnamefont {Constantinou}}, \bibinfo
  {author} {\bibfnamefont {K.}~\bibnamefont {Jansen}}, \bibinfo {author}
  {\bibfnamefont {A.}~\bibnamefont {Scapellato}}, \ and\ \bibinfo {author}
  {\bibfnamefont {F.}~\bibnamefont {Steffens}},\ }\bibfield  {title} {\enquote
  {\bibinfo {title} {{Transversity parton distribution functions from lattice
  QCD}},}\ }\href {\doibase 10.1103/PhysRevD.98.091503} {\bibfield  {journal}
  {\bibinfo  {journal} {Phys. Rev. D}\ }\textbf {\bibinfo {volume} {98}},\
  \bibinfo {pages} {091503} (\bibinfo {year} {2018}{\natexlab{a}})},\ \Eprint
  {http://arxiv.org/abs/1807.00232} {arXiv:1807.00232 [hep-lat]} \BibitemShut
  {NoStop}%
\bibitem [{\citenamefont {Chen}\ \emph {et~al.}(2016)\citenamefont {Chen},
  \citenamefont {Cohen}, \citenamefont {Ji}, \citenamefont {Lin},\ and\
  \citenamefont {Zhang}}]{Chen:2016utp}%
  \BibitemOpen
  \bibfield  {author} {\bibinfo {author} {\bibfnamefont {J.-W.}\ \bibnamefont
  {Chen}}, \bibinfo {author} {\bibfnamefont {S.~D.}\ \bibnamefont {Cohen}},
  \bibinfo {author} {\bibfnamefont {X.}~\bibnamefont {Ji}}, \bibinfo {author}
  {\bibfnamefont {H.-W.}\ \bibnamefont {Lin}}, \ and\ \bibinfo {author}
  {\bibfnamefont {J.-H.}\ \bibnamefont {Zhang}},\ }\bibfield  {title} {\enquote
  {\bibinfo {title} {{Nucleon Helicity and Transversity Parton Distributions
  from Lattice QCD}},}\ }\href {\doibase 10.1016/j.nuclphysb.2016.07.033}
  {\bibfield  {journal} {\bibinfo  {journal} {Nucl. Phys. B}\ }\textbf
  {\bibinfo {volume} {911}},\ \bibinfo {pages} {246} (\bibinfo {year}
  {2016})},\ \Eprint {http://arxiv.org/abs/1603.06664} {arXiv:1603.06664
  [hep-ph]} \BibitemShut {NoStop}%
\bibitem [{\citenamefont {Alexandrou}\ \emph
  {et~al.}(2018{\natexlab{b}})\citenamefont {Alexandrou}, \citenamefont
  {Bacchio}, \citenamefont {Cichy}, \citenamefont {Constantinou}, \citenamefont
  {Hadjiyiannakou}, \citenamefont {Jansen}, \citenamefont {Koutsou},
  \citenamefont {Scapellato},\ and\ \citenamefont
  {Steffens}}]{Alexandrou:2017dzj}%
  \BibitemOpen
  \bibfield  {author} {\bibinfo {author} {\bibfnamefont {C.}~\bibnamefont
  {Alexandrou}}, \bibinfo {author} {\bibfnamefont {S.}~\bibnamefont {Bacchio}},
  \bibinfo {author} {\bibfnamefont {K.}~\bibnamefont {Cichy}}, \bibinfo
  {author} {\bibfnamefont {M.}~\bibnamefont {Constantinou}}, \bibinfo {author}
  {\bibfnamefont {K.}~\bibnamefont {Hadjiyiannakou}}, \bibinfo {author}
  {\bibfnamefont {K.}~\bibnamefont {Jansen}}, \bibinfo {author} {\bibfnamefont
  {G.}~\bibnamefont {Koutsou}}, \bibinfo {author} {\bibfnamefont
  {A.}~\bibnamefont {Scapellato}}, \ and\ \bibinfo {author} {\bibfnamefont
  {F.}~\bibnamefont {Steffens}},\ }\bibfield  {title} {\enquote {\bibinfo
  {title} {{Computation of parton distributions from the quasi-PDF approach at
  the physical point}},}\ }\href {\doibase 10.1051/epjconf/201817514008}
  {\bibfield  {journal} {\bibinfo  {journal} {EPJ Web Conf.}\ }\textbf
  {\bibinfo {volume} {175}},\ \bibinfo {pages} {14008} (\bibinfo {year}
  {2018}{\natexlab{b}})},\ \Eprint {http://arxiv.org/abs/1710.06408}
  {arXiv:1710.06408 [hep-lat]} \BibitemShut {NoStop}%
\bibitem [{\citenamefont {Lin}\ \emph {et~al.}(2018)\citenamefont {Lin},
  \citenamefont {Chen}, \citenamefont {Ishikawa},\ and\ \citenamefont
  {Zhang}}]{Lin:2017ani}%
  \BibitemOpen
  \bibfield  {author} {\bibinfo {author} {\bibfnamefont {H.-W.}\ \bibnamefont
  {Lin}}, \bibinfo {author} {\bibfnamefont {J.-W.}\ \bibnamefont {Chen}},
  \bibinfo {author} {\bibfnamefont {T.}~\bibnamefont {Ishikawa}}, \ and\
  \bibinfo {author} {\bibfnamefont {J.-H.}\ \bibnamefont {Zhang}} (\bibinfo
  {collaboration} {LP3}),\ }\bibfield  {title} {\enquote {\bibinfo {title}
  {{Improved parton distribution functions at the physical pion mass}},}\
  }\href {\doibase 10.1103/PhysRevD.98.054504} {\bibfield  {journal} {\bibinfo
  {journal} {Phys. Rev. D}\ }\textbf {\bibinfo {volume} {98}},\ \bibinfo
  {pages} {054504} (\bibinfo {year} {2018})},\ \Eprint
  {http://arxiv.org/abs/1708.05301} {arXiv:1708.05301 [hep-lat]} \BibitemShut
  {NoStop}%
\bibitem [{\citenamefont {Fan}\ \emph {et~al.}(2018)\citenamefont {Fan},
  \citenamefont {Yang}, \citenamefont {Anthony}, \citenamefont {Lin},\ and\
  \citenamefont {Liu}}]{Fan:2018dxu}%
  \BibitemOpen
  \bibfield  {author} {\bibinfo {author} {\bibfnamefont {Z.-Y.}\ \bibnamefont
  {Fan}}, \bibinfo {author} {\bibfnamefont {Y.-B.}\ \bibnamefont {Yang}},
  \bibinfo {author} {\bibfnamefont {A.}~\bibnamefont {Anthony}}, \bibinfo
  {author} {\bibfnamefont {H.-W.}\ \bibnamefont {Lin}}, \ and\ \bibinfo
  {author} {\bibfnamefont {K.-F.}\ \bibnamefont {Liu}},\ }\bibfield  {title}
  {\enquote {\bibinfo {title} {{Gluon Quasi-Parton-Distribution Functions from
  Lattice QCD}},}\ }\href {\doibase 10.1103/PhysRevLett.121.242001} {\bibfield
  {journal} {\bibinfo  {journal} {Phys. Rev. Lett.}\ }\textbf {\bibinfo
  {volume} {121}},\ \bibinfo {pages} {242001} (\bibinfo {year} {2018})},\
  \Eprint {http://arxiv.org/abs/1808.02077} {arXiv:1808.02077 [hep-lat]}
  \BibitemShut {NoStop}%
\bibitem [{\citenamefont {Zhang}\ \emph {et~al.}(2019)\citenamefont {Zhang},
  \citenamefont {Ji}, \citenamefont {Sch\"afer}, \citenamefont {Wang},\ and\
  \citenamefont {Zhao}}]{Zhang:2018diq}%
  \BibitemOpen
  \bibfield  {author} {\bibinfo {author} {\bibfnamefont {J.-H.}\ \bibnamefont
  {Zhang}}, \bibinfo {author} {\bibfnamefont {X.}~\bibnamefont {Ji}}, \bibinfo
  {author} {\bibfnamefont {A.}~\bibnamefont {Sch\"afer}}, \bibinfo {author}
  {\bibfnamefont {W.}~\bibnamefont {Wang}}, \ and\ \bibinfo {author}
  {\bibfnamefont {S.}~\bibnamefont {Zhao}},\ }\bibfield  {title} {\enquote
  {\bibinfo {title} {{Accessing Gluon Parton Distributions in Large Momentum
  Effective Theory}},}\ }\href {\doibase 10.1103/PhysRevLett.122.142001}
  {\bibfield  {journal} {\bibinfo  {journal} {Phys. Rev. Lett.}\ }\textbf
  {\bibinfo {volume} {122}},\ \bibinfo {pages} {142001} (\bibinfo {year}
  {2019})},\ \Eprint {http://arxiv.org/abs/1808.10824} {arXiv:1808.10824
  [hep-ph]} \BibitemShut {NoStop}%
\bibitem [{\citenamefont {Alexandrou}\ \emph {et~al.}(2017)\citenamefont
  {Alexandrou}, \citenamefont {Cichy}, \citenamefont {Constantinou},
  \citenamefont {Hadjiyiannakou}, \citenamefont {Jansen}, \citenamefont
  {Panagopoulos},\ and\ \citenamefont {Steffens}}]{Alexandrou:2017huk}%
  \BibitemOpen
  \bibfield  {author} {\bibinfo {author} {\bibfnamefont {C.}~\bibnamefont
  {Alexandrou}}, \bibinfo {author} {\bibfnamefont {K.}~\bibnamefont {Cichy}},
  \bibinfo {author} {\bibfnamefont {M.}~\bibnamefont {Constantinou}}, \bibinfo
  {author} {\bibfnamefont {K.}~\bibnamefont {Hadjiyiannakou}}, \bibinfo
  {author} {\bibfnamefont {K.}~\bibnamefont {Jansen}}, \bibinfo {author}
  {\bibfnamefont {H.}~\bibnamefont {Panagopoulos}}, \ and\ \bibinfo {author}
  {\bibfnamefont {F.}~\bibnamefont {Steffens}},\ }\bibfield  {title} {\enquote
  {\bibinfo {title} {{A complete non-perturbative renormalization prescription
  for quasi-PDFs}},}\ }\href {\doibase 10.1016/j.nuclphysb.2017.08.012}
  {\bibfield  {journal} {\bibinfo  {journal} {Nucl. Phys. B}\ }\textbf
  {\bibinfo {volume} {923}},\ \bibinfo {pages} {394} (\bibinfo {year}
  {2017})},\ \Eprint {http://arxiv.org/abs/1706.00265} {arXiv:1706.00265
  [hep-lat]} \BibitemShut {NoStop}%
\bibitem [{\citenamefont {Chen}\ \emph {et~al.}(2018)\citenamefont {Chen},
  \citenamefont {Ishikawa}, \citenamefont {Jin}, \citenamefont {Lin},
  \citenamefont {Yang}, \citenamefont {Zhang},\ and\ \citenamefont
  {Zhao}}]{Chen:2017mzz}%
  \BibitemOpen
  \bibfield  {author} {\bibinfo {author} {\bibfnamefont {J.-W.}\ \bibnamefont
  {Chen}}, \bibinfo {author} {\bibfnamefont {T.}~\bibnamefont {Ishikawa}},
  \bibinfo {author} {\bibfnamefont {L.}~\bibnamefont {Jin}}, \bibinfo {author}
  {\bibfnamefont {H.-W.}\ \bibnamefont {Lin}}, \bibinfo {author} {\bibfnamefont
  {Y.-B.}\ \bibnamefont {Yang}}, \bibinfo {author} {\bibfnamefont {J.-H.}\
  \bibnamefont {Zhang}}, \ and\ \bibinfo {author} {\bibfnamefont
  {Y.}~\bibnamefont {Zhao}},\ }\bibfield  {title} {\enquote {\bibinfo {title}
  {{Parton distribution function with nonperturbative renormalization from
  lattice QCD}},}\ }\href {\doibase 10.1103/PhysRevD.97.014505} {\bibfield
  {journal} {\bibinfo  {journal} {Phys. Rev. D}\ }\textbf {\bibinfo {volume}
  {97}},\ \bibinfo {pages} {014505} (\bibinfo {year} {2018})},\ \Eprint
  {http://arxiv.org/abs/1706.01295} {arXiv:1706.01295 [hep-lat]} \BibitemShut
  {NoStop}%
\bibitem [{\citenamefont {Alexandrou}\ \emph
  {et~al.}(2021{\natexlab{b}})\citenamefont {Alexandrou}, \citenamefont
  {Cichy}, \citenamefont {Constantinou}, \citenamefont {Green}, \citenamefont
  {Hadjiyiannakou}, \citenamefont {Jansen}, \citenamefont {Manigrasso},
  \citenamefont {Scapellato},\ and\ \citenamefont
  {Steffens}}]{Alexandrou:2020qtt}%
  \BibitemOpen
  \bibfield  {author} {\bibinfo {author} {\bibfnamefont {C.}~\bibnamefont
  {Alexandrou}}, \bibinfo {author} {\bibfnamefont {K.}~\bibnamefont {Cichy}},
  \bibinfo {author} {\bibfnamefont {M.}~\bibnamefont {Constantinou}}, \bibinfo
  {author} {\bibfnamefont {J.~R.}\ \bibnamefont {Green}}, \bibinfo {author}
  {\bibfnamefont {K.}~\bibnamefont {Hadjiyiannakou}}, \bibinfo {author}
  {\bibfnamefont {K.}~\bibnamefont {Jansen}}, \bibinfo {author} {\bibfnamefont
  {F.}~\bibnamefont {Manigrasso}}, \bibinfo {author} {\bibfnamefont
  {A.}~\bibnamefont {Scapellato}}, \ and\ \bibinfo {author} {\bibfnamefont
  {F.}~\bibnamefont {Steffens}},\ }\bibfield  {title} {\enquote {\bibinfo
  {title} {{Lattice continuum-limit study of nucleon quasi-PDFs}},}\ }\href
  {\doibase 10.1103/PhysRevD.103.094512} {\bibfield  {journal} {\bibinfo
  {journal} {Phys. Rev. D}\ }\textbf {\bibinfo {volume} {103}},\ \bibinfo
  {pages} {094512} (\bibinfo {year} {2021}{\natexlab{b}})},\ \Eprint
  {http://arxiv.org/abs/2011.00964} {arXiv:2011.00964 [hep-lat]} \BibitemShut
  {NoStop}%
\bibitem [{\citenamefont {Alexandrou}\ \emph {et~al.}(2019)\citenamefont
  {Alexandrou}, \citenamefont {Cichy}, \citenamefont {Constantinou},
  \citenamefont {Hadjiyiannakou}, \citenamefont {Jansen}, \citenamefont
  {Scapellato},\ and\ \citenamefont {Steffens}}]{Alexandrou:2019lfo}%
  \BibitemOpen
  \bibfield  {author} {\bibinfo {author} {\bibfnamefont {C.}~\bibnamefont
  {Alexandrou}}, \bibinfo {author} {\bibfnamefont {K.}~\bibnamefont {Cichy}},
  \bibinfo {author} {\bibfnamefont {M.}~\bibnamefont {Constantinou}}, \bibinfo
  {author} {\bibfnamefont {K.}~\bibnamefont {Hadjiyiannakou}}, \bibinfo
  {author} {\bibfnamefont {K.}~\bibnamefont {Jansen}}, \bibinfo {author}
  {\bibfnamefont {A.}~\bibnamefont {Scapellato}}, \ and\ \bibinfo {author}
  {\bibfnamefont {F.}~\bibnamefont {Steffens}},\ }\bibfield  {title} {\enquote
  {\bibinfo {title} {{Systematic uncertainties in parton distribution functions
  from lattice QCD simulations at the physical point}},}\ }\href {\doibase
  10.1103/PhysRevD.99.114504} {\bibfield  {journal} {\bibinfo  {journal} {Phys.
  Rev. D}\ }\textbf {\bibinfo {volume} {99}},\ \bibinfo {pages} {114504}
  (\bibinfo {year} {2019})},\ \Eprint {http://arxiv.org/abs/1902.00587}
  {arXiv:1902.00587 [hep-lat]} \BibitemShut {NoStop}%
\bibitem [{\citenamefont {Zhang}\ \emph {et~al.}(2017)\citenamefont {Zhang},
  \citenamefont {Chen}, \citenamefont {Ji}, \citenamefont {Jin},\ and\
  \citenamefont {Lin}}]{Zhang:2017bzy}%
  \BibitemOpen
  \bibfield  {author} {\bibinfo {author} {\bibfnamefont {J.-H.}\ \bibnamefont
  {Zhang}}, \bibinfo {author} {\bibfnamefont {J.-W.}\ \bibnamefont {Chen}},
  \bibinfo {author} {\bibfnamefont {X.}~\bibnamefont {Ji}}, \bibinfo {author}
  {\bibfnamefont {L.}~\bibnamefont {Jin}}, \ and\ \bibinfo {author}
  {\bibfnamefont {H.-W.}\ \bibnamefont {Lin}},\ }\bibfield  {title} {\enquote
  {\bibinfo {title} {{Pion Distribution Amplitude from Lattice QCD}},}\ }\href
  {\doibase 10.1103/PhysRevD.95.094514} {\bibfield  {journal} {\bibinfo
  {journal} {Phys. Rev. D}\ }\textbf {\bibinfo {volume} {95}},\ \bibinfo
  {pages} {094514} (\bibinfo {year} {2017})},\ \Eprint
  {http://arxiv.org/abs/1702.00008} {arXiv:1702.00008 [hep-lat]} \BibitemShut
  {NoStop}%
\bibitem [{\citenamefont {Hua}\ \emph {et~al.}(2020)\citenamefont {Hua},
  \citenamefont {Chu}, \citenamefont {Sun}, \citenamefont {Wang}, \citenamefont
  {Xu}, \citenamefont {Yang}, \citenamefont {Zhang},\ and\ \citenamefont
  {Zhang}}]{Hua:2020gnw}%
  \BibitemOpen
  \bibfield  {author} {\bibinfo {author} {\bibfnamefont {J.}~\bibnamefont
  {Hua}}, \bibinfo {author} {\bibfnamefont {M.-H.}\ \bibnamefont {Chu}},
  \bibinfo {author} {\bibfnamefont {P.}~\bibnamefont {Sun}}, \bibinfo {author}
  {\bibfnamefont {W.}~\bibnamefont {Wang}}, \bibinfo {author} {\bibfnamefont
  {J.}~\bibnamefont {Xu}}, \bibinfo {author} {\bibfnamefont {Y.-B.}\
  \bibnamefont {Yang}}, \bibinfo {author} {\bibfnamefont {J.-H.}\ \bibnamefont
  {Zhang}}, \ and\ \bibinfo {author} {\bibfnamefont {Q.-A.}\ \bibnamefont
  {Zhang}},\ }\bibfield  {title} {\enquote {\bibinfo {title} {{Distribution
  Amplitudes of $K^*$ and $\phi$ at Physical Pion Mass from Lattice QCD}},}\
  }\href@noop {} {\  (\bibinfo {year} {2020})},\ \Eprint
  {http://arxiv.org/abs/2011.09788} {arXiv:2011.09788 [hep-lat]} \BibitemShut
  {NoStop}%
\bibitem [{\citenamefont {Wang}\ \emph {et~al.}(2020)\citenamefont {Wang},
  \citenamefont {Wang}, \citenamefont {Xu},\ and\ \citenamefont
  {Zhao}}]{Wang:2019msf}%
  \BibitemOpen
  \bibfield  {author} {\bibinfo {author} {\bibfnamefont {W.}~\bibnamefont
  {Wang}}, \bibinfo {author} {\bibfnamefont {Y.-M.}\ \bibnamefont {Wang}},
  \bibinfo {author} {\bibfnamefont {J.}~\bibnamefont {Xu}}, \ and\ \bibinfo
  {author} {\bibfnamefont {S.}~\bibnamefont {Zhao}},\ }\bibfield  {title}
  {\enquote {\bibinfo {title} {{$B$-meson light-cone distribution amplitude
  from Euclidean quantities}},}\ }\href {\doibase 10.1103/PhysRevD.102.011502}
  {\bibfield  {journal} {\bibinfo  {journal} {Phys. Rev. D}\ }\textbf {\bibinfo
  {volume} {102}},\ \bibinfo {pages} {011502} (\bibinfo {year} {2020})},\
  \Eprint {http://arxiv.org/abs/1908.09933} {arXiv:1908.09933 [hep-ph]}
  \BibitemShut {NoStop}%
\bibitem [{\citenamefont {Bhattacharya}\ \emph {et~al.}(2020)\citenamefont
  {Bhattacharya}, \citenamefont {Cichy}, \citenamefont {Constantinou},
  \citenamefont {Metz}, \citenamefont {Scapellato},\ and\ \citenamefont
  {Steffens}}]{Bhattacharya:2020cen}%
  \BibitemOpen
  \bibfield  {author} {\bibinfo {author} {\bibfnamefont {S.}~\bibnamefont
  {Bhattacharya}}, \bibinfo {author} {\bibfnamefont {K.}~\bibnamefont {Cichy}},
  \bibinfo {author} {\bibfnamefont {M.}~\bibnamefont {Constantinou}}, \bibinfo
  {author} {\bibfnamefont {A.}~\bibnamefont {Metz}}, \bibinfo {author}
  {\bibfnamefont {A.}~\bibnamefont {Scapellato}}, \ and\ \bibinfo {author}
  {\bibfnamefont {F.}~\bibnamefont {Steffens}},\ }\bibfield  {title} {\enquote
  {\bibinfo {title} {{Insights on proton structure from lattice QCD: The
  twist-3 parton distribution function $g_T(x)$}},}\ }\href {\doibase
  10.1103/PhysRevD.102.111501} {\bibfield  {journal} {\bibinfo  {journal}
  {Phys. Rev. D}\ }\textbf {\bibinfo {volume} {102}},\ \bibinfo {pages}
  {111501} (\bibinfo {year} {2020})},\ \Eprint
  {http://arxiv.org/abs/2004.04130} {arXiv:2004.04130 [hep-lat]} \BibitemShut
  {NoStop}%
\bibitem [{\citenamefont {Constantinou}(2021)}]{Constantinou:2020pek}%
  \BibitemOpen
  \bibfield  {author} {\bibinfo {author} {\bibfnamefont {M.}~\bibnamefont
  {Constantinou}},\ }\bibfield  {title} {\enquote {\bibinfo {title} {{The
  x-dependence of hadronic parton distributions: A review on the progress of
  lattice QCD}},}\ }\href {\doibase 10.1140/epja/s10050-021-00353-7} {\bibfield
   {journal} {\bibinfo  {journal} {Eur. Phys. J. A}\ }\textbf {\bibinfo
  {volume} {57}},\ \bibinfo {pages} {77} (\bibinfo {year} {2021})},\ \Eprint
  {http://arxiv.org/abs/2010.02445} {arXiv:2010.02445 [hep-lat]} \BibitemShut
  {NoStop}%
\bibitem [{\citenamefont {Cichy}\ and\ \citenamefont
  {Constantinou}(2019)}]{Cichy:2018mum}%
  \BibitemOpen
  \bibfield  {author} {\bibinfo {author} {\bibfnamefont {K.}~\bibnamefont
  {Cichy}}\ and\ \bibinfo {author} {\bibfnamefont {M.}~\bibnamefont
  {Constantinou}},\ }\bibfield  {title} {\enquote {\bibinfo {title} {{A guide
  to light-cone PDFs from Lattice QCD: an overview of approaches, techniques
  and results}},}\ }\href {\doibase 10.1155/2019/3036904} {\bibfield  {journal}
  {\bibinfo  {journal} {Adv. High Energy Phys.}\ }\textbf {\bibinfo {volume}
  {2019}},\ \bibinfo {pages} {3036904} (\bibinfo {year} {2019})},\ \Eprint
  {http://arxiv.org/abs/1811.07248} {arXiv:1811.07248 [hep-lat]} \BibitemShut
  {NoStop}%
\bibitem [{\citenamefont
  {Radyushkin}(2017{\natexlab{a}})}]{Radyushkin:2017cyf}%
  \BibitemOpen
  \bibfield  {author} {\bibinfo {author} {\bibfnamefont {A.~V.}\ \bibnamefont
  {Radyushkin}},\ }\bibfield  {title} {\enquote {\bibinfo {title}
  {{Quasi-parton distribution functions, momentum distributions, and
  pseudo-parton distribution functions}},}\ }\href {\doibase
  10.1103/PhysRevD.96.034025} {\bibfield  {journal} {\bibinfo  {journal} {Phys.
  Rev. D}\ }\textbf {\bibinfo {volume} {96}},\ \bibinfo {pages} {034025}
  (\bibinfo {year} {2017}{\natexlab{a}})},\ \Eprint
  {http://arxiv.org/abs/1705.01488} {arXiv:1705.01488 [hep-ph]} \BibitemShut
  {NoStop}%
\bibitem [{\citenamefont {Ma}\ and\ \citenamefont
  {Qiu}(2018{\natexlab{a}})}]{Ma:2014jla}%
  \BibitemOpen
  \bibfield  {author} {\bibinfo {author} {\bibfnamefont {Y.-Q.}\ \bibnamefont
  {Ma}}\ and\ \bibinfo {author} {\bibfnamefont {J.-W.}\ \bibnamefont {Qiu}},\
  }\bibfield  {title} {\enquote {\bibinfo {title} {{Extracting Parton
  Distribution Functions from Lattice QCD Calculations}},}\ }\href {\doibase
  10.1103/PhysRevD.98.074021} {\bibfield  {journal} {\bibinfo  {journal} {Phys.
  Rev. D}\ }\textbf {\bibinfo {volume} {98}},\ \bibinfo {pages} {074021}
  (\bibinfo {year} {2018}{\natexlab{a}})},\ \Eprint
  {http://arxiv.org/abs/1404.6860} {arXiv:1404.6860 [hep-ph]} \BibitemShut
  {NoStop}%
\bibitem [{\citenamefont {Ma}\ and\ \citenamefont
  {Qiu}(2018{\natexlab{b}})}]{Ma:2017pxb}%
  \BibitemOpen
  \bibfield  {author} {\bibinfo {author} {\bibfnamefont {Y.-Q.}\ \bibnamefont
  {Ma}}\ and\ \bibinfo {author} {\bibfnamefont {J.-W.}\ \bibnamefont {Qiu}},\
  }\bibfield  {title} {\enquote {\bibinfo {title} {{Exploring Partonic
  Structure of Hadrons Using ab initio Lattice QCD Calculations}},}\ }\href
  {\doibase 10.1103/PhysRevLett.120.022003} {\bibfield  {journal} {\bibinfo
  {journal} {Phys. Rev. Lett.}\ }\textbf {\bibinfo {volume} {120}},\ \bibinfo
  {pages} {022003} (\bibinfo {year} {2018}{\natexlab{b}})},\ \Eprint
  {http://arxiv.org/abs/1709.03018} {arXiv:1709.03018 [hep-ph]} \BibitemShut
  {NoStop}%
\bibitem [{\citenamefont {Sufian}\ \emph {et~al.}(2019)\citenamefont {Sufian},
  \citenamefont {Karpie}, \citenamefont {Egerer}, \citenamefont {Orginos},
  \citenamefont {Qiu},\ and\ \citenamefont {Richards}}]{Sufian:2019bol}%
  \BibitemOpen
  \bibfield  {author} {\bibinfo {author} {\bibfnamefont {R.~S.}\ \bibnamefont
  {Sufian}}, \bibinfo {author} {\bibfnamefont {J.}~\bibnamefont {Karpie}},
  \bibinfo {author} {\bibfnamefont {C.}~\bibnamefont {Egerer}}, \bibinfo
  {author} {\bibfnamefont {K.}~\bibnamefont {Orginos}}, \bibinfo {author}
  {\bibfnamefont {J.-W.}\ \bibnamefont {Qiu}}, \ and\ \bibinfo {author}
  {\bibfnamefont {D.~G.}\ \bibnamefont {Richards}},\ }\bibfield  {title}
  {\enquote {\bibinfo {title} {{Pion Valence Quark Distribution from Matrix
  Element Calculated in Lattice QCD}},}\ }\href {\doibase
  10.1103/PhysRevD.99.074507} {\bibfield  {journal} {\bibinfo  {journal} {Phys.
  Rev. D}\ }\textbf {\bibinfo {volume} {99}},\ \bibinfo {pages} {074507}
  (\bibinfo {year} {2019})},\ \Eprint {http://arxiv.org/abs/1901.03921}
  {arXiv:1901.03921 [hep-lat]} \BibitemShut {NoStop}%
\bibitem [{\citenamefont {Sufian}\ \emph {et~al.}(2020)\citenamefont {Sufian},
  \citenamefont {Egerer}, \citenamefont {Karpie}, \citenamefont {Edwards},
  \citenamefont {Jo\'o}, \citenamefont {Ma}, \citenamefont {Orginos},
  \citenamefont {Qiu},\ and\ \citenamefont {Richards}}]{Sufian:2020vzb}%
  \BibitemOpen
  \bibfield  {author} {\bibinfo {author} {\bibfnamefont {R.~S.}\ \bibnamefont
  {Sufian}}, \bibinfo {author} {\bibfnamefont {C.}~\bibnamefont {Egerer}},
  \bibinfo {author} {\bibfnamefont {J.}~\bibnamefont {Karpie}}, \bibinfo
  {author} {\bibfnamefont {R.~G.}\ \bibnamefont {Edwards}}, \bibinfo {author}
  {\bibfnamefont {B.}~\bibnamefont {Jo\'o}}, \bibinfo {author} {\bibfnamefont
  {Y.-Q.}\ \bibnamefont {Ma}}, \bibinfo {author} {\bibfnamefont
  {K.}~\bibnamefont {Orginos}}, \bibinfo {author} {\bibfnamefont {J.-W.}\
  \bibnamefont {Qiu}}, \ and\ \bibinfo {author} {\bibfnamefont {D.~G.}\
  \bibnamefont {Richards}},\ }\bibfield  {title} {\enquote {\bibinfo {title}
  {{Pion Valence Quark Distribution from Current-Current Correlation in Lattice
  QCD}},}\ }\href {\doibase 10.1103/PhysRevD.102.054508} {\bibfield  {journal}
  {\bibinfo  {journal} {Phys. Rev. D}\ }\textbf {\bibinfo {volume} {102}},\
  \bibinfo {pages} {054508} (\bibinfo {year} {2020})},\ \Eprint
  {http://arxiv.org/abs/2001.04960} {arXiv:2001.04960 [hep-lat]} \BibitemShut
  {NoStop}%
\bibitem [{\citenamefont {Ioffe}(1969)}]{Ioffe:1969kf}%
  \BibitemOpen
  \bibfield  {author} {\bibinfo {author} {\bibfnamefont {B.}~\bibnamefont
  {Ioffe}},\ }\bibfield  {title} {\enquote {\bibinfo {title} {{Space-time
  picture of photon and neutrino scattering and electroproduction cross-section
  asymptotics}},}\ }\href {\doibase 10.1016/0370-2693(69)90415-8} {\bibfield
  {journal} {\bibinfo  {journal} {Phys. Lett. B}\ }\textbf {\bibinfo {volume}
  {30}},\ \bibinfo {pages} {123} (\bibinfo {year} {1969})}\BibitemShut
  {NoStop}%
\bibitem [{\citenamefont {Braun}\ \emph {et~al.}(1995)\citenamefont {Braun},
  \citenamefont {Gornicki},\ and\ \citenamefont {Mankiewicz}}]{Braun:1994jq}%
  \BibitemOpen
  \bibfield  {author} {\bibinfo {author} {\bibfnamefont {V.}~\bibnamefont
  {Braun}}, \bibinfo {author} {\bibfnamefont {P.}~\bibnamefont {Gornicki}}, \
  and\ \bibinfo {author} {\bibfnamefont {L.}~\bibnamefont {Mankiewicz}},\
  }\bibfield  {title} {\enquote {\bibinfo {title} {{Ioffe - time distributions
  instead of parton momentum distributions in description of deep inelastic
  scattering}},}\ }\href {\doibase 10.1103/PhysRevD.51.6036} {\bibfield
  {journal} {\bibinfo  {journal} {Phys. Rev. D}\ }\textbf {\bibinfo {volume}
  {51}},\ \bibinfo {pages} {6036} (\bibinfo {year} {1995})},\ \Eprint
  {http://arxiv.org/abs/hep-ph/9410318} {arXiv:hep-ph/9410318} \BibitemShut
  {NoStop}%
\bibitem [{\citenamefont
  {Radyushkin}(2017{\natexlab{b}})}]{Radyushkin:2016hsy}%
  \BibitemOpen
  \bibfield  {author} {\bibinfo {author} {\bibfnamefont {A.}~\bibnamefont
  {Radyushkin}},\ }\bibfield  {title} {\enquote {\bibinfo {title}
  {{Nonperturbative Evolution of Parton Quasi-Distributions}},}\ }\href
  {\doibase 10.1016/j.physletb.2017.02.019} {\bibfield  {journal} {\bibinfo
  {journal} {Phys. Lett. B}\ }\textbf {\bibinfo {volume} {767}},\ \bibinfo
  {pages} {314} (\bibinfo {year} {2017}{\natexlab{b}})},\ \Eprint
  {http://arxiv.org/abs/1612.05170} {arXiv:1612.05170 [hep-ph]} \BibitemShut
  {NoStop}%
\bibitem [{\citenamefont {Polyakov}(1980)}]{Polyakov:1980ca}%
  \BibitemOpen
  \bibfield  {author} {\bibinfo {author} {\bibfnamefont {A.~M.}\ \bibnamefont
  {Polyakov}},\ }\bibfield  {title} {\enquote {\bibinfo {title} {{Gauge Fields
  as Rings of Glue}},}\ }\href {\doibase 10.1016/0550-3213(80)90507-6}
  {\bibfield  {journal} {\bibinfo  {journal} {Nucl. Phys. B}\ }\textbf
  {\bibinfo {volume} {164}},\ \bibinfo {pages} {171} (\bibinfo {year}
  {1980})}\BibitemShut {NoStop}%
\bibitem [{\citenamefont {Dotsenko}\ and\ \citenamefont
  {Vergeles}(1980)}]{Dotsenko:1979wb}%
  \BibitemOpen
  \bibfield  {author} {\bibinfo {author} {\bibfnamefont {V.~S.}\ \bibnamefont
  {Dotsenko}}\ and\ \bibinfo {author} {\bibfnamefont {S.~N.}\ \bibnamefont
  {Vergeles}},\ }\bibfield  {title} {\enquote {\bibinfo {title}
  {{Renormalizability of Phase Factors in the Nonabelian Gauge Theory}},}\
  }\href {\doibase 10.1016/0550-3213(80)90103-0} {\bibfield  {journal}
  {\bibinfo  {journal} {Nucl. Phys. B}\ }\textbf {\bibinfo {volume} {169}},\
  \bibinfo {pages} {527} (\bibinfo {year} {1980})}\BibitemShut {NoStop}%
\bibitem [{\citenamefont {Brandt}\ \emph {et~al.}(1981)\citenamefont {Brandt},
  \citenamefont {Neri},\ and\ \citenamefont {Sato}}]{Brandt:1981kf}%
  \BibitemOpen
  \bibfield  {author} {\bibinfo {author} {\bibfnamefont {R.~A.}\ \bibnamefont
  {Brandt}}, \bibinfo {author} {\bibfnamefont {F.}~\bibnamefont {Neri}}, \ and\
  \bibinfo {author} {\bibfnamefont {M.-a.}\ \bibnamefont {Sato}},\ }\bibfield
  {title} {\enquote {\bibinfo {title} {{Renormalization of Loop Functions for
  All Loops}},}\ }\href {\doibase 10.1103/PhysRevD.24.879} {\bibfield
  {journal} {\bibinfo  {journal} {Phys. Rev. D}\ }\textbf {\bibinfo {volume}
  {24}},\ \bibinfo {pages} {879} (\bibinfo {year} {1981})}\BibitemShut
  {NoStop}%
\bibitem [{\citenamefont {Craigie}\ and\ \citenamefont
  {Dorn}(1981)}]{Craigie:1980qs}%
  \BibitemOpen
  \bibfield  {author} {\bibinfo {author} {\bibfnamefont {N.~S.}\ \bibnamefont
  {Craigie}}\ and\ \bibinfo {author} {\bibfnamefont {H.}~\bibnamefont {Dorn}},\
  }\bibfield  {title} {\enquote {\bibinfo {title} {{On the Renormalization and
  Short Distance Properties of Hadronic Operators in {QCD}}},}\ }\href
  {\doibase 10.1016/0550-3213(81)90372-2} {\bibfield  {journal} {\bibinfo
  {journal} {Nucl. Phys. B}\ }\textbf {\bibinfo {volume} {185}},\ \bibinfo
  {pages} {204} (\bibinfo {year} {1981})}\BibitemShut {NoStop}%
\bibitem [{\citenamefont {Ishikawa}\ \emph {et~al.}(2017)\citenamefont
  {Ishikawa}, \citenamefont {Ma}, \citenamefont {Qiu},\ and\ \citenamefont
  {Yoshida}}]{Ishikawa:2017faj}%
  \BibitemOpen
  \bibfield  {author} {\bibinfo {author} {\bibfnamefont {T.}~\bibnamefont
  {Ishikawa}}, \bibinfo {author} {\bibfnamefont {Y.-Q.}\ \bibnamefont {Ma}},
  \bibinfo {author} {\bibfnamefont {J.-W.}\ \bibnamefont {Qiu}}, \ and\
  \bibinfo {author} {\bibfnamefont {S.}~\bibnamefont {Yoshida}},\ }\bibfield
  {title} {\enquote {\bibinfo {title} {{Renormalizability of quasiparton
  distribution functions}},}\ }\href {\doibase 10.1103/PhysRevD.96.094019}
  {\bibfield  {journal} {\bibinfo  {journal} {Phys. Rev. D}\ }\textbf {\bibinfo
  {volume} {96}},\ \bibinfo {pages} {094019} (\bibinfo {year} {2017})},\
  \Eprint {http://arxiv.org/abs/1707.03107} {arXiv:1707.03107 [hep-ph]}
  \BibitemShut {NoStop}%
\bibitem [{\citenamefont {Ji}\ \emph {et~al.}(2018)\citenamefont {Ji},
  \citenamefont {Zhang},\ and\ \citenamefont {Zhao}}]{Ji:2017oey}%
  \BibitemOpen
  \bibfield  {author} {\bibinfo {author} {\bibfnamefont {X.}~\bibnamefont
  {Ji}}, \bibinfo {author} {\bibfnamefont {J.-H.}\ \bibnamefont {Zhang}}, \
  and\ \bibinfo {author} {\bibfnamefont {Y.}~\bibnamefont {Zhao}},\ }\bibfield
  {title} {\enquote {\bibinfo {title} {{Renormalization in Large Momentum
  Effective Theory of Parton Physics}},}\ }\href {\doibase
  10.1103/PhysRevLett.120.112001} {\bibfield  {journal} {\bibinfo  {journal}
  {Phys. Rev. Lett.}\ }\textbf {\bibinfo {volume} {120}},\ \bibinfo {pages}
  {112001} (\bibinfo {year} {2018})},\ \Eprint
  {http://arxiv.org/abs/1706.08962} {arXiv:1706.08962 [hep-ph]} \BibitemShut
  {NoStop}%
\bibitem [{\citenamefont {Green}\ \emph {et~al.}(2018)\citenamefont {Green},
  \citenamefont {Jansen},\ and\ \citenamefont {Steffens}}]{Green:2017xeu}%
  \BibitemOpen
  \bibfield  {author} {\bibinfo {author} {\bibfnamefont {J.}~\bibnamefont
  {Green}}, \bibinfo {author} {\bibfnamefont {K.}~\bibnamefont {Jansen}}, \
  and\ \bibinfo {author} {\bibfnamefont {F.}~\bibnamefont {Steffens}},\
  }\bibfield  {title} {\enquote {\bibinfo {title} {{Nonperturbative
  Renormalization of Nonlocal Quark Bilinears for Parton Quasidistribution
  Functions on the Lattice Using an Auxiliary Field}},}\ }\href {\doibase
  10.1103/PhysRevLett.121.022004} {\bibfield  {journal} {\bibinfo  {journal}
  {Phys. Rev. Lett.}\ }\textbf {\bibinfo {volume} {121}},\ \bibinfo {pages}
  {022004} (\bibinfo {year} {2018})},\ \Eprint
  {http://arxiv.org/abs/1707.07152} {arXiv:1707.07152 [hep-lat]} \BibitemShut
  {NoStop}%
\bibitem [{\citenamefont {Orginos}\ \emph {et~al.}(2017)\citenamefont
  {Orginos}, \citenamefont {Radyushkin}, \citenamefont {Karpie},\ and\
  \citenamefont {Zafeiropoulos}}]{Orginos:2017kos}%
  \BibitemOpen
  \bibfield  {author} {\bibinfo {author} {\bibfnamefont {K.}~\bibnamefont
  {Orginos}}, \bibinfo {author} {\bibfnamefont {A.}~\bibnamefont {Radyushkin}},
  \bibinfo {author} {\bibfnamefont {J.}~\bibnamefont {Karpie}}, \ and\ \bibinfo
  {author} {\bibfnamefont {S.}~\bibnamefont {Zafeiropoulos}},\ }\bibfield
  {title} {\enquote {\bibinfo {title} {{Lattice QCD exploration of parton
  pseudo-distribution functions}},}\ }\href {\doibase
  10.1103/PhysRevD.96.094503} {\bibfield  {journal} {\bibinfo  {journal} {Phys.
  Rev. D}\ }\textbf {\bibinfo {volume} {96}},\ \bibinfo {pages} {094503}
  (\bibinfo {year} {2017})},\ \Eprint {http://arxiv.org/abs/1706.05373}
  {arXiv:1706.05373 [hep-ph]} \BibitemShut {NoStop}%
\bibitem [{\citenamefont {Braun}\ \emph {et~al.}(2019)\citenamefont {Braun},
  \citenamefont {Vladimirov},\ and\ \citenamefont {Zhang}}]{Braun:2018brg}%
  \BibitemOpen
  \bibfield  {author} {\bibinfo {author} {\bibfnamefont {V.~M.}\ \bibnamefont
  {Braun}}, \bibinfo {author} {\bibfnamefont {A.}~\bibnamefont {Vladimirov}}, \
  and\ \bibinfo {author} {\bibfnamefont {J.-H.}\ \bibnamefont {Zhang}},\
  }\bibfield  {title} {\enquote {\bibinfo {title} {{Power corrections and
  renormalons in parton quasidistributions}},}\ }\href {\doibase
  10.1103/PhysRevD.99.014013} {\bibfield  {journal} {\bibinfo  {journal} {Phys.
  Rev. D}\ }\textbf {\bibinfo {volume} {99}},\ \bibinfo {pages} {014013}
  (\bibinfo {year} {2019})},\ \Eprint {http://arxiv.org/abs/1810.00048}
  {arXiv:1810.00048 [hep-ph]} \BibitemShut {NoStop}%
\bibitem [{\citenamefont {Li}\ \emph {et~al.}(2021)\citenamefont {Li},
  \citenamefont {Ma},\ and\ \citenamefont {Qiu}}]{Li:2020xml}%
  \BibitemOpen
  \bibfield  {author} {\bibinfo {author} {\bibfnamefont {Z.-Y.}\ \bibnamefont
  {Li}}, \bibinfo {author} {\bibfnamefont {Y.-Q.}\ \bibnamefont {Ma}}, \ and\
  \bibinfo {author} {\bibfnamefont {J.-W.}\ \bibnamefont {Qiu}},\ }\bibfield
  {title} {\enquote {\bibinfo {title} {{Extraction of
  Next-to-Next-to-Leading-Order Parton Distribution Functions from Lattice QCD
  Calculations}},}\ }\href {\doibase 10.1103/PhysRevLett.126.072001} {\bibfield
   {journal} {\bibinfo  {journal} {Phys. Rev. Lett.}\ }\textbf {\bibinfo
  {volume} {126}},\ \bibinfo {pages} {072001} (\bibinfo {year} {2021})},\
  \Eprint {http://arxiv.org/abs/2006.12370} {arXiv:2006.12370 [hep-ph]}
  \BibitemShut {NoStop}%
\bibitem [{\citenamefont {Izubuchi}\ \emph {et~al.}(2018)\citenamefont
  {Izubuchi}, \citenamefont {Ji}, \citenamefont {Jin}, \citenamefont
  {Stewart},\ and\ \citenamefont {Zhao}}]{Izubuchi:2018srq}%
  \BibitemOpen
  \bibfield  {author} {\bibinfo {author} {\bibfnamefont {T.}~\bibnamefont
  {Izubuchi}}, \bibinfo {author} {\bibfnamefont {X.}~\bibnamefont {Ji}},
  \bibinfo {author} {\bibfnamefont {L.}~\bibnamefont {Jin}}, \bibinfo {author}
  {\bibfnamefont {I.~W.}\ \bibnamefont {Stewart}}, \ and\ \bibinfo {author}
  {\bibfnamefont {Y.}~\bibnamefont {Zhao}},\ }\bibfield  {title} {\enquote
  {\bibinfo {title} {{Factorization Theorem Relating Euclidean and Light-Cone
  Parton Distributions}},}\ }\href {\doibase 10.1103/PhysRevD.98.056004}
  {\bibfield  {journal} {\bibinfo  {journal} {Phys. Rev. D}\ }\textbf {\bibinfo
  {volume} {98}},\ \bibinfo {pages} {056004} (\bibinfo {year} {2018})},\
  \Eprint {http://arxiv.org/abs/1801.03917} {arXiv:1801.03917 [hep-ph]}
  \BibitemShut {NoStop}%
\bibitem [{\citenamefont {Radyushkin}(2018)}]{Radyushkin:2018cvn}%
  \BibitemOpen
  \bibfield  {author} {\bibinfo {author} {\bibfnamefont {A.}~\bibnamefont
  {Radyushkin}},\ }\bibfield  {title} {\enquote {\bibinfo {title} {{One-loop
  evolution of parton pseudo-distribution functions on the lattice}},}\ }\href
  {\doibase 10.1103/PhysRevD.98.014019} {\bibfield  {journal} {\bibinfo
  {journal} {Phys. Rev. D}\ }\textbf {\bibinfo {volume} {98}},\ \bibinfo
  {pages} {014019} (\bibinfo {year} {2018})},\ \Eprint
  {http://arxiv.org/abs/1801.02427} {arXiv:1801.02427 [hep-ph]} \BibitemShut
  {NoStop}%
\bibitem [{\citenamefont {Zhang}\ \emph {et~al.}(2018)\citenamefont {Zhang},
  \citenamefont {Chen},\ and\ \citenamefont {Monahan}}]{Zhang:2018ggy}%
  \BibitemOpen
  \bibfield  {author} {\bibinfo {author} {\bibfnamefont {J.-H.}\ \bibnamefont
  {Zhang}}, \bibinfo {author} {\bibfnamefont {J.-W.}\ \bibnamefont {Chen}}, \
  and\ \bibinfo {author} {\bibfnamefont {C.}~\bibnamefont {Monahan}},\
  }\bibfield  {title} {\enquote {\bibinfo {title} {{Parton distribution
  functions from reduced Ioffe-time distributions}},}\ }\href {\doibase
  10.1103/PhysRevD.97.074508} {\bibfield  {journal} {\bibinfo  {journal} {Phys.
  Rev. D}\ }\textbf {\bibinfo {volume} {97}},\ \bibinfo {pages} {074508}
  (\bibinfo {year} {2018})},\ \Eprint {http://arxiv.org/abs/1801.03023}
  {arXiv:1801.03023 [hep-ph]} \BibitemShut {NoStop}%
\bibitem [{\citenamefont {Chen}\ \emph {et~al.}(2021)\citenamefont {Chen},
  \citenamefont {Wang},\ and\ \citenamefont {Zhu}}]{Chen:2020ody}%
  \BibitemOpen
  \bibfield  {author} {\bibinfo {author} {\bibfnamefont {L.-B.}\ \bibnamefont
  {Chen}}, \bibinfo {author} {\bibfnamefont {W.}~\bibnamefont {Wang}}, \ and\
  \bibinfo {author} {\bibfnamefont {R.}~\bibnamefont {Zhu}},\ }\bibfield
  {title} {\enquote {\bibinfo {title} {{Next-to-Next-to-Leading Order
  Calculation of Quasiparton Distribution Functions}},}\ }\href {\doibase
  10.1103/PhysRevLett.126.072002} {\bibfield  {journal} {\bibinfo  {journal}
  {Phys. Rev. Lett.}\ }\textbf {\bibinfo {volume} {126}},\ \bibinfo {pages}
  {072002} (\bibinfo {year} {2021})},\ \Eprint
  {http://arxiv.org/abs/2006.14825} {arXiv:2006.14825 [hep-ph]} \BibitemShut
  {NoStop}%
\bibitem [{\citenamefont {Dokshitzer}(1977)}]{Dokshitzer:1977sg}%
  \BibitemOpen
  \bibfield  {author} {\bibinfo {author} {\bibfnamefont {Y.~L.}\ \bibnamefont
  {Dokshitzer}},\ }\bibfield  {title} {\enquote {\bibinfo {title} {{Calculation
  of the Structure Functions for Deep Inelastic Scattering and e+ e-
  Annihilation by Perturbation Theory in Quantum Chromodynamics.}}}\
  }\href@noop {} {\bibfield  {journal} {\bibinfo  {journal} {Sov. Phys. JETP}\
  }\textbf {\bibinfo {volume} {46}},\ \bibinfo {pages} {641} (\bibinfo {year}
  {1977})}\BibitemShut {NoStop}%
\bibitem [{\citenamefont {Gribov}\ and\ \citenamefont
  {Lipatov}(1972)}]{Gribov:1972rt}%
  \BibitemOpen
  \bibfield  {author} {\bibinfo {author} {\bibfnamefont {V.~N.}\ \bibnamefont
  {Gribov}}\ and\ \bibinfo {author} {\bibfnamefont {L.~N.}\ \bibnamefont
  {Lipatov}},\ }\bibfield  {title} {\enquote {\bibinfo {title} {{e+ e- pair
  annihilation and deep inelastic e p scattering in perturbation theory}},}\
  }\href@noop {} {\bibfield  {journal} {\bibinfo  {journal} {Sov. J. Nucl.
  Phys.}\ }\textbf {\bibinfo {volume} {15}},\ \bibinfo {pages} {675} (\bibinfo
  {year} {1972})}\BibitemShut {NoStop}%
\bibitem [{\citenamefont {Altarelli}\ and\ \citenamefont
  {Parisi}(1977)}]{Altarelli:1977zs}%
  \BibitemOpen
  \bibfield  {author} {\bibinfo {author} {\bibfnamefont {G.}~\bibnamefont
  {Altarelli}}\ and\ \bibinfo {author} {\bibfnamefont {G.}~\bibnamefont
  {Parisi}},\ }\bibfield  {title} {\enquote {\bibinfo {title} {{Asymptotic
  Freedom in Parton Language}},}\ }\href {\doibase
  10.1016/0550-3213(77)90384-4} {\bibfield  {journal} {\bibinfo  {journal}
  {Nucl. Phys. B}\ }\textbf {\bibinfo {volume} {126}},\ \bibinfo {pages} {298}
  (\bibinfo {year} {1977})}\BibitemShut {NoStop}%
\bibitem [{\citenamefont {Jo\'o}\ \emph
  {et~al.}(2019{\natexlab{a}})\citenamefont {Jo\'o}, \citenamefont {Karpie},
  \citenamefont {Orginos}, \citenamefont {Radyushkin}, \citenamefont
  {Richards}, \citenamefont {Sufian},\ and\ \citenamefont
  {Zafeiropoulos}}]{Joo:2019bzr}%
  \BibitemOpen
  \bibfield  {author} {\bibinfo {author} {\bibfnamefont {B.}~\bibnamefont
  {Jo\'o}}, \bibinfo {author} {\bibfnamefont {J.}~\bibnamefont {Karpie}},
  \bibinfo {author} {\bibfnamefont {K.}~\bibnamefont {Orginos}}, \bibinfo
  {author} {\bibfnamefont {A.~V.}\ \bibnamefont {Radyushkin}}, \bibinfo
  {author} {\bibfnamefont {D.~G.}\ \bibnamefont {Richards}}, \bibinfo {author}
  {\bibfnamefont {R.~S.}\ \bibnamefont {Sufian}}, \ and\ \bibinfo {author}
  {\bibfnamefont {S.}~\bibnamefont {Zafeiropoulos}},\ }\bibfield  {title}
  {\enquote {\bibinfo {title} {{Pion valence structure from Ioffe-time parton
  pseudodistribution functions}},}\ }\href {\doibase
  10.1103/PhysRevD.100.114512} {\bibfield  {journal} {\bibinfo  {journal}
  {Phys. Rev. D}\ }\textbf {\bibinfo {volume} {100}},\ \bibinfo {pages}
  {114512} (\bibinfo {year} {2019}{\natexlab{a}})},\ \Eprint
  {http://arxiv.org/abs/1909.08517} {arXiv:1909.08517 [hep-lat]} \BibitemShut
  {NoStop}%
\bibitem [{\citenamefont {Karpie}\ \emph {et~al.}(2021)\citenamefont {Karpie},
  \citenamefont {Orginos}, \citenamefont {Radyushkin},\ and\ \citenamefont
  {Zafeiropoulos}}]{Karpie:2021pap}%
  \BibitemOpen
  \bibfield  {author} {\bibinfo {author} {\bibfnamefont {J.}~\bibnamefont
  {Karpie}}, \bibinfo {author} {\bibfnamefont {K.}~\bibnamefont {Orginos}},
  \bibinfo {author} {\bibfnamefont {A.}~\bibnamefont {Radyushkin}}, \ and\
  \bibinfo {author} {\bibfnamefont {S.}~\bibnamefont {Zafeiropoulos}},\
  }\bibfield  {title} {\enquote {\bibinfo {title} {{The Continuum and Leading
  Twist Limits of Parton Distribution Functions in Lattice QCD}},}\ }\href@noop
  {} {\  (\bibinfo {year} {2021})},\ \Eprint {http://arxiv.org/abs/2105.13313}
  {arXiv:2105.13313 [hep-lat]} \BibitemShut {NoStop}%
\bibitem [{\citenamefont {Jo\'o}\ \emph {et~al.}(2020)\citenamefont {Jo\'o},
  \citenamefont {Karpie}, \citenamefont {Orginos}, \citenamefont {Radyushkin},
  \citenamefont {Richards},\ and\ \citenamefont {Zafeiropoulos}}]{Joo:2020spy}%
  \BibitemOpen
  \bibfield  {author} {\bibinfo {author} {\bibfnamefont {B.}~\bibnamefont
  {Jo\'o}}, \bibinfo {author} {\bibfnamefont {J.}~\bibnamefont {Karpie}},
  \bibinfo {author} {\bibfnamefont {K.}~\bibnamefont {Orginos}}, \bibinfo
  {author} {\bibfnamefont {A.~V.}\ \bibnamefont {Radyushkin}}, \bibinfo
  {author} {\bibfnamefont {D.~G.}\ \bibnamefont {Richards}}, \ and\ \bibinfo
  {author} {\bibfnamefont {S.}~\bibnamefont {Zafeiropoulos}},\ }\bibfield
  {title} {\enquote {\bibinfo {title} {{Parton Distribution Functions from
  Ioffe Time Pseudodistributions from Lattice Calculations: Approaching the
  Physical Point}},}\ }\href {\doibase 10.1103/PhysRevLett.125.232003}
  {\bibfield  {journal} {\bibinfo  {journal} {Phys. Rev. Lett.}\ }\textbf
  {\bibinfo {volume} {125}},\ \bibinfo {pages} {232003} (\bibinfo {year}
  {2020})},\ \Eprint {http://arxiv.org/abs/2004.01687} {arXiv:2004.01687
  [hep-lat]} \BibitemShut {NoStop}%
\bibitem [{\citenamefont {Bhat}\ \emph {et~al.}(2021)\citenamefont {Bhat},
  \citenamefont {Cichy}, \citenamefont {Constantinou},\ and\ \citenamefont
  {Scapellato}}]{Bhat:2020ktg}%
  \BibitemOpen
  \bibfield  {author} {\bibinfo {author} {\bibfnamefont {M.}~\bibnamefont
  {Bhat}}, \bibinfo {author} {\bibfnamefont {K.}~\bibnamefont {Cichy}},
  \bibinfo {author} {\bibfnamefont {M.}~\bibnamefont {Constantinou}}, \ and\
  \bibinfo {author} {\bibfnamefont {A.}~\bibnamefont {Scapellato}},\ }\bibfield
   {title} {\enquote {\bibinfo {title} {{Flavor nonsinglet parton distribution
  functions from lattice QCD at physical quark masses via the
  pseudodistribution approach}},}\ }\href {\doibase
  10.1103/PhysRevD.103.034510} {\bibfield  {journal} {\bibinfo  {journal}
  {Phys. Rev. D}\ }\textbf {\bibinfo {volume} {103}},\ \bibinfo {pages}
  {034510} (\bibinfo {year} {2021})},\ \Eprint
  {http://arxiv.org/abs/2005.02102} {arXiv:2005.02102 [hep-lat]} \BibitemShut
  {NoStop}%
\bibitem [{\citenamefont {Jo\'o}\ \emph
  {et~al.}(2019{\natexlab{b}})\citenamefont {Jo\'o}, \citenamefont {Karpie},
  \citenamefont {Orginos}, \citenamefont {Radyushkin}, \citenamefont
  {Richards},\ and\ \citenamefont {Zafeiropoulos}}]{Joo:2019jct}%
  \BibitemOpen
  \bibfield  {author} {\bibinfo {author} {\bibfnamefont {B.}~\bibnamefont
  {Jo\'o}}, \bibinfo {author} {\bibfnamefont {J.}~\bibnamefont {Karpie}},
  \bibinfo {author} {\bibfnamefont {K.}~\bibnamefont {Orginos}}, \bibinfo
  {author} {\bibfnamefont {A.}~\bibnamefont {Radyushkin}}, \bibinfo {author}
  {\bibfnamefont {D.}~\bibnamefont {Richards}}, \ and\ \bibinfo {author}
  {\bibfnamefont {S.}~\bibnamefont {Zafeiropoulos}},\ }\bibfield  {title}
  {\enquote {\bibinfo {title} {{Parton Distribution Functions from Ioffe time
  pseudo-distributions}},}\ }\href {\doibase 10.1007/JHEP12(2019)081}
  {\bibfield  {journal} {\bibinfo  {journal} {JHEP}\ }\textbf {\bibinfo
  {volume} {12}},\ \bibinfo {pages} {081} (\bibinfo {year}
  {2019}{\natexlab{b}})},\ \Eprint {http://arxiv.org/abs/1908.09771}
  {arXiv:1908.09771 [hep-lat]} \BibitemShut {NoStop}%
\bibitem [{\citenamefont {Fan}\ \emph {et~al.}(2021)\citenamefont {Fan},
  \citenamefont {Zhang},\ and\ \citenamefont {Lin}}]{Fan:2020cpa}%
  \BibitemOpen
  \bibfield  {author} {\bibinfo {author} {\bibfnamefont {Z.}~\bibnamefont
  {Fan}}, \bibinfo {author} {\bibfnamefont {R.}~\bibnamefont {Zhang}}, \ and\
  \bibinfo {author} {\bibfnamefont {H.-W.}\ \bibnamefont {Lin}},\ }\bibfield
  {title} {\enquote {\bibinfo {title} {{Nucleon gluon distribution function
  from 2 + 1 + 1-flavor lattice QCD}},}\ }\href {\doibase
  10.1142/S0217751X21500809} {\bibfield  {journal} {\bibinfo  {journal} {Int.
  J. Mod. Phys. A}\ }\textbf {\bibinfo {volume} {36}},\ \bibinfo {pages}
  {2150080} (\bibinfo {year} {2021})},\ \Eprint
  {http://arxiv.org/abs/2007.16113} {arXiv:2007.16113 [hep-lat]} \BibitemShut
  {NoStop}%
\bibitem [{\citenamefont {Egerer}\ \emph {et~al.}(2021)\citenamefont {Egerer},
  \citenamefont {Edwards}, \citenamefont {Orginos},\ and\ \citenamefont
  {Richards}}]{Egerer:2020hnc}%
  \BibitemOpen
  \bibfield  {author} {\bibinfo {author} {\bibfnamefont {C.}~\bibnamefont
  {Egerer}}, \bibinfo {author} {\bibfnamefont {R.~G.}\ \bibnamefont {Edwards}},
  \bibinfo {author} {\bibfnamefont {K.}~\bibnamefont {Orginos}}, \ and\
  \bibinfo {author} {\bibfnamefont {D.~G.}\ \bibnamefont {Richards}},\
  }\bibfield  {title} {\enquote {\bibinfo {title} {{Distillation at
  High-Momentum}},}\ }\href {\doibase 10.1103/PhysRevD.103.034502} {\bibfield
  {journal} {\bibinfo  {journal} {Phys. Rev. D}\ }\textbf {\bibinfo {volume}
  {103}},\ \bibinfo {pages} {034502} (\bibinfo {year} {2021})},\ \Eprint
  {http://arxiv.org/abs/2009.10691} {arXiv:2009.10691 [hep-lat]} \BibitemShut
  {NoStop}%
\bibitem [{\citenamefont {Edwards}\ \emph {et~al.}(2016)\citenamefont
  {Edwards}, \citenamefont {Jo\'o}, \citenamefont {Orginos}, \citenamefont
  {Richards},\ and\ \citenamefont {Winter}}]{jlab-wm-lanl}%
  \BibitemOpen
  \bibfield  {author} {\bibinfo {author} {\bibfnamefont {R.}~\bibnamefont
  {Edwards}}, \bibinfo {author} {\bibfnamefont {B.}~\bibnamefont {Jo\'o}},
  \bibinfo {author} {\bibfnamefont {K.}~\bibnamefont {Orginos}}, \bibinfo
  {author} {\bibfnamefont {D.}~\bibnamefont {Richards}}, \ and\ \bibinfo
  {author} {\bibfnamefont {F.}~\bibnamefont {Winter}},\ }\href@noop {}
  {\enquote {\bibinfo {title} {{U.S. 2+1 flavor clover lattice generation
  program}},}\ } (\bibinfo {year} {2016}),\ \bibinfo {note}
  {unpublished}\BibitemShut {NoStop}%
\bibitem [{\citenamefont {Borsanyi}\ \emph {et~al.}(2012)\citenamefont
  {Borsanyi} \emph {et~al.}}]{Borsanyi:2012zs}%
  \BibitemOpen
  \bibfield  {author} {\bibinfo {author} {\bibfnamefont {S.}~\bibnamefont
  {Borsanyi}} \emph {et~al.},\ }\bibfield  {title} {\enquote {\bibinfo {title}
  {{High-precision scale setting in lattice QCD}},}\ }\href {\doibase
  10.1007/JHEP09(2012)010} {\bibfield  {journal} {\bibinfo  {journal} {JHEP}\
  }\textbf {\bibinfo {volume} {09}},\ \bibinfo {pages} {010} (\bibinfo {year}
  {2012})},\ \Eprint {http://arxiv.org/abs/1203.4469} {arXiv:1203.4469
  [hep-lat]} \BibitemShut {NoStop}%
\bibitem [{\citenamefont {Yoon}\ \emph {et~al.}(2017)\citenamefont {Yoon} \emph
  {et~al.}}]{Yoon:2016jzj}%
  \BibitemOpen
  \bibfield  {author} {\bibinfo {author} {\bibfnamefont {B.}~\bibnamefont
  {Yoon}} \emph {et~al.},\ }\bibfield  {title} {\enquote {\bibinfo {title}
  {{Isovector charges of the nucleon from 2+1-flavor QCD with clover
  fermions}},}\ }\href {\doibase 10.1103/PhysRevD.95.074508} {\bibfield
  {journal} {\bibinfo  {journal} {Phys. Rev.}\ }\textbf {\bibinfo {volume}
  {D95}},\ \bibinfo {pages} {074508} (\bibinfo {year} {2017})},\ \Eprint
  {http://arxiv.org/abs/1611.07452} {arXiv:1611.07452 [hep-lat]} \BibitemShut
  {NoStop}%
\bibitem [{\citenamefont {Yoon}\ \emph {et~al.}(2016)\citenamefont {Yoon} \emph
  {et~al.}}]{Yoon:2016dij}%
  \BibitemOpen
  \bibfield  {author} {\bibinfo {author} {\bibfnamefont {B.}~\bibnamefont
  {Yoon}} \emph {et~al.},\ }\bibfield  {title} {\enquote {\bibinfo {title}
  {{Controlling Excited-State Contamination in Nucleon Matrix Elements}},}\
  }\href {\doibase 10.1103/PhysRevD.93.114506} {\bibfield  {journal} {\bibinfo
  {journal} {Phys. Rev.}\ }\textbf {\bibinfo {volume} {D93}},\ \bibinfo {pages}
  {114506} (\bibinfo {year} {2016})},\ \Eprint
  {http://arxiv.org/abs/1602.07737} {arXiv:1602.07737 [hep-lat]} \BibitemShut
  {NoStop}%
\bibitem [{\citenamefont {Morningstar}\ and\ \citenamefont
  {Peardon}(2004)}]{Morningstar:2003gk}%
  \BibitemOpen
  \bibfield  {author} {\bibinfo {author} {\bibfnamefont {C.}~\bibnamefont
  {Morningstar}}\ and\ \bibinfo {author} {\bibfnamefont {M.~J.}\ \bibnamefont
  {Peardon}},\ }\bibfield  {title} {\enquote {\bibinfo {title} {{Analytic
  smearing of SU(3) link variables in lattice QCD}},}\ }\href {\doibase
  10.1103/PhysRevD.69.054501} {\bibfield  {journal} {\bibinfo  {journal} {Phys.
  Rev. D}\ }\textbf {\bibinfo {volume} {69}},\ \bibinfo {pages} {054501}
  (\bibinfo {year} {2004})},\ \Eprint {http://arxiv.org/abs/hep-lat/0311018}
  {arXiv:hep-lat/0311018} \BibitemShut {NoStop}%
\bibitem [{\citenamefont {Allton}\ \emph {et~al.}(1993)\citenamefont {Allton}
  \emph {et~al.}}]{Allton:1993wc}%
  \BibitemOpen
  \bibfield  {author} {\bibinfo {author} {\bibfnamefont {C.~R.}\ \bibnamefont
  {Allton}} \emph {et~al.} (\bibinfo {collaboration} {UKQCD}),\ }\bibfield
  {title} {\enquote {\bibinfo {title} {{Gauge invariant smearing and matrix
  correlators using Wilson fermions at Beta = 6.2}},}\ }\href {\doibase
  10.1103/PhysRevD.47.5128} {\bibfield  {journal} {\bibinfo  {journal} {Phys.
  Rev. D}\ }\textbf {\bibinfo {volume} {47}},\ \bibinfo {pages} {5128}
  (\bibinfo {year} {1993})},\ \Eprint {http://arxiv.org/abs/hep-lat/9303009}
  {arXiv:hep-lat/9303009} \BibitemShut {NoStop}%
\bibitem [{\citenamefont {Peardon}\ \emph {et~al.}(2009)\citenamefont
  {Peardon}, \citenamefont {Bulava}, \citenamefont {Foley}, \citenamefont
  {Morningstar}, \citenamefont {Dudek}, \citenamefont {Edwards}, \citenamefont
  {Joo}, \citenamefont {Lin}, \citenamefont {Richards},\ and\ \citenamefont
  {Juge}}]{Peardon:2009gh}%
  \BibitemOpen
  \bibfield  {author} {\bibinfo {author} {\bibfnamefont {M.}~\bibnamefont
  {Peardon}}, \bibinfo {author} {\bibfnamefont {J.}~\bibnamefont {Bulava}},
  \bibinfo {author} {\bibfnamefont {J.}~\bibnamefont {Foley}}, \bibinfo
  {author} {\bibfnamefont {C.}~\bibnamefont {Morningstar}}, \bibinfo {author}
  {\bibfnamefont {J.}~\bibnamefont {Dudek}}, \bibinfo {author} {\bibfnamefont
  {R.~G.}\ \bibnamefont {Edwards}}, \bibinfo {author} {\bibfnamefont
  {B.}~\bibnamefont {Joo}}, \bibinfo {author} {\bibfnamefont {H.-W.}\
  \bibnamefont {Lin}}, \bibinfo {author} {\bibfnamefont {D.~G.}\ \bibnamefont
  {Richards}}, \ and\ \bibinfo {author} {\bibfnamefont {K.~J.}\ \bibnamefont
  {Juge}} (\bibinfo {collaboration} {Hadron Spectrum}),\ }\bibfield  {title}
  {\enquote {\bibinfo {title} {{A Novel quark-field creation operator
  construction for hadronic physics in lattice QCD}},}\ }\href {\doibase
  10.1103/PhysRevD.80.054506} {\bibfield  {journal} {\bibinfo  {journal} {Phys.
  Rev.}\ }\textbf {\bibinfo {volume} {D80}},\ \bibinfo {pages} {054506}
  (\bibinfo {year} {2009})},\ \Eprint {http://arxiv.org/abs/0905.2160}
  {arXiv:0905.2160 [hep-lat]} \BibitemShut {NoStop}%
\bibitem [{\citenamefont {Bali}\ \emph {et~al.}(2016)\citenamefont {Bali},
  \citenamefont {Lang}, \citenamefont {Musch},\ and\ \citenamefont
  {Schäfer}}]{Bali:2016lva}%
  \BibitemOpen
  \bibfield  {author} {\bibinfo {author} {\bibfnamefont {G.~S.}\ \bibnamefont
  {Bali}}, \bibinfo {author} {\bibfnamefont {B.}~\bibnamefont {Lang}}, \bibinfo
  {author} {\bibfnamefont {B.~U.}\ \bibnamefont {Musch}}, \ and\ \bibinfo
  {author} {\bibfnamefont {A.}~\bibnamefont {Schäfer}},\ }\bibfield  {title}
  {\enquote {\bibinfo {title} {{Novel quark smearing for hadrons with high
  momenta in lattice QCD}},}\ }\href {\doibase 10.1103/PhysRevD.93.094515}
  {\bibfield  {journal} {\bibinfo  {journal} {Phys. Rev.}\ }\textbf {\bibinfo
  {volume} {D93}},\ \bibinfo {pages} {094515} (\bibinfo {year} {2016})},\
  \Eprint {http://arxiv.org/abs/1602.05525} {arXiv:1602.05525 [hep-lat]}
  \BibitemShut {NoStop}%
\bibitem [{\citenamefont {Edwards}\ \emph {et~al.}(2011)\citenamefont
  {Edwards}, \citenamefont {Dudek}, \citenamefont {Richards},\ and\
  \citenamefont {Wallace}}]{Edwards:2011jj}%
  \BibitemOpen
  \bibfield  {author} {\bibinfo {author} {\bibfnamefont {R.~G.}\ \bibnamefont
  {Edwards}}, \bibinfo {author} {\bibfnamefont {J.~J.}\ \bibnamefont {Dudek}},
  \bibinfo {author} {\bibfnamefont {D.~G.}\ \bibnamefont {Richards}}, \ and\
  \bibinfo {author} {\bibfnamefont {S.~J.}\ \bibnamefont {Wallace}},\
  }\bibfield  {title} {\enquote {\bibinfo {title} {{Excited state baryon
  spectroscopy from lattice QCD}},}\ }\href {\doibase
  10.1103/PhysRevD.84.074508} {\bibfield  {journal} {\bibinfo  {journal} {Phys.
  Rev. D}\ }\textbf {\bibinfo {volume} {84}},\ \bibinfo {pages} {074508}
  (\bibinfo {year} {2011})},\ \Eprint {http://arxiv.org/abs/1104.5152}
  {arXiv:1104.5152 [hep-ph]} \BibitemShut {NoStop}%
\bibitem [{\citenamefont {Dudek}\ and\ \citenamefont
  {Edwards}(2012)}]{Dudek:2012ag}%
  \BibitemOpen
  \bibfield  {author} {\bibinfo {author} {\bibfnamefont {J.~J.}\ \bibnamefont
  {Dudek}}\ and\ \bibinfo {author} {\bibfnamefont {R.~G.}\ \bibnamefont
  {Edwards}},\ }\bibfield  {title} {\enquote {\bibinfo {title} {{Hybrid Baryons
  in QCD}},}\ }\href {\doibase 10.1103/PhysRevD.85.054016} {\bibfield
  {journal} {\bibinfo  {journal} {Phys. Rev. D}\ }\textbf {\bibinfo {volume}
  {85}},\ \bibinfo {pages} {054016} (\bibinfo {year} {2012})},\ \Eprint
  {http://arxiv.org/abs/1201.2349} {arXiv:1201.2349 [hep-ph]} \BibitemShut
  {NoStop}%
\bibitem [{\citenamefont {Thomas}\ \emph {et~al.}(2012)\citenamefont {Thomas},
  \citenamefont {Edwards},\ and\ \citenamefont {Dudek}}]{Thomas:2011rh}%
  \BibitemOpen
  \bibfield  {author} {\bibinfo {author} {\bibfnamefont {C.~E.}\ \bibnamefont
  {Thomas}}, \bibinfo {author} {\bibfnamefont {R.~G.}\ \bibnamefont {Edwards}},
  \ and\ \bibinfo {author} {\bibfnamefont {J.~J.}\ \bibnamefont {Dudek}},\
  }\bibfield  {title} {\enquote {\bibinfo {title} {{Helicity operators for
  mesons in flight on the lattice}},}\ }\href {\doibase
  10.1103/PhysRevD.85.014507} {\bibfield  {journal} {\bibinfo  {journal} {Phys.
  Rev. D}\ }\textbf {\bibinfo {volume} {85}},\ \bibinfo {pages} {014507}
  (\bibinfo {year} {2012})},\ \Eprint {http://arxiv.org/abs/1107.1930}
  {arXiv:1107.1930 [hep-lat]} \BibitemShut {NoStop}%
\bibitem [{\citenamefont {Maiani}\ \emph {et~al.}(1987)\citenamefont {Maiani},
  \citenamefont {Martinelli}, \citenamefont {Paciello},\ and\ \citenamefont
  {Taglienti}}]{Maiani:1987by}%
  \BibitemOpen
  \bibfield  {author} {\bibinfo {author} {\bibfnamefont {L.}~\bibnamefont
  {Maiani}}, \bibinfo {author} {\bibfnamefont {G.}~\bibnamefont {Martinelli}},
  \bibinfo {author} {\bibfnamefont {M.~L.}\ \bibnamefont {Paciello}}, \ and\
  \bibinfo {author} {\bibfnamefont {B.}~\bibnamefont {Taglienti}},\ }\bibfield
  {title} {\enquote {\bibinfo {title} {{Scalar Densities and Baryon Mass
  Differences in Lattice {QCD} With Wilson Fermions}},}\ }\href {\doibase
  10.1016/0550-3213(87)90078-2} {\bibfield  {journal} {\bibinfo  {journal}
  {Nucl. Phys. B}\ }\textbf {\bibinfo {volume} {293}},\ \bibinfo {pages} {420}
  (\bibinfo {year} {1987})}\BibitemShut {NoStop}%
\bibitem [{\citenamefont {Capitani}\ \emph {et~al.}(2012)\citenamefont
  {Capitani}, \citenamefont {Della~Morte}, \citenamefont {von Hippel},
  \citenamefont {Jager}, \citenamefont {Juttner}, \citenamefont {Knippschild},
  \citenamefont {Meyer},\ and\ \citenamefont {Wittig}}]{Capitani:2012gj}%
  \BibitemOpen
  \bibfield  {author} {\bibinfo {author} {\bibfnamefont {S.}~\bibnamefont
  {Capitani}}, \bibinfo {author} {\bibfnamefont {M.}~\bibnamefont
  {Della~Morte}}, \bibinfo {author} {\bibfnamefont {G.}~\bibnamefont {von
  Hippel}}, \bibinfo {author} {\bibfnamefont {B.}~\bibnamefont {Jager}},
  \bibinfo {author} {\bibfnamefont {A.}~\bibnamefont {Juttner}}, \bibinfo
  {author} {\bibfnamefont {B.}~\bibnamefont {Knippschild}}, \bibinfo {author}
  {\bibfnamefont {H.~B.}\ \bibnamefont {Meyer}}, \ and\ \bibinfo {author}
  {\bibfnamefont {H.}~\bibnamefont {Wittig}},\ }\bibfield  {title} {\enquote
  {\bibinfo {title} {{The nucleon axial charge from lattice QCD with controlled
  errors}},}\ }\href {\doibase 10.1103/PhysRevD.86.074502} {\bibfield
  {journal} {\bibinfo  {journal} {Phys. Rev. D}\ }\textbf {\bibinfo {volume}
  {86}},\ \bibinfo {pages} {074502} (\bibinfo {year} {2012})},\ \Eprint
  {http://arxiv.org/abs/1205.0180} {arXiv:1205.0180 [hep-lat]} \BibitemShut
  {NoStop}%
\bibitem [{\citenamefont {Gao}\ \emph {et~al.}(2020)\citenamefont {Gao},
  \citenamefont {Jin}, \citenamefont {Kallidonis}, \citenamefont {Karthik},
  \citenamefont {Mukherjee}, \citenamefont {Petreczky}, \citenamefont
  {Shugert}, \citenamefont {Syritsyn},\ and\ \citenamefont
  {Zhao}}]{Gao:2020ito}%
  \BibitemOpen
  \bibfield  {author} {\bibinfo {author} {\bibfnamefont {X.}~\bibnamefont
  {Gao}}, \bibinfo {author} {\bibfnamefont {L.}~\bibnamefont {Jin}}, \bibinfo
  {author} {\bibfnamefont {C.}~\bibnamefont {Kallidonis}}, \bibinfo {author}
  {\bibfnamefont {N.}~\bibnamefont {Karthik}}, \bibinfo {author} {\bibfnamefont
  {S.}~\bibnamefont {Mukherjee}}, \bibinfo {author} {\bibfnamefont
  {P.}~\bibnamefont {Petreczky}}, \bibinfo {author} {\bibfnamefont
  {C.}~\bibnamefont {Shugert}}, \bibinfo {author} {\bibfnamefont
  {S.}~\bibnamefont {Syritsyn}}, \ and\ \bibinfo {author} {\bibfnamefont
  {Y.}~\bibnamefont {Zhao}},\ }\bibfield  {title} {\enquote {\bibinfo {title}
  {{Valence parton distribution of the pion from lattice QCD: Approaching the
  continuum limit}},}\ }\href {\doibase 10.1103/PhysRevD.102.094513} {\bibfield
   {journal} {\bibinfo  {journal} {Phys. Rev. D}\ }\textbf {\bibinfo {volume}
  {102}},\ \bibinfo {pages} {094513} (\bibinfo {year} {2020})},\ \Eprint
  {http://arxiv.org/abs/2007.06590} {arXiv:2007.06590 [hep-lat]} \BibitemShut
  {NoStop}%
\bibitem [{\citenamefont {Buckley}\ \emph {et~al.}(2015)\citenamefont
  {Buckley}, \citenamefont {Ferrando}, \citenamefont {Lloyd}, \citenamefont
  {Nordstr\"om}, \citenamefont {Page}, \citenamefont {R\"ufenacht},
  \citenamefont {Sch\"onherr},\ and\ \citenamefont {Watt}}]{Buckley:2014ana}%
  \BibitemOpen
  \bibfield  {author} {\bibinfo {author} {\bibfnamefont {A.}~\bibnamefont
  {Buckley}}, \bibinfo {author} {\bibfnamefont {J.}~\bibnamefont {Ferrando}},
  \bibinfo {author} {\bibfnamefont {S.}~\bibnamefont {Lloyd}}, \bibinfo
  {author} {\bibfnamefont {K.}~\bibnamefont {Nordstr\"om}}, \bibinfo {author}
  {\bibfnamefont {B.}~\bibnamefont {Page}}, \bibinfo {author} {\bibfnamefont
  {M.}~\bibnamefont {R\"ufenacht}}, \bibinfo {author} {\bibfnamefont
  {M.}~\bibnamefont {Sch\"onherr}}, \ and\ \bibinfo {author} {\bibfnamefont
  {G.}~\bibnamefont {Watt}},\ }\bibfield  {title} {\enquote {\bibinfo {title}
  {{LHAPDF6: parton density access in the LHC precision era}},}\ }\href
  {\doibase 10.1140/epjc/s10052-015-3318-8} {\bibfield  {journal} {\bibinfo
  {journal} {Eur. Phys. J. C}\ }\textbf {\bibinfo {volume} {75}},\ \bibinfo
  {pages} {132} (\bibinfo {year} {2015})},\ \Eprint
  {http://arxiv.org/abs/1412.7420} {arXiv:1412.7420 [hep-ph]} \BibitemShut
  {NoStop}%
\bibitem [{\citenamefont {Akiyama}\ \emph {et~al.}(2019)\citenamefont {Akiyama}
  \emph {et~al.}}]{Akiyama:2019bqs}%
  \BibitemOpen
  \bibfield  {author} {\bibinfo {author} {\bibfnamefont {K.}~\bibnamefont
  {Akiyama}} \emph {et~al.} (\bibinfo {collaboration} {Event Horizon
  Telescope}),\ }\bibfield  {title} {\enquote {\bibinfo {title} {{First M87
  Event Horizon Telescope Results. IV. Imaging the Central Supermassive Black
  Hole}},}\ }\href {\doibase 10.3847/2041-8213/ab0e85} {\bibfield  {journal}
  {\bibinfo  {journal} {Astrophys. J. Lett.}\ }\textbf {\bibinfo {volume}
  {875}},\ \bibinfo {pages} {L4} (\bibinfo {year} {2019})},\ \Eprint
  {http://arxiv.org/abs/1906.11241} {arXiv:1906.11241 [astro-ph.GA]}
  \BibitemShut {NoStop}%
\bibitem [{\citenamefont {Backus}\ and\ \citenamefont
  {Gilbert}(1968)}]{Backus}%
  \BibitemOpen
  \bibfield  {author} {\bibinfo {author} {\bibfnamefont {G.}~\bibnamefont
  {Backus}}\ and\ \bibinfo {author} {\bibfnamefont {F.}~\bibnamefont
  {Gilbert}},\ }\bibfield  {title} {\enquote {\bibinfo {title} {{The Resolving
  Power of Gross Earth Data}},}\ }\href {\doibase
  10.1111/j.1365-246X.1968.tb00216.x} {\bibfield  {journal} {\bibinfo
  {journal} {Geophysical Journal International}\ }\textbf {\bibinfo {volume}
  {16}},\ \bibinfo {pages} {169} (\bibinfo {year} {1968})},\ \Eprint
  {http://arxiv.org/abs/https://academic.oup.com/gji/article-pdf/16/2/169/5891044/16-2-169.pdf}
  {https://academic.oup.com/gji/article-pdf/16/2/169/5891044/16-2-169.pdf}
  \BibitemShut {NoStop}%
\bibitem [{\citenamefont {Asakawa}\ \emph {et~al.}(2001)\citenamefont
  {Asakawa}, \citenamefont {Hatsuda},\ and\ \citenamefont
  {Nakahara}}]{Asakawa:2000tr}%
  \BibitemOpen
  \bibfield  {author} {\bibinfo {author} {\bibfnamefont {M.}~\bibnamefont
  {Asakawa}}, \bibinfo {author} {\bibfnamefont {T.}~\bibnamefont {Hatsuda}}, \
  and\ \bibinfo {author} {\bibfnamefont {Y.}~\bibnamefont {Nakahara}},\
  }\bibfield  {title} {\enquote {\bibinfo {title} {{Maximum entropy analysis of
  the spectral functions in lattice QCD}},}\ }\href {\doibase
  10.1016/S0146-6410(01)00150-8} {\bibfield  {journal} {\bibinfo  {journal}
  {Prog. Part. Nucl. Phys.}\ }\textbf {\bibinfo {volume} {46}},\ \bibinfo
  {pages} {459} (\bibinfo {year} {2001})},\ \Eprint
  {http://arxiv.org/abs/hep-lat/0011040} {arXiv:hep-lat/0011040} \BibitemShut
  {NoStop}%
\bibitem [{\citenamefont {Karpie}\ \emph {et~al.}(2019)\citenamefont {Karpie},
  \citenamefont {Orginos}, \citenamefont {Rothkopf},\ and\ \citenamefont
  {Zafeiropoulos}}]{Karpie:2019eiq}%
  \BibitemOpen
  \bibfield  {author} {\bibinfo {author} {\bibfnamefont {J.}~\bibnamefont
  {Karpie}}, \bibinfo {author} {\bibfnamefont {K.}~\bibnamefont {Orginos}},
  \bibinfo {author} {\bibfnamefont {A.}~\bibnamefont {Rothkopf}}, \ and\
  \bibinfo {author} {\bibfnamefont {S.}~\bibnamefont {Zafeiropoulos}},\
  }\bibfield  {title} {\enquote {\bibinfo {title} {{Reconstructing parton
  distribution functions from Ioffe time data: from Bayesian methods to Neural
  Networks}},}\ }\href {\doibase 10.1007/JHEP04(2019)057} {\bibfield  {journal}
  {\bibinfo  {journal} {JHEP}\ }\textbf {\bibinfo {volume} {04}},\ \bibinfo
  {pages} {057} (\bibinfo {year} {2019})},\ \Eprint
  {http://arxiv.org/abs/1901.05408} {arXiv:1901.05408 [hep-lat]} \BibitemShut
  {NoStop}%
\bibitem [{\citenamefont {Liang}\ \emph {et~al.}(2020)\citenamefont {Liang},
  \citenamefont {Draper}, \citenamefont {Liu}, \citenamefont {Rothkopf},\ and\
  \citenamefont {Yang}}]{Liang:2019frk}%
  \BibitemOpen
  \bibfield  {author} {\bibinfo {author} {\bibfnamefont {J.}~\bibnamefont
  {Liang}}, \bibinfo {author} {\bibfnamefont {T.}~\bibnamefont {Draper}},
  \bibinfo {author} {\bibfnamefont {K.-F.}\ \bibnamefont {Liu}}, \bibinfo
  {author} {\bibfnamefont {A.}~\bibnamefont {Rothkopf}}, \ and\ \bibinfo
  {author} {\bibfnamefont {Y.-B.}\ \bibnamefont {Yang}} (\bibinfo
  {collaboration} {XQCD}),\ }\bibfield  {title} {\enquote {\bibinfo {title}
  {{Towards the nucleon hadronic tensor from lattice QCD}},}\ }\href {\doibase
  10.1103/PhysRevD.101.114503} {\bibfield  {journal} {\bibinfo  {journal}
  {Phys. Rev. D}\ }\textbf {\bibinfo {volume} {101}},\ \bibinfo {pages}
  {114503} (\bibinfo {year} {2020})},\ \Eprint
  {http://arxiv.org/abs/1906.05312} {arXiv:1906.05312 [hep-ph]} \BibitemShut
  {NoStop}%
\bibitem [{\citenamefont {Hansen}\ \emph {et~al.}(2017)\citenamefont {Hansen},
  \citenamefont {Meyer},\ and\ \citenamefont {Robaina}}]{Hansen:2017mnd}%
  \BibitemOpen
  \bibfield  {author} {\bibinfo {author} {\bibfnamefont {M.~T.}\ \bibnamefont
  {Hansen}}, \bibinfo {author} {\bibfnamefont {H.~B.}\ \bibnamefont {Meyer}}, \
  and\ \bibinfo {author} {\bibfnamefont {D.}~\bibnamefont {Robaina}},\
  }\bibfield  {title} {\enquote {\bibinfo {title} {{From deep inelastic
  scattering to heavy-flavor semileptonic decays: Total rates into multihadron
  final states from lattice QCD}},}\ }\href {\doibase
  10.1103/PhysRevD.96.094513} {\bibfield  {journal} {\bibinfo  {journal} {Phys.
  Rev. D}\ }\textbf {\bibinfo {volume} {96}},\ \bibinfo {pages} {094513}
  (\bibinfo {year} {2017})},\ \Eprint {http://arxiv.org/abs/1704.08993}
  {arXiv:1704.08993 [hep-lat]} \BibitemShut {NoStop}%
\bibitem [{\citenamefont {Accardi}\ \emph {et~al.}(2016)\citenamefont
  {Accardi}, \citenamefont {Brady}, \citenamefont {Melnitchouk}, \citenamefont
  {Owens},\ and\ \citenamefont {Sato}}]{Accardi:2016qay}%
  \BibitemOpen
  \bibfield  {author} {\bibinfo {author} {\bibfnamefont {A.}~\bibnamefont
  {Accardi}}, \bibinfo {author} {\bibfnamefont {L.}~\bibnamefont {Brady}},
  \bibinfo {author} {\bibfnamefont {W.}~\bibnamefont {Melnitchouk}}, \bibinfo
  {author} {\bibfnamefont {J.}~\bibnamefont {Owens}}, \ and\ \bibinfo {author}
  {\bibfnamefont {N.}~\bibnamefont {Sato}},\ }\bibfield  {title} {\enquote
  {\bibinfo {title} {{Constraints on large-$x$ parton distributions from new
  weak boson production and deep-inelastic scattering data}},}\ }\href
  {\doibase 10.1103/PhysRevD.93.114017} {\bibfield  {journal} {\bibinfo
  {journal} {Phys. Rev. D}\ }\textbf {\bibinfo {volume} {93}},\ \bibinfo
  {pages} {114017} (\bibinfo {year} {2016})},\ \Eprint
  {http://arxiv.org/abs/1602.03154} {arXiv:1602.03154 [hep-ph]} \BibitemShut
  {NoStop}%
\bibitem [{\citenamefont {Martin}\ \emph {et~al.}(2009)\citenamefont {Martin},
  \citenamefont {Stirling}, \citenamefont {Thorne},\ and\ \citenamefont
  {Watt}}]{Martin:2009iq}%
  \BibitemOpen
  \bibfield  {author} {\bibinfo {author} {\bibfnamefont {A.}~\bibnamefont
  {Martin}}, \bibinfo {author} {\bibfnamefont {W.}~\bibnamefont {Stirling}},
  \bibinfo {author} {\bibfnamefont {R.}~\bibnamefont {Thorne}}, \ and\ \bibinfo
  {author} {\bibfnamefont {G.}~\bibnamefont {Watt}},\ }\bibfield  {title}
  {\enquote {\bibinfo {title} {{Parton distributions for the LHC}},}\ }\href
  {\doibase 10.1140/epjc/s10052-009-1072-5} {\bibfield  {journal} {\bibinfo
  {journal} {Eur. Phys. J. C}\ }\textbf {\bibinfo {volume} {63}},\ \bibinfo
  {pages} {189} (\bibinfo {year} {2009})},\ \Eprint
  {http://arxiv.org/abs/0901.0002} {arXiv:0901.0002 [hep-ph]} \BibitemShut
  {NoStop}%
\bibitem [{\citenamefont {Harland-Lang}\ \emph {et~al.}(2015)\citenamefont
  {Harland-Lang}, \citenamefont {Martin}, \citenamefont {Motylinski},\ and\
  \citenamefont {Thorne}}]{Harland-Lang:2014zoa}%
  \BibitemOpen
  \bibfield  {author} {\bibinfo {author} {\bibfnamefont {L.~A.}\ \bibnamefont
  {Harland-Lang}}, \bibinfo {author} {\bibfnamefont {A.~D.}\ \bibnamefont
  {Martin}}, \bibinfo {author} {\bibfnamefont {P.}~\bibnamefont {Motylinski}},
  \ and\ \bibinfo {author} {\bibfnamefont {R.~S.}\ \bibnamefont {Thorne}},\
  }\bibfield  {title} {\enquote {\bibinfo {title} {{Parton distributions in the
  LHC era: MMHT 2014 PDFs}},}\ }\href {\doibase 10.1140/epjc/s10052-015-3397-6}
  {\bibfield  {journal} {\bibinfo  {journal} {Eur. Phys. J. C}\ }\textbf
  {\bibinfo {volume} {75}},\ \bibinfo {pages} {204} (\bibinfo {year} {2015})},\
  \Eprint {http://arxiv.org/abs/1412.3989} {arXiv:1412.3989 [hep-ph]}
  \BibitemShut {NoStop}%
\bibitem [{\citenamefont {Hou}\ \emph {et~al.}(2021)\citenamefont {Hou} \emph
  {et~al.}}]{Hou:2019efy}%
  \BibitemOpen
  \bibfield  {author} {\bibinfo {author} {\bibfnamefont {T.-J.}\ \bibnamefont
  {Hou}} \emph {et~al.},\ }\bibfield  {title} {\enquote {\bibinfo {title} {{New
  CTEQ global analysis of quantum chromodynamics with high-precision data from
  the LHC}},}\ }\href {\doibase 10.1103/PhysRevD.103.014013} {\bibfield
  {journal} {\bibinfo  {journal} {Phys. Rev. D}\ }\textbf {\bibinfo {volume}
  {103}},\ \bibinfo {pages} {014013} (\bibinfo {year} {2021})},\ \Eprint
  {http://arxiv.org/abs/1912.10053} {arXiv:1912.10053 [hep-ph]} \BibitemShut
  {NoStop}%
\bibitem [{\citenamefont {Ball}\ \emph {et~al.}(2015)\citenamefont {Ball} \emph
  {et~al.}}]{Ball:2014uwa}%
  \BibitemOpen
  \bibfield  {author} {\bibinfo {author} {\bibfnamefont {R.~D.}\ \bibnamefont
  {Ball}} \emph {et~al.} (\bibinfo {collaboration} {NNPDF}),\ }\bibfield
  {title} {\enquote {\bibinfo {title} {{Parton distributions for the LHC Run
  II}},}\ }\href {\doibase 10.1007/JHEP04(2015)040} {\bibfield  {journal}
  {\bibinfo  {journal} {JHEP}\ }\textbf {\bibinfo {volume} {04}},\ \bibinfo
  {pages} {040} (\bibinfo {year} {2015})},\ \Eprint
  {http://arxiv.org/abs/1410.8849} {arXiv:1410.8849 [hep-ph]} \BibitemShut
  {NoStop}%
\bibitem [{\citenamefont {Ball}\ \emph {et~al.}(2017)\citenamefont {Ball} \emph
  {et~al.}}]{Ball:2017nwa}%
  \BibitemOpen
  \bibfield  {author} {\bibinfo {author} {\bibfnamefont {R.~D.}\ \bibnamefont
  {Ball}} \emph {et~al.} (\bibinfo {collaboration} {NNPDF}),\ }\bibfield
  {title} {\enquote {\bibinfo {title} {{Parton distributions from
  high-precision collider data}},}\ }\href {\doibase
  10.1140/epjc/s10052-017-5199-5} {\bibfield  {journal} {\bibinfo  {journal}
  {Eur. Phys. J. C}\ }\textbf {\bibinfo {volume} {77}},\ \bibinfo {pages} {663}
  (\bibinfo {year} {2017})},\ \Eprint {http://arxiv.org/abs/1706.00428}
  {arXiv:1706.00428 [hep-ph]} \BibitemShut {NoStop}%
\bibitem [{\citenamefont {Moffat}\ \emph {et~al.}(2021)\citenamefont {Moffat},
  \citenamefont {Melnitchouk}, \citenamefont {Rogers},\ and\ \citenamefont
  {Sato}}]{Moffat:2021dji}%
  \BibitemOpen
  \bibfield  {author} {\bibinfo {author} {\bibfnamefont {E.}~\bibnamefont
  {Moffat}}, \bibinfo {author} {\bibfnamefont {W.}~\bibnamefont {Melnitchouk}},
  \bibinfo {author} {\bibfnamefont {T.}~\bibnamefont {Rogers}}, \ and\ \bibinfo
  {author} {\bibfnamefont {N.}~\bibnamefont {Sato}},\ }\bibfield  {title}
  {\enquote {\bibinfo {title} {{Simultaneous Monte Carlo analysis of parton
  densities and fragmentation functions}},}\ }\href@noop {} {\  (\bibinfo
  {year} {2021})},\ \Eprint {http://arxiv.org/abs/2101.04664} {arXiv:2101.04664
  [hep-ph]} \BibitemShut {NoStop}%
\bibitem [{\citenamefont {Martin}\ \emph {et~al.}(2010)\citenamefont {Martin},
  \citenamefont {Stirling}, \citenamefont {Thorne},\ and\ \citenamefont
  {Watt}}]{Martin:2010db}%
  \BibitemOpen
  \bibfield  {author} {\bibinfo {author} {\bibfnamefont {A.~D.}\ \bibnamefont
  {Martin}}, \bibinfo {author} {\bibfnamefont {W.~J.}\ \bibnamefont
  {Stirling}}, \bibinfo {author} {\bibfnamefont {R.~S.}\ \bibnamefont
  {Thorne}}, \ and\ \bibinfo {author} {\bibfnamefont {G.}~\bibnamefont
  {Watt}},\ }\bibfield  {title} {\enquote {\bibinfo {title} {{Heavy-quark mass
  dependence in global PDF analyses and 3- and 4-flavour parton
  distributions}},}\ }\href {\doibase 10.1140/epjc/s10052-010-1462-8}
  {\bibfield  {journal} {\bibinfo  {journal} {Eur. Phys. J. C}\ }\textbf
  {\bibinfo {volume} {70}},\ \bibinfo {pages} {51} (\bibinfo {year} {2010})},\
  \Eprint {http://arxiv.org/abs/1007.2624} {arXiv:1007.2624 [hep-ph]}
  \BibitemShut {NoStop}%
\bibitem [{\citenamefont {Radyushkin}(2019)}]{Radyushkin:2019owq}%
  \BibitemOpen
  \bibfield  {author} {\bibinfo {author} {\bibfnamefont {A.~V.}\ \bibnamefont
  {Radyushkin}},\ }\bibfield  {title} {\enquote {\bibinfo {title} {{Generalized
  parton distributions and pseudodistributions}},}\ }\href {\doibase
  10.1103/PhysRevD.100.116011} {\bibfield  {journal} {\bibinfo  {journal}
  {Phys. Rev. D}\ }\textbf {\bibinfo {volume} {100}},\ \bibinfo {pages}
  {116011} (\bibinfo {year} {2019})},\ \Eprint
  {http://arxiv.org/abs/1909.08474} {arXiv:1909.08474 [hep-ph]} \BibitemShut
  {NoStop}%
\bibitem [{\citenamefont {Dulat}\ \emph {et~al.}(2016)\citenamefont {Dulat},
  \citenamefont {Hou}, \citenamefont {Gao}, \citenamefont {Guzzi},
  \citenamefont {Huston}, \citenamefont {Nadolsky}, \citenamefont {Pumplin},
  \citenamefont {Schmidt}, \citenamefont {Stump},\ and\ \citenamefont
  {Yuan}}]{Dulat:2015mca}%
  \BibitemOpen
  \bibfield  {author} {\bibinfo {author} {\bibfnamefont {S.}~\bibnamefont
  {Dulat}}, \bibinfo {author} {\bibfnamefont {T.-J.}\ \bibnamefont {Hou}},
  \bibinfo {author} {\bibfnamefont {J.}~\bibnamefont {Gao}}, \bibinfo {author}
  {\bibfnamefont {M.}~\bibnamefont {Guzzi}}, \bibinfo {author} {\bibfnamefont
  {J.}~\bibnamefont {Huston}}, \bibinfo {author} {\bibfnamefont
  {P.}~\bibnamefont {Nadolsky}}, \bibinfo {author} {\bibfnamefont
  {J.}~\bibnamefont {Pumplin}}, \bibinfo {author} {\bibfnamefont
  {C.}~\bibnamefont {Schmidt}}, \bibinfo {author} {\bibfnamefont
  {D.}~\bibnamefont {Stump}}, \ and\ \bibinfo {author} {\bibfnamefont {C.~P.}\
  \bibnamefont {Yuan}},\ }\bibfield  {title} {\enquote {\bibinfo {title} {{New
  parton distribution functions from a global analysis of quantum
  chromodynamics}},}\ }\href {\doibase 10.1103/PhysRevD.93.033006} {\bibfield
  {journal} {\bibinfo  {journal} {Phys. Rev. D}\ }\textbf {\bibinfo {volume}
  {93}},\ \bibinfo {pages} {033006} (\bibinfo {year} {2016})},\ \Eprint
  {http://arxiv.org/abs/1506.07443} {arXiv:1506.07443 [hep-ph]} \BibitemShut
  {NoStop}%
\bibitem [{\citenamefont {Bali}\ \emph {et~al.}(2018)\citenamefont {Bali},
  \citenamefont {Braun}, \citenamefont {Gl\"a\ss{}le}, \citenamefont
  {G\"ockeler}, \citenamefont {Gruber}, \citenamefont {Hutzler}, \citenamefont
  {Korcyl}, \citenamefont {Sch\"afer}, \citenamefont {Wein},\ and\
  \citenamefont {Zhang}}]{Bali:2018spj}%
  \BibitemOpen
  \bibfield  {author} {\bibinfo {author} {\bibfnamefont {G.~S.}\ \bibnamefont
  {Bali}}, \bibinfo {author} {\bibfnamefont {V.~M.}\ \bibnamefont {Braun}},
  \bibinfo {author} {\bibfnamefont {B.}~\bibnamefont {Gl\"a\ss{}le}}, \bibinfo
  {author} {\bibfnamefont {M.}~\bibnamefont {G\"ockeler}}, \bibinfo {author}
  {\bibfnamefont {M.}~\bibnamefont {Gruber}}, \bibinfo {author} {\bibfnamefont
  {F.}~\bibnamefont {Hutzler}}, \bibinfo {author} {\bibfnamefont
  {P.}~\bibnamefont {Korcyl}}, \bibinfo {author} {\bibfnamefont
  {A.}~\bibnamefont {Sch\"afer}}, \bibinfo {author} {\bibfnamefont
  {P.}~\bibnamefont {Wein}}, \ and\ \bibinfo {author} {\bibfnamefont {J.-H.}\
  \bibnamefont {Zhang}},\ }\bibfield  {title} {\enquote {\bibinfo {title}
  {{Pion distribution amplitude from Euclidean correlation functions: Exploring
  universality and higher-twist effects}},}\ }\href {\doibase
  10.1103/PhysRevD.98.094507} {\bibfield  {journal} {\bibinfo  {journal} {Phys.
  Rev. D}\ }\textbf {\bibinfo {volume} {98}},\ \bibinfo {pages} {094507}
  (\bibinfo {year} {2018})},\ \Eprint {http://arxiv.org/abs/1807.06671}
  {arXiv:1807.06671 [hep-lat]} \BibitemShut {NoStop}%
\bibitem [{\citenamefont {Bali}\ \emph {et~al.}(2019)\citenamefont {Bali},
  \citenamefont {Braun}, \citenamefont {B\"urger}, \citenamefont {G\"ockeler},
  \citenamefont {Gruber}, \citenamefont {Hutzler}, \citenamefont {Korcyl},
  \citenamefont {Sch\"afer}, \citenamefont {Sternbeck},\ and\ \citenamefont
  {Wein}}]{Bali:2019dqc}%
  \BibitemOpen
  \bibfield  {author} {\bibinfo {author} {\bibfnamefont {G.~S.}\ \bibnamefont
  {Bali}}, \bibinfo {author} {\bibfnamefont {V.~M.}\ \bibnamefont {Braun}},
  \bibinfo {author} {\bibfnamefont {S.}~\bibnamefont {B\"urger}}, \bibinfo
  {author} {\bibfnamefont {M.}~\bibnamefont {G\"ockeler}}, \bibinfo {author}
  {\bibfnamefont {M.}~\bibnamefont {Gruber}}, \bibinfo {author} {\bibfnamefont
  {F.}~\bibnamefont {Hutzler}}, \bibinfo {author} {\bibfnamefont
  {P.}~\bibnamefont {Korcyl}}, \bibinfo {author} {\bibfnamefont
  {A.}~\bibnamefont {Sch\"afer}}, \bibinfo {author} {\bibfnamefont
  {A.}~\bibnamefont {Sternbeck}}, \ and\ \bibinfo {author} {\bibfnamefont
  {P.}~\bibnamefont {Wein}} (\bibinfo {collaboration} {RQCD}),\ }\bibfield
  {title} {\enquote {\bibinfo {title} {{Light-cone distribution amplitudes of
  pseudoscalar mesons from lattice QCD}},}\ }\href {\doibase
  10.1007/JHEP08(2019)065} {\bibfield  {journal} {\bibinfo  {journal} {JHEP}\
  }\textbf {\bibinfo {volume} {08}},\ \bibinfo {pages} {065} (\bibinfo {year}
  {2019})},\ \bibinfo {note} {[Addendum: JHEP 11, 037 (2020)]},\ \Eprint
  {http://arxiv.org/abs/1903.08038} {arXiv:1903.08038 [hep-lat]} \BibitemShut
  {NoStop}%
\bibitem [{\citenamefont {Segovia}\ \emph {et~al.}(2014)\citenamefont
  {Segovia}, \citenamefont {Chang}, \citenamefont {Clo\"et}, \citenamefont
  {Roberts}, \citenamefont {Schmidt},\ and\ \citenamefont
  {Zong}}]{Segovia:2013eca}%
  \BibitemOpen
  \bibfield  {author} {\bibinfo {author} {\bibfnamefont {J.}~\bibnamefont
  {Segovia}}, \bibinfo {author} {\bibfnamefont {L.}~\bibnamefont {Chang}},
  \bibinfo {author} {\bibfnamefont {I.~C.}\ \bibnamefont {Clo\"et}}, \bibinfo
  {author} {\bibfnamefont {C.~D.}\ \bibnamefont {Roberts}}, \bibinfo {author}
  {\bibfnamefont {S.~M.}\ \bibnamefont {Schmidt}}, \ and\ \bibinfo {author}
  {\bibfnamefont {H.-s.}\ \bibnamefont {Zong}},\ }\bibfield  {title} {\enquote
  {\bibinfo {title} {{Distribution amplitudes of light-quark mesons from
  lattice QCD}},}\ }\href {\doibase 10.1016/j.physletb.2014.02.006} {\bibfield
  {journal} {\bibinfo  {journal} {Phys. Lett. B}\ }\textbf {\bibinfo {volume}
  {731}},\ \bibinfo {pages} {13} (\bibinfo {year} {2014})},\ \Eprint
  {http://arxiv.org/abs/1311.1390} {arXiv:1311.1390 [nucl-th]} \BibitemShut
  {NoStop}%
\bibitem [{\citenamefont {Fukaya}\ \emph {et~al.}(2020)\citenamefont {Fukaya},
  \citenamefont {Hashimoto}, \citenamefont {Kaneko},\ and\ \citenamefont
  {Ohki}}]{Fukaya:2020wpp}%
  \BibitemOpen
  \bibfield  {author} {\bibinfo {author} {\bibfnamefont {H.}~\bibnamefont
  {Fukaya}}, \bibinfo {author} {\bibfnamefont {S.}~\bibnamefont {Hashimoto}},
  \bibinfo {author} {\bibfnamefont {T.}~\bibnamefont {Kaneko}}, \ and\ \bibinfo
  {author} {\bibfnamefont {H.}~\bibnamefont {Ohki}},\ }\bibfield  {title}
  {\enquote {\bibinfo {title} {{Towards fully nonperturbative computations of
  inelastic $\ell N$ scattering cross sections from lattice QCD}},}\ }\href
  {\doibase 10.1103/PhysRevD.102.114516} {\bibfield  {journal} {\bibinfo
  {journal} {Phys. Rev. D}\ }\textbf {\bibinfo {volume} {102}},\ \bibinfo
  {pages} {114516} (\bibinfo {year} {2020})},\ \Eprint
  {http://arxiv.org/abs/2010.01253} {arXiv:2010.01253 [hep-lat]} \BibitemShut
  {NoStop}%
\bibitem [{\citenamefont {Karpie}\ \emph {et~al.}(2018)\citenamefont {Karpie},
  \citenamefont {Orginos},\ and\ \citenamefont
  {Zafeiropoulos}}]{Karpie:2018zaz}%
  \BibitemOpen
  \bibfield  {author} {\bibinfo {author} {\bibfnamefont {J.}~\bibnamefont
  {Karpie}}, \bibinfo {author} {\bibfnamefont {K.}~\bibnamefont {Orginos}}, \
  and\ \bibinfo {author} {\bibfnamefont {S.}~\bibnamefont {Zafeiropoulos}},\
  }\bibfield  {title} {\enquote {\bibinfo {title} {{Moments of Ioffe time
  parton distribution functions from non-local matrix elements}},}\ }\href
  {\doibase 10.1007/JHEP11(2018)178} {\bibfield  {journal} {\bibinfo  {journal}
  {JHEP}\ }\textbf {\bibinfo {volume} {11}},\ \bibinfo {pages} {178} (\bibinfo
  {year} {2018})},\ \Eprint {http://arxiv.org/abs/1807.10933} {arXiv:1807.10933
  [hep-lat]} \BibitemShut {NoStop}%
\bibitem [{\citenamefont {Golub}\ and\ \citenamefont {Pereyra}(1973)}]{Golub}%
  \BibitemOpen
  \bibfield  {author} {\bibinfo {author} {\bibfnamefont {G.~H.}\ \bibnamefont
  {Golub}}\ and\ \bibinfo {author} {\bibfnamefont {V.}~\bibnamefont
  {Pereyra}},\ }\bibfield  {title} {\enquote {\bibinfo {title} {{The
  Differentiation of Pseudo-Inverses and Nonlinear Least Squares Problems Whose
  Variables Separate}},}\ }\href {\doibase 10.1137/0710036} {\bibfield
  {journal} {\bibinfo  {journal} {SIAM Journal on Numerical Analysis}\ }\textbf
  {\bibinfo {volume} {10}},\ \bibinfo {pages} {413} (\bibinfo {year}
  {1973})}\BibitemShut {NoStop}%
\bibitem [{\citenamefont {Brodsky}\ and\ \citenamefont
  {Farrar}(1973)}]{Brodsky:1973kr}%
  \BibitemOpen
  \bibfield  {author} {\bibinfo {author} {\bibfnamefont {S.~J.}\ \bibnamefont
  {Brodsky}}\ and\ \bibinfo {author} {\bibfnamefont {G.~R.}\ \bibnamefont
  {Farrar}},\ }\bibfield  {title} {\enquote {\bibinfo {title} {{Scaling Laws at
  Large Transverse Momentum}},}\ }\href {\doibase 10.1103/PhysRevLett.31.1153}
  {\bibfield  {journal} {\bibinfo  {journal} {Phys. Rev. Lett.}\ }\textbf
  {\bibinfo {volume} {31}},\ \bibinfo {pages} {1153} (\bibinfo {year}
  {1973})}\BibitemShut {NoStop}%
\bibitem [{\citenamefont {Del~Debbio}\ \emph {et~al.}(2021)\citenamefont
  {Del~Debbio}, \citenamefont {Giani}, \citenamefont {Karpie}, \citenamefont
  {Orginos}, \citenamefont {Radyushkin},\ and\ \citenamefont
  {Zafeiropoulos}}]{DelDebbio:2020rgv}%
  \BibitemOpen
  \bibfield  {author} {\bibinfo {author} {\bibfnamefont {L.}~\bibnamefont
  {Del~Debbio}}, \bibinfo {author} {\bibfnamefont {T.}~\bibnamefont {Giani}},
  \bibinfo {author} {\bibfnamefont {J.}~\bibnamefont {Karpie}}, \bibinfo
  {author} {\bibfnamefont {K.}~\bibnamefont {Orginos}}, \bibinfo {author}
  {\bibfnamefont {A.}~\bibnamefont {Radyushkin}}, \ and\ \bibinfo {author}
  {\bibfnamefont {S.}~\bibnamefont {Zafeiropoulos}},\ }\bibfield  {title}
  {\enquote {\bibinfo {title} {{Neural-network analysis of Parton Distribution
  Functions from Ioffe-time pseudodistributions}},}\ }\href {\doibase
  10.1007/JHEP02(2021)138} {\bibfield  {journal} {\bibinfo  {journal} {JHEP}\
  }\textbf {\bibinfo {volume} {02}},\ \bibinfo {pages} {138} (\bibinfo {year}
  {2021})},\ \Eprint {http://arxiv.org/abs/2010.03996} {arXiv:2010.03996
  [hep-ph]} \BibitemShut {NoStop}%
\bibitem [{\citenamefont {Bringewatt}\ \emph {et~al.}(2021)\citenamefont
  {Bringewatt}, \citenamefont {Sato}, \citenamefont {Melnitchouk},
  \citenamefont {Qiu}, \citenamefont {Steffens},\ and\ \citenamefont
  {Constantinou}}]{Bringewatt:2020ixn}%
  \BibitemOpen
  \bibfield  {author} {\bibinfo {author} {\bibfnamefont {J.}~\bibnamefont
  {Bringewatt}}, \bibinfo {author} {\bibfnamefont {N.}~\bibnamefont {Sato}},
  \bibinfo {author} {\bibfnamefont {W.}~\bibnamefont {Melnitchouk}}, \bibinfo
  {author} {\bibfnamefont {J.-W.}\ \bibnamefont {Qiu}}, \bibinfo {author}
  {\bibfnamefont {F.}~\bibnamefont {Steffens}}, \ and\ \bibinfo {author}
  {\bibfnamefont {M.}~\bibnamefont {Constantinou}},\ }\bibfield  {title}
  {\enquote {\bibinfo {title} {{Confronting lattice parton distributions with
  global QCD analysis}},}\ }\href {\doibase 10.1103/PhysRevD.103.016003}
  {\bibfield  {journal} {\bibinfo  {journal} {Phys. Rev. D}\ }\textbf {\bibinfo
  {volume} {103}},\ \bibinfo {pages} {016003} (\bibinfo {year} {2021})},\
  \Eprint {http://arxiv.org/abs/2010.00548} {arXiv:2010.00548 [hep-ph]}
  \BibitemShut {NoStop}%
\bibitem [{\citenamefont {Stanzione}\ \emph {et~al.}(2020)\citenamefont
  {Stanzione}, \citenamefont {West}, \citenamefont {Evans}, \citenamefont
  {Minyard}, \citenamefont {Ghattas},\ and\ \citenamefont {Panda}}]{frontera}%
  \BibitemOpen
  \bibfield  {author} {\bibinfo {author} {\bibfnamefont {D.}~\bibnamefont
  {Stanzione}}, \bibinfo {author} {\bibfnamefont {J.}~\bibnamefont {West}},
  \bibinfo {author} {\bibfnamefont {R.~T.}\ \bibnamefont {Evans}}, \bibinfo
  {author} {\bibfnamefont {T.}~\bibnamefont {Minyard}}, \bibinfo {author}
  {\bibfnamefont {O.}~\bibnamefont {Ghattas}}, \ and\ \bibinfo {author}
  {\bibfnamefont {D.~K.}\ \bibnamefont {Panda}},\ }\bibfield  {title} {\enquote
  {\bibinfo {title} {Frontera: The evolution of leadership computing at the
  national science foundation},}\ }in\ \href {\doibase 10.1145/3311790.3396656}
  {\emph {\bibinfo {booktitle} {Practice and Experience in Advanced Research
  Computing}}},\ \bibinfo {series and number} {PEARC '20}\ (\bibinfo
  {publisher} {Association for Computing Machinery},\ \bibinfo {address} {New
  York, NY, USA},\ \bibinfo {year} {2020})\ p.\ \bibinfo {pages}
  {106–111}\BibitemShut {NoStop}%
\bibitem [{\citenamefont {Towns}\ \emph {et~al.}(2014)\citenamefont {Towns},
  \citenamefont {Cockerill}, \citenamefont {Dahan}, \citenamefont {Foster},
  \citenamefont {Gaither}, \citenamefont {Grimshaw}, \citenamefont {Hazlewood},
  \citenamefont {Lathrop}, \citenamefont {Lifka}, \citenamefont {Peterson},
  \citenamefont {Roskies}, \citenamefont {Scott},\ and\ \citenamefont
  {Wilkins-Diehr}}]{xsede}%
  \BibitemOpen
  \bibfield  {author} {\bibinfo {author} {\bibfnamefont {J.}~\bibnamefont
  {Towns}}, \bibinfo {author} {\bibfnamefont {T.}~\bibnamefont {Cockerill}},
  \bibinfo {author} {\bibfnamefont {M.}~\bibnamefont {Dahan}}, \bibinfo
  {author} {\bibfnamefont {I.}~\bibnamefont {Foster}}, \bibinfo {author}
  {\bibfnamefont {K.}~\bibnamefont {Gaither}}, \bibinfo {author} {\bibfnamefont
  {A.}~\bibnamefont {Grimshaw}}, \bibinfo {author} {\bibfnamefont
  {V.}~\bibnamefont {Hazlewood}}, \bibinfo {author} {\bibfnamefont
  {S.}~\bibnamefont {Lathrop}}, \bibinfo {author} {\bibfnamefont
  {D.}~\bibnamefont {Lifka}}, \bibinfo {author} {\bibfnamefont {G.~D.}\
  \bibnamefont {Peterson}}, \bibinfo {author} {\bibfnamefont {R.}~\bibnamefont
  {Roskies}}, \bibinfo {author} {\bibfnamefont {J.}~\bibnamefont {Scott}}, \
  and\ \bibinfo {author} {\bibfnamefont {N.}~\bibnamefont {Wilkins-Diehr}},\
  }\bibfield  {title} {\enquote {\bibinfo {title} {Xsede: Accelerating
  scientific discovery},}\ }\href {\doibase 10.1109/MCSE.2014.80} {\bibfield
  {journal} {\bibinfo  {journal} {Computing in Science \& Engineering}\
  }\textbf {\bibinfo {volume} {16}},\ \bibinfo {pages} {62} (\bibinfo {year}
  {2014})}\BibitemShut {NoStop}%
\bibitem [{\citenamefont {Edwards}\ and\ \citenamefont
  {Joo}(2005)}]{Edwards:2004sx}%
  \BibitemOpen
  \bibfield  {author} {\bibinfo {author} {\bibfnamefont {R.~G.}\ \bibnamefont
  {Edwards}}\ and\ \bibinfo {author} {\bibfnamefont {B.}~\bibnamefont {Joo}}
  (\bibinfo {collaboration} {SciDAC, LHPC, UKQCD}),\ }\bibfield  {title}
  {\enquote {\bibinfo {title} {{The Chroma software system for lattice QCD}},}\
  }\href {\doibase 10.1016/j.nuclphysbps.2004.11.254} {\bibfield  {journal}
  {\bibinfo  {journal} {Nucl. Phys. B Proc. Suppl.}\ }\textbf {\bibinfo
  {volume} {140}},\ \bibinfo {pages} {832} (\bibinfo {year} {2005})},\ \Eprint
  {http://arxiv.org/abs/hep-lat/0409003} {arXiv:hep-lat/0409003} \BibitemShut
  {NoStop}%
\bibitem [{\citenamefont {Clark}\ \emph {et~al.}(2010)\citenamefont {Clark},
  \citenamefont {Babich}, \citenamefont {Barros}, \citenamefont {Brower},\ and\
  \citenamefont {Rebbi}}]{Clark:2009wm}%
  \BibitemOpen
  \bibfield  {author} {\bibinfo {author} {\bibfnamefont {M.~A.}\ \bibnamefont
  {Clark}}, \bibinfo {author} {\bibfnamefont {R.}~\bibnamefont {Babich}},
  \bibinfo {author} {\bibfnamefont {K.}~\bibnamefont {Barros}}, \bibinfo
  {author} {\bibfnamefont {R.~C.}\ \bibnamefont {Brower}}, \ and\ \bibinfo
  {author} {\bibfnamefont {C.}~\bibnamefont {Rebbi}},\ }\bibfield  {title}
  {\enquote {\bibinfo {title} {{Solving Lattice QCD systems of equations using
  mixed precision solvers on GPUs}},}\ }\href {\doibase
  10.1016/j.cpc.2010.05.002} {\bibfield  {journal} {\bibinfo  {journal}
  {Comput. Phys. Commun.}\ }\textbf {\bibinfo {volume} {181}},\ \bibinfo
  {pages} {1517} (\bibinfo {year} {2010})},\ \Eprint
  {http://arxiv.org/abs/0911.3191} {arXiv:0911.3191 [hep-lat]} \BibitemShut
  {NoStop}%
\bibitem [{\citenamefont {Babich}\ \emph {et~al.}(2010)\citenamefont {Babich},
  \citenamefont {Clark},\ and\ \citenamefont {Joo}}]{Babich:2010mu}%
  \BibitemOpen
  \bibfield  {author} {\bibinfo {author} {\bibfnamefont {R.}~\bibnamefont
  {Babich}}, \bibinfo {author} {\bibfnamefont {M.~A.}\ \bibnamefont {Clark}}, \
  and\ \bibinfo {author} {\bibfnamefont {B.}~\bibnamefont {Joo}},\ }\bibfield
  {title} {\enquote {\bibinfo {title} {{Parallelizing the QUDA Library for
  Multi-GPU Calculations in Lattice Quantum Chromodynamics}},}\ }in\ \href@noop
  {} {\emph {\bibinfo {booktitle} {{SC 10 (Supercomputing 2010)}}}}\ (\bibinfo
  {year} {2010})\ \Eprint {http://arxiv.org/abs/1011.0024} {arXiv:1011.0024
  [hep-lat]} \BibitemShut {NoStop}%
\bibitem [{\citenamefont {Winter}\ \emph {et~al.}(2014)\citenamefont {Winter},
  \citenamefont {Clark}, \citenamefont {Edwards},\ and\ \citenamefont
  {Jo\'o}}]{Winter:2014dka}%
  \BibitemOpen
  \bibfield  {author} {\bibinfo {author} {\bibfnamefont {F.~T.}\ \bibnamefont
  {Winter}}, \bibinfo {author} {\bibfnamefont {M.~A.}\ \bibnamefont {Clark}},
  \bibinfo {author} {\bibfnamefont {R.~G.}\ \bibnamefont {Edwards}}, \ and\
  \bibinfo {author} {\bibfnamefont {B.}~\bibnamefont {Jo\'o}},\ }\bibfield
  {title} {\enquote {\bibinfo {title} {{A Framework for Lattice QCD
  Calculations on GPUs}},}\ }in\ \href {\doibase 10.1109/IPDPS.2014.112} {\emph
  {\bibinfo {booktitle} {{28th IEEE International Parallel and Distributed
  Processing Symposium}}}}\ (\bibinfo {year} {2014})\ \Eprint
  {http://arxiv.org/abs/1408.5925} {arXiv:1408.5925 [hep-lat]} \BibitemShut
  {NoStop}%
\bibitem [{\citenamefont {Jo\'o}\ \emph {et~al.}(2013)\citenamefont {Jo\'o},
  \citenamefont {Kalamkar}, \citenamefont {Vaidyanathan}, \citenamefont
  {Smelyanskiy}, \citenamefont {Pamnany}, \citenamefont {Lee}, \citenamefont
  {Dubey},\ and\ \citenamefont {Watson}}]{Joo:2013lwm}%
  \BibitemOpen
  \bibfield  {author} {\bibinfo {author} {\bibfnamefont {B.}~\bibnamefont
  {Jo\'o}}, \bibinfo {author} {\bibfnamefont {D.~D.}\ \bibnamefont {Kalamkar}},
  \bibinfo {author} {\bibfnamefont {K.}~\bibnamefont {Vaidyanathan}}, \bibinfo
  {author} {\bibfnamefont {M.}~\bibnamefont {Smelyanskiy}}, \bibinfo {author}
  {\bibfnamefont {K.}~\bibnamefont {Pamnany}}, \bibinfo {author} {\bibfnamefont
  {V.~W.}\ \bibnamefont {Lee}}, \bibinfo {author} {\bibfnamefont
  {P.}~\bibnamefont {Dubey}}, \ and\ \bibinfo {author} {\bibfnamefont
  {W.}~\bibnamefont {Watson}},\ }\bibfield  {title} {\enquote {\bibinfo {title}
  {{Lattice QCD on Intel\textregistered{} Xeon Phi Coprocessors}},}\ }\href
  {\doibase 10.1007/978-3-642-38750-0_4} {\bibfield  {journal} {\bibinfo
  {journal} {Lect. Notes Comput. Sci.}\ }\textbf {\bibinfo {volume} {7905}},\
  \bibinfo {pages} {40} (\bibinfo {year} {2013})}\BibitemShut {NoStop}%
\bibitem [{\citenamefont {Jo{\'o}}\ \emph {et~al.}(2016)\citenamefont
  {Jo{\'o}}, \citenamefont {Kalamkar}, \citenamefont {Kurth}, \citenamefont
  {Vaidyanathan},\ and\ \citenamefont {Walden}}]{optimising}%
  \BibitemOpen
  \bibfield  {author} {\bibinfo {author} {\bibfnamefont {B.}~\bibnamefont
  {Jo{\'o}}}, \bibinfo {author} {\bibfnamefont {D.~D.}\ \bibnamefont
  {Kalamkar}}, \bibinfo {author} {\bibfnamefont {T.}~\bibnamefont {Kurth}},
  \bibinfo {author} {\bibfnamefont {K.}~\bibnamefont {Vaidyanathan}}, \ and\
  \bibinfo {author} {\bibfnamefont {A.}~\bibnamefont {Walden}},\ }\bibfield
  {title} {\enquote {\bibinfo {title} {Optimizing wilson-dirac operator and
  linear solvers for intel® knl},}\ }in\ \href@noop {} {\emph {\bibinfo
  {booktitle} {High Performance Computing}}},\ \bibinfo {editor} {edited by\
  \bibinfo {editor} {\bibfnamefont {M.}~\bibnamefont {Taufer}}, \bibinfo
  {editor} {\bibfnamefont {B.}~\bibnamefont {Mohr}}, \ and\ \bibinfo {editor}
  {\bibfnamefont {J.~M.}\ \bibnamefont {Kunkel}}}\ (\bibinfo  {publisher}
  {Springer International Publishing},\ \bibinfo {address} {Cham},\ \bibinfo
  {year} {2016})\ pp.\ \bibinfo {pages} {415--427}\BibitemShut {NoStop}%
\end{thebibliography}%

\end{document}